\documentclass[12pt,a4paper]{report} 

\pdfoutput=1
\usepackage[margin=3cm]{geometry}

\usepackage{setspace}

\usepackage[numbers]{natbib}
\usepackage[affil-it]{authblk}

\usepackage{amsmath}
\usepackage{amsthm}
\usepackage[english]{babel}
\usepackage{graphicx}
\usepackage{amsfonts} 
\usepackage{amssymb} 
\usepackage{float}
\usepackage{times}
\usepackage{tensor}
\usepackage{mathtools}
\usepackage{color}
\usepackage{mathrsfs}
\usepackage{bbold}
\usepackage{bigints}
\usepackage{paralist}

\usepackage{pdflscape}
\usepackage{rotating}
\usepackage{multirow}
\usepackage{array}
\usepackage{adjustbox}

\usepackage{wrapfig}

\usepackage[small,labelfont=bf,up,textfont=it,up]{caption} 

\usepackage{titlesec} 

\usepackage{mdframed}
\usepackage{hyperref}

\numberwithin{equation}{chapter}

\newcommand{\deriv}[1]{\frac{d}{d #1}}
\newcommand{\DERIV}[2]{\frac{d #1}{d #2}}
\newcommand{\pd}[1]{\frac{\partial}{\partial #1}}
\newcommand{\PD}[2]{\frac{\partial #1}{\partial #2}}

\newcommand{\half}[0]{\frac{1}{2}}

\newcommand{\fd}[1]{\frac{\delta}{\delta #1}}

\newcommand{\de}{\partial}
\newcommand{\lqu}{\textrm`}
\newcommand{\rqu}{\textrm'}
\newcommand{\QU}[1]{[\lqu #1 \rqu]}

\newcommand{\oppr}[3]{\ensuremath{\left \langle \left. #1\right.  \right| #2 \left| \left. #3 \right. \right \rangle}}
\newcommand{\ket}[1] {\ensuremath {\left| #1 \right\rangle }}

\newcommand{\p}  	{\prime}

\newcommand{\del}	{\partial}
\newcommand{\supfour}	{~^{(4)}\!}
\newcommand{\phat}[0]	{\hat{p}}
\newcommand{\qhat}[0]	{\hat{q}}
\newcommand{\gtilde}[0]	{\tilde{g}}
\newcommand{\pitilde}[0]{\tilde{\pi}}

\newcommand{\curlyH}	{\mathscr{H}}

\newcommand{\rootg}	{\sqrt{g}}

\newcommand{\gdot}	{\dot{g}}
\newcommand{\gbar}	{\bar{g}}

\newcommand{\phibar}	{\bar{\phi}}
\newcommand{\phidot}	{\dot{\phi}}

\newcommand{\pphi}[1]	{{p_{\phi_{#1}}}}
\newcommand{\pphibar}[1]{{\bar{p}}_{\phi_{#1}}}
\newcommand{\pphisq}	{p_\phi^2}

\newcommand{\tk}	{2\kappa}
\newcommand{\zero}	{{(0)}}
\newcommand{\one}	{{(1)}}
\newcommand{\two}	{{(2)}}

\newcommand{\Rbar}	{\bar{R}}

\title{Gravitation and cosmology with York time \vspace{2cm}} 
\author{Philipp Roser \vspace{1cm}}
\affil{Department of Physics and Astronomy,\\ Clemson University, Kinard Laboratory,\\ Clemson, SC 29631-0978, USA}
\date{\vspace{-0.5cm}}		

\begin{document}
\hypersetup{pageanchor=false} 
\baselineskip16pt

\begin{titlepage}
 \maketitle
 \thispagestyle{empty}
\end{titlepage} 

\pagenumbering{roman}
\setcounter{page}{1}

\chapter*{}
\begin{doublespace}
\begin{centering}
In partial fulfillment of the requirements for the degree of\\DOCTOR OF PHILOSOPHY\\in\\PHYSICS\\at\\Clemson University, Clemson, South Carolina, USA.\\ \vspace{5cm}
\begin{tabular}{ll}
 Committee chair:\hspace{12pt} & Antony Valentini \\
 Committee members:\hspace{12pt} & Murray Daw, Dieter Hartmann, Bradley Meyer
\end{tabular}

\end{centering}
\end{doublespace}



\chapter*{Abstract}

Despite decades of inquiry an adequate theory of `quantum gravity' has remained elusive, in part due to the absence of data that would guide the search and in part due to technical difficulties, prominently among them the `problem of time'. The problem is a result of the attempt to quantise a classical theory with temporal reparameterisation and refoliation invariance such as general relativity.

One way forward is therefore the breaking of this invariance via the identification of a preferred foliation of spacetime into parameterised spatial slices. In this thesis we argue that a foliation into slices of constant extrinsic curvature, parameterised by `York time', is a viable contender. We argue that the role of York time in the initial-value problem of general relativity as well as a number of the parameter's other properties make it the most promising candidate for a physically preferred notion of time. 

A Hamiltonian theory describing gravity in the York-time picture may be derived from general relativity by `Hamiltonian reduction', a procedure that eliminates certain degrees of freedom --- specifically the local scale and its rate of change --- in favour of an explicit time parameter and a functional expression for the associated Hamiltonian. In full generality this procedure is impossible to carry out since the equation that determines the Hamiltonian cannot be solved using known methods.

However, it is possible to derive explicit Hamiltonian functions for cosmological scenarios (where matter and geometry is treated as spatially homogeneous). Using a perturbative expansion of the unsolvable equation enables us to derive a quantisable Hamiltonian for cosmological perturbations on such a homogeneous background. We analyse the (classical) theories derived in this manner and look at the York-time description of a number of cosmological processes.

We then proceed to apply the canonical quantisation procedure to these systems and analyse the resulting quantum theories. We discuss a number of conceptual and technical points, such as the notion of volume eigenfunctions and the absence of a momentum representation as a result of the non-canonical commutator structure. While not problematic in a technical sense, the conceptual problems with canonical quantisation are particularly apparent when the procedure is applied in cosmological contexts.

In the final part of this thesis we develop a new quantisation method based on configu\-ration-space trajectories and a dynamical configuration-space Weyl geometry. There is no wavefunction in this type of quantum theory and so many of the conceptual issues do not arise. We outline the application of this quantisation procedure to gravity and discuss some technical points. The actual technical developments are however left for future work.

We conclude by reviewing how the York-time Hamiltonian-reduced theory deals with the problem of time. We place it in the wider context of a search for a theory of quantum gravity and briefly discuss the future of physics if and when such a theory is found.




\chapter*{Acknowledgements}

First and foremost, I am indebted to my doctoral advisor Antony Valentini at Clemson University for his help in all matters from technical to literary, and his support in enabling me to pursue the big questions that fascinate me. I am also grateful to Lucien Hardy for many discussions on foundations of quantum theory, and for his hospitality at Perimeter Institute during 2012/2013, where a non-trivial amount of the research in this thesis (especially chapter 11) was carried out. I furthermore wish to thank Lee Smolin, who in early 2012 pointed me toward the literature on the then very recent developments in Shape Dynamics, and Rafael Sorkin, whose critical and novel way to look at questions in both quantum theory and gravity has contributed to my own approach. I owe much of my appreciation of the issues of gravity, quantum theory and quantum theory of gravity to my education in philosophy of physics at Oxford University, in particular to Oliver Pooley, Harvey Brown, Simon Saunders and David Wallace.

I am also grateful to the Clemson University Department of Physics and Astronomy and my PhD committee to allow me to pursue my work via somewhat unconventional arrangements: first, through financial support in the form of a minimal assistantship during my time at Perimeter Institute, and second by allowing me to spend the last two years working from my home, some three thousand kilometres from campus.

Finally, I wish to thank my wife, Jemma, for her patience.

\newpage
\setcounter{tocdepth}{1} 
\tableofcontents

\newpage
\listoffigures
\newpage
\listoftables

\newpage

\pagenumbering{arabic}
\chapter{Introduction} \label{chap:Introduction}

\section{Quantum theory and gravity}

Cosmology, the study of the history of universe, has a unique role in the investigation of physical reality. Unlike, for example, particle physics or condensed-matter theory it cannot be tested experimentally in the laboratory. All evidence must come from observation of the one and only universe to which we have access. To make matters worse, all our observations are made roughly from a single point in space and time. 

Yet despite these limitations, cosmology --- in particular the study of the very early moments in the history of the universe --- is not only fascinating in itself but also exceptionally useful in that it allows us to explore physics at extremely high energies and on very small scales, where both gravitational and quantum effects are expected to play a crucial role. That is, through observations of remnants of this very early period in the history of the cosmos, we can hope to probe physics that is irreproducible in the lab to any degree of practicality, if not in principle.\footnote{Since it is impossible to have the entire universe inside a piece of apparatus that is itself part of the universe.}

The observational investigation of the early universe is uniquely challenging, but also uniquely rewarding.

Aside from technical difficulties (to which we come shortly) it is this lack of experimental access that has led to the elusiveness of a complete and testable theory of quantum gravity, a theory that is to describe physics under the conditions that existed in the immediate aftermaths of the big bang,\footnote{Throughout this thesis I will use the term `big bang' to refer to the initial singularity with which the universe is thought to have started (unless quantum gravity were to tell us otherwise). A particle physicist on the other hand would probably place the `big bang' at a point when the universe had a size of roughly the Planck scale (where classical general relativity is sure to break down), while some cosmologists would use the term to describe a point in time as far forward as the end of inflation (since everything before that is highly speculative).} and that is to reduce to the testable limits of classical general relativity and quantum field theory in the appropriate regimes. The elusive theory of quantum gravity therefore has a unique historical status: Advances in electromagnetism during the nineteenth century were propelled forwards by observing phenomena such as static electricity and the action of magnetic forces, thermodynamics by observing the behaviour of gases and other fluids. During the early twentieth century the mystery of spectral lines and other atomic phenomena led, ultimately, to quantum mechanics. Even general relativity, undoubtedly one of the most elegant theoretical developments in the history of physics, was not primarily a quest for mathematical beauty but one to explain the advancing perihelion of mercury, for example.\footnote{In fact, later in life Einstein revised his own history, stating that mathematical beauty had been the driving force behind his development of general relativity. The documentation of the period 1910-1916 tells, however, a different story \citep{Smolin2015_Einstein1915}. Einstein's motivation was almost entirely experimental.} And while quantum field theory was developed with the idea in mind of making quantum mechanics compatible with the Lorentz-invariance of special relativity, experimental progress kept up with theoretical predictions.\footnote{Perhaps a notable exception is the Higgs boson, which was only observed some forty years after having been predicted theoretically.} On the other hand, the search for quantum gravity is driven not by experimental facts in need of explanation but by a desire for a unified description of phenomena, each of which is already well described by an appropriate theory such as general relativity or quantum field theory. 

The fact that the quest for quantum gravity is driven by theory rather than experiment is exactly the reason why experimentally corroborating any particular quantum-gravity contender is difficult; if appropriate experiments were available in the lab we would likely already have performed them, observed new phenomena and would have had empirical evidence to guide our search.

That no accepted theory of quantum gravity exists is not due to a lack of contenders. Proposals are numerous, albeit few are sufficiently developed theoretically to justifiably claim the status of a self-consistent theory with concrete predictions. The proposed theories differ radically even in the answers they give to most fundamental questions: Is spacetime fundamental or is there a true differentiation between space and time? Are space and time infinitely divisible? Is there a physically fundamental measure of time? Is time even physical at all, and if so, what does that mean? Is everything dynamical or is there a fixed `background' structure that serves as the arena of the universe's dynamics?

The reason for this plethora of theoretical frameworks is the failure of the `standard' procedure when applied to gravity: the canonical quantisation of the classical gravitational field of general relativity. In the present work we will spend some time gaining an understanding of why this method fails. Suffice to say at this stage, both the success of canonical quantisation with anything \emph{but} the gravitational field, as well as the procedure's failure \emph{with} it are in their own way remarkable.

A priori the idea of canonical quantisation appears ludicrous: Begin with a `classical' theory describing entities such point particles, or fields, or rigid objects, where the chosen classical theory is experimentally adequate on sufficiently large scales and sufficiently low energies. Now perform a series of ad-hoc mathematical steps to arrive at a new theory whose ontology (that is, the physical entities that purport to make up the world) is radically different from that of the classical theory (if at all well-defined). Finally establish a link with what is measurable by introducing highly questionable postulates concerning the outcomes of `experiments'. 

This new `quantum' theory is now supposed to be understood as being more fundamental than the classical theory from which it originated, the latter being considered a limit of the former under appropriate conditions. It appears that the reasoning is backward, with the fundamental being derived from the approximation. Yet the phenomenal experimental success of this `recipe' in all areas of physics \emph{except} gravity is truly remarkable. When applied to the most fundamental classical theory known, the fields comprising the standard model of particle physics, the resultant quantum theory has been corroborated experimentally to an extremely high degree of accuracy.

However, when the same procedure is applied to the metric field $g_{\mu\nu}$, a $(0,2)$-tensor classically used to represent the gravitational field (or, more accurately, to the spatial metric $g_{ab}$ on a suitably chosen space-like `slice' taken from a foliation of the spacetime), it fails spectacularly. In particular, the resultant theory implies that the universe is frozen, that no dynamical processes are possible at all. This is one facet of what has been dubbed the `problem(s) of time', which we will be discussing in some detail in chapter \ref{chap:problemoftime}.

At the heart of the problem lies the general covariance of the classical theory, that is, the fact that the dynamical equations of general relativity are covariant under arbitrary coordinate transformations on spacetime. General covariance is not a physically meaningful property of a theory. Any (classical) theory can be expressed in a generally covariant form, a criticism going back almost a century \citep{Kretschmann1917}. However, applying the quantisation recipe to distinct forms of the same classical theory does not necessarily yield physically equivalent quantum theories. This fact is central to the line of inquiry pursued in this thesis and will be discussed further in future chapters.

In general relativity one can consider four independent generators of coordinate transformations at each point in spacetime, three in space and one in time. These are the gauge freedoms of the theory at any one point in spacetime. Each of these gauge freedoms leads to a constraint equation, totalling three so-called `momentum constraints' and one `Hamiltonian constraint'\footnote{Gauge constraints usually generate (non-physical) gauge transformations (`generate' referring to motion in the configuration space determined by the Poisson bracket of the constraint and the configuration variable). However, it has been argued that the Hamiltonian constraint described real \emph{physical} change \citep{BarbourFoster2008}. We briefly survey the argument at the end of section \ref{sec:VanishingH}.} stemming from the three spatial and one temporal gauge freedom respectively. The problem of time is associated with the latter.

Understanding how the `frozen' dynamics follow from the temporal gauge freedom is possible without reference to the full theory of general relativity. In fact, a simple particle model suffices, provided its laws (that is, its action) is given in a \emph{time-reparameterisation invariant} form. This time-reparameterisation invariance serves as a toy model for the temporal gauge freedom of general relativity. It is however important to emphasise that in general relativity there are other facets to the problem of time which do not appear in the case of a particle model, primarily due to the fact that while in the toy model there is a single gauge degree of freedom determining the temporal parameterisation, there is one such degree \emph{per spacetime point} in general relativity. As a consequence in general relativity it is not only that the sequence of spatial slices constituting the history of the universe may be freely relabelled, but it is not even clear how the slicing should be done in the first place.

Short of more radical proposals, two `fixes' come to mind. First, one may consider partially \emph{gauge-fixing} general relativity. Here one specifies the temporal gauge, thereby choosing a unique slicing and a unique labelling of those slices. This eliminates the Hamiltonian constraint and hence the frozen dynamics. A second approach is to construct an entirely new theory which does not possess the slicing ambiguity in the first place and has a fixed temporal parameterisation, provided the phenomenology of general relativity can be recovered sufficiently closely in order not to contradict past observations.\footnote{It is not strictly necessary that all predictions of general relativity are recovered by such a theory, but only those that have actually been tested. Discrepancies would allow for the ability to distinguish between the two theories in future experiments.} One such contender is Shape Dynamics, a theory whose dynamical variables are those describing a conformal three-geometry (rather than the non-conformal four-geometry of general relativity). Shape dynamics matches the phenomenology of general relativity for spacetimes that can be foliated into globally hyperbolic slices (for example, a spatially flat Friedmann universe) but differs in its prediction for what to expect on the other side of a black-hole event horizon, for example. It turns out that in the presence of such a hyperbolic slicing Shape Dynamics and general relativity have a gauge overlap. That is, there is an appropriate gauge fixing (a choice of coordinate description) for each such that the resultant two gauge-fixed theories are identical.

From the perspective of general relativity this gauge fixing consists of splitting spacetime into slices of \emph{constant mean extrinsic curvature} (CMC), a concept we will introduce in some detail later. The slices are labelled by `York time', named after James York in whose work on the initial-value problem of general relativity in the early 1970s this parameter had special significance \citep{York1972, ChoquetBruhatYork1980}. In a cosmological setting York time is proportional to the negative of the Hubble parameter. This choice has a number of convenient properties which we will describe in chapter \ref{chap:Yorktime}. The choice corresponds to the gauge overlap between Shape Dynamics and general relativity.

In the light of the problem of time, performing such a gauge fixing is an obvious strategy. However, it is nonetheless radical in that it constitutes an explicit abandonment of general spacetime covariance in favour of a more limited spatial covariance. Physically speaking, it assumes the existence of a fundamental slicing of spacetime into space and time, at least if this gauge choice is meant to be more than a convenient coordinate choice for general relativity. Since we intend to construct a quantum theory from this choice, and since different gauge fixings of the same theory may well lead to phenomenologically distinct quantum theories after application of the quantisation recipe (see chapter \ref{chap:choiceoftime}), the choice is indeed more than one of coordinates.

Abandoning spacetime in favour of space and time might appear to run in the face of the last one hundred years of developments in physics, not least quantum field theory whose development was in part motivated by the desire to construct a Lorentz-covariant theory of particles. However, at closer inspection it is not, in fact, as far-fetched as it might appear. One reason is that in quantum theory the notion of `time' takes on a fundamentally different role. Consider the quantisation of field theories. The quantisation must be performed with a specific choice of temporal parameter. It is only after the quantisation is complete that one can show that the chosen splitting of space and time (here in a Lorentzian setting with flat spacetime rather than the curved spacetime of general relativity) is, in fact, undetectable and that no superluminal `signalling' is possible. That is, spacetime covariance is recovered at the phenomenological rather than fundamental level. Further, covariant quantisation of the electro-magnetic field (or equally fields of the strong and electro-weak sector of the standard model) leads to `ghosts', unphysical modes in the field which are nonetheless essential to the mathematical structure of such a formulation. Once again these do not appear at the level of the phenomenology, but their status is somewhat of an open question on the ontological level if the quantum field theory is treated with any degree of physical realism (which it rarely is in practice). A fundamental choice of space and time would alleviate this concern.

A very different reason to suspect the existence of a fundamental distinction between space and time is found in cosmology. It is well established that the observable universe is highly homogeneous and isotropic at sufficiently large scales. 
Over distances of hundreds of megaparsecs matter and radiation are very evenly distributed throughout the universe. However, this homogeneity is not generally invariant but only holds in one particular foliation of spacetime (up to small perturbations with scales smaller than the scale on which the homogeneity holds).

Quantum theory demands an unambiguous notion of simultaneity. This is because time is treated very differently from any other variable in standard quantum theory. 

The departure from a fundamentally `relativistic' physics is radical, but not unreasonable. In fact, even at the classical level one can very naturally arrive at a field theory in which the universal light-cone, that is, the constancy of the speed of light and the fact that this `speed limit' appears to apply to all fields of matter equally, is a \emph{result} rather than a postulate of the theory. Starting from the desire to implement a truly `relational' viewpoint of dynamics this has been achieved as part of the Shape Dynamics program. However, for the emergence of the classical light cone a slightly simpler theory suffices, dubbed `Relativity without relativity' by its creators \citep{BarbourFosterOMurchadha2002}.

In order to arrive at a theory describing the evolution of CMC slices we will perform a procedure called Hamiltonian reduction. It provides a method to arrive at a physical (that is, non-vanishing) Hamiltonian for the foliation-fixed theory starting from the gauge-general Hamiltonian of general relativity in its Arnowitt-Deser-Misner (ADM) formulation \citep{ADM1962}. The ADM Hamiltonian vanishes in virtue of being the sum of four constraints. In general, the Hamiltonian reduction cannot be completed analytically since it involves solving a highly non-trivial (and non-linear) differential equation \citep{ChoquetBruhatYork1980}. However, at sufficiently large scales on which the universe is spatially homogeneous the procedure is relatively straightforward. For example, for a cosmological model with a set of scalar fields the equation reduces to a cubic, which can be readily solved (see chapter \ref{chap:Friedmann}). At the level of perturbations it is more intricate and a perturbative solution to the problem and the discussion of the resulting perturbation theory will constitute chapters \ref{chap:perttheoprelims} and \ref{chap:perttheoformalism}.

The reduced-Hamiltonian theory, unlike the original constraint system of general relativity, can be quantised without (too much) difficulty, although we will encounter some subtleties in the process. How then is the quantisation to be carried out?

It is important to remember that no quantisation method is intrinsically superior to another. Ultimately empirical data corroborates the contenders for the correct quantum-gravitational theory and eliminates incorrect ones. If two theories, constructed from the same classical theory using different methods of quantisation, are identical in their prediction, one may be preferable for its conceptual clarity. However, realists must proceed with care. The inference of facts about the world's underlying ontology necessarily involves a certain degree of `philosophy' and one must be careful when choosing between two empirically equivalent but ontologically distinct theories. This is not to say that such reasoning is irrational, although many physicists throughout the twentieth century and beyond have taken this attitude.\footnote{More accurately, there are two distinct points here. First, there are schools of thought that claim or at least imply that any form of reasoning about the underlying reality of a theory is meaningless. Historically the best known such group were the logical positivists, most prominently represented by the famous Vienna Circle. In a similar vein, arguably, falls the rising popularity of attempting to formulate quantum mechanics in terms of partial knowledge and information with an `epistemic' wavefunction (concerning knowledge rather than physical facts), an approach appearing attractive to some at least in part due to the progress in quantum computing and quantum information theory. The other point we are referring to here is the unfortunate tendency of many physicists to look down on any form of philosophical reasoning, in some cases using the word in an almost (and sometimes outright) derogatory manner.} In fact, here we wish to argue that any fundamental physical theory must make statements about what the basic ontological entities are with which the theory is concerned. Lacking such information a theory is more akin to a prescription for predicting experimental outcomes, without or with only very little explanatory value.

Yet it is no hyperbole to say that the last century has been dominated by confusion with regards to the correct `interpretation' of quantum mechanics, even in the context of, say, quantised non-relativistic particle models.\footnote{It would be an apt if slightly too general observation to criticise much of the work that has been done in the foundations of quantum mechanics for being too focussed on non-relativistic models. While mathematically and conceptually easier to handle, it is by no means clear that any insight gained in this realm generalises. On the contrary, as we will argue, insight into the correct `interpretation' might be gained from insights in quantum gravity.} We chose to use quotation marks around the term since it is rather inaccurate. What is generally counted among such interpretations includes theories with potential experimental discrepancies (such as dynamical-collapse theories and non-equilibrium pilot-wave theory), as well as `interpretations' that are in fact no such thing but an explicit lack thereof (so-called operational pictures). 

Given the extent to which the quantum field theory of the standard model of particle physics has been tested it is highly doubtful that discrepancies in the predictions of laboratory measurement outcomes by different formulations will be detected in applications relating to conventional (that is, non-gravitational) quantum field theory. On the other hand, given the fundamental incompatibility of the standard quantum formalism with general relativity, one or both of them must be revised. This might plausibly lead to diverging phenomenologies in appropriate regimes, in particular in cosmological observations. One is therefore justified in the hope that cosmological observation may help narrow down the set of viable proposals for theories of quantum gravity, which in turn may contain indications for the correct formulation, or interpretation, of quantum theory. 

We already touched on how our theory of gravity may be modified in agreement with astronomical observations, for example in the form of Shape Dynamics instead of general relativity or by considering the existence of a physically fundamental notion of time. We shortly turn towards a discussion of appropriate formulations of quantum theory.

Before doing so however we must emphasise that there is no guarantee that \emph{any} choice of classical theory and quantisation method will succeed. Given the ludicrousness of the quantisation method, it is entirely plausible to expect that the correct theory of quantum gravity can only be attained `from scratch', that is, by direct construction from fundamental principles and not via quantisation at all. One example of such an approach is that of causal sets (\citep{Sorkin2003}and references therein; for the original proposal, see \citep{BombelliLeeMeyerSorkin1987}) where the spacetime four-manifold is only considered a smooth approximation to a discrete set of spacetime elements that are related via a partial ordering.\footnote{Depending on the particulars of the approach, matter fields are either defined \emph{on} those spacetime elements or the elements are \emph{all} there is and matter fields are emergent via a Kaluza-Klein formalism  \citep{Kaluza1921,Klein1926,OverduinWesson1997}. One should also add that even the causal-set approach does not completely abandon the notion of quantisation. It employs a `sum-over-histories' method in its construction.} However, given the success of quantisation in other areas, the attempt to obtain an adequate theory of quantum gravity in the `conventional' manner is at least plausible.

It is not entirely clear what the `standard' method of quantisation (that is, \emph{canonical} quantisation) is beyond the relationship between mathematical quantities, even leaving aside the option of path-integral quantisation completely. Formally, canonical quantisation is the construction of an operator algebra from a Poisson algebra of classical functions by `promoting' variables and functions of variables to operators \citep{Thiemann2007}. This operator algebra may then be \emph{represented} as operators acting on complex functions on a Hilbert space. Quantum mechanics is then itself about relationships between `observables',\footnote{Some authors \citep[section 3.2.4]{Rovelli2007} suggest that in a relativistic setting even classical mechanics is about the relationship of observables.} a misleading term used to refer to the set of those operators that are self-adjoint on the chosen Hilbert space. It is difficult to see how this formalism alone leads to the physical reality of everyday objects or even that of particles or fields. At best, it describes some structural properties thereof.

In slightly less mathematics-focussed approaches `canonical quantisation' is assumed to also include the existence of a `quantum state' or `wavefunction', although the latter term is often reserved to refer specifically to the quantum state when represented in the position basis of the Hilbert space. This wavefunction is then considered to represent the physical state of the universe or at least the system under consideration and the action of operators on this wavefunction is used to extract information about the likelihood of measurement outcomes (hence the name `observables'), albeit without being clear what exactly constitutes a `measurement'. The time evolution of the wavefunction is determined by the action of a particular operator, the quantum Hamiltonian. If `measurement' is understood as referring to a physical process (as it must or its meaning is entirely obscure), then it is unclear why this process is supposed to have a special status and is not itself encoded in the wavefunction and its evolution. This is particularly problematic in the context of cosmology where the entirety of the universe, including anything capable of performing a measurement, is supposed to be described by the `universal wavefunction'. But even in non-cosmological application the concept is problematic and a large literature has been dedicated to various facets of the `problem of measurement' (see \citep{Wallace2008} for a relatively recent review with many references).

In the minimalist sense canonical quantisation is a purely mathematical procedure and does not constitute the construction of a physical theory. If the term is used in a broader sense it is an incomplete if not inconsistent recipe to construct one. Either way, by itself it is insufficient to arrive at an ontologically meaningful theory, although it may describe certain mathematical structures underlying such a theory. In order to arrive at a more satisfactory picture canonical quantisation must therefore be supplemented with a set of physical principles that relate the operator algebra to physical reality. The abstract notion of measurement does not suffice. 

One can separate proposals to overcome the overcome the measurement problem into two groups, those that are purely \emph{interpretative} of the quantum formalism and those that add or change ontological and dynamical elements of the theory. Among the former one counts the Everett `many-worlds' interpretation (\citep{Everett1956}, though see \citep[][sec.~4]{Wallace2008} for a review and discussion), and Quantum Bayesianism \citep{FuchsMerminSchack2014} and other `psi-epistemic' (the quantum state describes a state of information or knowledge) approaches. Among the latter are dynamical collapse model such as those of Ghirardi, Rimini and Weber \citep{GRW1986} and de~Broglie-Bohm pilot-wave theory \citep{deBroglie1928,Bohm1952,Holland1993}. The list is by no means exhaustive and this is not the place for detailed critiques of these ideas, which by themselves would easily fill several doctoral theses.

To us, de~Broglie-Bohm theory will be of particular interest. The reason is not that it is a picture that is conceptually or otherwise more appealing than its contenders (although it arguably is, as I have argued elsewhere \citep{RoserMScThesis}), but that it provides mathematical tools that are unavailable in other formulations. These tools are configuration-space trajectories just like they exist in classical physics, although they are governed by non-classical dynamics. Specifically, the trajectories may be used to provide a new way to deal with quantum cosmological perturbations.\footnote{There is no reason to think that the method we use here does not generalise to other applications, although this will be the subject of future work.} 

Does the unique availability of mathematical methods imply that de~Broglie-Bohm theory deserves greater credence? In the process of constructing the quantum cosmological perturbation theory via the quantisation of a reduced-Hamiltonian system we will see hints towards the plausibility of de~Broglie-Bohm. Yet in the absence of a full, non-perturbative theory of quantum gravity that explicitly relies on quantum trajectories any definite conclusion of the sort would be rash.

However, it is possible to develop a quantum theory from a classical starting point based entirely on the notion of such configuration-space trajectories and without reference to a wavefunction. The exploration of such a framework and its tentative application to cosmology will form the final part of this thesis. A wavefunction-free formulation of quantisation may be particularly suited to gravity and to the quantum theory of the universe as a whole. Whatever its conceptual benefits, whether or not such an approach can really lead to a fully satisfactory theory of quantum gravity remains to be seen. In part \ref{part4} we sketch the way forward.


\section{Outline of this thesis}

This thesis is divided into three parts. Part \ref{part1} introduces the relevant background and motivates the line of inquiry pursued in this thesis. In part \ref{part2} we develop the classical theory of York-time cosmology and in part \ref{part3} the corresponding quantum theory. Chapters \ref{chap:problemoftime} and \ref{chap:Yorktime} for the most part review material (though hopefully introduce some new perspectives), while chapters \ref{chap:choiceoftime} (at least section \ref{sec:choiceoftimesignificance}), \ref{chap:Friedmann}, \ref{chap:cosmext}, \ref{chap:perttheoprelims}, \ref{chap:perttheoformalism}, \ref{chap:QuantFriedmann}, \ref{chap:quantcosmpert}, \ref{chap:traj} and \ref{chap:cosmtraj} contain primarily original work. 

Most of the quantitative work has been published. The contents of chapters \ref{chap:Friedmann} and \ref{chap:QuantFriedmann} (with the exception of section \ref{sec:inflation}) has been published in ref.~\citep{RoserValentini2014a}, that of chapter \ref{chap:cosmext} in ref.~\citep{Roser2015CosmExtension}, sections \ref{sec:ClassKasner} and \ref{sec:quantKasner} in ref.~\citep{Roser2015a}, while sections \ref{sec:inflation}, \ref{sec:cosmhist},\ref{sec:modefreezing} and \ref{sec:modereentry} correspond to the content of ref.~\citep{RoserValentini2016}. The work on perturbation theory proper (chapter \ref{chap:perttheoformalism} and section \ref{sec:quantperts}) is contained in ref.~\citep{Roser2015b}, while that on trajectory-based quantisation methods (chapter \ref{chap:traj}) is presented in ref.~\citep{Roser2015_TrajGeometry}.

We begin by setting the scene. Chapter \ref{chap:problemoftime} begins by considering the classical laws governing a system of particles with the aim to express them in a time-parameterisation invariant manner. That is, we wish to describe the dynamics in a way that is invariant under the transformation $t\rightarrow T(t)$, the transformation from Newtonian time $t$ to an arbitrary (smooth monotonic) function $T(t)$ thereof. Classically this turns out to be not difficult. However, we find that it is a necessary condition of time-reparameterisation invariance that the value of the Hamiltonian function vanishes, $H=0$. This `Hamiltonian constraint' leads to difficulties after quantisation however, where it seems to imply that the quantum dynamics is frozen. This is a facet of the `problem of time'. We leave the particle system behind and review relevant concepts in general relativity, in particular the `$3+1$' formalism of Arnowitt, Deser and Misner. The formalism has properties resembling those of the parameterised particle model and once again leads to frozen dynamics after quantisation. However, there are other facets of the problem of time, related to the spatio-temporal symmetries of general relativity.

In chapter \ref{chap:choiceoftime} we review the idea of Hamiltonian reduction, a method to `pick' a particular time parameter as physically fundamental and derive an associated non-vanishing Hamiltonian, eliminating the problem of time. We argue that while in the classical theory the choice of time is purely aesthetic, without consequences for observation, quantisation with distinct choices of time may lead to diverging phenomenologies, implying that if indeed there is a physical time, the list of contenders can be narrowed down empirically. We illustrate this using a particle model.

Finally we introduce York time in chapter \ref{chap:Yorktime}, a particular choice of time parameter in general relativity which is a plausible candidate for an underlying physically fundamental time. We discuss its role in the initial-value problem of general relativity, perhaps the strongest motivation for taking York time seriously, before looking at some of its formal and physical properties, specifically the Poisson structure of the reduced variables, the behaviour of the York-time foliation around singularities and how various aspects of the problem of time are resolved. 

Hamiltonian reduction cannot be performed explicitly in the case of full general relativity without additional symmetry assumptions since in this case the Hamiltonian constraint cannot be solved explicitly. In part \ref{part2} we perform the procedure for various physically relevant scenarios in which solving the Hamiltonian constraint \emph{is} possible due to the presence of symmetries or at least approximate symmetries. 

In chapter \ref{chap:Friedmann} we consider the simplest possible cosmological scenario, a homogeneous and isotropic (`Friedmann-Lema\^itre') universe. This simplification leaves only one geometric degree of freedom: scale. This one degree of freedom is `absorbed' into the notion of time through the Hamiltonian reduction, so that only matter degrees of freedom remain, for which we consider one or more scalar fields. The model, though simple, is nonetheless of relevance. First, it serves to illustrate some of the features of the reduced-Hamiltonian method in practice. Second, the model will describe the dynamics of the homogeneous background on which cosmological perturbation theory is defined. Furthermore, aside from the simplified matter content used here, the Friedmann-Lema\^itre universe does, in fact, provide a good approximation to our actual universe. We conclude the chapter with a discussion of cosmological inflation in terms of the York-time description.

In chapter \ref{chap:cosmext} we discuss an important, if unexpected implication of considering York time the physically fundamental time parameter: the history of the universe should (depending on the global geometry of space, that is, whether the universe is closed, flat or open) be extended beyond what is normally regarded as temporal infinity $t=\infty$. This point corresponds to a finite value of $T$ and \emph{a priori} there is no reason for $T$ to `end' here. However, there are also concrete physical reasons why the quantum theory based on York time requires this extension for consistency. We consider the conditions on various scalar-field potentials, with particular emphasis on potentials favoured as candidates for inflation, for a smooth transition from `our' side of cosmological history to the extension. 

Chapter \ref{chap:perttheoprelims} begins the development of the theory of cosmological perturbations in York time. First we consider a simple model (a homogeneous but anisotropic model classically equivalent to the `Kasner' models) that explores the unconventional Poisson structure, that is, the fact that the geometric reduced variables are only `almost' canonical. We then examine phenomena that are well understood in conventional perturbation theory, such as the freezing of modes as their physical wavelengths grow to scales larger than the Hubble radius, in the context of York time cosmology. 

Finally, drawing on the preliminary insights gained in the last chapter, in chapter \ref{chap:perttheoformalism} we develop the formalism for cosmological perturbation theory based on York-time Hamiltonian reduction, discuss a variety of related conceptual points and aim to obtain a physical understanding of the theory.

Chapter \ref{chap:QuantFriedmann} begins part \ref{part3} the development of York-time quantum cosmology. We consider the quantum theory of the cosmological `background' by canonical quantisation of the theory developed in chapter \ref{chap:Friedmann}, explore the dynamics and draw some conclusions for different scalar-field potentials. We also discuss some of the more subtle technical points of the quantisation procedure.

In the next chapter, chapter \ref{chap:quantcosmpert}, we undertake the project of canonically quantising cosmological perturbation theory as developed in chapter \ref{chap:perttheoprelims}. First, we consider the quantised anisotropic minisuperspace model explored earlier. This quantum theory can be solved exactly, that is, we are able to find the Hamiltonian eigenvalues and eigenfunctions, for example. Following quantisation the non-canonicity of the Poisson brackets has some curious consequences relating to the nature of momenta. In particular, (1) the theory does not have a momentum representation and (2) the momenta are not observables (are not Hermitian). We discuss the implication thereof. We then move on to the quantisation of perturbation theory proper and discuss a number of its features. Because of its complexity, no explicit solutions can be found and explicit connection with observation will likely require numerical work.

Chapter \ref{chap:traj} considers an alternative to canonical quantisation. After all, the problem of time is the result of the application of the canonical quantisation recipe to general relativity. In the rest of the thesis we only modified general relativity by selecting a physically preferred choice of foliation. In this chapter on the other hand we consider a modification of the quantisation recipe by `quantising' via the mutual interaction of an infinite set of configuration-space trajectories. The particular approach proposed here, we argue, is preferable over similar quantisation methods developed in recent years for certain technical reasons that we explain in detail.

The approach of chapter \ref{chap:traj} is understood for simple finite-dimensional particle models. Its application to gravity remains an open question. In chapter \ref{chap:cosmtraj} we discuss the application of the quantisation scheme developed in the last chapter to gravitational field theories and consider some cosmological examples. The mathematical development here remains mostly schematic. Given that this line of inquiry is still in its infancy, these last two chapters are to be understood as a speculative guide and the beginning of a new direction of research rather than a presentation of established results.

Finally, in chapter \ref{chap:conclusions} we summarise what has been accomplished in this work and discuss its implications and open questions, as well as future directions of research.

\part{Setting the scene}\label{part1}
\chapter{The problem of time}\label{chap:problemoftime}

\textit{In this chapter we explore the problems that arise when attempting to quantise a time-reparameterisation and refoliation invariant theory such as general relativity. These `problems of time' are perhaps the most pressing difficulty when formulating a theory of quantum gravity.}

\section{Time-reparameterisation invariance} \label{sec:TRepInvariance}
Much of modern physics describes the physical world through the use of \emph{variables}. These are mathematical quantities --- often real or complex numbers, or entities representable as ordered sets thereof (such as vectors, matrices and so on), or occasionally more complicated objects with different mathematical behaviours (such Grassmann numbers) --- which are understood to have some physical interpretation, are subject to certain laws expressible in the form of equations limiting the values the variables can take and establishing relations between values at different (often infinitesimally separated) points in time. We leave a more careful characterisation of the relationship between the \emph{mathematical} entities and \emph{physical} reality to philosophical discourse. 

The variables and equations that provide the best known description of our actual world, the quantum state vector on the space of field variables of the standard model of particle physics and, somewhat speculatively, of some appropriate theory of gravity along the lines of general relativity, are so complicated that explicit calculations are only possible for highly symmetric situations (such as Friedmann-Lema\^itre cosmology) or through perturbative approaches (such as particle scattering theory or cosmological perturbation theory). It is therefore often beneficial to look at non-realistic, simple, finite-dimensional models in order to understand individual physical features of the more complicated realistic theory. The implications of \emph{time-reparameterisation invariance}, that is, roughly, the irrelevance of the choice of temporal parameter to the form of the equations, is no exception.

While the notion of a `variable' is fairly clear, at least from a practical perspective, the notion of `time' is not. Greater questions aside, one can ask: Should time $t$ itself be a variable? Suppose we have a classical system of $n$ particles of identical mass $m$, moving in three-dimensional space, whose instantaneous configuration may therefore be described by a $3n$-dimensional vector in configuration space. The set of all points that describe the particle configuration at some time forms a trajectory in this configuration space. This trajectory may well cross itself any (finite or infinite) number of times. Suppose instead we consider the \emph{extended} configuration space, that is, the $3n$-dimensional configuration space together with an extra dimension, time. Time $t$ is treated as a further variable. The trajectory in this $3n+1$-dimensional space never crosses itself (since every instant of time only occurs once). Furthermore, with time $t$ explicitly included in the information given by the vector in the extended configuration space, the trajectory may be parameterised by any choice of parameter $\tau$, not necessarily the value of $t$. 

The natural question raised by this picture is whether or not the physical laws (equations) that govern which trajectories are dynamically allowed (that is, are solutions to the equations of motion) can be expressed in terms of the arbitrarily chosen parameter $\tau$. For simplicity, suppose the system is conservative (or \emph{scleronomic} in Boltzmann's language), that is, its dynamics has no explicit time dependence. Consider the form of the action when parameterised by $\tau$ rather than $t$:\footnote{In this and the next few paragraphs we follow roughly the presentation in Lanczos beautiful book \citep{Lanczos1949} on the variational principles of classical mechanics, chapter V, section 6.}
\begin{align} \text{Action }
 &= \int_{t_1}^{t_2} dt\; L\big(q_1, \dots, q_{3n}; \dot{q}_1, \dots, \dot{q}_{3n} \big) \notag\\
 &= \int_{\tau_1}^{\tau_2} d\tau\, t^\p\; L\left(q_1, \dots, q_{3n}; \frac{q_1^\p}{t^\p},\dots, \frac{q_{3n}^\p}{t^\p}\right), \label{eq:2.1-parameterisedaction}
\end{align}
where $\dot{q}_i=dq_i/dt$, $q^\p=dq_i/d\tau$ and $t^\p=dt/d\tau$. We limit our considerations to Lagrangians without second or higher-order time-derivatives. This still encompasses the majority of physically relevant Lagrangians and indeed all those relevant for the present work. Note that the variable $t$ appears only in the form $t^\p$. It is \emph{kinosthenic}, or \emph{cyclic} in Helmholtz's terminology. It is then a basic theorem of classical mechanics that its conjugate momentum, $p_t$ is a constant of motion. Bearing in mind that the Lagrangian in the action in $\tau$-form is $t^\p L$ rather than merely $L$, this momentum is
\begin{align} p_t
 &= \PD{(t^\p L)}{t^\p} = L-\left(\sum\limits_{i=1}^{3n}\PD{L}{\dot{q}_i}\frac{q_i^\p}{t^{\p2}}\right)t^\p = L -\sum\limits_{i=1}^{3n}p_i\dot{q}_i \notag\\
 &= -\left(\sum_{i=1}^{3n}p_i\dot{q}_i - L\right). \label{eq:2.1-ptisminusH}
\end{align}
The momentum $p_t$ conjugate to time is just the negative of the Hamiltonian of the system. This insight will form the basis of the `reduced Hamiltonian formalism' we will employ below. At present, we note that the fact that $p_t$ is a constant of motion therefore implies that the total energy of the system is, too. This constitutes a proof that the energy in conservative systems is conserved, although this was not our primary objective here.

Another consequence of $p_t$ being a constant of motion is that $t$ may be eliminated entirely from the variational problem defined by this Lagrangian. This effectively constitutes a partial integration (hence a partial solution) of the Lagrangian equations of motion. In our case this eliminates any reference to the original notion of Newton's absolute time $t$. The general procedure to eliminate kinosthenic variables can be found in most textbooks on classical mechanics (or e.g.\ ref.~\citep[ch.\ 5, sec.\ 4]{Lanczos1949}). In brief, the procedure is as follows.

Suppose variable $q_k$ only appears as `$\dot{q}_k$' in the Lagrangian. In this case the Euler-Lagrange equation for $q_k$ is simply the statement that the conjugate momentum $p_k\equiv \partial L/\partial \dot{q}_k$ is constant, $p_k=c_k$. One introduces a modified Lagrangian,
\begin{equation} \bar{L}\equiv L-c_k\dot{q}_k.\end{equation}
Now solve the original Euler-Lagrange equation of $q_k$, $\partial L/\partial \dot{q}_k = c_k$ for $\dot{q}_k$ and substitute the result into $\bar{L}$ to arrive at a variational problem leading to dynamics equivalent to the original but without dependence on the kinosthenic variable.

Let us apply this procedure to the variable $t$ in order to eliminate any reference to it. The momentum conjugate to $t$ is a constant, 
\begin{equation}p_t=-E. \label{eq:2.1-ptisminusE}\end{equation} 
The modified action is
\begin{equation} \bar{A} = \int \bar{L} = \int_{\tau_1}^{\tau_2} d\tau\; \big(Lt^\p-p_tt^\p\big) 
					= \int_{\tau_1}^{\tau_2} d\tau\; (L-p_t)t^\p 
					=\int_{\tau_1}^{\tau_2} d\tau\; \sum\limits_{i=1}^{3n} p_i\dot{q}_it^\p , \label{eq:2.1-ModifiedAction}
\end{equation}
in which we must eliminate $t^\p$ via equation \ref{eq:2.1-ptisminusE}. Assuming that the Lagrangian has the usual form $L=K.E.-V$, note that the integrand on the right-hand side of eq.\ \ref{eq:2.1-ModifiedAction} is just twice the kinetic energy,
\begin{equation} K.E. = \tfrac12\sum\limits_{i=1}^{3n} p_i\dot{q}_it^\p = \tfrac12 \sum\limits_{i=1}^{3n} m \dot{q}_i^2 t^\p 
		      =\tfrac12\sum\limits_{i=1}^{3n} m\left(\frac{q^{\p}_i}{t^\p}\right)^2 t^\p
		      =\frac{1}{2t^\p}\sum\limits_{i=1}^{3n} m q_i^{\p2}. \label{eq:2.1-KineticEnergy}
\end{equation}
Having chosen to focus on Lagrangians of the usual form, eq.\ \ref{eq:2.1-ptisminusH} combined with eq.\ \ref{eq:2.1-ptisminusE} now gives the familiar relation
\begin{equation} E = K.E.+V(\vec{q}), \label{eq:2.1-EnergyBalance}\end{equation}
where the argument $(\vec{q})$ is shorthand for $(q_1,\dots,q_{3n})$, so that eq.\ \ref{eq:2.1-KineticEnergy} implies
\begin{equation} t^\p = \sqrt{\frac{\sum_{i=1}^{3n} m q_i^{\p2}}{2(E-V)}} \end{equation}
and the resulting action is (after another use of eq.\ \ref{eq:2.1-EnergyBalance}),
\begin{equation} \bar{A} = \int_{\tau_1}^{\tau_2} d\tau\; \sqrt{(E-V)\cdot\sum_{i=1}^{3n} \tfrac12 m q_i^{\p2}}. \label{eq:2.1-Barboursqrtaction}\end{equation}
This action is of \emph{geodesic form}: it may be written as
\begin{equation} \bar{A}= \int \sqrt{\sum_{i=1,j=1}^{3n,3n}G_{ij}(\vec{q}) dq_i dq_j},\qquad G_{ij}(\vec{q}) = \tfrac12\big(E-V(\vec{q})\big)\delta_{ij}, \end{equation}
where $\delta_{ij}$ is the Kronecker delta function. That is, the configuration space may be considered to be equipped with a metric $G_{ij}(\vec{q})$ and the variational problem is one of minimising the path length in configuration space with respect to that metric. We will meet actions of this form throughout this thesis.

The action principle \ref{eq:2.1-Barboursqrtaction} has been used by Barbour \citep{Barbour2011,BarbourBertotti1982} as a starting point for time-reparameterisation invariant mechanics. He justifies this by arguing that the `kinetic metric' $G_{ij}$ is the natural choice of metric on the configuration space. We chose an alternative route for its derivation by promoting Newtonian time $t$ to a variable and parameterising the configuration-space trajectory by an arbitrary monotonic parameter, identifying $t$ as kinosthenic and hence being able to eliminate it, resulting in the geodesic action \ref{eq:2.1-Barboursqrtaction}. The Euler-Lagrange equations for this action are
\begin{equation} \deriv{\tau}\left(\sqrt{\frac{E-V(\vec{q})}{\sum_{k}\frac12 m \left(\DERIV{q_k}{\tau}\right)^2}}\, m\DERIV{q_i}{\tau}\right) 
				= -\sqrt{\frac{\sum_{k}\frac12 m \left(\DERIV{q_k}{\tau}\right)^2}{E-V(\vec{q})}}\PD{V}{q_i}.
\end{equation}
These equations are dramatically simplified if we choose $\tau$ such that $\sum_{k}\frac12 m (\DERIV{q_k}{\tau})^2=E-V(\vec{q})$. This is, of course, the case when $\tau=t$, where one recovers the familiar Newtonian equations of motion. The Newtonian time parameter is special: when the dynamical laws are expressed with respect to $t$ they take their simplest form. However, we are not forced to make that choice or ascribe any fundamental meaning to $t$. But given that $t$ has this simplifying property it is no surprise that Newton and his followers believed in the existence of an `absolute' time.\footnote{Newton's reasoning was, of course, considerably more subtle and relied on philosophical and theological considerations as well \citep{NewtonPrincipia,LeibnizClarkeCorrespondence}. Furthermore, the Lagrangian formalism had not yet been invented, although it is interesting to speculate what he might have thought of the above line of argument.}

\section{Vanishing Hamiltonians and frozen dynamics} \label{sec:VanishingH}

It is an obvious next step to try to understand time-reparameterisation invariance in the Hamiltonian picture. However, promoting Newtonian time $t$ to a variable and parameterising the configuration-space trajectory instead via an arbitrary parameter $\tau$ leads to a difficulty: The Hamiltonian vanishes. The Hamiltonian\footnote{We use a subscript $\tau$ to identify the Hamiltonian as the one defined with respect to the arbitrary parameter and not, say, the Hamiltonian $H_t$ defined with respect to Newtonian time $t$, which is given by $H_t=-p_t$.} is defined as
\begin{equation} H_\tau = p_tt^\p + \sum_{i=1}^{3n}\PD{(Lt^\p)}{q_i^\p}q_i^\p - Lt^\p  \label{eq:2.2-Htau}\end{equation}
since the Lagrangian is $Lt^\p$ and the set of configuration variables is $(t,q_i,\dots,q_{3n})$. But from eq.\ \ref{eq:2.1-ptisminusH} and $\dot{q}_i=q_i^\p/t^\p$ it follows immediately that the right-hand side vanishes,
\begin{equation} H_\tau = 0. \label{eq:2.2-Htauvanishes}\end{equation}
This result may be understood heuristically, although as with all heuristic explanations care must be taken not to take this for being more rigorous than it is.\footnote{It is not an uncommon occurrence for heuristic explanations to become the canon that is repeated and propagated without question. A prime example is undoubtedly Hawking radiation, whose supposed heuristic explanation by spontaneous pair production near a black-hole horizon has very little to do with the rigorous derivation via the inequivalence of inertial frames at the horizon and at infinity. The heuristic version goes back to a paragraph in Hawking's original paper \citep{Hawking1974}, though it is doubtful that the author intended this idea to be popularised in the way it did.} The heuristic picture is as follows. A Hamiltonian function generates (via the Poisson brackets) motion of the system in phase space that corresponds to the physical evolution of the system in time. But $H_\tau$ is to generate motion for \emph{all} possible choices of $\tau$. For example, if $H_\tau$ generates the motion for one choice $\tau=\tau_1$, exactly the same Hamiltonian $H_\tau$ is to generate the motion for another choice, $\tau=\tau_2$ given by $\tau_2=2\tau_1$, that is, a choice where the parameter-time `passes' twice as fast. But this is inconsistent unless the Hamiltonian vanishes.

Eq.\ \ref{eq:2.1-ptisminusH} followed from the \emph{parameterised form} of the action \ref{eq:2.1-parameterisedaction}. The vanishing of the Hamiltonian is a direct result of the definition of the momenta (specifically $p_t$) for an action of this form and hence is a primary constraint in the language of Dirac \citep{Dirac1964}. It is the so-called \emph{Hamiltonian constraint}. As an aside, note that the other momenta $p_1,\dots,p_{3n}$ conjugate to the original variables $q_1,\dots,q_{3n}$ are independent of the choice of time since
\begin{equation} p_i^{(t)} \equiv \PD{L}{\dot{q}_i}=\PD{L}{q_i^\p}t^\p = \PD{(Lt^\p)}{q_i^\p} \equiv p_i^{(\tau)}.\end{equation}

The result that the Hamiltonian vanishes remains unchanged if one eliminates the cyclic variable $t$ first. Starting with the the Lagrangian given by expression \ref{eq:2.1-Barboursqrtaction} (the Lagrangian of Barbour and Bertotti \citep{BarbourBertotti1982}) instead, the momenta are
\begin{equation} p_i^{BB} = \pd{q_i^\p}\sqrt{(E-V)\sum_{k=1}^{3n}\tfrac12m q_k^{\p2}} = \frac12\sqrt{\frac{E-V}{\sum_{k=1}^{3n}\frac12 mq_k^{\p2}}}\cdot mq_i^\p, \end{equation}
in virtue of which the associated Hamiltonian vanishes,
\begin{equation} H_\tau^{BB} = \sum_{i=1}^{3n} p_i^{BB} q_i^\p - \sqrt{(E-V)\sum_{i=1}^{3n}\tfrac12 mq_i^{\p2}} =0.\end{equation}

The Hamiltonian constraint does not pose any problem for the dynamics, at least classically. For example in a conventional particle system with $L=K.E.-V$ and $H_t=E=K.E.+V=-p_t$ it merely says that the total energy $E$ is zero. Since the dynamics does not depend on the value of $V$ but only its gradient (is invariant under a shift in the potential $V\rightarrow V+$const.), whether or not $E=0$ is not detectable and the choice of potential and resulting dynamics is effectively unrestricted. 

The Hamiltonian constraint does however pose a problem for quantisation. During canonical quantisation the canonical variables are `promoted' to operators, which act on `states', that is, complex functions $\psi$ over the configuration space of the system. The constraint equations become equations of operators, which may either be identically satisfied at the operator level or, if not, select a subspace of states that are considered `physical' in virtue of satisfying this quantum constraint equation, rather than mere mathematical artifacts. We will encounter both cases in this thesis. This is not the place to review the systematic treatment of constraints in the classical and quantum theory, which is due to Dirac \citep{Dirac1964}. A comprehensive discussion of constraint systems can be found in ref.~\citep{Sundermeyer1982}.

Accordingly, physical states in time-reparameterisation invariant systems, that is, systems with a Hamiltonian constraint $H_\tau=0$, must satisfy
\begin{equation} \hat{H}_\tau \,\psi_{phys} = 0, \label{eq:2.2-particleWdW}\end{equation}
where $\hat{H}_{\tau}$ is the Hamiltonian operator obtained in the usual manner by replacing $q_i$ and $p_i$ by operators $\hat{q}_i$ and $\hat{p}_i$ respectively in the functional expression for $H_\tau$. In the context of general relativity the analogue equation is known as the Wheeler-deWitt equation, which we will encounter in section \ref{sec:PoTinGR}. The implication of equation \ref{eq:2.2-particleWdW} is that the Schr\"odinger equation for physical states is then
\begin{equation} i\hbar\PD{\psi_{phys}}{\tau} = \hat{H}_\tau\psi_{phys} = 0. \end{equation}
The state vector does not evolve and the quantum dynamics is `frozen'. This static result is known as the `problem of time', although in the context of general relativity there are other, related issues summarised under this label (see section \ref{sec:OtherProblemsOfTime}).

The above treatment of canonical quantisation does, of course, not constitute a rigorous discussion. As we discussed in chapter \ref{chap:Introduction}, a more rigorous approach would be formulated in terms of an operator algebra whose commutator structure is inherited from the Poisson algebra of the classical theory. For a thorough treatment, see ref.~\citep{Thiemann2007}. Suffice to say, the problem of time is unaffected.

The Hamiltonians $H_\tau$ and $H_\tau^{BB}$ were derived by extending the set of configuration variables to include Newtonian time $t$. One might consider this counterintuitive to the stated goal that is the elimination of an external time parameter. Is this the cause of the problem?

Not so. Return instead to the parameterised action \ref{eq:2.1-parameterisedaction}. Suppose we wish to eliminate any dependence on $t$, which only appears in the form of its derivative $t^\p$. If $t$ is to be meaningless in the desired dynamics then $t^\p$ merely denotes some arbitrary non-negative\footnote{What about $\DERIV{t}{\tau}=0$? This would suggest that an extended interval of $\tau$ would correspond to a single point on the $t$-timeline. A physical interpretation of this is not immediately apparent. However, later on we will discover examples with such a relationship between two choices of time parameter, for example, York time with de~Sitter space.} function, which we will call $N(\tau)$. Its counterpart in general relativity, which we will meet in section \ref{sec:PoTinGR}, is referred to as the `lapse' function and we also adopt this nomenclature here. Our proposed action is then
\begin{equation} \text{Action }=\int_{\tau_1}^{\tau_2} d\tau\; N\;L\left(q_1,\dots,q_{3n};\frac{q_1^\p}{N},\dots,\frac{q_{3n}^\p}{N}\right) \label{eq:2.2-parameterisedactionwithN}.\end{equation}
Having `forgotten' the origin of $N(\tau)$, this is now an arbitrary function that encodes the freedom of choice with regards to the temporal parameter. 

In many texts the argument proceeds by treating $N$ like another dynamical variable. One then introduces its conjugate momentum $p_N$, which must vanish since $dN/d\tau$ does not occur in the action. Such variables do not introduce dynamical equations of motion but so-called primary constraints. The easiest way to see this is to write the action in \emph{canonical form}, that is, introduce conjugate momenta and perform a Legendre transformation of the integrand,
\begin{equation} \text{Canonical action }=\int_{\tau_1}^{\tau_2} d\tau\; \Bigg[\sum_{i=1}^{3n} p_iq_i^\p - H(q,p)\Bigg],\end{equation}
where $H(q,p)$ is a function such that the expression matches the original Lagrangian up to total derivatives and is, of course, the Hamiltonian of the system. The new variable $N$ does not contribute to the sum since $p_N$ vanishes. Note that the canonical action presents a new variational problem with $6n$ variables $\{\vec{q},\vec{p}\}$ and its Euler-Lagrange equations are just Hamilton's equation. The Hamiltonian formalism can therefore be understood as a special case of the Lagrangian formalism, namely limited to problems in which the Lagrangian takes the form of a sum made up of pairs of variables one of which (the `$q$') appears as a time derivative while the other one (the `$p$') does not, minus a function of the variables only (without time derivatives). This result is independent of any notion of reparameterisation invariance and the nature of the temporal parameter. It applies just the same to mechanics in terms of standard Newtonian time \citep{Lanczos1949}. In the case of the action \ref{eq:2.2-parameterisedactionwithN}, the canonical action is
\begin{equation} \text{Canonical action } = \int_{\tau_1}^{\tau_2} d\tau\; \Bigg[\sum_{i=1}^{3n} p_iq_i^\p - N\curlyH(\vec{q},\vec{p})\Bigg], \end{equation}
that is, the Hamiltonian is
\begin{equation} H(N,\vec{q},\vec{p}) = N\cdot\curlyH(\vec{q},\vec{p}) = N\cdot\left(\sum_{i=1}^{3n}\tfrac{1}{2m}p_i^2-V(\vec{q})\right). \end{equation}
The Euler-Lagrange equation for $N$ then implies that 
\begin{equation} \curlyH(\vec{q},\vec{p}) = 0,\end{equation}
which is exactly the Hamiltonian constraint obtained above.

The reasoning of the last paragraph --- that is, the reasoning of most texts reviewing the subject --- is, on the whole, correct. The one step that may be objectionable is the very first: Why should $N$ be treated as a dynamical variable? It is a function of $\tau$ but has no obvious physical meaning. The point we wish to make here is that it is, in fact, not necessary to adopt $N$ as a dynamical variable. First note that the form of the canonical action is independent of this step. The system is expected to follow the `path of least action' for \emph{any} choice of $N$ (since our dynamics is to be independent of that choice), which is akin to demanding that it be extremised with respect to $N$. That is, the variational principle for $N$ is justified on physical grounds, not because the function is adopted to the pool of variables that are subject to the well-known Lagrangian machinery. The formal consequence, namely the appearance of the Hamiltonian constraint, is the same as in the standard argument. Here we merely took issue with its justification.

The derivation of the Hamiltonian constraint (and the resulting frozen dynamics following canonical quantisation) via the lapse function $N$ is, in fact, the way the problem is usually understood in the context of general relativity. There treating $N$ as a variable may more justified depending on one's perspective: The lapse is effectively the information contained in the time-time entry of the spacetime metric $\supfour g_{\mu\nu}$ (i.e.\ $\mu=\nu=0$) and so from a spacetime point of view is indeed one of the variables. On the other hand, if a `3+1' formalism is considered fundamental (that is, a slice of space evolving in time), then $N$ should not be considered a physical variable. We will discuss these perspectives and the contrast between them in greater detail below.

Below we will also show that the problem of time may be addressed by supposing the existence of a physically meaningful time parameter. Formally this is done via a so-called `reduced Hamiltonian formalism', where the choice of physical time $T$ is extracted from the physical variables and a corresponding (non-vanishing) Hamiltonian function is derived. At this stage however it is important to acknowledge that this is by no means the only possibility. In particular, the procedure of canonical quantisation---itself questionable as we argued in chapter \ref{chap:Introduction}---may be abandoned in favour of perhaps some other method to arrive at a quantum theory with the appropriate classical limit (corresponding to the classical system with which we started) but without the suffering from the problem of time. This more ambitious endeavour is began in chapter \ref{chap:traj}.

{
\begin{mdframed}

\setlength{\parindent}{10pt}

\small
\bigskip
\textbf{The relational particle dynamics of Barbour et al.}\bigskip
 
The elimination of a theory dependent on $t$ in favour of a reparameterisation-invariant description was in part motivated by the fact that $t$ had no physical role (although it did lead to the simplest equations of motion). Yet $t$ is not the only variable that has that property. Any gauge degree of freedom should similarly be eliminated. In the context of Newtonian particle theories, for example, concepts such as absolute position and orientation are physically meaningless unless reference to a notion of `absolute space' is made. But since absolute space plays no role in the physics, its introduction is unwarranted. For a system of $N$ particles in three dimensions the real number of degrees of freedom is therefore not $3N$ but $3N-6$, the coordinates of all the particles minus the centre-of-mass position (three degrees of freedom) and the overall orientation in three-dimensional space (three further degrees of freedom, for example in the form of Euler angles). 

Eliminating unphysical degrees of freedom of this kind motivated Barbour and various collaborators to develop the `shape dynamics'\footnote{\label{footnote:2-BarbourBoxSD} Without capitalisation we use this term to refer to any theory, which aims to eliminate unphysical degrees of freedom as described. When capitalised, `Shape Dynamics' refers to a particular theory of gravity based on three-dimensional conformal geometries evolving in time, developed in the last six years or so \citep[e.g.][]{GomesGrybKoslowski2011,GomesKoslowski2011,BarbourKoslowskiMercati2013ProbOfTime,Mercati2014}. The terminology (shape dynamics versus Shape Dynamics) is my choice. In the literature context and author usually determines which is meant.} of particles and fields \citep[e.g.][]{Barbour2011,BarbourBertotti1982,Barbour2003SIGPD, AndersonBarbourFosterOMurchadha2003, BarbourFosterOMurchadha2002}. They include not only the six degrees of freedom mentioned but also the overall scale (only \emph{ratios} of distances are considered physical), so that a total of seven degrees of freedom are eliminated. Why this leaves the dynamics of `shape' is best illustrated by an example: Consider three particles, effectively forming the three vertices of a triangle, na\"ively having nine degrees of freedom. However, the orientation of the plane of the triangle in space is physically irrelevant (using up three degrees of freedom), as is the position of the centre of mass on the plane (two more degrees of freedom) and its orientation in the plane (one degree of freedom). Finally, the overall size of the triangle is presumed physically irrelevant (so that geometrically similar triangles are considered just representations of a single physical triangle), eliminating a seventh degree of freedom. The remaining two degrees of freedom determine only the ratios of the three sides to one another --- in other words the \emph{shape} of the triangle.

Formulating the dynamics explicitly in terms of physical quantities only is difficult. Instead these ideas are implemented via a procedure called `best matching'. The idea is this: two system configurations differing only by the value of variables considered unphysical are consequently themselves physically equivalent. At each instant one can therefore freely `slide' the configuration-space point representing the system up and down these gauge orbits. Now choose at each instant the gauge variables such that the resulting system (the triangle, for example) matches that of the previous instant most closely. This minimisation procedure (one is minimising the difference in Newtonian representation between temporally adjacent system configurations (triangles)) implies that certain boundary terms in the action are forced to vanish, resulting in constraints.\footnote{We will use a very similar construction in the development of a new wavefunction-free quantisation procedure in chapter \ref{chap:traj}.} In particular, the constraints are that the momenta conjugate to the unphysical degrees of freedom vanish, namely
\begin{alignat*}{4}
 \vec{P} &\equiv&\; \sum_{I=1}^{n} m_I\dot{\vec{x}}_I 			&=&\; 0 &&\qquad \text{(Total linear momenta)}\\
 \vec{L} &\equiv&\; \sum_{I=1}^{n} M_I \vec{x}_I\times\dot{\vec{x}}_I	&=&\; 0 &&\qquad \text{(Total angular momenta)}\\
 D	 &\equiv&\; 2\sum_{I=1}^{n} M_I \vec{x}_I\cdot\dot{\vec{x}}_I	&=&\; 0 &&\qquad \text{(`Dilational' momentum)}.
\end{alignat*}
In addition, the elimination of an specific time parameter introduces the Hamiltonian constraint. Applied to a universe made up of particles, the consequences are that this universe has zero total energy (due to the Hamiltonian constraint), zero total linear momentum, zero angular momentum and zero dilational momentum. The value of energy is not detectable anyway since shifting the potential by a constant has no dynamical effect. The value of the total linear momentum is also undetectable unless Galilean relativity were violated. The total value of angular momentum is however detectable, at least in principle (a rotating reference frame can be dynamically distinguished from a non-rotating one from within, and a similar result holds for the dilational momentum. Hence this theory would be falsifiable via measurement of the universe's total angular or dilational momentum if indeed we lived in a universe whose fundamental building blocks are point particles in three-dimensional space.

A question here is however whether specifying initial data independent of the gauge variables is sufficient to uniquely determine the future evolution of the physical variables. If it is, this would arguably constitute an implementation of Mach's Principle, a concept that has received ample philosophical discussion over the last one hundred years but disagreement persists over what exactly the principle is.

One can apply analogous ideas to gravitational field theories, which led to a series of papers on ideas such as conformal geometrodynamics \citep[e.g.][]{BarbourFosterOMurchadha2002, AndersonBarbourFosterOMurchadha2003} and ultimately Shape Dynamics (see footnote \ref{footnote:2-BarbourBoxSD}).

Finally, the question arises whether the Hamiltonian constraint itself can be the generator of physical transformations, since all the other constraints relating to the total momenta do not, and according to Dirac's classification scheme \citep{Dirac1964} the constraint is primary and first-class, normally associated with physically irrelevant motion. Barbour and Foster \citep{BarbourFoster2008} have however pointed out that the argument relies on a absolute time being the independent variable. This assumptions is explicitly violated in time-reparameterisation invariant systems. The Hamiltonian constraint may therefore be a generator of physical change, not just movement along a gauge orbit.

\end{mdframed}
}

\section{The problem of time in general relativity}\label{sec:PoTinGR}

\subsection{General relativity}

General relativity is a classical theory that describes gravity. General relativity postulates a \emph{space-time}, represented by a four-dimensional (pseudo-)Riemannian manifold. The fundamental physical variables are encoded in the \emph{spacetime metric} $\supfour g_{\mu\nu}(x)$, that is, ten real numbers for each point of spacetime in the form of a four-by-four symmetric matrix.\footnote{Throughout this thesis we will use the raised prefix `$\supfour$' to denote quantities relating to four-dimensional space-\emph{time} (e.g.\ $\supfour g_{\mu\nu}$, $\supfour R$, ,,,), while quantities concerning three-dimensional \emph{space} are written without a prefix (e.g.\ $g_{ab}$, $R$, ...). Greek indices assume values $0,1,2,3$ with $0$ representing the temporal direction, while Latin indices take values $1,2,3$, ranging over spatial directions only. Except where otherwise indicated (e.g.\ section \ref{sec:ClassKasner}) we follow the usual Einstein summation convention, assuming a sum over repeated indices.} The metric encodes the \emph{geometry}, that is, the notion of length and angles by defining the scalar product between two four-vectors $A,B$ at a spacetime point $x$,
\begin{equation} (A\cdot B)(x) = \supfour g_{\mu\nu}(x) A^\mu(x) B^\nu(x). \end{equation}
The length of a vector is then given by $|A|=\sqrt{A\cdot A}$. The equations governing the behaviour of the metric are the ten independent Einstein Field Equations (EFEs),
\begin{equation} \supfour R_{\mu\nu} - \tfrac12\supfour R \supfour g_{\mu\nu} = \frac{8\pi G}{c^4}\supfour T_{\mu\nu}, \end{equation}
where $\supfour R$ and $\supfour R_{\mu\nu}$ are, respectively, the scalar and Ricci curvature derived from $\supfour g_{\mu\nu}$, and $\supfour T_{\mu\nu}$ is the stress-energy tensor of the appropriate matter fields, $G$ being the gravitational constant. For a full introduction to general relativity we refer the reader to \citep{MisnerThorneWheeler1973} or another book of their choice. In what follows we will assume the use of units such that the speed of light $c$ is unity.

The number of geometric degrees of freedom is however less than ten. This is because the entries in the matrix $\supfour g_{\mu\nu}$ are coordinate dependent. In other words, a single \emph{geometry} corresponds to an infinity of possible \emph{coordinate representations} of that geometry. The EFEs take the same form in all coordinate systems. That is, they are \emph{generally covariant}, whence the name `general relativity'. Furthermore, every measurable quantity must be independent of the choice of coordinate system. 

The precise physical meaning of general covariance and its relation to notions such as `background independence' (the absence of non-dynamical structures) and diffeomorphism invariance (roughly, that the theory possesses a certain number of continuous symmetries) is, oddly, still somewhat contentious. See ref.~\citep{Pooley2015} for a recent contribution and references therein. In brief: Coordinates have no physical content. They are labels of points in space and time, that is, methods of \emph{describing} physical entities. Whether or not the description of the physics (that is, equations and their formal solutions) is unchanged when the labelling of points is changed is a question concerned entirely with our description of the physics, not the physics itself. General covariance is a property of our description, not physical reality. Indeed any theory can be cast into a fully covariant form.\footnote{For example, to obtain standard Klein-Gordon scalar-field theory in Minkowski spacetime, promote spacetime derivatives $\partial_\mu$ in the Klein-Gordon equation to covariant derivatives $\nabla_\mu$ and introduce an additional equation to determine the spacetime properties,
\begin{equation*} R_{\mu\nu\rho\sigma} = 0.\end{equation*}} This, in effect, was a point raised as early as 1917 by Kretschmann \citep{Kretschmann1917} as a response to the undue importance some researchers --- including, arguably, Einstein himself --- attached to general relativity being general covariant.

The Einstein Field Equations follow from an action principle that is simply the extremisation of the scalar four-curvature, the so-called Einstein-Hilbert action,
\begin{equation} \text{EH action }= \tfrac12 M_{Pl}^2\;\int\;d^4x\;\left(\sqrt{|\supfour g|}\;\supfour R\;+L_{matter}\right), \label{eq:2.3-EHAction}\end{equation}
where $L_{matter}$ is the four-covariant Lagrangian of the matter content and may include a cosmological constant $\Lambda$.\footnote{Whether or not this is conceptually appropriate depends on the proposed origin of $\Lambda$ \citep[see e.g.\ ][]{BianchiRovelli2010}. Either way, mathematically it is unproblematic.}

In equation \ref{eq:2.3-EHAction} $M_{Pl}$ denotes the reduced Planck mass. Its relationship to the cosmological constant $G$ is
\begin{equation} M_{Pl} = \sqrt{\frac{\hbar c}{8\pi G}}  \label{eq:2.3-PlanckMass} \end{equation}
but throughout the rest of this thesis we choose units such that $\hbar=1$ and $c=1$. We refrain from setting $M_{Pl}=1$ (or equivalently $8\pi G=1$) since it can at times be insightful to see the dependence of the dynamics on these constants explicitly. We have to choose if we wish to make reference to $M_{Pl}$ or $G$ in our derivations. While this is ultimately only an aesthetic choice, the two options do suggest slightly different physical insights: how quantities and dynamics depend on the Planck scale $M_{Pl}$, which presumably encodes where our classical theories break down entirely, versus how they depend on the `strength of gravity' $G$. Since ultimately we are interested in the construction of a theory of quantum gravity, we choose to make reference to $M_{Pl}$ in this thesis.

\subsection{The Arnowitt-Deser-Misner formalism}

The \emph{spacetime} description of general relativity given above is widely accepted as the correct way to interpret general relativity. Some particular solution of the Einstein Field Equations is a description of a block universe, the geometry (and value of matter fields) at all points in spacetime. It is not a sequence of configurations labelled by a time parameter. This makes a Hamiltonian formulation of the theory impossible since a Hamiltonian is to be understood as a generator of time translations, and this in turn means that the system cannot be canonically quantised. Early attempts to construct a Hamiltonian formulation of general relativity culminated in a series of papers by Arnowitt, Deser and Misner in the late fifties and early sixties, in particular \citep{ADM1962}. Their approach became known as the Arnowitt-Deser-Misner (or ADM) formalism and forms the basis of much of the subsequent work on canonical quantum gravity. Other terminology used is the `3+1' or `Cauchy' formulation of general relativity.

For a na\"ive approach to a Hamiltonian formulation of general relativity one might propose the following: Choose some coordinate system that covers the region of spacetime for which we wish to arrive at a Hamiltonian description. Slices of constant $t$ automatically provide a foliation of the spacetime into spatial slices. Two points in space on slices labelled by $t_1$ and $t_2$ respectively are considered the same point if and only if they have the same spatial coordinate. Now construct a Hamiltonian density (by some appropriate method) that adequately describes the evolution of all field values (matter field and the spatial geometry induced on the slice) at all spatial points from slice $t$ to $t+\Delta t$.

Such an approach, or a variant thereof, would be, however, deeply problematic. First, whether or not this can be carried out depends crucially on the original choice of coordinate system. For example, the spatial coordinates must cover the same ranges on each temporal slice, which is by no means guaranteed. Second, even if the procedure can be carried out, the Hamiltonian theory has lost a crucial feature of general relativity: its general covariance. 

Part of the feat of ADM was to come up with a formulation that retains the full general covariance at the Hamiltonian level. That is, the formulation has to be general enough to allow for any choice of foliation, and any choice of coordinates on each slice, after the Hamiltonian has been derived, but no such choice must be made prior to it. In what follows we will not reproduce the full derivation, which can be found in any textbook or set of lecture notes on the subject (e.g.\ \citep{MisnerThorneWheeler1973}) but merely highlight critical points.

The first step is the decomposition of the spacetime metric $\supfour g_{\mu\nu}$ into a spatial metric $g_{ab}$ that represents the geometry on a given spatial slice and four functions, which establish the spatio-temporal relationships between one slice and infinitesimally nearby ones and points thereon. The decomposition is as follows:
\begin{equation}
   \supfour g_{00} = - (N^2-N_iN^i), \qquad   \supfour g_{0a}= N_a, \qquad  \supfour g_{ab}=g_{ab}.
\end{equation}
The function $N$ is the lapse function and will be shown to indeed correspond to a close analogy to the notion of lapse explored in section \ref{sec:VanishingH}. The three functions $N_a$ are the `shift', which relate the coordinate values of points of one slice with those of infinitesimally close ones. For reference, we also note the expressions for the inverse metric,
\begin{equation}
 \supfour g^{00} = - N^{-2},\qquad	\supfour g^{0a} = \frac{N^a}{N^2}, \qquad	\supfour g^{ab}= g^{ab}-\frac{N^aN^b}{N^2},
\end{equation}
and for the spacetime volume element,
\begin{equation} \sqrt{|det (\supfour g_{\mu\nu})|} = N\sqrt{det (g_{ij})} = N\sqrt{g}, \end{equation}
where the last equality is purely notational, $g\equiv det (g_{ij})$.

The lapse $N$ receives its name from the fact that it denotes how much proper time $d\tau$ lapses per interval of coordinate time $dt$ in the direction in spacetime that is orthogonal to the embedded spatial slice. This interpretation relies on the existence of the entire spacetime, in which spatial slices have geometrical relations to one another. This is a natural assumption if we consider the four-dimensional spacetime picture to be fundamental and our `3+1' formalism to be merely a re-description. On the other hand, if we consider a picture in which the geometric properties of a spatial slice evolve in time (in conjunction with matter fields defined on the slice), then this interpretation is nonsensical as there is no notion of orthogonality to a slice since there is no embedding. In this case $N$ (and $N_a$) must be considered free functions without geometric content, although it is the appearance of these functions in the formalism that ultimately allows the reconstruction of a spacetime picture by `welding' spatial slices together in an appropriate manner. The situation is mostly analogous to that in section \ref{sec:VanishingH}, where $N$ may be understood to encode a relationship of the arbitrary time parameter and Newtonian time if and only if Newtonian time is considered to have a meaning in the first place, that is, if and only if the parameterisation-invariant formalism is not considered the original starting point. The situation is not entirely analogous however because not being able to ascribe $N$ with an obvious interpretation is due to slightly different reasons: In the particle formalism it was due to the `non-existence' of Newtonian time, while here it is due to the absence of the notion of orthogonality. In the ADM formalism $N$ relates the arbitrary coordinate time to \emph{proper} time rather than some `background' time \`a la Newton. We will discuss the notion of proper time in section \ref{sec:YTproperties}.

Whether the ADM formalism is considered to describe the relation between spatial slices and geometric variables thereon or rather the evolution of a single slice also determines whether $N$ and $N_a$ should be considered dynamical variables. In the former case they describe geometric properties between slices and the identification of points from slice to slice, and can therefore be considered geometric variables, namely just a transformation on the time-time and time-space components of $\supfour g_{\mu\nu}$. In the latter case $N$ and $N_a$ are simply arbitrary function without dynamical meaning and therefore should not be treated as variables. However, just like in the case of parameterised particle mechanics there is no distinction between the resulting equations and dynamics, merely between the lines of reasoning used to arrive at them.

I am not aware that this point concerning the ambiguous nature of $N$ has been made explicitly in the literature, although at least some authors are clearly aware of it. For example, Unruh and Wald \citep{UnruhWald1989} introduce $N$ and $N_a$ as arbitrary functions on the spatial slice ``which have the interpretation of lapse function and shift vector in the spacetime constructed from the time evolution.''

In the space-time picture the shift $N_a$ relates the identification of points by spatial coordinates on neighbouring spatial slices. In particular, they denote how the point identified by a particular spatial coordinate value $x$ is shifted from the point that lies orthogonally to the future of the point identified by $x$ on the infinitesimally previous slice. This implies that unlike the lapse $N$ the shift $N_a$ depends on the purely spatial coordinates chosen on the slice. As a consequence, the choice of how to split spacetime into space and time fixes $N$, but $N_a$ is only determined once a coordinate system is chosen \emph{on} the slice. If one therefore makes some choice of physical time, such as the `York time' we will choose later in this work, the shift remains undetermined. In other words, because the way spacetime is separated into space and time is more physically meaningful than the choice of spatial coordinates (in particular with regards to subsequent quantisation), the lapse and shift are not exactly analogous to one another. 

In order to establish the dynamics of the ADM picture, we must rewrite the Einstein-Hilbert action \ref{eq:2.3-EHAction} in terms of the new variables. This is not fundamentally difficult, although the decomposition of the scalar 4-curvature $\supfour R$ into a scalar 3-curvature $R$ and other terms is somewhat involved algebraically. Helpful here are the Gauss-Codazzi relations (see ref.~\citep{MisnerThorneWheeler1973}).
The final result is the ADM action, 
\begin{equation} \text{ADM Action } = \int \,dt\,d^3x\; N\rootg\;\left(R+g^{ac}g^{bd}K_{ab}K_{cd}-K^2\right) + \text{ matter terms}. \label{eq:2.3-ADMAction}\end{equation}
All terms in the Lagrangian density are understood to be functions of time and space. The term $K_{ab}$ denotes the \emph{extrinsic curvature} and $K\equiv g^{ab}K_{ab}$ its trace. We will explain their meaning momentarily. We first wish to remark however on the form of this Lagrangian density, which matches the usual `kinetic-minus-potential' form where the terms involving $K_{ab}$ constitute the kinetic term (which is furthermore a squared first-order time derivative, as will become clear) and the term $-R$, which depends only on the geometric variables $g_{ab}$ and not their time derivatives, takes the role of the potential. 

The extrinsic curvature $K_{ab}$ describes how the spatial slice is embedded in the spacetime manifold in relation to other, temporally infinitesimally separated slices. An isolated slice cannot be said to have an extrinsic curvature, so at least at this stage we must still assume that the spacetime picture is the fundamental one and the `3+1' formalism only a re-description. Formally the extrinsic curvature may be defined as the Lie derivative of the spatial metric along the normalised vector $\hat{n}$ orthogonal to the spatial slice,
\begin{equation} K_{ab} \equiv -\tfrac12\mathcal{L}_{\hat{n}}g_{ab} = -\tfrac12\mathcal{L}_{\hat{n}}\supfour g_{ab} 
			  = -\tfrac12\left(\hat{n}^\mu\del_\mu\supfour g_{ab} + \supfour g_{a\mu}\partial_b\hat{n}^\mu -\supfour g_{\mu b}\partial_a\hat{n}^\mu \right).
\end{equation}
Another possible expression is in terms of a covariant derivative, $K_{ab}=-\nabla_b\hat{n}_a$. In more useful and intuitive terms, the extrinsic curvature denotes the fractional rate of deformation of a volume element---or equivalently, the rate of change of $g_{ab}$---with respect to proper time $\tau$, orthogonal to the spatial slice,
\begin{equation} K_{ab} = -\frac12\left(\PD{g_{ab}}{\tau} - \frac{\nabla_a N_b}{N} - \frac{\nabla_b N_a}{N}\right),\end{equation}
or in terms of an arbitrary coordinate time $t$,
\begin{equation} K_{ab} = -\frac{1}{2N}\left(\PD{g_{ab}}{t} - \nabla_a N_b - \nabla_b N_a\right). \end{equation}
The terms involving the shift $N_a$ ensure orthogonality since the time derivative alone would measure the rate of change of the metric along the line connecting points of equal spatial-coordinate value, which may be shifted relative to the point orthogonally `above' the starting point. In many scenarios one is able to set the shift to zero, which constitutes a partial gauge fixing of the spatial coordinates (a partial coordinate choice) and indeed we will do so below since it significantly simplifies the algebra without significant loss of insight.

The scalar extrinsic curvature $K=g^{ab}K_{ab}$ is the rate of fractional shrinkage of volume with respect to proper time. In a homogeneous and isotropic cosmology like the one we will consider below this implies that $K$ is nothing but the Hubble parameter up to a negative constant.

It is important to emphasise the importance of an embedding spacetime for the meaningfulness of $K_{ab}$. At the same time, however, it is not essential for $K_{ab}$ to have this geometric interpretation. Instead it is possible to view $K_{ab}$ merely as a convenient way to express the kinetic term of the ADM action.

The other point of note is that the ADM action \ref{eq:2.3-ADMAction} is already in \emph{parameterised form}, analogous to the particle action \ref{eq:2.2-parameterisedactionwithN} we obtained in section \ref{sec:VanishingH}. There we had to `manually' introduce an arbitrary time parameter and treat Newtonian time as one of the dynamical variables. Here this step is unnecessary. The reason is that in general relativity the time parameter is \emph{already} dynamical, or at least its rate of change (given by the lapse $N$ or equivalently the temporal components of the 4-metric) is. In the particle model there was no notion of shift since it was not a field theory and the variables had discrete labels, but otherwise the situation is exactly analogous. Consequently, as we will see shortly, the ADM formalism also suffers from the problem of time.

In order to make this apparent, first introduce the geometric momenta,
\begin{equation} \pi^{ab} = \PD{\mathcal{L}}{\dot{g}_{ab}} =  \sqrt{g}\left(Kg^{ab}-K^{ab}\right), \label{eq:2.3-ADMmomenta}\end{equation}
where the Lagrangian density $\mathcal{L}$ denotes the integrand of the action \ref{eq:2.3-ADMAction} and $\dot{g}_{ab}=dg_{ab}/dt$. The action can then be written in canonical form,
\begin{equation} \text{Canon.\ ADM action }=  \int \,dt\,d^3x\; \bigg[ \pi^{ab}\dot{g}_{ab} + \text{`}\phi p_\phi\text{'} - N\curlyH -N_a\curlyH^a \bigg],  \label{eq:2.3-CanonicalADMAction}\end{equation}
where `$\phi p_\phi$' stands for the canonical kinetic term of the matter content (the $\dot{q}\cdot p$ terms that characterise the canonical action integral) and
\begin{align}
 \curlyH   &= -\tfrac12 \rootg M_{Pl}^2 R + \frac{2}{M_{Pl}^2\rootg}\left(g_{ac}g_{bd}-\tfrac12 g_{ab}g_{cd}\right)\pi^{ab}\pi^{cd} + \curlyH_{matter}, \label{eq:2.3-HConstraintExpression} \\
 \curlyH^a &= -2\nabla_b\pi^{ab} + \curlyH^a_{matter}, \label{eq:2.3-MomConstraintExpression}
\end{align}
with $\curlyH_{matter}$ and $\curlyH^a_{matter}$ being place-holders for the terms appearing as a result of the matter fields, which we are leaving general for now. It is common terminology to call $\curlyH$ and $\curlyH_a$ the Hamiltonian constraint and momentum constraint respectively, although this is somewhat confusing since the literal `constraints' are the demand that these expressions vanish, 
\begin{align}
  \curlyH   &= 0 \label{eq:2.3-HConstraint} \\
  \curlyH^a &= 0 \label{eq:2.3-MomConstraint},
\end{align}
not the name for the expressions themselves.

The reason that they do vanish is analogous to the situation in the particle model of sections \ref{sec:TRepInvariance}, \ref{sec:VanishingH}. Either one takes the spacetime view, in which case $N$ and $N_a$ are dynamical variables and then the Hamiltonian and momentum constraints follow from their respective variation, or, in the space-in-time picture, they are just arbitrary functions but the action is required to be extremised for any choice of $N$ and $N_a$, in which case the constraints also follow.

Note that each constraint describes the vanishing of a \emph{density} and therefore constitutes four constraint \emph{at each spatial point}. An equivalent set of constraint could be imposed,
\begin{align}
 \int d^3x\; \xi^0\curlyH &=0 		\label{eq:2.3-HConstraintXiForm}\\ 
 \int d^3x\; \xi^a\curlyH_a &= 0 	\label{eq:2.3-MomConstraintXiForm}
\end{align}
for any choice of the four functions $\xi^0(x)$, $\xi^a(x)$. The expression on the left-hand side have another role: The momentum constraint expressions \ref{eq:2.3-MomConstraintXiForm} generate canonical transformations that result from spatial diffeomorphisms on the slice, that is, infinitesimal changes in the spatial coordinates. Expression \ref{eq:2.3-HConstraintXiForm} is the generator of canonical transformations corresponding to diffeomorphisms of the spacetime metric (constructed from $g_{ab}$, $\pi^{ab}$, $N$ and $N_a$ if spacetime is not presupposed), although there are subtleties with this interpretation since the correspondence only holds if the field equations are satisfied \citep{UnruhWald1989}. On the phase-space hypersurface on which all constraints are satisfied however, the set of expressions \ref{eq:2.3-MomConstraintXiForm}, \ref{eq:2.3-HConstraintXiForm} suffice to generate all the spacetime diffeomorphisms of general relativity.

\subsection{The problem of time in the ADM formalism}

From equation \ref{eq:2.3-CanonicalADMAction} we can read off the ADM Hamiltonian,
\begin{equation} H = \int d^3x\; \Big(N_a\curlyH^a + N\curlyH\Big), \label{eq:2.3-ADMHamiltonian} \end{equation}
which vanishes in virtue of eqs.\ \ref{eq:2.3-HConstraint}, \ref{eq:2.3-MomConstraint} or eqs.\ \ref{eq:2.3-HConstraintXiForm}, \ref{eq:2.3-MomConstraintXiForm} (simply set $\xi_a=N_a$, $\xi_0=N$). 
According to the canonical quantisation procedure one now `promotes' the canonical variables to operators, which are acting on `states', as we discussed in the introduction (chapter \ref{chap:Introduction}). The quantum Hamiltonian given by expression \ref{eq:2.3-ADMHamiltonian} with $g_{ab}$ and $\pi^{ab}$ replaced by $\hat{g}_{ab}$ and $\hat{\pi}^{ab}$ respectively (and similar replacements are made for the matter variables) determines the time evolution of those states. Operator-expressions derived from the constraints now select which states are considered physically possible. The Hamiltonian and momentum constraints \ref{eq:2.3-HConstraintXiForm}, \ref{eq:2.3-MomConstraintXiForm} therefore lead to the conditions
\begin{align}
 \left[\int d^3x\; \xi^0\hat{\curlyH}\right] \Psi_{phys} 	&= 0 \label{eq:2.3-QuantumHConstraintXiForm}\\
 \left[\int d^3x\; \xi^a\hat{\curlyH}_a\right] \Psi_{phys}	&= 0.\label{eq:2.3-QuantumMomConstraintXiForm}
\end{align}
These constraints ensure that $\Psi$ is invariant under the same set of spacetime diffeomorphisms as the classical theory. In particular, \ref{eq:2.3-QuantumMomConstraintXiForm} ensures that $\Psi$ only depends on the spatial three-geometry and not its metric representation. This means $\Psi$ is a function on \emph{superspace}, the space of three-geometries of the manifold, each of which is an equivalence class of metrics related via spatial diffeomorphisms. An analogous interpretation of \ref{eq:2.3-QuantumHConstraintXiForm} is not possible since its meaning concerns the relationship between quantum states at different times. In conjunction with \ref{eq:2.3-QuantumMomConstraintXiForm} the effect is that acting with a spacetime diffeomorphism in the spacetime constructed from the spatial slice at different times on the argument of the state $\Psi$ does not change it.

Equation \ref{eq:2.3-QuantumHConstraintXiForm} is known as the \emph{Wheeler-deWitt} equation. It is the equation central to the problem of time, more so than eq.\ \ref{eq:2.3-QuantumMomConstraintXiForm}. The reason is that in the Hamiltonian \ref{eq:2.3-ADMHamiltonian} and equivalently in its quantum operator form
\begin{equation} \hat{H} = \int d^3x\; \Big(N_a\hat{\curlyH}^a +N\hat{\curlyH}\Big) \end{equation}
one can set $N_a=0$ by an appropriate choice of coordinates but not $N$, since $N=0$ would lead to singular evolution---no proper time would pass during a finite coordinate-time interval. That is, the contribution of the momentum constraint to the Hamiltonian essentially reduces to a matter of coordinates, the Hamiltonian constraint does not. Equivalently the Wheeler-deWitt equation may be written as one constraint per spacetime point,
\begin{equation} \hat{\curlyH} \Psi_{phys} =0, \label{eq:2.3-QuantumHConstraint} \end{equation}
in analogy with the classical case.

In virtue of the Wheeler-deWitt equation \ref{eq:2.3-QuantumHConstraintXiForm} and of \ref{eq:2.3-QuantumMomConstraintXiForm}, the Hamiltonian annihilates physical states,
\begin{equation} \hat{H}\Psi_{phys} =0, \end{equation}
and this once again makes the Schr\"odinger equation imply a frozen dynamics,
\begin{equation} i\hbar\pd{t}\Psi_{phys} = 0.\end{equation}
If $\Psi$ is presumed to describe the physical state of the universe, which is the case in quantum theory, then no evolution is possible. This is not only a formal problem but it appears to outright contradict the fact that the state of the universe is evidently changing. This constitutes the problem of time in quantum geometrodynamics, the quantum theory of the dynamics of spatial geometry as derived from the ADM formalism, or at least one aspect of it. There are other issues arising from the spacetime-diffeomorphism invariance (that is, the ability to choose one's foliation freely and parameterise the slices at will), which we will discuss in the next section.

\section{Other problems of time} \label{sec:OtherProblemsOfTime}

The problem of frozen dynamics is only one of a number of difficulties and questions that arises when quantising (or even just preparing to quantise) a classical theory that is refoliation invariant. A number of these issues are essentially related to the fact that two theories that are classically equivalent --- or indeed two methods of expressing the same theory (for example, once in a four-covariant and once in a `gauge-fixed' form) --- do not necessarily lead to equivalent quantum theories when the rules of quantisation are applied. Many of these problems do not have counterparts in a finite-dimensional models such as the particle model studied above.

The classification of the following list is primarily due to Isham \citep{Isham1992}, Kucha\v{r} \citep{Kuchar2011} and Anderson \citep{Anderson2012}. However, a number of issues are linked and cannot strictly be separated, and in some cases a solution to one issue would also constitute a solution to another. Anderson fittingly calls them `facets' of the problem of time.

\paragraph{The global problem of time.}
It is not in general guaranteed that a system allows for a time parameter that can be defined consistently everywhere. In particular, there are scenarios where no function of the physical variables is always monotonic. It is therefore not guaranteed that a global notion of time can be constructed.

\paragraph{The (thin) sandwich problem.}
Suppose data of the intrinsic spatial metric (such that it satisfies the momentum and Hamiltonian constraints) is given on two (possibly infinitesimally) separated hypersurfaces, each of which is considered to be `space at a given time'. It is unclear whether (or to what degree) this specification of the boundary data uniquely determines the spacetime geometry. The reconstruction of a spacetime from a series of spatial slices requires knowledge of the lapse and shift functions in addition to the intrinsic variables on each slice. The problem is already a classical one. 

\paragraph{Spacetime reconstruction problem.}
In the quantum version of the sandwich problem (presumably one would have to give a wave functional over spatial geometries as data for the initial and final time) it is not even clear if and in what manner a spacetime geometry (or functional over spacetime geometries) can be constructed at all. This touches on the much bigger question on the role of spacetime in quantum gravity. 

\begin{figure}[htb]
 \label{fig:differentfoliations}
 \includegraphics[width=0.8\linewidth]{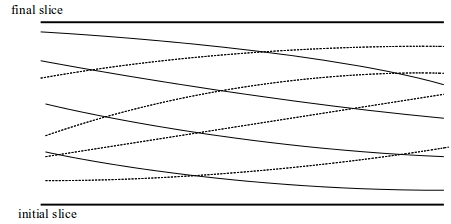}
 \caption[Different ways of foliating spacetime]{Different ways of foliating the spacetime into spatial slices while keeping the initial and final slice (the `bread' of the sandwich) fixed. The sandwich problem and spacetime-reconstruction problem concern determination of the foliation from data at the end points, while the functional-evolution problem asks whether a quantum state (over geometries) on the initial slice uniquely determines the quantum state on the final slice irrespective of the intermediate foliation. At least the classical analogue, the question of the independence of the classical evolution on the foliation, can be answered affirmatively.}
\end{figure}

\paragraph{Independence of evolution on the foliation.}
The (more or less) converse of the sandwich problem is: Suppose an initial and final spatial surface is chosen and the initial data is fixed. If one then evolves the data via two different intermediate foliations, does one arrive at the same final data? Classically it can be shown that this is guaranteed by the first-class nature of the constraints (they Poisson commute with the Hamiltonian) \citep{KucharTorre1991} (and so there is not actually a problem). But for a quantum theory this is not clear:

\paragraph{The functional-evolution problem.}
In particular, evolving an initial state $\Psi_{init}[g]$ via two different intermediate foliations to a final state $\Psi_{fin}[g]$, it is not guaranteed that the two final states obtained in this manner are the same. In other words, are the quantum Hamiltonian and momentum constraints all that is needed to determine consistent evolution of a state?

\paragraph{Multiple-choice problem.}
This is essentially the issue that forces us to carefully consider how to identify a physical time parameter. Different choices, while associated with classically equivalent theories, lead to non-equivalent quantum theories. 

\paragraph{} 
These problems are mostly concerned with the relationship between space (or spatial geometry) evolving through time and spacetime, in particular in a quantum theory. Therefore, if we are willing to abandon the idea that spacetime is in some some fundamental, or that it must play some role in the quantum theory, these problems are really no problems at all. In fact, these problems arguably hint at the fact that the notion of spacetime ought to be abandoned in the quantum theory. A spacetime picture can be constructed in the classical theory in virtue of the constraints. The construction is rather compelling --- so much so that one may be forgiven to consider it more fundamental than the theory of spatial geometry evolving through time. However, ultimately the problems of time hint at the fact that this may be a mistake. 

That said, in the absence of a fundamental notion of spacetime it would be desirable to have a better understanding of why the constraints `conspire' in such a way that a spacetime construction is possible at the classical level. More accurately, why is the (quantum) theory that describes our universe such that its classical limit has a structure (namely, the necessary symmetries) that allows for the construction of spacetime? Of course, any question in physics beginning with `Why?' is notoriously difficult to answer and at best an answer consists of a reduction of the problem to more basic and simpler statements (for example, in the form of a simple physical principle). Nonetheless some reduction of the sort would be highly desirable in order to do away with the crutch of spacetime in explaining why the laws of gravity are what they are. I am not at this stage aware of any work on the subject, either by physicists or philosophers.

\chapter{Physical time}\label{chap:choiceoftime}

\textit{In this chapter we explain how the existence of a physically fundamental notion of time would alleviate the problem of time. We explain how to derive via `Hamiltonian reduction' a theory with a fundamental notion of time from a classically equivalent theory with time-reparameterisation invariance. We provide an example of how two classically equivalent theories with distinct choices of time may lead to distinguishable quantum theories.}

\section{Choosing a physical time: Hamiltonian reduction} \label{sec:Hamreduction}

There are numerous ways to overcome or avoid the problem of time. Very broadly speaking, these may be divided into two categories: approaches that modify the quantisation formalism such that the vanishing classical Hamiltonian does not lead to frozen dynamics in the quantum theory, and approaches that modify the classical theory in order to avoid the vanishing Hamiltonian, although some ideas may fall into both categories. The former require for the most part a radical departure from what is usually understood to constitute the framework of quantum theory. In chapters \ref{chap:traj} and \ref{chap:cosmtraj} we explore a new method of quantisation, although also in conjunction with a preferred spacetime foliation.

Among the latter type of approaches is what is perhaps the simplest idea of them all, the choice of a fundamental time parameter. In the case of relativity, this includes two separate steps: (1) the choice of foliation, that is, how to split spacetime into space and time, and (2) the choice of parameterisation of slices. In the case of the time-reparameterisation invariant particle model of sections \ref{sec:TRepInvariance} and \ref{sec:VanishingH} only the second step is necessary. 

Two claims constitute such a proposal, which a proponent will have to justify. First, there is a claim of \emph{existence}. The proponent claims that there is a physically preferred foliation and time parameter. Second, the proponent identifies a \emph{particular choice} of foliation and time parameter and must explain why this is a plausible choice from the infinite number of options. Of course, it is a perfectly viable position to make the first claim and be agnostic with regards to the second. In fact, as conscientious scientists we do not, of course, subscribe to any such proposal in particular. Rather, we consider this solution to the problem of time plausible (it is no theoretical impossibility) and therefore worth of investigation, but we do not believe that any particular choice of time parameter is correct until we have empirical or other\footnote{It is highly questionable what other form of evidence there could be. Clearly a theory has to pass basic tests of mathematical and physical consistency. However, any claims beyond this must be viewed very critically. In fact, the tendency by a few physicists (specifically, some string theorists \citep{Dawid2013}) to argue for non-empirical selection criteria for theories has prompted a critique by Ellis and Silk \citep{EllisSilk2014}, which in turn has led to a fair degree of argument within the high-energy theoretical-physics community over the last year or two. For a very recent and insightful clarification on the subject, see ref.~\citep{Rovelli2016NonEmpiricalDangers}.} evidence to support such a claim. However, in order to guide our search for plausible choices of time parameter we must invoke some theoretical ideas and since scientific inquiry is carried out by fallible human beings our respective evaluation of particular proposals may vary.\footnote{We are not only fallible with regards to our ability for analytical assessment of theoretical proposals but also, for example, with regards to our biases towards familiar concepts (what we understand and built our career on appears more plausible) and the biases of our funding agency (proposals more likely to receive grant money suddenly become a lot more plausible).} 
We will discuss a particular choice of time parameter, York time, in the next chapter (chapter~\ref{chap:Yorktime}).

The basic idea is then as follows. Choose a time parameter that serves as the physical time, derive a (non-vanishing) Hamiltonian that describes the dynamics with respect to this parameter and quantise based on this new Hamiltonian. In this section we will aim to understand the procedure that allows us to derive the non-vanishing, \emph{physical} Hamiltonian based on a choice of physical time parameter.  The procedure to do so is called \emph{Hamiltonian reduction} and the resulting theory with the choice of York time is the central topic of investigation in this thesis.

Of course, a crucial question is whether or not the choice of time has any measurable consequence. That is, is the claim that some particular choice is the correct physical time parameter falsifiable? Fortunately, the answer appears to be affirmative at least in principle. In section \ref{sec:choiceoftimesignificance} we will illustrate this using a simple two-particle model. 

The central idea of Hamiltonian reduction is that the Hamiltonian is the generator of transformations in time. In general, the quantity that generates transformations in a variable $q_i$ is the momentum $p_i$ conjugate to that variable. Therefore, if the physical time $T$ is one of those variables (and this is what we mean by $T$ being a `physical' time), then motion with respect to $T$ is generated by the momentum $P_T$ that is conjugate to $T$. The physical Hamiltonian is therefore $H_T\equiv -P_T$, where the minus sign is due to convention. If $T$ is not one of the `$q_i$', then one must first perform an appropriate change of variable to make it so. Depending on one's choice of $T$, performing such a change of variable may not always be possible---York time is a notable example of this---and one must instead introduce $T$ and $P_T$ as additional variables in conjunction with constraints. Furthermore, the remaining \emph{reduced} variables may then obey a non-standard Poisson structure. We will see this explicitly in later sections.

The physical Hamiltonian, or equivalently the momentum $P_T$, must be expressed as a function of the other variables, so that equations of motion may be derived. Of course, this assumes that $P_T$ is fully determined by the remaining variables. This it is in virtue of the Hamiltonian constraint, which provides exactly the one condition (or one condition per spacetime point in the case of field theories, in which one obtains the Hamiltonian density via this method) that implies that having fixed all but one of the set of canonical variables, that last one is determined. Hence one has to \emph{solve} the Hamiltonian constraint for $P_T$ as a function of the other variables (including $T$). This gives the desired functional expression for $H_T$. The existence of the Hamiltonian constraint is guaranteed by the time-reparameterisation invariance of the original dynamical description, in other words, the parameterised form of the action. If the system is not parameterised, it cannot be deparameterised. 

It is by no means guaranteed that the Hamiltonian constraint is analytically solvable. In fact, the full Hamiltonian constraint of general relativity is not. This is the reason why we are restricted to minisuperspace models and perturbative solutions in later chapters.

The physical time $T$ must be chosen in such a way that $T$ is monotonic. Otherwise $P_T$ may become singular and at best the reduced-Hamiltonian description is limited to time intervals during which $T$ is monotonic. In the case of York time problems arise if one attempts to describe a pure de~Sitter space, in which $T$ is constant over an extended interval of conventional cosmological time $t$, for example. 

Let us apply the Hamiltonian reduction procedure to the particle model of chapter \ref{chap:problemoftime}. The simplest possibility is a return to Newtonian time. The vanishing Hamiltonian $H_\tau$ has the form \ref{eq:2.2-Htau}, which is just
\begin{equation} H_\tau = p_t t^\prime + \left(\sum_{i+1}^{3n} p_i\DERIV{q_i}{t}-L\right) t^\prime. \end{equation}
The Hamiltonian constraint $H_\tau=0$ is then solved straightforwardly for $H_t=-p_t$ and one obtains
\begin{equation} H_t(q_1,\dots,q_{3n}, p_1,\dots,p_{3n},t) = \sum_{i+1}^{3n} p_i\DERIV{q_i}{t}-L, \end{equation}
that is, the conventional Hamiltonian of such a finite-dimensional model.

A less trivial example would be to pick one of the particles to function as a `clock'. In other words, have some function of its position and momentum correspond to the physical time $T$. We will explore this choice for a particular two-particle model in section \ref{sec:choiceoftimesignificance}, where we use it to illustrate that two quantum theories, derived via canonical quantisation from a pair of physically equivalent parameterised and deparameterised classical theories respectively, are not themselves observationally equivalent.

{
\begin{mdframed}
\setlength{\parindent}{10pt}

\small
\bigskip
\textbf{Proof: Classically the parameterised and deparameterised theories are equivalent}\bigskip
 
So far we have merely asserted that the dynamics generated by the reduced Hamiltonian $H_T$ are identical with those generated by the vanishing Hamiltonian $H_\tau$ of the parameterised description. We justified the reduction procedure by the fact that the `momentum' conjugate to `time' is the Hamiltonian, but it is not necessarily clear that the functional expression obtained by solving the Hamiltonian constraint for $H_T\equiv-P_T$ should indeed lead to a set of physically equivalent equations. 

I am not aware of such a proof anywhere in the literature, although frequently it seems to be tacitly assumed that the reduced and unreduced dynamics are identical. Nonetheless I do not doubt that workers in the field (such as Arnowitt, Deser and Misner, or Choquet-Bruhat and York) were aware of a proof such as the one below or similar and that for one reason or another it was just never written down in their publications (to my knowledge).

For easier readability we consider the case of a finite-dimensional model such as the particle model of chapter \ref{chap:problemoftime}. We also assume that the proposed physical time is identifiable with one of the variables, $T=q_{3n+1}$. This can either be achieved by a suitable chosen canonical transformation or by the introduction of an auxiliary variable together with an additional constraint. (The latter would be required, for example, in the case of York time, as we will see in the anisotropic minisuperspace model discussed in section \ref{sec:ClassKasner}.) This also means that the Poisson brackets are the standard canonical ones after deparamterisation and Hamilton's equations take their usual form. As a result the short proof given here is not fully general, although a generalisation is readily possible if rephrased in terms of Poisson brackets.

Let us abbreviate functional dependence by writing $f(\vec{q})$ for $f(q_1, ...,q_{3n})$ and so on. The parameterised system satisfies the Hamiltonian constraint,
\begin{equation*} H_{\tau}(\vec{q},\vec{p},T,-H_T) =0\end{equation*} 
where $T$ and $-H_T$ are just another canonically conjugate pair of variables and have not yet been given any special significance. The Hamiltonian constraint also generates motion in the arbitrary time parameter $\tau$,
\begin{alignat*}{4}
 \DERIV{q_i}{\tau} &= \{q_i,H_{\tau}\} = \PD{H_{\tau}}{p_i},  \qquad &\DERIV{T}{\tau} &=\{T,H_{\tau}\} = \PD{H_{\tau}}{(-H_T)},\\
 \DERIV{p_i}{\tau} &= \{p_i,H_{\tau}\} = -\PD{H_{\tau}}{q_i}, \qquad &\DERIV{H_T}{\tau} &= \{H_T,H_{\tau}\} = \PD{H_{\tau}}{T}. \\
\end{alignat*}
The Hamiltonian-reduction procedure now requires us to solve $H_{\tau}=0$ for $H_T$ as a function of the other variables. Suppose we do so and obtain $H_T(\vec{q},\vec{p},T)$. Then define new function $G(\vec{q},\vec{p},T)$ as
\begin{equation*} G(\vec{q},\vec{p},T)\equiv H_{\tau}\big(\vec{q},\vec{p},T,-H_T(\vec{q},\vec{p},T)\big), \end{equation*}
that is, as the function $H_{\tau}$ with the functional expression of $H_T(\vec{q},\vec{p},T)$ substituted in for the variable $H_T$. By the nature of this construction,
\begin{equation*} G(\vec{q},\vec{p},T) \equiv 0 \end{equation*}
identically, that is, for all values of $\vec{q}$, $\vec{p}$ and $T$, and so it is also the case that
\begin{equation*} \PD{G}{q_i} = 0,\qquad \PD{G}{p_i}=0\quad \text{and} \quad \PD{G}{T}=0.\end{equation*}
One can now express $\DERIV{q_i}{\tau}$ in terms of $G(\vec{q},\vec{p},T)$,
\begin{equation*}\DERIV{q_i}{\tau} = \PD{H_{\tau}}{p_i} = \PD{G}{p_i}-\PD{H_{\tau}}{H_T}\PD{H_T}{p_i}=-\PD{H_{\tau}}{H_T}\PD{H_T}{p_i} = \DERIV{T}{\tau}\PD{H_T}{p_i}.\end{equation*}
But also, by the chain rule, since $T$ is a suitably monotonic parameter by hypothesis,
\begin{equation*} \DERIV{q_i}{\tau} = \DERIV{q_i}{T}\DERIV{T}{\tau}, \end{equation*}
so that Hamilton's equation for the deparameterised dynamics in $T$ follows,
\begin{equation*} \DERIV{q_i}{T} = \PD{H_T}{p_i}.\end{equation*}
The derivation of the equations for $\DERIV{p_i}{T}$ is exactly analogous.

The steps can also be easily inverted to obtain a proof in the other direction, so that full equivalence is established.
\end{mdframed}
}

\section{The meaning of the reduced Hamiltonian} \label{sec:meaningofHamiltonian}

A Hamiltonian is two things: First, it is a \emph{function} of a set of variables representing the physical degrees of freedom of a dynamical system, and possibly of time. Second, the \emph{numerical value} of the Hamiltonian has some physical interpretation, usually `energy'. That is, having arrived at solutions to Hamilton's equations one can substitute the values taken by the variables at some time back into the Hamiltonian function to obtain a number. In many conventional cases the dynamics is independent of $t$, which implies that the value of the Hamiltonian, the energy, is constant. It is doubtful that the concept of energy would even be considered physically meaningful were it not usually conserved. That is, we can talk of a `flow of energy' from one part of a system to another exactly because overall it is conserved and can therefore be conceptualised as something that is passed around. This interpretation is then applied to cases where the the numerical value of the Hamiltonian is not conserved and we speak of energy being `lost' or `gained', usually representing interactions with the world outside the system under consideration. In other words, even here energy is taken to be conserved overall but the appearance of energy gain or loss is due to our attempt to describe a system interacting with the rest of the universe as if it were not. Yet the assumption is that if we were to provide a complete description, then the Hamiltonian would once again be time independent.\footnote{In gravitational system the notion of energy is more complicated since here the energy density is part of the dynamics itself, co-determining the stress-energy tensor and hence via the Einstein equations the curvature of space-time. The two notions are can be roughly identified with each other physically if we place a finite-dimensional particle system into a gravitational field. This description is not entirely consistent though and the two notions of energy should really be kept separate, ideally using different terminology.}

This assumption is justified. The conventional variable $t$ has no physical meaning in itself. While to Newton it was an absolute `background' time at the very foundation of the structure of the physical world, today we may view $t$ as the one choice of time parameters that brings the physical laws into their simplest form (see our discussion in chapter \ref{chap:problemoftime}) and that can be `read off' most easily from the time evolution of systems such as pendula or the motion of the planets (ephemeris time). Having no physical significance, the dynamics cannot be $t$-dependent except in cases where a $t$-dependent term (usually the potential) is used to represent evolution outside the system under investigation.

In a formalism with a physical time $T$ however the situation is very different. The interpretation of the numerical value of the Hamiltonian is not `energy' but whatever the physical meaning of $-P_T$, the negative of the momentum conjugate to $T$, since this is how the physical Hamiltonian was defined. Hence if the position of some particle is taken for $T$, then the same particle's momentum is identified with $-H_T$. In general reduced Hamiltonians are time dependent and their numerical values are not conserved in time. In the case of York time, which we will explore in later chapters, the physical meaning of the associated Hamiltonian density is the local spatial volume element and the full Hamiltonian is the volume of the universe, both of which vary in time.

When quantising a reduced classical system, the eigenvalues and eigenfunctions of the Hamiltonian are consequently not `energy eigenvalues' and `energy eigenfunctions' but also have physical interpretations depending on the meaning of the Hamiltonian, such as `volume eigenvalues' and so on in the case of York time.


\section{Significance of the choice of time}\label{sec:choiceoftimesignificance}

If one takes a realist view of scientific inquiry --- the idea that our theories not only provide models useful for predictions but in some way are also a reflection of at least some properties of the world itself, properties that are `really there' --- then theories with different choices of physical time are manifestly distinct as their ontology is manifestly distinct, even if their empirical predictions are identical. A physicist who rejects a scientific realism on the other hand would disagree. To them an empirical distinction, that is, a different prediction for observation, is required for two theories to be distinct. 

However, even a realist should find a situation where there are two or more competing theories with identical predictions troubling. It would imply that a fundamental feature of reality, namely, in our case, what the true physical time parameter is, is undetectable in principle. The question of which theory is correct cannot be answered scientifically and becomes one of metaphysics rather than physics. In some cases one of a set of empirically equivalent theories may be strongly preferable for other reasons, such as mathematical or ontological simplicity. However, here one invokes \emph{a priori} assumptions about what reality is really like: that it is ontologically simple or is governed by laws whose mathematical form is concise. 

While such assumptions, whether explicit or implicit, have undoubtedly propelled forward progress in physics at multiple occasions in history, to elevate them to ontological tie-breakers for observationally equivalent theories is rash. Furthermore, the occasions that spring to mind, such as the departure from geocentrism, the unification of electricity and magnetism, special and general relativity, the electro-weak unification and so on, all were at least in part observationally motivated and verified. Mathematical and ontological simplicity may have been ideas nudging the scientists' minds in the right direction but ultimately these theories made new predictions which could be subsequently experimentally corroborated. By hypothesis, this would not be the case for equivalent theories differing in their choices of fundamental time.

At the end of the last section we showed that classically the reduced-Hamiltonian theory is equivalent to the parameterised theory of some particular system. This raises the question whether or not quantum theories constructed via quantisation of these classical theories are themselves identical. In fact, that the choice of time does not matter even when composing the quantum theory is a tacit assumption frequently made in particular in the context of cosmological perturbation theory. Here one has to make a `gauge choice', meaning a choice of coordinates describing the small perturbations on the homogeneous cosmological background. Changing the gauge leaves the background description undisturbed (since the transformation is `small') but transforms the representation of the perturbations. A choice must be made (although ultimately measurable quantities must turn out to be gauge invariant) because the four-diffeomorphism invariance of general relativity implies that some of the degrees of freedom are merely coordinate freedom, although there is no matter of fact of which ones these are. The gauge choice not only concerns spatial coordinates on a slice but also small perturbations of the slice itself (small `bumps' in the foliation), so that it involves small deformations in the time parameter. Classically this is known to be indeed only a choice of description and not physical but it is then usually assumed without discussion that the gauge choice is also irrelevant for the physical properties of the quantum theory obtained through quantisation of the perturbations.

Such an assumption is problematic. For example, different choices of gauge will lead to Hilbert spaces whose degrees of freedom have different physical interpretations and it is entirely unclear that the quantum dynamics is physically equivalent when moving from one Hilbert space to another (the functional-evolution facet of the problem of time in the context of cosmological perturbation theory). Different choices of time lead to different reduced Hilbert spaces. A general argument against the equivalence has been presented, for example, in ref.~ \citep{Malkiewicz2015}. Here we will consider a concrete two-particle model which illustrates that different choices of time can indeed lead to observationally distinct theories.\footnote{There are, strictly speaking, a number of separate issues: (1) Can the canonical quantum theories based on two distinct foliations be empirically separated? (2) Can quantum theories based on identical foliations but distinct parameterisations be distinguished? (3) Can quantised reduced-Hamiltonian theories be distinguished from quantisations of the unreduced theories with gauge choices corresponding to the same foliation but different parameterisation? This last question is the one we answer in this section. A full exploration of all three of these and the specific conditions that lead to empirical distinguishability is left for future work.} This saves the realist from having to rely on metaphysics, although there may remain classes of time parameters which lead to equivalent physics, so that observation might only provide the possibility of a partial determination of the true physical time. 

Nonetheless it is encouraging to see that we might ultimately be able to detect what this fundamental physical time is if indeed there is one. The observation would necessarily be one of particular quantum phenomena. In particular, this model suggests that differences may arise with regards to observations relating to the physical degree of freedom that has been chosen as time. If there may be other differences not directly related to the choice of time itself remains uncertain. Furthermore, a theory with a fundamental choice of time not only competes with other choices of time but also other, radically different theories of quantum gravity. How far observations can disentangle all such choices is an open question and will depend on the details of the various theories.

Consider then a system of two particles, each with mass $m=1$, and suppose they are constrained to a one-dimensional `box' (or infinite well) but are otherwise free. Denote the position of the particles by $x$ and $y$ respectively and let the box be of length $\pi$,
\begin{equation} V(x,y) = -U \text{ if } x\in[0,\pi], y\in[0,\pi], \qquad \infty \text{ otherwise}, \end{equation}
where $U>0$ is a constant. We consider two descriptions of this system. First, the conventional `textbook' description using Newtonian time $t$. Second, a deparameterised description using one of the two particles as a clock. The parameterised action for this system is
\begin{align} S
 &= \int d\tau\;N\left(\frac{\dot{x}^2}{2N^2}+\frac{\dot{y}^2}{2N^2} -V\right) \notag\\
 &= \int d\tau\,\left(\dot{x}p_x+\dot{y}p_y - NH_\tau\right),
\end{align}
where the second line is the canonical form. The momenta are
\begin{equation} p_x = \frac{1}{N}\DERIV{x}{\tau},\qquad p_y=\frac{1}{N}\DERIV{y}{\tau} \end{equation}
and the function $H_\tau$, whose vanishing constitutes the Hamiltonian constraint, is
\begin{equation} H_\tau = \tfrac12p_x^2+\tfrac12p_y^2+V =0. \end{equation}
As we discussed in chapter \ref{chap:problemoftime} the Hamiltonian constraint may be considered the equation of motion for $N$. Note that the Hamiltonian constraint has no well-defined solution if one particle is outside the box, since this would require 
\begin{equation} \tfrac12p_x^2 + \tfrac12p_y^2 =-\infty. \end{equation}
Inside the box the Hamiltonian constraint implies that the total kinetic energy is fixed to the constant $U$,
\begin{equation}\tfrac12p_x^2+\tfrac12p_y^2=U.\end{equation}
This equation illustrates why we chose to set the bottom of the well to a negative finite constant rather than zero. Had we chosen the latter option, the particles would have been forced to be at rest.

We begin now with the classical analysis in terms of Newtonian time $t$. Here we simply set $N=1$ and relabel `$\tau$' as `$t$'. This would be the textbook description of this scenario, although we also retain the Hamiltonian constraint, which now constrains the total energy to be equal to zero.\footnote{Dropping the Hamiltonian constraint in order to have the system as it would actually appear in a textbook indifferent to parameterisation invariance would not change the conclusions drawn from the analysis of this section.} The Newtonian Hamiltonian is
\begin{equation} H_t = \tfrac12p_x^2+\tfrac12p_y^2-U \end{equation}
while the two particles are inside the well, which is guaranteed by the Hamiltonian constraint.\footnote{It is a noteworthy if somewhat pedantic point that without the Hamiltonian constraint one must additionally impose the requirement that the initial conditions are such that the particles are inside the well, or equivalently that the total energy is finite. In that case, they can never leave the well. Otherwise starting with a particle outside the well would constitute perfectly reasonable boundary conditions. } Hamilton's equations are just those of a free particle,
\begin{equation}\deriv{t}p_x =0,\;\deriv{t}p_y=0,\qquad \deriv{t}x=p_x,\;\deriv{t}y=p_y, \label{eq:3.2-tHamiltonsequations}\end{equation}
excluding the instances when a particle hits a wall. For the purposes of this model we are content not to treat collisions and instead limit the applicability of our analysis to a time interval within which no collision occurs. If we wanted to treat collisions, we would need to treat the well as the limit of a continuous potential, for example by expressing $V$ as a Fourier series terminated after $n$ terms for the dynamical analysis before taking the limit $n\rightarrow\infty$. The solutions of the equations \ref{eq:3.2-tHamiltonsequations} are, of course, trivial.

Now consider that we instead choose a physical time by considering the particle described by $y$ a `clock'. Specifically, define $T$ and its conjugate momentum $P_T$ as
\begin{equation} T\equiv \frac{p_y}{2y}, \qquad P_T=-y^2.\end{equation}
This choice will not only illustrate the point that the two resulting quantum theories are observationally distinct, but it also is a choice of time that is remarkably similar mathematically to York time in minisuperspace models such as the ones explored in chapters \ref{chap:Friedmann} and \ref{chap:QuantFriedmann}, and sections \ref{sec:ClassKasner} and \ref{sec:quantKasner}. The fact that $T$ and $P_T$ do indeed form a conjugate pair may be shown either by deriving their Poisson brackets (using the known Poisson structure for $y$ and $p_y$), or by considering the canonical kinetic term $\DERIV{T}{\tau}P_T$ and showing that it is equal to $\DERIV{y}{\tau}p_y$ up to a total derivative. We now perform the coordinate transformation $(x,p_x,y,p_y)\rightarrow(x,p_x,T,P_T)$, that is, we substitute $T$ and $P_T$ for $y$ and $p_y$ in $H_\tau$, 
\begin{equation} H_\tau = \tfrac12p_x^2+\tfrac12\cdot 4T^2(-P_T)+V,  \end{equation}
from which we can then solve the Hamiltonian constraint $H_\tau=0$ for $P_T$ in terms of the other variables and derive the physical Hamiltonian according to the procedure described in section \ref{sec:Hamreduction},
\begin{equation}H_T(x,p_x,T)\equiv-P_T(x,p_x,T)=-\frac{1}{4T^2}(p_x^2+2V).\end{equation}
This leads to the Hamilton's equations
\begin{equation} \deriv{T}x=-\frac{1}{2T^2}p_x,\qquad \deriv{T}p_x = 0.\end{equation}
The second of these might have been predicted since $x$ and $p_x$ have been untouched by the introduction of $T$ and $P_T$ and $\deriv{t}p_x=0$ above (something no change in time parameter chan change--if it is constant, it is constant no matter how time is measured). The equations of motion can be solved:
\begin{equation} x=\frac{p_x^0}{2T}+x_0 = \frac{p_x^0}{p_y^0}y+x_0 \end{equation}
with $p_x^0$, $p_y^0$ and $x_0$ being constants determined by the initial conditions. We know that the dynamics must be equivalent in virtue of the proof given in the previous section, but one can also verify explicitly that the equations in $t$ and $T$ are indeed physically equivalent.

Let us now canonically quantise the two descriptions. We are going to be agnostic about the underlying quantum ontology and simply perform the standard procedure of canonical quantisation of which we provided a discussion in chapter \ref{chap:Introduction}. In terms of $t$ the quantum system is well known from any introductory textbook on quantum mechanics with the exception of inclusion of the Hamiltonian constraint. The quantum Hamiltonian operator is
\begin{equation} \hat H_t = \tfrac12\hat p_x^2 +\tfrac12\hat p_y^2 +V \end{equation}
where the momentum operators take their usual form in the position basis. In order to solve the Hamiltonian eigenequation, consider the ansatz $\psi=\psi_x(x)\psi_y(y)$, which leads to the equation\footnote{For the purposes of this section only, we include the constant $\hbar$ explicitly.}
\begin{equation} -\frac{\hbar^2}{2}\partial_x^2\psi_x = (E_x-V_x)\psi_x \end{equation}
for $x$-eigenfunction $\psi_x$ and $x$-eigenvalue $E_x$, and an analogous equation for $\psi_y$ and $E_y$, with $V_x+V_y=V$. The infinite-well potential manifests in boundary conditions $\psi_x(0)=\psi_x(\pi)=0$, with which the equation gives solutions (labelled by $k\in\mathbb{N}$)
\begin{equation} \psi_x^k = \sqrt{\frac2\pi}\sin(kx).\end{equation}
An anologous results holds for $\psi_y$. For a Hamiltonian eigenstate $\psi^{nm}(x,y)=\pi_x^n(x)\psi_y^m(y)$ the corresponding energy eigenvalue is
\begin{equation} E = \frac{\hbar^2}{2}(n^2+m^2)-U .\end{equation}
This essentially completes the standard textbook analysis. We also consider the presence of the Hamiltonian constraint. Constraints in a classical system manifest in the form of conditions on the physical wavefunction. The quantum Hamiltonian constraint states that physical states satisfy $\hat{H}_t\psi_{phys}=0$, or
\begin{equation} -\frac{\hbar^2}{2}(\partial_x^2+\partial_y^2)\psi_{phys} = U\psi_{phys}, \end{equation}
while states not satisfying this condition are considered mathematical artifacts without physical relevance. Expressed in the Hamiltonian eigenbasis,
\begin{equation} \psi_{phys} = \sum_{n,m}c_{nm}\psi_x^n\psi_y^m, \label{eq:3.2-generalstate}\end{equation}
this condition becomes
\begin{equation} \sum_{n,m} c_{nm}\frac{\hbar^2}{2}(n^2+m^2) = U. \end{equation}
This has the rather peculiar implication that there are physical Hamiltonian eigenstates (corresponding to having only one non-vanishing coefficient $c_{nm}$) only if $U$ is equal to one of a subset of integer multiples of $\hbar^2/2$. The physical meaning of this is somewhat of an open question but since the infinite square well is not a realistic model of our universe this peculiarity may not concern us for our purposes. The quantum constraint is effectively a Wheeler-deWitt-like equation and implies that the dynamics is frozen

We alluded earlier that phenomenological differences arise when considering measurements relating to observables associated with the physical time $T$, that is, the $y$-particle. To see this, consider the expectation value $\langle y^2\rangle$. For a general physical state \ref{eq:3.2-generalstate} this is
\begin{equation} \oppr{\psi}{y^2}{\psi} = \sum_k\sum_{m}|c_{km}|^2\cdot\left(\frac{\pi^2}{3}-\frac{1}{2m^2}\right) 
					  + \sum_k\sum_{l\neq m} c_{kl}^\ast c_{km} (-1)^{l+m}\cdot\frac{8lm}{(l^2-m^2)^2}.
\end{equation}
The term inside the sum in the first part of this expression is what one has for Hamiltonian eigenstates, the second term corresponds to the cross-term contribution only present in non-eigenstates. In general $c_{kl}^\ast c_{km}$ has a time-dependent overall phase factor, so that the second sum itself is time-dependent, while the first is constant. However, suppose that $U$ is such that there exist physical energy eigenstates, that is, $U$ is an integer multiple of $\hbar/2$. Then for such states the second sum vanishes and there exist physically allowed states in which $\langle y^2\rangle$ is time-independent.

Now quantise the classical theory with physical time $T$. We will show that there are no states in this theory such that $\langle y^2\rangle$ is constant in time, illustrating disagreement with the quantised theory in $t$. Here the configuration space of the system is one-dimensional since only $x$ remains as a variable, while $y$ and its momentum have been `absorbed' into the notion of time. Denoting the eigenvalue of the Hamiltonian by $F$, the Hamiltonian eigenequation $\hat{H}_T\psi_x=F\psi_x$ is
\begin{equation}-\frac{1}{2T^2}\left(\tfrac12p_x^2+V\right)\psi_x=F\psi_x. \end{equation}
This is exactly analogous to the equation in the quantisation of the $t$-description except that there is an additional time-dependent pre-factor. Hence the eigenfunctions are the same as $\psi_x$ obtained above,
\begin{equation} \psi_x^n = \sqrt{\frac2\pi}\sin(nx),\qquad n\in\mathbb{N},\end{equation}
(and vanishing outside the wel) and the eigenvalues are
\begin{equation} F_n = -\frac{1}{2T^2}\left(\frac{\hbar^2n^2}{2}-U\right).\end{equation}
The Hamiltonian constraint has been eliminated as part of the Hamiltonian reduction and as a result is identically satisfied as is easy to check.

Above we calculated the expectation value $\langle y^2\rangle$ and showed that at least for some choices of $U$ there are states in which it is constant in time. Here $\langle y^2\rangle$ is, according to our discussion in section \ref{sec:meaningofHamiltonian}, the Hamiltonian expectation value, since the physical meaning of the Hamiltonian was defined as $H_T=-P_T=y^2$. For a general state
\begin{equation}\psi = \sum_n c_n\psi_x^n\end{equation}
we find (using the orthogonality of the eigenstates)
\begin{equation} \oppr{\psi}{y^2}{\psi} = -\frac{1}{2T^2}\left(\sum_n |c_n|^2\frac{\hbar^2 n^2}{2}-U\right),\end{equation}
which is time dependent for \emph{all} states since the part in parentheses is not. This is in manifest disagreement with the range of possible expectation values resulting from the quantisation of the $t$-description.

There remains a question of how this would be measurable in practice given the above model. One might think it possible to simply wait a sufficiently long time such that the expectation values are sufficiently different to be easily distinguishable with good probability from a single measurement, although one is of course limited by the limited applicability of the model to the time frame between classical bounces. Furthermore, a measurement process should be included dynamically, not considered external, since it would have to be part of the Hamiltonian-reduction process. Furthermore, it is arguably impossible to test the expectation value via repeated measurements of systems prepared in the same state since in the reduced formalism there is only one time variable, not one per copy of the system. Physical time is always extracted for the universe as a whole, not multiple times within a universe, and multiple copies of a system are necessarily part of the same universe if one claims to be able to make any statistical inference from them.

These considerations suggest that great care has to be taken on how to actually distinguish between the two quantum theories in practice, in particular in a more realistic model. However, another way one might hope to establish the distinction experimentally with a single measurement is by simply measuring $y^2$ to high precision. For any value of $T$ other than a discrete set of values the possible outcomes of such a measurement do not match the possible values of measurements obtained in the $t$ formalism. Hence one would be able to say with great probability whether or not the measured outcome could possibly be one that is allowed in the $t$-formalism or not. 

Despite these subtleties the above illustration is sufficient to show that one may plausibly expect experimental differences in the respective quantum theories with different physical time parameters. One might be hopeful that such a time parameter might therefore be detectable for our universe if indeed there is one. Considering York-time quantum gravity is then not a purely formal exercise but may ultimately lead to concretely different observable physics, at least in principle.


\chapter{York time}\label{chap:Yorktime}

\textit{In this chapter we introduce York time, arguably the most promising candidate for a physically fundamental notion of time. We explain its role in the initial-value problem of general relativity and discuss some of its other properties and how it would address various questions about the nature of quantum gravity.}

\section{The initial-value problem of general relativity}\label{sec:IVP}

In the last chapter we discovered that different choices of physical time may lead to phenomenologically distinct theories following quantisation. In principle, therefore, one can hope that observation will ultimately constrain the list of temporal candidates, or rule out such an approach to quantum gravity all together. However, a priori there are an infinite number of ways to carry out the time-space split. We must therefore apply theoretical considerations in order to identify viable candidates. Roughly speaking one may divide proposals into three categories: 
\begin{enumerate}
 \item Intrinsic time: the time parameter depends only on the configuration variables within a spatial slice, that is, the metric $g_{ab}$. Furthermore, the parameter ought to be invariant under spatial diffeomorphisms, so that the parameter depends really only on the geometry, that is, an equivalence class of metrics. One simple example in cosmology would be to use a monotonic function of the scale factor $a$ as a time parameter. For example, Misner \citep{Misner1969c} proposed the use of $-\ln a$ to describe cosmological history.
 \item Extrinsic time: the time parameter depends only on geometrical degrees of freedom but this includes the relationship between infinitesimally separated slices, so that it may be a function of $g_{ab}$ and the geometric momenta $\pi^{ab}$. Once again the parameter ought to be invariant under spatial diffeomorphisms, which may vary from slice to slice. York time is an example of an extrinsic time parameter.
 \item Matter time: the time parameter is extracted from the properties of matter. Whether some particular matter field and if so which, or perhaps a combination of all matter should be used to formulate the parameter is entirely open. Some illustrative proposals using dust have been made in refs.~ \citep{BrownKuchar1995} and \citep{HusainPawlowski2012}. 
\end{enumerate}
In principle one could define a time parameter also by mixing geometric and matter degrees of freedom, although it is difficult to see what proposal of this sort would have any theoretical basis. Nonetheless it cannot be ruled out.

With a plethora of choices available and a finite number of hours in a researcher's life, how should one select a candidate parameter for further investigation? The most promising route is to look at the structure of the theory, general relativity, itself. The first point to note is that the physical degrees of freedom of general relativity are not the variables themselves but only equivalence classes thereof. That is, general relativity is generally covariant, meaning (roughly) the equations are identical for different choices of coordinates. This coordinate freedom is not physical and our choice of time, as already indicated above, should not depend on the chosen coordinates. It should not depend, that is, on the metric representation of the geometry.

In addition to the dynamical equations that govern how the spatial geometry changes over time the space of solutions is subject to four constraints, the three momentum constraints \ref{eq:2.3-MomConstraint} and the Hamiltonian constraint \ref{eq:2.3-HConstraint}. One consequence of the existence of such constraints is that the \emph{initial-value problem} of general relativity is non-trivial. The initial-value problem is the task to specify a full set of Cauchy data on a spatial slice such that the evolution is uniquely determined at all times thereafter (and before). In practice this means specifying values for $g_{ab}(x)$ and the geometric momentum $\pi^{ab}(x)$ at every point $x$ on a spatial slice. Doing so uniquely determines their values on all other slices via the evolution equations. This holds true no matter how the spacetime is foliated to the future of the initial slice since $g_{ab}(x,t)$ and $\pi^{ab}(x,t)$ being fixed for one foliation means they are determined for any foliation as such a change corresponds to a mere spacetime coordinate transformation.

For many other theories the initial-value problem is simple. For example, in a system with $n$ particles one simply specifies the momentum and position of each particle. If the total energy is constrained (to zero, for example, as the result of time-reparameterisation invariance) then we can imagine placing $n-1$ particles as we like (depending on the form of the potential function there might be a maximum total energy for those $n-1$ particles that we have to observe) and then the position and/or momentum of the $n$th particle is at least partially determined by the constraint. The future evolution of the system is then fully determined and the initial-value problem is solved.

In general relativity there is a catch, however. There is no obvious straightforward way to specify initial data $g_{ab}(x)$ and $\pi^{ab}(x)$ such that all the constraints are satisfied. General relativity is a field theory with an infinity of degrees of freedom and the constraints are sufficiently non-trivial\footnote{An (artificial) example of sufficiently trivial constraints would be constraints that are super-local, meaning they do not involve spatial derivatives. In this case a finite set of data could be found at each point individually and only continuity from point to point is required.} that no simple mechanical construction is possible.

A beautiful solution to the initial-value problem was found by James York in the early 1970s \citep{York1971, York1972, York1973, ChoquetBruhatYork1980}. Specifically he showed that only the conformal part of the metric and conjugate momentum, meaning the `scale-free' or `shape-only' geometric degrees of freedom describing the anisotropy of space, may be specified, and only on slices of constant extrinsic curvature. The local notion of scale and its rate of change are then determined by the constraints. The details are as follows below. A reader not interested in the technical details may choose to skip to the non-technical discussion at the end of this section, taking note however of equation \ref{eq:4.1-YorkTime}.

Recall that the Hamiltonian and momentum constraints are respectively the vanishing of
\begin{align}
 \curlyH   &= -\tfrac12 \rootg M_{Pl}^2 R + \frac{2}{M_{Pl}^2\rootg}\left(g_{ac}g_{bd}-\tfrac12 g_{ab}g_{cd}\right)\pi^{ab}\pi^{cd} + \curlyH_{matter}, \tag{\ref{eq:2.3-HConstraintExpression}} \\
 \curlyH^a &= -2\nabla_b\pi^{ab} + \curlyH^a_{matter}. \tag{\ref{eq:2.3-MomConstraintExpression}}
\end{align}
Following \citep{York1973}, the first step in specifying a full set of data that satisfies the constraints is to construct data satisfying the momentum constraints only. To do so, specify some arbitrary functions for the metric and momentum, $\bar{g}_{ab}(x)$ and $\bar{\pi}^{ab}(x)$. From these one can construct uniquely a set of data that satisfies the momentum constraints by decomposing it into pure-trace, longitudinal and transverse-traceless parts at each spatial point. Consider for the sake of generality a tensor $\psi^{ab}$ and write
\begin{equation} \psi^{ab} = \psi^{ab}_{TT} + \psi^{ab}_L + \psi^{ab}_{Tr}. \label{eq:4.1-TTLTrdecomposition}\end{equation}
The trace contribution is determined by single degree of freedom $\psi$, 
\begin{equation} \psi^{ab}_{Tr} = \tfrac13\psi g^{ab},\qquad \psi=g_{cd}\psi^{cd} \label{eq:4.1-tensortracepart}\end{equation}
while the longitudinal contribution may be expressed in terms of a vector $W^a$,
\begin{equation} \psi^{ab}_L \equiv (LW)^{ab}\equiv \nabla^aW^b + \nabla^bW^a-\tfrac23 g^{ab}\nabla_cW^c, \label{eq:4.1-tensorlongpart}\end{equation}
where indices are raised and lowered with the spatial metric $g_{ab}$ and its inverse $g^{ab}$ as usual, and $\nabla_a$ denotes the spatial covariant derivative associated with the given metric. The transverse-traceless part is defined as what is leftover to satisfy equation \ref{eq:4.1-TTLTrdecomposition}, although the fact that $\psi_{TT}^{ab}$ really is transverse and traceless is still to be shown. We assume throughout that all tensor quantities are $C^\infty$. 

It is easy to show that $\psi^{ab}_{TT}$ is indeed traceless since $Tr(\psi)=Tr(\psi_{Tr})$ from eq.\ \ref{eq:4.1-tensortracepart} and $Tr(\psi_L)\equiv0$ for any $W^a$ in virtue of its definition \ref{eq:4.1-tensorlongpart}. The non-trivial part of the construction is to choose $W^a$ such that $\psi_{TT}$ is indeed transverse, 
\begin{equation} \nabla_b\psi_{TT}^{ab} = 0. \label{eq:4.1-transverse} \end{equation}
This effectively constitutes the requirement
\begin{equation} \nabla_b(LW)^{ab} = -\nabla_b\psi_{TF}^{ab}, \end{equation}
where $\psi_{TF}^{ab}\equiv\psi^{ab}-\tfrac13\psi g^{ab}$ denotes the tracefree part of $\psi^{ab}$. Introducing the operator $D$ defined by
\begin{equation} (DW)^a = -\nabla_b(LW)^{ab} \end{equation}
one can show that $D$ is linear, second-order, positive-definite and Hermitian. The first two properties are easily seen by inspection, while the latter two may be shown by integrating by parts a global scalar product over the manifold, giving
\begin{equation} (W,DW) \equiv \int W_a (DW)^a = \tfrac12\int (LW)_{ab}(LW)^{ab} \equiv \tfrac12 (LW,LW) \geq0, \end{equation}
and for any two vectors $V^a$, $W^b$ one can show by integrating by parts twice that
\begin{equation} (DV,W) = (V,DW) \end{equation}
illustrating Hermiticity. The case $LW=0$ is a special case, implying that either $W=0$ or $W$ is a conformal Killing vector field. In either case they do not affect the possible solutions for $W^a$ (details are found in ref.~\citep{York1973}). These properties of $D$ imply that solutions exist and can be found by considering eigenfunctions of $D$. The solutions are, formally,
\begin{equation} W^a = \left(D^{-1}(-\nabla\cdot\psi_{TF})\right)^a, \end{equation}
where the inverse $D^{-1}$ of $D$ exists in virtue of the same properties, although a solution $W^a$ is unique only up to conformal Killing vector fields. Thus, strictly speaking, one obtains an equivalence class of vectors $W^a$ for any one initially selected tensor $\psi^{ab}$. The tensor $(LW)^{ab}$, which enters the definition of $\psi_{TT}^{ab}$ is however unique for any one such equivalence class. Having solved for $W^a$, the transverse traceless part $\psi_{TT}^{ab}$ now has those properties provably by construction. 

Applying this method to the initially chosen momentum data $\bar{\pi}^{ab}$ helps one to find the part $\pi^{ab}$ of $\bar{\pi}^{ab}$ that satisfies the momentum constraint with respect to the metric $\bar{g}_{ab}$, as the momentum constraint is effectively a requirement of transverseness. 

Suppose then we have a set of data $(g_{ab}(x),\pi^{ab}(x))$ satisfying the momentum constraints but not yet the Hamiltonian constraint. Consider a conformal transformation
\begin{equation} g_{ab}\rightarrow\tilde{g}_{ab}=\phi^4g_{ab},\qquad \pi^{ab}\rightarrow\tilde{\pi}^{ab} =\phi^{-4}\pi^{ab}, \label{eq:4.1-conftrans}\end{equation}
where $\phi(x)$ is an arbitrary smooth function on the slice. The momentum constraint is invariant under such a transformation, that is, it is satisfied by the data $(\tilde{g}_{ab},\tilde{\pi}^{ab})$ if it is satisfied by $(g_{ab},\pi^{ab})$, if the slice is \emph{maximal}. A slice is maximal if and only if the trace of the momentum vanishes everywhere, $\pi\equiv Tr(\pi^{ab})\equiv g_{ab}\pi^{ab} = 0$.

Assume then that the data is indeed given for a maximal slice. The momentum constraints are assumed to be satisfied but the Hamiltonian constraint, in general, is not. The goal is to perform a conformal transformation \ref{eq:4.1-conftrans} such that the transformed data $(\tilde{g}_{ab},\tilde{\pi}^{ab})$ does satisfy the Hamiltonian constraint. If this can be done, then the transformed data would indeed satisfy all constraints and the initial-value problem would be solved, at least for data on maximal slices. Substitute therefore the transformed data in the form $(\phi^4g_{ab},\phi^{-4}\pi^{ab})$ into the Hamiltonian constraint, which can now be understood as an equation determining the correct choice of $\phi$ in terms of $g_{ab}$ and $\pi^{ab}$, namely
\begin{equation} 8\nabla^2\phi+M_{Pl}^{-2}g^{-1}\pi_{ab}\pi_{ab}\phi^{-7}-M_{Pl}^2R\phi=0. \label{eq:4.1-Lichnerowicz}\end{equation}
This equation is known as the Lichnerowicz equation. Having solved this equation for $\phi$ one has explicit data satisfying both the momentum and Hamiltonian constraints.

The method described in the last two paragraphs assumes that the data is given on a maximal slice. This restriction may be weakened to slices of constant mean curvature (constant scalar extrinsic curvature), that is, slices on which the parameter
\begin{equation} T\equiv \frac{2\pi}{3\sqrt{g}} \label{eq:4.1-YorkTime}\end{equation}
is constant ($\pi$ here still denotes the trace of $\pi^{ab}$, not the constant relating geometric properties of circles). This requires an extra step, the replacement of $\pi^{ab}$ by its traceless part
\begin{equation} \sigma^{ab}\equiv \pi^{ab}-\tfrac13\pi g^{ab}.\end{equation}
Since $\nabla_ag^{ab}=0$, one has that if $\pi^{ab}$ satisfies the momentum constraint ($\nabla_a\pi^{ab}=0$), then so does $\sigma^{ab}$ ($\nabla_a\sigma^{ab}=0$). The crucial aspect of a slice of constant mean curvature is that the decomposition \ref{eq:4.1-TTLTrdecomposition} is invariant under conformal transformations.\footnote{The distinction between maximal and constant-mean-curvature slices is essentially that in the former case the trace part $\pi^{ab}_{Tr}$ as defined in \ref{eq:4.1-TTLTrdecomposition} vanishes, while in the latter case the part has to be subtracted `manually' to arrive at the transverse traceless part.} One proceeds as for the case of maximal slicing by substituting the transformed data $(\tilde{g}_{ab}, \tilde{\sigma}^{ab})$ into the Hamiltonian, which now yields the \emph{extended Lichnerowicz equation},
\begin{equation} 8\nabla^2\phi+M_{Pl}^{-2}g^{-1}\pi_{ab}\pi_{ab}\phi^{-7}-M_{Pl}^2R\phi -\tfrac38 M_{Pl}^{-2}T^2\phi^5 =0. \label{eq:4.1-extendedLichnerowicz} \end{equation}
For $T=0$ one recovers the (unextended) Lichnerowicz equation \ref{eq:4.1-Lichnerowicz} as expected.

There are two key points that follow from this solution to the initial-value problem. First, the degrees of freedom that may be freely specified are the \emph{conformal} geometric degrees of freedom, that is, the variables determining local `shape' (or angles together with anisotropy), and not the local absolute scale. Second, the initial data can only be given on a a slice of constant scalar extrinsic curvature, or equivalently of constant $T$.

If we take the idea of a theory of gravity described by three-dimensional space whose geometry evolves through time (rather than the four-covariant `spacetime' picture) seriously, then the latter fact strongly suggests that slices of constant $T$ are slices of constant time, so that the foliation on which the initial-value problem can be solved is indeed the foliation that corresponds to stacking of spaces at consecutive instances. For if physical time corresponded to a different time variable, that is, if the reconstruction of spacetime from the space-through-time theory were not a reconstruction from a constant-mean-curvature foliation, then as a consequence initial data could not be specified at a single instance in time. This would pose a major conundrum for the notion of what determines the dynamics of a physical system.

Accepting that the correct physical foliation is one of constant $T$, two question remain. First, is $T$ itself a viable notion of time? In other words, is it monotonic? And is the constant-mean-curvature slicing unique? 

We will consider the question of uniqueness in the next section. Regarding the first of these, note first that there are solutions to the equations of general relativity that do not permit a global constant-mean-curvature slicing. One example is any solution with a closed timelike loop. Such solutions of general relativity would therefore not constitute solutions to our theory of gravity described by space evolving through time. This inequivalence is likely unproblematic since such solutions of general relativity do not as far as we know correspond to our actual universe and the theories are therefore phenomenologically equivalent given what we already know about our universe. For solutions of general relativity that do allow a constant-mean-curvature slicing (such as universes with closed or asymptotically flat cosmologies) one can indeed show as a result of the field equations that $T$ is non-decreasing.\footnote{This is in general true provided a sufficiently well-behaved matter content is assumed.}

This leaves only the possibility that $T$ is constant across multiple slices. One example of such a scenario is de~Sitter space, where homogeneous space expands at a constant fractional rate. De~Sitter space often serves as a first approximation for the behaviour of the universe during inflation, which we will treat in detail in section \ref{sec:inflation}. Fortunately there de~Sitter is only an approximation and a higher-order calculation shows that $T$ is, in fact, increasing during this cosmological period, albeit slowly (with respect to conventional `cosmological' time). A true de~Sitter space would pose issues for considering $T$ as a time variable. However, once again we know that our universe is not described by a pure de~Sitter space, so that the viability of the York parameter $T$ is as a time parameter is not threatened. 

The second insight gained from the initial-value problem was that `scale' (and its conjugate momentum) are not freely specifiable but are determined by the Hamiltonian constraint. This suggests that these variables do not have the same physical status as the other, shape-determining variables. Indeed, the (fractional) rate of change of scale is just the value of the York parameter on the slice, or the answer to the question `what time is it?' The scale itself (expressed as the local volume element $\sqrt{g}$) is the momentum conjugate to the time $T$.

Considering $T$ as our physical time parameter implies then that $\sqrt{g}$ is equal to the physical Hamiltonian density, as discussed in section \ref{sec:meaningofHamiltonian}. The status of volume is therefore analogous to that of `energy' in conventional theories; not a physical variable (although arguably not void of physical meaning) and given by the numerical value of the Hamiltonian. Exploring this idea will constitute the majority of the remainder of this thesis.

In our argument there is one caveat. While it is true that York's solution to the initial-value problem gives reasons to take the proposal of York time as physical time seriously, it is unproven that the initial-value problem can indeed not be solved in another manner, in particular with another foliation. It is the case that no other solutions are known\footnote{although there were attempts along different lines preceding York, for example due to Baierlein, Sharp and Wheeler \citep{BaierleinSharpWheeler1962}}, but at least the possibility of other solutions has not been ruled out.

York time and constant-mean-curvature slicing do however also have a number of other properties that support the plausibility of their fundamental physical significance. This will be the subject of the next section.


\section{Properties of York time}\label{sec:YTproperties}

The York parameter $T$ plays a central role in the initial-value problem in that it is to our knowledge only surfaces of constant $T$ on which the problem can be solved. The notion of local scale and its rate of change is `extracted' from among the physical variables. Taking the implication of this seriously, we argued, means to consider the York parameter as the true physical time and its conjugate momentum as the Hamilton density, whose physical interpretation is that of local volume.

York time however does have other properties as the result of which it stands out from among the candidates for physical time. Suppose we were unaware of York's work on the initial-value problem and we wished to identify a physical time based on a few simple theoretical considerations. First, we wish to extract it from the geometric degrees of freedom only (no matter-dependent definitions) and furthermore we believe that it should only depend on the \emph{local} degrees of freedom, that is, `what time is it?' should be answerable without reference to the properties of other, distant points in space. The latter criterion enables subsequently the identification of a local Hamiltonian density. So physical time and the Hamiltonian density at $x$ are functions of $g_{ab}(x)$ and $\pi^{ab}(x)$ only. 

Further, it is reasonable to suppose that the time parameter be identified as an isotropic quantity. There is nothing that leads us to believe that there is any sort of fundamental anisotropy in the laws of the universe. This then implies that any of the `shape' degrees of freedom, which determine the notion of angles, are not to be used. Only the overall scale and its rate of change (or conjugate momentum), that is, the metric determinant $g=\det(g_{ab})$ and the momentum trace $\pi=g_{ab}\pi^{ab}$, remain.

This still leaves various options. For example, one might consider the local scale $\sqrt{g}$ to be time and its conjugate momentum to be the Hamilton, essentially a reversal of the roles we propose. The issue here is however that in general $\sqrt{g}$ does not change monotonically according to the laws of general relativity. Disallowing such solutions would be far more restricting than disallowing spacetimes that cannot be foliated in the manner of constant mean curvature, although in cosmological contexts (assuming a homogeneous universe) such a time parameter may be viable and has been considered in the literature (by Misner, for example, \citep{Misner1969c}). The requirement of monotonicity rules out the vast majority of possible functions of $g$ and $\pi$.

I cannot say with certainty that only constant mean curvature slicing guarantees monotonicity, although I am unaware of any other, non-equivalent foliation of the sort. However, the simplicity of the interpretation of the Hamiltonian is uniquely compelling. So, even if there are other options, York time must be considered a favourite among them based only on a few theoretical principles.

In the last section we also left unanswered the question whether the constant-mean-curvature slicing of a spacetime that forms a solution of the field equations of general relativity is unique. We already discussed that not all such spacetimes are, in fact, constant-mean-curvature slicable and those that are not would not be solutions of our theory of York-time geometrodynamics. However, if the spacetime is slicable by a constant-mean-curvature foliation, can there be more than one such foliation?

The last question has been answered in the negative by Marsden and Tipler \citep{MarsdenTipler1980}. They established that the constant-mean-curvature slicing of a spacetime, if indeed there is one, is unique, at least for a subset of solutions of general relativity. Specifically, these are the cases of \emph{closed} universes, that is, those that, roughly speaking, experience a `big crunch' after some finite proper time. However, their argument also applies to universes which are asymptotically flat, that is, where space in the infinite distance approaches a Euclidean geometry. 

\begin{wrapfigure}{R}{0.4\linewidth} \label{fig:QadirWheeler}
\includegraphics[width=\linewidth]{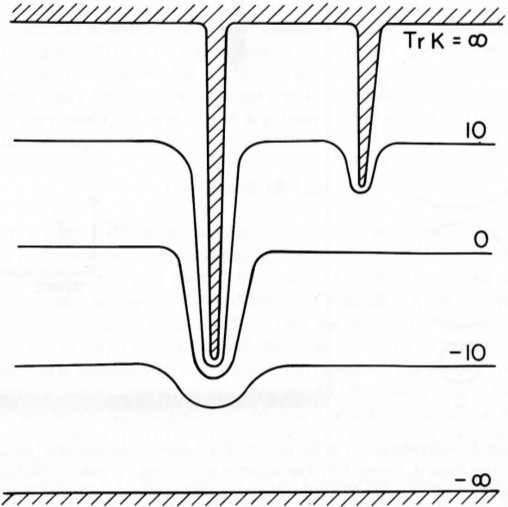}
\caption[Surfaces of constant York time around singularities]{Surfaces of finite constant $T$ wrap around singularities, which are all part of a single final singularity at $T=\infty$. The image is taken from ref.\ \citep{QadirWheeler1985}.}
\end{wrapfigure}

Another encouraging aspect of constant-mean-curvature slicing is the shape such a foliation takes in the presence of singularities. In the 1980s Qadir and Wheeler \citep{QadirWheeler1985} discovered that the scalar extrinsic curvature $K$, or equivalently $-T$, approaches infinity as the singularity is approached. This applies equally to a final singularity that is the `big crunch' in universes that have such a singularity and to local singularities such as black holes. As a result slices of constant $T$ `wrap around' a singularity and all black holes are part of the single, final singularity found at $T=\infty$.\footnote{As in all parts of this thesis, a slice is itself a three-dimensional object with non-zero volume, and `wrap around' is to be understood as a three-dimensional (hyper)slice `wrapping' in the fourth dimension.} Consequently there are no geometric singularities on any slice labelled by a finite $T$.\footnote{However: In a universe that expands forever (and $T\rightarrow0$ as $t\rightarrow\infty$ in the homogeneous approximation), what is the shape of the York foliation then? In the vicinity of a singularity the cosmological properties are irrelevant and $T\rightarrow\infty$ but in other parts of the universe $T$ ends at zero. This apparent contradiction motivates us to propose a cosmological extension for such universes. This we discuss in detail in chapter \ref{chap:cosmext}.}

This completely preempts a difficult conceptual issue that any theory of quantum gravity has to address. The issue concerns the existence of singularities. The first question is whether or not singularities, which are clearly solutions of the classical theory, are also part of the quantum theory. If yes, then what is the quantum description of such singularity? If not, then how do singularities arise as classical approximations? I am not aware that any approach of quantum gravity has answered these questions fully and with certainty.

The York slicing avoids classical singularities entirely since they are `banished' to $T=\pm\infty$. Hence the quantum geometrodynamics, the quantum theory of spatial geometries, can be defined on those slices without concern regarding the recovery of classical singularities in any limit.

{
\begin{mdframed}
 \setlength{\parindent}{10pt}

\small
\bigskip
\textbf{Proper time}\bigskip

The attentive reader will have noticed that the notion of constant extrinsic curvature and York time was introduced with reference to another, more familiar notion: proper time. In particular, the physical understanding of the scalar extrinsic curvature is the local fractional rate of growth of volume, where `rate of growth' refers to growth with respect to proper time. Does this reliance on another notion of time, proper time, spell doom to our idea to consider York time a physically fundamental time parameter? What about proper time as a physically fundamental time parameter? And what is this proper time anyway?

Our endeavour is not doomed. Referring to proper time provided a physical intuition for the York parameter but in its formal definition in terms of the geometric momenta $\pi^{ij}$ and the metric $g_{ij}$ no such reference was made. The expression for the momenta, when traced back to the original action principle, involved the lapse $N$, whose own physical interpretation is the amount of proper time passing per coordinate time, but recall that mathematically $N$ merely encodes the freedom to choose any monotonic time parameter. Its physical meaning is not 
relevant for this, nor for the procedure of Hamiltonian reduction, which will be used to derive the dynamical equations. The notion of York time does not, in fact, rely on the idea of proper time at all. 

In fact, if we take the view that our theory of gravity is fundamentally one of the time evolution of a spatial geometry and not just a `3+1' description of a space\emph{time} geometry, then the lapse $N$ and proper time $\tau$ were never more than a crutch that allowed us to come up with a theory that is phenomenologically equivalent with general relativity anyway.

Proper time itself does not constitute a viable fundamental time parameter. For one, it does not actually define a single preferred foliation. At any point on a slice chosen in any desirably way one can define the notion of proper time (of a comoving observer), but `what proper time it is' is not a notion that can be answered in an absolute manner. A `slice of constant proper time' is not a meaningful notion unless there already is another such slice that is assumed to be such and relative to which absolute values of proper time can be assigned. That is, a constant-proper-time foliation is determined by choosing a single slice, but how to choose that first slice is not unique. Proper time further lacks a relation to what constitutes, arguably, the physical variables given the initial-value problem.

Another issue proper time faces is the criticism made by Valentini \citep{Valentini1996} that we discussed above.

Nonetheless, it is difficult to deny that the notion of proper time must have some direct physical significance. After all, it is exactly the time parameter that is measured by a clock,
\begin{equation*} \Delta t_{clock} = \int_{path} d\tau = \int_{path} \sqrt{dt^2-g_{ij}dx^idx^j}. \end{equation*}
A clock may be considered a `hodometer of spacetime' \citep{Brown2005}, measuring the Lorentzian length of the spacetime path taken akin to the mileage counter of an automobile. Clocks are physical systems governed by dynamical equations depending on the exact constituents of the clock. It is a remarkable conspiracy that all clocks --- no matter whether consisting of pendula, vibrating atoms or systems of orbiting bodies --- `tick' at the same rate. There are two views one can take: either this conspiracy shows that there is an underlying spacetime structure that the clocks `sense'; or it is exactly this conspiracy that allows a \emph{description} in terms of spacetime, but the conspiracy is purely a matter of matter dynamics.

The first view is the one commonly presented in the majority of textbooks on relativity, although too often clocks (and rods) are simply considered as primitive entities that --- almost by definition --- display the spacetime structure. Opening many a textbook one finds diagrams of coordinate grids with clocks drawn at each intersection, or similar. This apparently compelling view is called into question however when we try to understand the details by which actual clocks sense this spacetime structure. We are forced to consider the interaction between dynamics and spacetime, but this then is just a rephrasing of the second view. The notion of spacetime can be invoked because the universality of the dynamics allows for a geometric description but the clock conspiracy is ultimately a question of dynamics. 

This universality is essentially due to the way the gravitational field represented by the spacetime metric $\supfour g_{\mu\nu}$ is coupled to the matter fields via \emph{minimal coupling}. (Co-)vectorial quantities (such as spatial derivatives of fields) are summed, for example in the kinetic terms of a scalar field, via contraction of the indices by the metric. The spatial derivatives are furthermore modified to be covariant, giving them the right transformation properties that the gravitational field may be given a geometric interpretation.

The reliance on notions that ultimately require dynamical treatment, such as rods and clocks, as primitive is as old as relativity itself. Early in Einstein's celebrated 1905 paper \emph{Zur Elektrodynamik bewegter K\"orper} \citep{Einstein1905_Elektrodynamik} he writes,
\begin{quote}
 ``Like all electrodynamics, the theory to be developed here is based on the kinematics of a rigid body, since the assertions of any such theory have to do with the relations among rigid bodies (coordinate systems), clocks, and electromagnetic processes.''
\end{quote}
Few authors\footnote{Many years later Einstein himself however became uneasy with his approach to special relativity as is evident from his autobiographical notes, in which he calls the primitive treatment of rods and clocks ``inconsistent'' \citep{EinsteinAutobiographicalNotes}.} have encouraged a dynamical understanding of relativistic effects such as clock dilation and length contraction without relying on spacetime geometry as a primitive explanation. One notable exception is Bell, who provides a beautiful discussion of the deformation of a moving atom in terms of electrodynamics in ``How to teach special relativity'', published as something of an oddity among his collection of papers on quantum mechanics \citep{Bell1987}. The most careful discussion of the principles and assumption of special relativity of which I am aware is due to Brown \citep{Brown2005}, who criticises the operationalist viewpoint of rods and clocks exemplified by the above passage. Here we cannot do the subject justice. Our point is merely that proper time, which can be understood as a geometric notion in the context of spacetime, cannot be separated from the dynamics of physical systems.

To consider spacetime conceptually prior to the dynamics of matter would, one may argue, be putting the cart before the horse.\footnote{This statement is intentionally a reference to a claim by Balashov and Janssen \citep{BalashovJanssen2003}, who defend exactly the opposite view, although in the context of Minkowski's non-dynamical geometry. The paper is heavily criticised by Brown, who argues that ``[t]he appropriate structure [to define a kinematics for mechanics] is Minkowksi geometry \emph{precisely because} the laws of physics of the non-gravitational interactions are Lorentz covariant'' (original italics) \citep[][p.\ 133]{Brown2005}. } Proper time is intrinsically dynamical, not an aspect of spatial or spatio-temporal structure, and is therefore not obviously a viable choice of fundamental time parameter.

\end{mdframed}
}

\section{The reduced variables and their Poisson structure}\label{sec:Poisson}

Hamiltonian reduction is a mathematical procedure that reduces the number of config\-uration-space degrees of freedom by one and the number of phase-space degrees of freedom by two (one `position' and one `momentum'). If the variables extracted to serve as time and as the Hamiltonian (or Hamiltonian density) are simply one pair of canonical variables chosen from among the full set of canonical pairs, then the Poisson structure of the set of remaining variables is unchanged. On the other hand, if the notion of physical time and Hamiltonian are definable only in terms of multiple variables and there is no canonical transformation that avoids this, then the number of reduced variables is \emph{not} smaller than the number of original degrees of freedom. Instead the reduced variables must satisfy a new pair of constraints, so that the number of degrees of freedom is nonetheless reduced. That these constraints remain satisfied with time irrespective of the particular form of the Hamiltonian is however guaranteed by modified Poisson brackets. The new Poisson structure is obtained directly from the Poisson structure of the original variables and some basic properties of Poisson brackets.

The content of the preceding paragraph, describing abstractly how a non-canonical Poisson structure arises, may appear obscure. A simple example helps to clarify this mathematical process.

Consider a system described by three pairs of canonical variables $Q_i,P^i$, $i=1,2,3$. These may describe the position and momentum of three particles confined to move in one dimension, for example. Let its (unreduced) Hamiltonian be the usual one (setting the particle mass to $m=1$), 
\begin{equation} H_t=\tfrac12\sum_i P^{i2} + V(Q_1,Q_2,Q_3), \end{equation}
where the subscript $t$ denotes the fact that this Hamiltonian describes evolution with respect to $t$. We demand time-reparameterisation invariance (see chapter \ref{chap:problemoftime}), so that the system obeys a Hamiltonian constraint,
\begin{equation} H_t = 0.\end{equation}
Since the set of variables $\{Q_1,Q_2,Q_3,,P^1,P^2,P^3\}$ is by hypothesis canonical and therefore obeys the usual Poisson structure,
\begin{equation} \{Q_i,P^j\} = \delta_i^j, \end{equation}
with all other Poisson brackets vanishing. 

Define for convenience the quantities
\begin{equation} G \equiv Q_1Q_2Q_3,\qquad \Pi = \sum_i Q_iP^i,\end{equation}
somewhat resembling the metric determinant $g$ and the trace of the geometric momentum, $\pi=g_{ij}\pi^{ij}$ respectively. In fact, the model used here to illustrate the appearance of non-canonical Poisson brackets describes a homogeneous but anisotropic spatially flat universe and will be developed in full in section \ref{sec:ClassKasner}. We choose a physical time with respect to which to perform a Hamiltonian reduction, namely 
\begin{equation} T\equiv \frac{2\Pi}{3\sqrt{G}},\qquad P_T \equiv -\sqrt{G}.\end{equation}
The choice is obviously analogous to York time and its conjugate momentum. However, for our present purposes (to derive the Poisson structure of the reduced variables) the exact form does not matter, only that they are functions of the `scale' variables: $T=T(G,\Pi)$, $P_T=P_T(G,\Pi)$. 

There is no canonical transformation from the set of variables $\{Q_1,Q_2,Q_3,P_1,P_2,P_3\}$ to another set $\{G,\tilde{Q}_1,\tilde{Q}_2,\Pi,\tilde{P_1},\tilde{P_2}\}$ such that the latter includes $G$ and $\Pi$. This implies that the set of reduced variables is not just an unreduced set of canonical variables minus one canonical pair. Instead it is another set of six variables from which the `scale' is extracted via constraints. The most obvious way to do this is to define the reduced variables as
\begin{equation} q_i \equiv G^{-\frac13}Q_i,\qquad p^i\equiv G^\frac13(P^i-\tfrac13\Pi/Q_i). \label{eq:4.3-modelreducedvars}\end{equation}
The new variables configuration variables $q_i$ obey the constraint
\begin{equation} q_1q_2q_3 = 1 \label{eq:4.3-modelconstraint1}\end{equation}
identically in virtue of their definition. However, if we `kick away the ladder' and treat the variables $q_i$ as fundamental, that is, if we `forget' how we arrived at the reduced description, then the constraint is a genuine one. The variables also obey
\begin{equation} \sum_i q_ip^i =0,\end{equation}
which ensures that the first constraint remains satisfied with time evolution irrespective of the exact form of the Hamiltonian. The constraint also holds identically in virtue of the definition of the momenta $p^i$. In fact, the form of the momenta \ref{eq:4.3-modelreducedvars} is exactly such that any motion generated by the momenta is tangent to the constraint surface defined by eq.\ \ref{eq:4.3-modelconstraint1}. Other definitions, such as $\tilde{p}^i\equiv G^\frac13P^i$, would not have this property. We will discuss this in more detail in section \ref{sec:ClassKasner}. 

For now we are only interested in the derivation of the Poisson structure for the set~$\{q_1,q_2,q_3,p^1,p^2,p^3\}$. The procedure to find the new Poisson structure (such that the reduced dynamics is physically equivalent with the unreduced dynamics) is as follows: First write the Poisson bracket to be determined in terms of the original variables. Second, use the original Poisson structure together with general properties of Poisson brackets to evaluate the selected bracket in terms of the old variables. Rewrite the result in terms of the new reduced variables. This ensures consistency of the new Poisson structure with the old. 

For illustrative purposes we wish to determine $\{q_i,p^j\}$. Following the procedure just described we write,
\begin{equation} \{q_i,p^j\} = \left\{ (Q_1Q_2Q_3)^{-\frac13}Q_i\;,\;(Q_1Q_2Q_3)^{\frac13}\left(P^j-\tfrac13\frac{\sum_k Q_kP^k}{Q_j}\right) \right\}. \end{equation}
Now use the property of Poisson brackets that for a function $F(Q_1,Q_2,Q_3)$ it is true that
\begin{equation} \{F,P^j\} = \sum_k\PD{F}{Q_k}\{Q_k,P^j\}.\end{equation}
This follows from distributivity of the Poisson brackets and the for our purposes reasonable assumption that $F$ is sufficiently well behaved to have a series expansion. Evaluating the Poisson bracket in this manner gives (after rewriting the right-hand side in terms of the new reduced variables):
\begin{equation} \{q_i,p^j\} = \delta_i^j-\frac{q_i}{3q_j}.\label{eq:4.3-modelqpPB}\end{equation}
Proceeding analogously with the other Poisson brackets, one finds
\begin{align}
 \{q_i,q_j\} &= 0 \label{eq:4.3-modelqqPB}\\
 \{p^i,p^j\} &= \frac{p^i}{3q_j}-\frac{p^j}{3q_i}. \label{eq:4.3-modelppPB}
\end{align}
The existence of the second term in the expression \ref{eq:4.3-modelqpPB} and the non-vanishing of the right-hand side of eq.~\ref{eq:4.3-modelppPB} are the reason the new reduced variables are not canonical. We will return to this model, which is a description of the geometrodynamics of a homogeneous but anisotropic universe, in chapter~\ref{chap:perttheoprelims} before proceeding to consider the analogous Poisson brackets of the perturbation theory.

In contrast with this finite-dimensional model, the full theory of York-time geometrodynamics is a field theory. The theory is, however, local, so that Poisson brackets vanish if compared between two distinct spatial points. Otherwise the Poisson brackets one obtains (proceeding in the same manner as we did above) are very similar:
\begin{align}
 \{\tilde{g}_{ab}(x),\tilde{g}_{cd}(y)\}     &= 0 \label{eq:4.3-ggPB}\\
 \{\tilde{g}_{ab}(x),\tilde{\pi}^{cd}(y)\}   &= \left(\delta_a^{(c}\delta_b^{d)}-\tfrac13\tilde{g}_{ab}(x)\tilde{g}^{cd}(x)\right)\delta^3(x-y) \label{eq:4.3-gpiPB}\\
 \{\tilde{\pi}^{ab}(x),\tilde{\pi}^{cd}(y)\} &= 
	  \frac13\Big(\tilde{g}^{cd}(x)\tilde{\pi}^{ab}(x)-\tilde{g}^{ab}(x)\tilde{\pi}^{cd}(x)\Big)\delta^3(x-y). \label{eq:4.3-pipiPB}
\end{align}
Here $\tilde{g}_{ab}$ and $\tilde{\pi}^{ab}$ denote the reduced variables and $x$ and $y$ denote points on the spatial slice. The argument of $\tilde{g}_{ab}$ and $\tilde{\pi}^{ab}$ on the right-hand sides has been written as `$x$', although we might as well have written `$y$' since the expressions are non-vanishing only if the two selected spatial points are one and the same. 

If we consider the universe to be homogeneous and additionally restrict the metric to have a form such that off-diagonal elements vanish, the expressions \ref{eq:4.3-ggPB} - \ref{eq:4.3-pipiPB} reduce to expressions \ref{eq:4.3-modelqpPB} - \ref{eq:4.3-modelppPB}.

In section \ref{sec:ClassKasner} we discuss the meaning and consequences of the non-canonical Poisson structure further, and its implication for quantisation in section \ref{sec:quantKasner}.

\section{Cosmological history with York time}\label{sec:cosmhist}

A conventional description of cosmological history in terms of cosmological time $t$ may be criticised for a variety of reasons: The universe began at some finite time $t_0$ whose value is essentially arbitrary. In other words, only intervals $\Delta t$ matter. However, at the same time the universe might not end at a finite $t$, so that the cosmological timeline is semi-infinite. Furthermore, different eras of interest vary hugely (by dozens of orders of magnitude) in their duration. In particular, much of the early history of the cosmos took place within the first $10^{-33}$ (inflation) or even $10^{-43}$ (Planck era) of a second.\footnote{Changing units to, say, Planck units leads to absurdly large numbers for later eras instead.}

There is nothing intrinsically wrong with the description in $t$. One might, however, consider it lacking in an aesthetic sense. Cosmological time $t$ is a highly unnatural choice of time parameter when discussing the very early universe.

In 1969 Misner proposed that the quantity $-\ln a$ --- or (more or less equivalently) the logarithm of the temperature in the homogeneous approximation --- would provide a time parameter with which to give an account of cosmic history that is more adequate than the conventional cosmological time $t$ \citep{Misner1969c}. The new parameter would avoid the absurdly small numbers needed to describe early epochs. Misner wrote: `The universe is meaningfully infinitely old, since infinitely many things have happened since the beginning.' 

In terms of York time it is the early history that takes up the longest period $\Delta T$. In fact, just as in the case of Misner's parameter $-\ln a$, the `beginning' lies in the infinite past and unlike in the conventional description in terms of $t$ there is no notion of `before the Big Bang'. The Planck era stretches from $T=-\infty$ to just before the onset of inflation. Recent data \citep{Planck2015_ConstraintsOnInflation} leads to an upper bound on the Hubble parameter during inflation of $3.6\times10^{-5}M_{Pl}$, so that in reduced Planck units (where $M_{Pl}=1$) this number also gives a rough estimate of $T$. 

Following inflation and reheating, the energy density then becomes dominated by radiation until $t\sim10^4$ years. At this point matter becomes dominant and the Hubble parameter is of order $10^{-12}$s$^{-1}$, or $10^{-53}$ in reduced Planck units, which is already within the final moments of the universe when considered in terms of York time. Evidence suggests that today, at $T\sim H\sim 10^{-61}$, we are nearing a possibly final era dominated by a cosmological constant.

The York time description clearly does not do away with absurdly large order-of-magnitude differences in the way Misner might have envisioned. Rather, periods of large $T$-duration are of short $t$-duration and vice versa. The infinite `York age' of the universe might however be considered an aesthetic advantage.

One might speculate whether York time really `ends' at $T=0$ or whether cosmic history continues and our description ought to be extended. Indeed, if $T$ does have fundamental physical significance, then it is unwarranted to conclude that the universe would end then simply because the usual parameter $t$ ceases to describe such speculative future eras, although problems may arise if physical quantities are not well behaved during the transition to $T>0$. The existence of a cosmological constant also raises questions. In such a case the conventionally considered timeline never reaches $T=0$ but stops slightly short since $H(t)\rightarrow\Lambda$ (up to constants) as $t\rightarrow\infty$ rather than approaching zero. An analysis concerning the viability or indeed necessity of such an extension will be performed in chapter \ref{chap:cosmext}.

\begin{figure}
  \includegraphics[width=\textwidth]{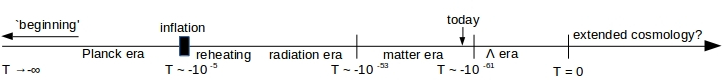}
  \caption[The cosmological timeline with York time]{The cosmological timeline with York time. The entire known cosmological history takes place within $10^{-5}$ from the end of the universe. If there is a cosmological constant then cosmological history ends just short of $T=0$. The figure is taken from ref.~\citep{RoserValentini2016} by the author.}
\end{figure}

\section{York time as a solution to the problem of time}\label{sec:YTandIsham}

In an important review on the subject of the problem of time and approaches to overcome it, Isham \citep{Isham1992} identifies the key questions that any theory of quantum gravity must answer. The points raised are (verbatim):
\begin{enumerate}
 \item the status of the concept of probability and the extent to which it is conserved;
 \item the status of the associated concepts of causality and unitarity;
 \item the time-honoured debate about whether quantum gravity should be approached via a canonical, or a covariant, quantisation scheme;
 \item the extent to which spacetime is a meaningful concept;
 \item the extent to which classical geometrical concepts can, or should, be maintained in the quantum theory;
 \item the way in which our classical world emerged from some primordial quantum event at the big-bang;
 \item the whole question of the interpretation of quantum theory and, in particular, the domain of applicability of the conventional Copenhagen view.
\end{enumerate}
In this section we wish to briefly address what answer quantum geometrodynamics based on York time and constant-mean-curvature slicing gives to each of these. Some answers are not determined by this change of the classical starting point from four-diffeomorphic general relativity to a space-through-time theory with spatial diffeomorphisms only, but will depend on one's favoured formulation of quantum mechanics. At this stage we will remain agnostic about the latter question, although the reader will undoubtedly become aware of our favoured approach. 

York-time quantum geometrodynamics has a clear notion of time and simultaneity across space, even if these simultaneity surfaces --- the foliation --- is not classically detectable. Hence the quantum theory is not plagued by the question of how to define quantum states and similar issues. A quantum state $\Psi$ is a complex function over the configuration space of the theory, which is just \emph{conformal superspace}, the space of all equivalence classes of spatial geometries where two geometries are in the same class if they are related via a local conformal transformation $g_{ab}(x)\rightarrow\phi^4(x)g_{ab}(x)$. 

The notion of probability is therefore well defined and conserved as much as this is the case in any theory. Rigorously defining probability densities here requires defining a measure on the configuration space, which has an uncountably infinite number of dimensions. These complications are however inherent to all field theories and not generally considered an issue. The main point here is that in York-time quantum geometrodynamics instants of time are clearly-defined concepts and no additional issue arises.

Since the foliation is fixed, the quantum state of the universe evolves just like in other more conventional systems as the state space is well defined. One quantum state follows another, so if one understands causality in the sense of a temporal ordering with one state being a necessary and/or sufficient condition for another state to arise at a later stage, then the notion of causality makes sense. However, in general I do not find the introduction of such a notion helpful. For example, since the evolution of the quantum state is deterministic and linear (given by a functional Schr\"odinger equation) if the universe is described by a state $\Psi_1$ at time $T_1$, this is as much necessary and/or sufficient for $\Psi_2$ to be the state at a later time $T_2$ as is the converse. So being necessary and/or sufficient is surely not what the idea of `causality' is trying to express. But then without any temporal aspect to the notion of causality, how is one to understand the question `is A a cause of B?' The question is problematic independently of any notion of space, time, spacetime or quantum theory and should be left to philosophers to discuss,\footnote{The debate over what constitutes causation goes back at least as far as Hume \citep{Hume1748}, and more recently has been tackled by philosophers such as Lewis \citep[e.g.][]{Lewis1973}, Armstrong \citep[e.g.][]{Armstrong1983}, Mellor \citep[e.g.][]{Mellor1995} and others too numerous to list.} although even there one may ask how far the exercise of defining the notion of causality really is more than one of semantics. Unitarity on the other hand is as well defined as in any field theory. One may have to impose certain boundary conditions at spatial infinity, for example, for unitarity and well-defined probability but there is no additional issue from the fact that the theory is considered `gravitational' or `geometrical'.

The canonical-versus-covariant debate has clearly been answered since four-covariance is explicitly broken and only the more trivial covariance under spatial diffeomorphisms remains. The approach to quantisation here is canonical, although in chapters \ref{chap:traj} and \ref{chap:cosmtraj} we suggest an alternative approach, which is neither canonical nor covariant.

Is spacetime a meaningful concept in York-time quantum geometrodynamics? This is a more difficult question to answer. In the classical theory spacetime may be constructed from a sequence of spatial slices, that is, from the classical state of the space and its matter content at a continuous progression of instants. In other words, it may be constructed from the trajectory of the system (the universe) in conformal superspace. In standard quantum theory no such trajectories exist and such a construction is not possible. If on the other hand one introduces quantum trajectories as in de~Broglie-Bohm theory or the proposal in chapter \ref{chap:traj} of this thesis, for example, then spacetime may once again be constructed. However, even in this case this constructed spacetime is not `all there is' since the wavefunction (in de~Broglie-Bohm) that drives the evolution is a complex function over the entire conformal superspace. It is also an open question whether or not a quantum equivalent of the four-covariant field equations can be written down, that is, a spacetime description of the quantum dynamics that is detached from the York foliation.\footnote{If such a feat were accomplished, could one kick away the ladder of York time all together?}

Similarly the notion of the possibility of a geometric interpretation depends on the chosen approach to quantum theory. For any particular trajectory the geometry is clearly defined since each point on the trajectory exactly determines the conformal geometry (and the conformal factor, that is the scale, may be recovered via the numerical value of the Hamiltonian density at the respective point in space, which is a function of the dynamical `shape' variables and was obtained in the Hamiltonian reduction that produced the classical theory in the first place). We discuss this in more detail in chapter \ref{chap:perttheoprelims}. In the absence of trajectories one still has a complex function over conformal spatial geometries, so the concept is not lost, although it is impossible to associate a particular geometry with the universe at any one instant of time.

Instead of considering the emergence of some primordial quantum event at the big bang, York-time quantum geometrodynamics denies the existence of such an event since the big bang is simply a name of the infinite past, $T=-\infty$. One can, of course, instead ask about properties of the universe prior to some very early time such as the Planck time, but even here the quantum evolution is at all times well defined. The York-time approach to quantum gravity gives no explanation of a beginning because the universe simply has none. It is infinitely old. There are however a series of questions one can ask about the early universe, such as how the classical approximation of the universe that we observe today can be extracted from the dynamics and whether one can expect such a classical limit to arise independently of the boundary conditions or whether some special assumptions must be made about the quantum state of the early universe (a `quantum past-hypothesis'?) and how that state might be explained.

Finally, as Isham's wording indicates, the `whole question of the interpretation of quantum theory' is just that. It is a whole separate question. Irrespective of the choice of York time however certain things can be said. For example, by definition (of the universe) there is no notion of an external observer who could perform a measurement. Hence any approach to quantum theory that relies on such a notion is inconsistent. Second, there is only one universe, so the notion of probability cannot be understood in a frequentist manner. Similarly, in de~Broglie-Bohm theory, there is no notion of an ensemble density $\rho$,\footnote{In the approach proposed in chapter \ref{chap:traj} the situation is different since there is, ontologically speaking, an infinity of `classical' universes (defined by classical variables, although not evolving according to classical laws). However, here one really ought to view the entire collection as `the universe', so the exception is really one of nomenclature rather than substance.} so also the idea of quantum equilibrium versus non-equilibrium is meaningless when applied to the entire universe, although it may apply to subsystems of which many copies exist within the universe, such as hydrogen atoms or some primordial particle species. The conventional Copenhagen view is not applicable although it may be possible to show, with an appropriate definition of what constitutes a measurement dynamically, how measurement processes within the universe appear to obey Copenhagen quantum mechanics. There is nothing fundamental about these processes though. Copenhagen quantum mechanics is problematic irrespective of the application of quantum theory to the entire universe, although the latter certainly makes some of the issues more apparent.

In summary, the answers York-time quantum geometrodynamics gives to the conceptual questions of quantum gravity engendered by the problem of time are on the whole relatively conservative in that quantum gravity is reduced to a conventional field theory via the choice of a preferred slicing. Even issues particular to gravitational theories, such as the question concerning geometric singularities, are avoided via the particular foliation chosen. In this limited sense the proposal made in this thesis is anything but revolutionary and arguably less exciting than other proposals that present more radical departures from the gravitational and matter theories we understand. Observation must be the ultimate arbiter.

\part{Classical York-time cosmology}\label{part2}

\chapter{The homogeneous isotropic universe with York time}\label{chap:Friedmann}

\textit{In this chapter we explore the Friedmann-Lema\^itre universe in the context of York time and perform Hamiltonian reduction explicitly in the case of a minisuperspace model with a scalar field. We also discuss cosmological inflation in the York-time description. The contents of this chapter were published in ref.~\citep{RoserValentini2014a}, with the exception of section \ref{sec:inflation}, which was presented in ref.~\citep{RoserValentini2016}.}

\section{The role of cosmology in gravitational theories}\label{sec:roleofcosm}

The equations of the full theory of general relativity are too complicated to allow us to visualise or otherwise intuitively grasp some given solution thereof. It is, of course, not strictly necessary to have an intuition or visualisation of such solutions in order to understand the theory or arrive at new insights, purely formal or otherwise. However, it would be insincere to state that intuitions play no role in how we as human beings with limited mental capabilities (and with brain functions primarily evolved to ensure survival and procreation rather than consider mathematical abstractions) develop and indeed advance comprehension of theories in physics. 

Intuitive insights into a theory are gained via simplification. For example, in Newtonian mechanics one might consider a small number of rigid, perhaps pointlike bodies and neglect the rest of the universe or model its influence as an external field. Indeed, this is how we spend much of our early physics education. In some cases the simple scenario considered may, in fact, provide a relatively good approximation to a real-world system. This is also the case in general relativity and reduced-Hamiltonian theories derived therefrom. To consider a \emph{cosmology} of a gravitational field theory is to impose a large number of symmetries (such as spatial homogeneity) on the solution space, reducing the degrees of freedom that characterise the physics to a finite number. The remaining dynamics is then no longer a field theory. Such a model is also commonly called a \emph{minisuperspace} model since superspace --- the infinite-dimensional configuration space of the gravitational field --- is reduced to a `smaller' finite-dimensional subspace. This terminology is perhaps less ambiguous than `cosmology', which is used differently in other contexts. 

Cosmologies (or minisuperspace models) can, in general, be visualised because of their simpler mathematical structure and therefore allow for an intuitive grasp of some of the implications of the equations of the gravitational theory. Furthermore, observational evidence strongly suggests that the universe is highly homogeneous at sufficiently large scales, so that at least some cosmological models give solutions that approximate the actual universe.

Considering cosmologies is clearly worth our time. However, one cannot expect to gain a full understanding of a gravitational theory by looking at cosmologies. Some features may be hidden by the symmetries. For example, in this chapter we explore a model universe that is homogeneous and isotropic (a `Friedmann-Lema\^itre' cosmology). However, as we will see, this hides the non-canonical form of the Poisson brackets of the reduced variables since these are related to the anisotropic degrees of freedom. In contrast, in section \ref{sec:ClassKasner} we develop an anisotropic (but still homogeneous) cosmology where their presence becomes apparent, before proceeding to the perturbation theory itself in the rest of chapters \ref{chap:perttheoprelims} and \ref{chap:perttheoformalism}.

Minisuperspace approximations also hide a number of the `problems of time' discussed in section \ref{sec:OtherProblemsOfTime}. Fortunately for our illustrative purposes the problem of the frozen dynamics remains since it is a result of reparameterisation invariance rather than anything strictly related to field theories (in fact, we introduced it in chapter \ref{chap:problemoftime} first in the context of finite dimensional theories). Yet evidently one cannot rashly generalise from the cosmological theory to the full theory.

Later we also quantise the minisuperspace models (chapters \ref{chap:QuantFriedmann} and section \ref{sec:quantKasner}). Doing so is comparatively simple since the classical theories are by design finite dimensional. This can be insightful as it gives us a feel for some of the features of the theory of quantum gravity obtained via quantisation of the Hamiltonian-reduced theory. Strictly speaking however there is a missing step if the insights are to be of any value: One first must have reason to believe that the cosmological (dimensionally reduced / symmetrised) limit of the quantised theory of gravity is the same or at least sufficiently similar to the theory resulting from quantising an already cosmological (dimensionally reduced / symmetrised) classical theory. In other words, one must have reason to believe that quantising the cosmology gives the same result as `cosmologising' the quantum theory. This proviso ought to be kept in mind.

\section[Friedmann-Lema\^itre cosmology `translated' into York time]{\texorpdfstring{Friedmann-Lema\^itre cosmology `translated' into\\ York time}{Friedmann-Lema\^itre cosmology `translated' into York time}}\label{sec:handsonYTcosm}

Our purpose here is to explore the cosmology that is the York-time reduced gravitational theory when restricted to spatially homogeneous and isotropic solutions. We learned in chapter \ref{chap:choiceoftime} that at least classically a reduced theory is equivalent to the original theory. We therefore expect to recover the standard results of classical Friedmann-Lema\^itre cosmology expressed in the `language' of York time. We can exploit this in order to check that our procedure is correct. To do so, in this section we `translate' the standard results of Friedmann-Lema\^itre cosmology, usually expressed in cosmological time $t$ or conformal time $\eta$, into the York-time description. Only later in section \ref{sec:classFriedmann} do we perform the Hamiltonian reduction explicitly and derive and solve the resulting equations.

Consider then a homogeneous and isotropic universe. For the matter content we choose a single scalar field $\phi$ with potential $V(\phi)$ for simplicity. The Einstein field equations reduce to a set of equations equivalent to the so-called Friedmann-Lema\^itre\footnote{Frequently the homogeneous isotropic model is at least in part attributed to Robertson and Walker and the equations are referred to as the FLRW or FRW (omitting Lema\^itre) equations. However, the development of this model is due to Friedmann \citep{Friedmann1922,Friedmann1924} and separately Lema\^itre \citep{Lemaitre1933}. The contribution of Robertson and Walker was to write influential review articles on the subject a few years later, hence our chosen nomenclature.} equations:
\begin{align}
 \frac{\dot{a}^2}{a^2}+\frac{k}{a^2}&=\frac{1}{3 M_{Pl}^2}\rho \label{eq:5.2-CosmoFriedmann1}\\
 \frac{\ddot{a}}{a}&=-\frac{1}{6M_{Pl}^2}(\rho+3\mathcal{P}). \label{eq:5.2-CosmoFriedmann2}
\end{align}
The first of these corresponds to the time-time component of the Einstein field equations and the second to a combination of the space-space and time-time components. Here a dot denotes differentiation with respect to cosmological time $t$, corresponding to a choice of lapse $N_t=1$. The parameter $k$ takes values $0,+1,-1$ for a flat, closed and open universe respectively. The energy density is denoted by $\rho$ and for a scalar field takes the form $\rho=\half\dot{\phi}^2+V(\phi)$. The pressure $\mathcal{P}$ is given by $\mathcal{P}=\half\dot{\phi}^2-V(\phi)$. A cosmological constant may be included formally as a constant contribution to $V(\phi)$. 

Using $T=-2M_{Pl}^2\frac{\dot{a}}{a}$, written in terms of York time, equation (\ref{eq:5.2-CosmoFriedmann1}) becomes
\begin{equation} \frac{T^2}{4M_{Pl}^4}+\frac{k}{a^2} = \frac{1}{3M_{Pl}^2}\rho, \label{eq:5.2-GeneralFriedmann1}\end{equation}
which is purely algebraic if $\rho$ does not contain any time-derivatives, as is the case for a dust-dominated universe, or if we have a field for which appropriate ``slow-roll'' conditions --- for example during an inflationary phase --- are satisfied, in which case we can ignore the kinetic term. In cases where the equation is algebraic, it does not describe a time \emph{evolution} (how a quantity changes from one instant to the next) but merely a time \emph{progression} (what a quantity is at any given time). This possible reduction of order in return for explicit time dependence will be a key feature we encounter in the reduced Hamiltonian formalism below. Note though that if $\rho$ does contain time-derivatives as it is the case for a general scalar field, then we must convert the derivative with respect to $t$ into one with respect to $T$ in order to avoid reference to $t$. In general this requires determination of the York lapse $N_T$, defined by $dt=N_TdT$ (since $N_t=1$). 

As elsewhere in this thesis, let a prime denote differentiation with respect to $T$. It is $N_T^{-1}a^\prime=\dot{a}$, which together with $T=-2M_{Pl}^2\frac{\dot{a}}{a}$ implies $N_T^{-1}=-\frac{1}{2M_{Pl}^2} T\frac{a}{a^\prime}$. Having determined the lapse in this manner, we can replace $\dot{\phi}$ by $N_T^{-1}\phi^\prime$ explicitly, for example. Since $a^\prime$ appears in the lapse, the resultant form of the first Friedmann-Lema\^itre equation is not algebraic. In the simplest case, namely a free scalar field ($V=0$) in a flat ($k=0$) universe, it takes the form $\frac{\dot{a}^2}{a^2}=\frac{1}{6M_{Pl}^2}\dot{\phi}^2$, which gives 
\begin{equation}\frac{a^{\prime2}}{a^2}=\frac{1}{6M_{Pl}^2}\phi^{\prime2} \label{eq:5.2-freeFriedmann1}\end{equation} 
after multiplying both sides by the $N_T^{-2}$. In this particular case, since the equation is homogeneous in time derivatives (all terms have two first-order derivatives) explicit knowledge of $N_T$ would not have been necessary.

In order to `translate' the second Friedmann-Lema\^itre equation to York time, one observes that
\begin{equation}N_T^{-1}=\DERIV{T}{t}=-2M_{Pl}^2\deriv{T}\frac{\dot{a}}{a}=-2M_{Pl}^2\frac{\ddot{a}}{a}+\frac{1}{2M_{Pl}^2}T^2 \end{equation}
and combines this with the previous expression for $N_T$ to obtain
\begin{equation}\frac{\ddot{a}}{a}=\frac{1}{4M_{Pl}^4}\left(T^2+\frac{aT}{a^\prime}\right). \label{eq:5.2-adoubledot}\end{equation}

For the scalar field the equation \ref{eq:5.2-CosmoFriedmann2}, after some appropriate rearrangement, takes the form
\begin{equation} \frac{a^\prime}{a}=-\frac{T}{T^2+\frac{2}{3}M_{Pl}^2(\rho+3\mathcal{P})}, \label{eq:5.2-GeneralFriedmann2}\end{equation}
which is a first order equation even after replacement of $\dot{\phi}$ by $N_T^{-1}\phi^\prime$. Here the reduction of order in return for explicit time dependence is general, provided $\rho$ does not contain second or higher order time derivatives, a property that holds true in most conceivable physical examples. Alternatively this equation could have been obtained by explicit differentiation using $\frac{d^2}{dt^2}=N_T^{-1}\deriv{T}\left(N_T^{-1}\deriv{T}\right)$ together with $N_T^{-1}=-\frac{1}{2M_{Pl}^2} T\frac{a}{a^\prime}$.

For a scalar field the last relevant equation is the Klein-Gordon equation. In terms of $t$ this is given by 
\begin{equation}0=\ddot{\phi}+3\frac{\dot{a}}{a}\dot{\phi}+\PD{V}{\phi}(\phi). \label{eq:5.2-CosmoKG}\end{equation} 
Expressed in terms of dependence on $T$ rather than $t$ this equation can be written in the form
\begin{equation} \label{eq:5.2-YorkKG}
0 = \phi^{\prime\prime}+\frac{\phi^\prime}{T}+4\frac{a^\prime}{a}\phi^\prime-\frac{a^{\prime\prime}}{a^\prime}\phi^\prime
	      +\frac{4M_{Pl}^4}{T^2}\frac{a^{\prime2}}{a^2}\PD{V}{\phi}(\phi)
\end{equation}
where extra terms have arisen from the differentiation of the lapse in $\ddot{\phi}=N_T^{-1}\deriv{T}(N_T^{-1}\DERIV{\phi}{T})$.

We have transformed the Friedmann-Lema\^itre and Klein-Gordon equations from the language of cosmological time $t$ to that of York time $T$. Since the classical dynamics are of the standard theory and the reduced-Hamiltonian theory developed below are provably equivalent, the `translated' equations found in this section give us the ability to check the results of the reduction procedure.

\section[Reduced-Hamiltonian Friedmann-Lema\^itre cosmology]{\texorpdfstring{Reduced-Hamiltonian Friedmann-Lema\^itre\\ cosmology}{Reduced-Hamiltonian Friedmann-Lema\^itre cosmology}}\label{sec:classFriedmann}

\subsection{Derivation of the Hamiltonian}\label{subsec:classFriedmannHamiltonian}

In the previous section we arrived at equations for the dynamics of a homogeneous isotropic universe by a transformation of the time coordinate from cosmological time $t$ to York time $T$. However, the equations --- in particular the Klein-Gordon equation in $T$, eq.~\ref{eq:5.2-YorkKG} --- is rather awkward to use. In fact, these equations are not the `natural' equations to describe York-time dynamics. In this section we begin with an action principle for the Friedmann-Lema\^itre universe and perform a Hamiltonian reduction to arrive at a different set of equations, which forms a more natural description (although is, of course, equivalent to the equations above). 

The action principle with which we start is simply the Einstein-Hilbert action for an homogeneous isotropic universe, that is, where the spatial metric takes the form
\begin{equation} g_{ab}(t) = a^2(t)\,\gamma_{ab}, \end{equation}
where $\gamma_{ab}$ is constant and in spherical coordinates given by the diagonal matrix 
\begin{equation} \gamma_{ab} =  \ensuremath{\left[\begin{array}{ccc} (1-kr^2)^{-1} & 0 & 0 \\ 0 & r^2 & 0 \\ 0 & 0 & r^2\,\sin^2\theta\end{array}\right]}_{ab}, \end{equation} 
for $k=1$ (spatially closed), $k=0$ (spatially flat) or $k=-1$ (spatially open). In the flat case ($k=0$) one can also use Cartesian coordinates, in which case $\gamma_{ab}$ is just the identity matrix.

The only geometric variable is the scale $a$, which is only a function of time and not of spatial position. This makes the Friedmann-Lema\^itre universe a special case in that the York-time Hamiltonian reduction, in which the notion of scale is `absorbed' into the time parameter, reduces the number of geometrical dynamical variables to zero and only leaves the degrees of freedom pertaining to matter. For simplicity here we consider a single scalar field. One could envision a scenario with no matter content at all except for a cosmological constant. In this case the reduction would lead to an expression for the Hamiltonian that is only a function of time and directly encodes the volume of the universe. There would be no dynamics, merely time passing.

With a scalar field $\phi$ with a potential $V(\phi)$ the action takes the form
\begin{align}\label{eq:5.3-action}
 \text{Action } &= \int d\tau\;Na^3\left(L_{grav}+L_\phi\right) \notag\\
		&=\tfrac12M_{Pl}^2\int d\tau\;\left(6Nka-\frac{6}{N}a\dot{a}^2\right) +\int d\tau\;\left(\frac{a^3}{2N}\dot{\phi}^2 - Na^3V(\phi)\right),
\end{align}
where $N$ is, of course, the lapse associated with the arbitrary temporal parameter $\tau$ and $k\in\{0,\pm1\}$ encodes the global properties of space. A dot refers to differentiation with respect to $\tau$ (and not cosmological time $t$ as elsewhere).

In order to obtain this expression from the Einstein-Hilbert action one must integrate a spatial constant (essentially the instantaneous value of $a$) over all space. This just introduces an overall constant factor. However, unless the universe is spatially closed ($k=1$) this integral is technically infinite, so one must choose a `normalisation volume'. Here we use a normalisation such that the volume is given by $a^3$ without any numerical factor. For the flat case, on which we will restrict our focus shortly, the choice of normalisation has no effect and a numerical factor $C$ drops out at the level of the Hamiltonian constraint when written in terms of York variables (since York time $T$ and its conjugate momentum $P_T$ are then rescaled by $1/C$ and $C$ respectively). For a closed universe the appropriate expression for the volume is $2\pi^2a^3$ rather than $a^3$.

The standard Friedmann-Lema\^itre equations in cosmological time $t$ (eqs.~\ref{eq:5.2-CosmoFriedmann1}, \ref{eq:5.2-CosmoFriedmann2}) can be obtained by variation of $N$ and $a$, setting $N=1$ in the resulting Euler-Lagrange equations and identifying the time coordinate with cosmological time $t$. The Klein-Gordon equation (eq.~\ref{eq:5.2-CosmoKG}) in cosmological time is obtained via variation of $\phi$, with $N=1$.

The canonical momenta conjugate to $a$ and $\phi$ are
\begin{equation}
 p_a= -\frac{6M_{Pl}^2}{N}a\dot{a},\qquad  p_\phi= \frac{a^3}{N}\dot{\phi}.
\end{equation}
Using these to rewrite the action in canonical form gives
\begin{equation} A = \int dt\left[\dot{a}p_a+\dot{\phi}p_\phi-N(\curlyH_{grav}+\curlyH_\phi)\right] \end{equation}
where 
\begin{equation} \curlyH_{grav}+\curlyH_\phi = -\frac{1}{12M_{Pl}^2}\frac{p_a^2}{a}-3kaM_{Pl}^2+\frac{p_\phi^2}{2a^3}+a^3V(\phi). \end{equation}
Variation of $N$ implies the Hamiltonian constraint, 
\begin{equation}\curlyH_{grav}+\curlyH_\phi=0,\end{equation} 
which is the origin of many of the problems in canonical quantum gravity (see the discussion in chapter \ref{chap:problemoftime}).

We now begin with the Hamiltonian reduction. Recall that the procedure consists of first transforming the Hamiltonian constraint from the variables $(a,p_a)$, encoding scale and its rate of change, to York time $T$ and its conjugate momentum $P_T$, then solving it for $P_T$ as a function of $T$ and the other variables. York time and its momentum (eq.~\ref{eq:4.1-YorkTime}) take the form,
\begin{equation} T\equiv p_a/3a^2, \qquad P_T=-a^3 \end{equation}
in the Friedmann-Lema\^itre universe. It is easy to verify that indeed $P_T\dot{T}=p_a\dot{a}$ up to a total derivative (which does not change the equations of motion). With the change of variables $(a,p_a)\rightarrow(T,P_T)$ the Hamiltonian constraint is now
\begin{equation}\label{eq:5.3-THconstraint} \curlyH_{grav}+\curlyH_\phi= -\frac{p_\phi^2}{2P_T}+\frac{3}{4M_{Pl}^2}T^2P_T+ 3kM_{Pl}^2P_T^\frac{1}{3} - P_TV(\phi) =0. \end{equation}
This is the equation we must now solve for $-P_T$, which will be identified with our reduced (physical) Hamiltonian $H_T$. It is the minisuperspace counterpart to the Lichnerowicz equation, written in terms of the York variables. The equation is seen to be a depressed cubic in $P_T^{-2/3}$, which may be solved by standard methods, giving
\begin{equation}\label{eq:5.3-physicalHamiltonian}
 H = -p_\phi\left[U+3\left(\frac{1}{4}U^2-C_k^2\right)^\frac{1}{3}
	    \left(\left(\half U+C_k\right)^\frac{1}{3}+\left(\half U-C_k\right)^\frac{1}{3}\right)\right]^{-\half},
\end{equation}
where we define 
\begin{align}
  U &=U(\phi,T)\equiv \tfrac{3}{2}M_{Pl}^{-2}T^2-2V(\phi) \label{eq:5.3-defofU}\\
  C_k &=C_k(\phi,p_\phi,T)\equiv\sqrt{\tfrac{1}{4}U^2-(2kM_{Pl}^2)^3p_\phi^{-2}}
\end{align}
for notational convenience. This solution is not valid for $P_T=0$, that is, the singular case $a=0$. The physical interpretation of the numerical value of $H$ --- as discussed in chapter \ref{chap:choiceoftime} --- is that of a `volume' (although in the flat and open case only volume ratios are meaningful).

In principle there is no reason not to work with this Hamiltonian, except that it is algebraically rather unwieldy, as are the derived Hamilton's equations. We therefore choose to consider the case of a spatially flat cosmology ($k=0$), which simplifies the expression significantly,\footnote{The results of this section were published in \citep{RoserValentini2014a}. However, there we erroneously wrote $p_\phi$ rather than $|p_\phi|$ in the numerator. This had some consequences for the resulting analysis. The error was corrected in \citep{Roser2015CosmExtension}. There are implications of this for the rigorous quantisation of the model. This will be discussed in chapter~\ref{chap:QuantFriedmann}.} 
\begin{equation} H_{k=0}= \pm(p_\phi^2/U)^{\half}= \pm\frac{|p_\phi|}{\sqrt{\frac{3}{2M_{Pl}^2}T^2-2V(\phi)}}.\label{eq:5.3-k0Hamiltonian}\end{equation}
In fact, the flat case is physically the most relevant as current data suggests that the universe is, in fact, spatially flat. Recent data leads to an effective energy-density contribution of curvature given by  $\Omega_k=-0.011\pm0.012$ or $-0.014\pm0.017$ depending on the exact choice of data and dark-energy model \citep{WMAP2007} (flatness corresponds to $\Omega_k=0$). For the remainder of this section we will focus on the flat case and we will drop the subscript `$k=0$' when denoting the Hamiltonian.

Since the Poisson structure of the non-geometric degrees of freedom is the usual canonical ones, Hamilton's equations for this Hamiltonian are
\begin{align}
 \phi^\prime &= \pm\frac{p_\phi/|p_\phi|}{\sqrt{\frac{3}{2M_{Pl}^2}T^2-2V(\phi)}} \label{eq:5.3-phiprime} \\
 p_\phi^\prime &= \mp \frac{|p_\phi|}{(\frac{3}{2M_{Pl}^2}T^2-2V(\phi))^\frac32}\PD{V}{\phi}. \label{eq:5.3-pphiprime}
\end{align}
The sign ambiguity arises from the fact that both sign choices for $H$ are formal solutions to the Hamiltonian constraint. The physical interpretation of the numerical value of $H$ is `volume' (since $H\equiv-P_T=\sqrt{g}\propto a^3$), from which one can readily infer that $H>0$ constitutes the physically relevant choice. Aside from this interpretation, mathematically speaking the choice is arbitrary. The set of solutions for $(\phi(T),p_{\phi}(T))$ is identical except for a reflection in phase space (swapping the signs in the corresponding functions $p_\phi(T)$). Henceforth we assume the positive solution $H\ge0$, although we emphasise that this is only relevant for the physical interpretation, not the mathematical development. 

One may be concerned that the radicand $U=\frac32M_{Pl}^{-2}T^2-2V(\phi)$ in the denominator of the Hamiltonian \ref{eq:5.3-k0Hamiltonian} (and \ref{eq:5.3-phiprime}, \ref{eq:5.3-pphiprime}) may take non-positive values. Fortunately, this is never the case. The easiest way to see this is to realise that the positivity of the radicand is equivalent to the positivity of the kinetic energy. That is, using $T=-2M_{Pl}^2\dot{a}/a$ it is easy to show that $\frac32M_{Pl}^{-2}T^2>2V$ if and only if $\dot{a}^2/a^2-(3M_{Pl}^2)^{-1}V>0$, which by the first Friedmann-Lema\^itre equation (\ref{eq:5.2-CosmoFriedmann1}) is exactly the condition for the kinetic term $\dot{\phi}^2$ to be positive. That the Friedmann-Lema\^itre equations are obeyed follows from the equivalence of the reduced-Hamiltonian and the standard formalism, which can be derived as follows.

\subsection{Explicit equivalence between the reduced and standard formalism in the case of the homogeneous isotropic universe}\label{subsec:classFriedmannEquivalence}

We already showed the equivalence of reduced-Hamiltonian and unreduced dynamics in generality in section \ref{sec:Hamreduction}. Nonetheless it is insightful to see the equivalence explicitly for the present cosmological scenario.

The Friedmann-Lema\^itre equations with a scalar field $\phi$ in a spatially flat universe ($k=0$) written in terms of York time $T$ are
\begin{align}
 \tfrac14M_{Pl}^{-4}T^2 &=(3M_{Pl}^2)^{-1}\left(\half N_T^{-2}\phi^{\p2}+V(\phi)\right) \label{eq:5.3-F1T}\\
 \frac{a^\p}{a} &= -\frac{T}{T^2+\frac43M_{Pl}^2\left(N_T^2\phi^{\p2}-V(\phi)\right)}, \label{eq:5.3-F2T}
\end{align}
where the inverse York lapse is $N_T^{-1}=-\frac12M_{Pl}^{-2} T a/a^\p$. These are eqs.\ \ref{eq:5.2-GeneralFriedmann1} and \ref{eq:5.2-GeneralFriedmann2} with $\rho=\half\dot{\phi}^2+V(\phi)=\half N_T^{-2}\phi^{\p2}+V(\phi)$ and $\mathcal{P}=\half\dot{\phi}^2-V(\phi)=\half N_T^{-2}\phi^{\p2}-V(\phi)$. To show equivalence we must be able to recover these equations from the reduced Hamiltonian dynamics, that is, eqs.\ \ref{eq:5.3-phiprime} and \ref{eq:5.3-pphiprime} together with the interpretation of the numerical value of the Hamiltonian as `volume', $a^3=H$, with $H$ given by eq.\ \ref{eq:5.3-k0Hamiltonian}.

Begin by noting that $a^3=H(\phi,p_\phi,T)$ is to hold at all times, and so $da^3/dT=dH/dT$. Using eq.\ \ref{eq:5.3-k0Hamiltonian}, the right-hand side becomes the sum of three terms, two of which cancel (since $\DERIV{H}{T}=\PD{H}{T}$ for a Hamiltonian function $H$, or equivalently by using the equations of motion \ref{eq:5.3-phiprime} and \ref{eq:5.3-pphiprime}), leaving 
\begin{equation}3a^3\frac{a^\prime}{a} = \frac{T}{T^2-\frac43M_{Pl}^2V(\phi)}\cdot \frac{p_\phi}{\sqrt{\frac32M_{Pl}^{-2}T^2-2V(\phi)}}.\end{equation}
The second factor on the right-hand side is just the expression for the negative of the Hamiltonian $H$, which cancels with the $a^3$ term on the left, up to a sign, so that
\begin{equation} \frac{a^\prime}{a}=-\frac{T}{3T^2-4M_{Pl}^2V(\phi)}. \label{eq:5.3-keyeqinapp}\end{equation}
Using this and eq.\ \ref{eq:5.3-phiprime} implies after a little algebra
\begin{equation} \tfrac32M_{Pl}^{-2}T^2 = \tfrac14M_{Pl}^{-4}T^2\frac{a^2}{a^{\prime2}}+2V(\phi), \end{equation}
which, after division by $6M_{Pl}^2$ and identification of the lapse $N_T$ yields eq.\ \ref{eq:5.3-F1T}.

In order to derive eq.\ \ref{eq:5.3-F2T} we return to eq.\ \ref{eq:5.3-keyeqinapp}, writing it in the form
\begin{equation}\frac{a^\p}{a}=-\frac{T}{T^2-\frac43M_{Pl}^{2}V(\phi) + \left(2T^2-\frac83M_{Pl}^2V(\phi)\right)}. \end{equation}
Using eq.\ \ref{eq:5.3-F1T} and the expression for $N_T$ to eliminate the term in parentheses in favour of $\frac43M_{Pl}^2N_T^{-2}\phi^{\prime2}$ then gives the desired result.

The Klein-Gordon equation (\ref{eq:5.2-YorkKG}) is recovered by differentiation of eqs.\ \ref{eq:5.3-keyeqinapp} and \ref{eq:5.3-phiprime} with respect to $T$ and elimination of the term involving $\PD{V}{\phi}$. Subsequent use of eq.\ \ref{eq:5.3-phiprime} and \ref{eq:5.3-F1T} then allows rearrangement into the form of eq.\ \ref{eq:5.2-YorkKG}.

Note the `holistic' manner in which the dynamics is equivalent. All dynamical equations and the interpretation $a^3=H$ were necessary together in order to derive all of the standard Friedmann-Lema\^itre and Klein-Gordon equations (although these are not the natural equations with which to work in practice in the reduced Hamiltonian formalism).

\subsection{Evolution of the free scalar field}\label{subsec:FriedmannFreeScalar}

In order to discuss some of the features of cosmological Hamiltonian dynamics in York time, we begin by discussing the free massless case, setting $V(\phi)=0$. Dynamics with a general potential is treated in section \ref{subsec:FriedmannWithPotential} below. Mathematically this case is rather trivial. However, it is our first actual application of the reduced-Hamiltonian formalism in cosmology and helps to illustrate the role of the various physical quantities and a number of features of such a reduced theory in practice.

The equations of motion are
\begin{align}
 \phi^\p &= \sqrt{\tfrac32}M_{Pl} \frac{sign(p_\phi)}{|T|}  \label{eq:5.3-phiprimefree}\\
 p_\phi^\p &= 0. \label{eq:5.3-pphiprimefree}
\end{align}
An interesting feature of these equations is that the evolution of the field value $\phi(T)$ depends only on the sign of the momentum, and not its magnitude. With the momentum a constant, this sign does not change, so that eq.\ \ref{eq:5.3-phiprimefree} is easily solved. The physically viable range for $T$ is $T\in(-\infty,0)$ for a spatially flat universe, so that $|T|=-T$ in all cases (unless we consider the extension discussed in chapter \ref{chap:cosmext}). Leaving the sign of $p_\phi$ unspecified, we solve
\begin{equation} \phi^\p = \pm\sqrt{\tfrac32}M_{Pl}\frac{1}{-T} \end{equation}
to find
\begin{equation} \phi(T) = \phi_0 \mp\sqrt{\tfrac32}M_{Pl} \ln\left(\frac{|T|}{|T_0|}\right),  \end{equation}
where $\phi_0=\phi(T_0)$ is an integration constant that has to be fixed by boundary data at some reference time $T_0$. Throughout the history of the universe the field changes monotonically from positive to negative infinity or vice versa (depending on the sign of $p_\phi$),how to approach such problems in practice
\begin{align}
 \phi(T\rightarrow-\infty) &\longrightarrow\mp\infty \\
 \phi(T\rightarrow0) &\longrightarrow\pm\infty.
\end{align}

The physical interpretation of the value of the Hamiltonian is volume. In order to make this explicit within the reduced-Hamiltonian framework, we can re-introduce the scale factor $a$, now \emph{defining} its third power as the value of the Hamiltonian $H(T)$,
\begin{equation} a^3(T)\equiv H(T) \equiv H\big(\phi(T),p_\phi(T),T\big). \label{eq:5.3-reintroducescalefactor}\end{equation}
That is, we consider the reduced formalism the fundamental theory and `kick away the ladder' used to derived it. The scale factor $a$ has no natural role in the reduced formalism and is merely introduced as a way to connect the results to the standard picture. In no way does $a$ have any dynamical role however. The time evolution of the scale factor is then easily calculated:
\begin{equation} 3a^3\frac{a^\p}{a} = -\sqrt{\frac32}M_{Pl}\frac{|p_\phi|}{T^2} \end{equation}
and therefore (using $H(T)=a^3$ again)
\begin{equation} \frac{a^\p}{a} = -\frac{1}{3T} = \frac{1}{3|T|}. \label{eq:5.3-afreesolution}\end{equation}
As expected, the universe grows from zero volume at $T\rightarrow-\infty$ to infinite volume as $T\rightarrow0$. Eq.\ \ref{eq:5.3-afreesolution} is independent of the arbitrariness in the volume normalisation and in the definition of the scale factor $a$.

\subsection{Evolution of a scalar field with a potential}\label{subsec:FriedmannWithPotential}

In the presence of a field potential $V(\phi)$ the full equations \ref{eq:5.3-phiprime} and \ref{eq:5.3-pphiprime} must be considered. Once again the rate of evolution $\phi^\p$ of the field depends only on the sign of the momentum $p_\phi$, not its magnitude.\footnote{This is not a general feature however. In the presence of multiple scalar fields $\phi_A$ with corresponding momenta $p_\phi^A$ the numerator on the right-hand side of the corresponding equation of motion for $\phi_A$ reads $p_{phi}^A/(\sum_B p_{\phi}^{B2})^\frac12)$, so that at least the ratios of the momenta matter, although $\phi_A^\p$ is invariant under a uniform rescaling of all momenta by the same factor.} However, the momentum $p_\phi$ is not constant and can change sign, leading to a discontinuity in $\phi^\p$. Let us call such events `turning points' and examine them more closely.

If $\phi^\p$ or equivalently $\dot{\phi}$ is considered to pass through zero at the turning point (for example, if $\phi^\prime$ is treated as some appropriate limit of a sequence of continuous functions), one finds that at such points $\frac32M_{Pl}^{-2}T^2-2V(\phi)=0$ (one may infer from the first Friedmann equation \ref{eq:5.3-F1T} and $T=-2M_{Pl}^2\frac{\dot{a}}{a}$ that $\dot{\phi}^2\propto\frac32M_{Pl}^{-2}T^2-2V(\phi)$, which also guarantees the non-negativity of the radicand as discussed above). It appears that the expressions for the Hamiltonian \ref{eq:5.3-k0Hamiltonian} and the rates of change of field and momentum \ref{eq:5.3-phiprime} and \ref{eq:5.3-pphiprime} become singular.

Since we consider York time to be fundamental, let us frame our discussion entirely in terms of $T$ without reference to $t$ or the Friedmann-Lema\^itre equations. A turning point is defined by a change in sign of $p_\phi$, that is, by $p_\phi = 0$. Note that unlike $\phi^\prime$, the momentum $p_\phi$ changes continuously, so its vanishing at the turning point is established without reference to any kind of limiting procedure. But then the expression for the Hamiltonian \ref{eq:5.3-k0Hamiltonian} implies that either $H=0$ (and therefore $a^3=0$, constituting an initial or final singularity), or (if the turning point is to occur at finite volume) that $\frac32M_{Pl}^{-2}T^2-2V(\phi)\rightarrow0$ as the turning point is approached, specifically $(\frac32M_{Pl}^{-2}T^2-2V(\phi))^\half\sim |p_\phi|$. The implication is that $\phi^\p\rightarrow0$ with opposite signs for approach of the turning point from the future and past.

\begin{wrapfigure}{R}{0.4\linewidth}
 \includegraphics[width=\linewidth]{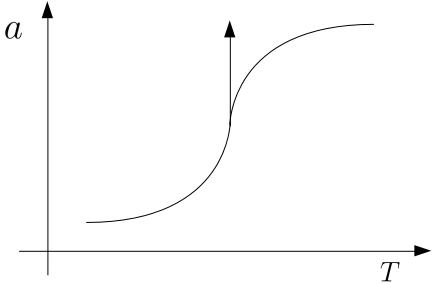}
 \caption[The scale factor near scalar-field `turning points']{Schematic illustration of the evolution of the scale factor near a turning point. The rate of growth becomes infinite at a single point but there is no discontinuity in the scale factor $a$.}
 \label{turningpointgraphic}
\end{wrapfigure}
In order to understand the nature of these turning points, consider the `interpretation' of the Hamiltonian as volume, $H\propto a^3$. Using expression \ref{eq:5.3-k0Hamiltonian} and differentiating with respect to $T$ gives
\begin{equation} \frac{a^\p}{a}=-\frac{T}{3T^2-4M_{Pl}^2V(\phi)}, \label{eq:5.3-Yubble}\end{equation}
so if the turning point is to occur at finite volume (and therefore finite $T$), we see that $a^\prime/a\rightarrow\infty$ as we approach the turning point. At the turning point the fractional rate of expansion (with respect to York time) diverges, although the total fractional growth remains finite and there is no discontinuity in $a$ (see fig.\ \ref{turningpointgraphic}). This corresponds to a de~Sitter phase (of infinitesimal duration in terms of cosmological time), where the fractional rate of expansion is constant and the York lapse diverges, $N_T\rightarrow\infty$. Note that even in a universe with a single scalar field, where such turning points are relatively generic, they do no necessarily occur. Their existence depends on the form of the potential $V(\phi)$.

Finding general solutions $\phi(T)$ is difficult even for comparatively simple forms of the potential $V(\phi)$. However, one can draw some conclusions about the evolution for certain regimes corresponding to different cosmological periods. Of course, the present model containing only a single scalar field does not adequately describe our actual universe and so its `epochs' do not match what we know about actual cosmological history. 

In the present model it is plausible to differentiate between an `early' and a 'late' epoch, respectively characterised by $|T|\gg1$ and $|T|\ll1$ and qualitatively different from one another with regards to which one of the two terms $\frac32M_{Pl}^{-2}T^2$ and $2V(\phi)$ dominates. Sufficiently early the term in $T^2$ dominates and the field behaves like the free field discussed in section \ref{subsec:FriedmannFreeScalar} unless $\phi(T)$ behaves such that $V(\phi)$ grows at least as fast as $T^2$ as one goes backwards in time. In the free theory the evolution of the field is characterised by $\phi(T)\sim\ln|T|$, so that if $\frac32M_{Pl}^{-2}T^2\gg|2V(\phi)|$ is satisfied at some point in time and $V(\phi)$ is bounded by some polynomial in $\phi$, then it is also satisfied at all earlier times. Indeed if $V(\phi)$ being at most polynomial is a sufficient condition for such a sufficiently early time to exist. If the model were considered as offering a partial description of our universe --- for example, if $\phi$ is considered as the field driving inflation --- then one might wonder how viable candidates for the inflaton potential fare. For an overview over such contenders based on recent observation, see ref.\ \citep{MartinEtAl2013}.

The end of this early epoch is reached when $2V(\phi)=\alpha^2\cdot\frac32M_{Pl}^{-2}T^2$, where $0<\alpha^2<1$ is chosen to suit one's desired level of accuracy ($\alpha=1$ would correspond to the end of the universe, $T=0$ as the end of the early period). For concreteness, consider a simple viable potential, such as a simple mass term, instantiating so-called large-field inflation for parameter $p=2$ (section A.3 in \citep{MartinEtAl2013}), that is, 
\begin{equation}V(\phi)=M^4(\phi/M_{Pl})^2,\end{equation} 
where $M$ is a constant corresponding to a mass scale (related to the chosen CMB normalisation if the proposal of $\phi$ as the inflaton field in our universe is taken seriously). For a first (and rather rough) approximation, let us assume that the free theory remains valid until equality is reached. Then the boundary $T_B$ between epochs is given by the equation
\begin{equation} \alpha^2\cdot \tfrac32M_{Pl}^{-2} T_B^2 = 2M^4\left(\frac{\sqrt{2}\ln(-T_B)}{\sqrt{3}}\right)^2
\end{equation}
whose solution is 
\begin{equation}T_B=-\exp\left[-w\left(\frac{3\alpha}{\sqrt{8}M^2M_{Pl}}\right)\right]\end{equation} 
where $w(x)$ is the Lambert Product-Log function.

For any potential that is strictly non-negative, as is the case in the chosen example, the free evolution forms a lower bound for the (magnitude of) the scalar-field velocity $\phi^\prime$ as well as the fractional rate of change of the scale factor. That is, the existence of a positive potential accelerates the rate of change of $\phi$ as well as the expansion of the universe in comparison to the free theory. This applies to all epochs.

In the late epoch $2V(\phi)$ becomes comparable to $\frac32M_{Pl}^{-2}T^2$ in magnitude (by definition of the onset of the late epoch). This implies that the denominator in eq.\ \ref{eq:5.3-k0Hamiltonian} becomes small faster. Thus the interpretation $a^3=H$ implies that the existence of a potential causes an increase in expansion when compared to the free theory. 

\section{Inflation}\label{sec:inflation}
Before concluding our treatment of the homogeneous isotropic universe in the York-time reduced-Hamiltonian formalism we return briefly to the classical theory and consider a particular cosmological process that plausibly characterised part of the early history of our actual universe: inflation.

The theory of inflation poses an immediate puzzle for a description in terms of York time. Inflation is characterised by a de~Sitter phase, that is, a phase during which the rate of spatial expansion (with respect to proper time) is exponential. The Hubble parameter is constant and therefore so is the York parameter as the two are proportional. During this `era' the universe appears to expand by a finite amount while no time passes. The York-time description appears singular. 

The puzzle is resolved if we recall that in a realistic inflationary phase the Hubble parameter is constant only to a first approximation and is, in fact, decreasing, albeit only by a small amount. The implication is that an inflationary period corresponds to a very short time interval and furthermore that the York-time description must use a higher order of approximation than is conventionally used in the standard picture.

Conventional slow-roll inflation assumes that during this period the matter content of the universe is dominated by a scalar field, whose contribution to the energy density $\rho = \frac12\dot{\phi}^2+V(\phi)$ is in turn dominated by the field's potential function $V(\phi)$. (Recall that $\dot{\phi}$ denotes the derivative of $\phi$ with respect to conventional cosmological time $t$, or, more or less equivalently, proper time.) Specifically, one assumes the `slow-roll conditions'
\begin{equation} \tfrac12\dot{\phi}^2\ll V(\phi),\qquad |\ddot{\phi}|\ll\left|\PD{V}{\phi}\right|, \end{equation}
while the dynamics is determined by the Friedmann and Klein-Gordon equations. Provided the slow-roll conditions hold, in terms of York time $T=-2M_{Pl}^2H$ the Friedmann equation \ref{eq:5.2-CosmoFriedmann1} becomes
\begin{equation}\tfrac34 M_{Pl}^{-2} T^2\approx V(\phi), \end{equation}
which suggests that the numerical value of the potential effectively provides a measure of York time. Approximating the potential to be roughly constant is therefore not a viable description. 

Instead one proceeds to calculate the next-order, time-dependent correction to $H$, and hence $T$. At `zeroth' order one ignores the contribution of the kinetic term to $\rho$, so that equation \ref{eq:5.2-CosmoFriedmann1} becomes 
\begin{equation} H \approx \sqrt{V(\phi)/3M_{Pl}^2}, \label{eq:5.4-FriedmannZerothOrder}\end{equation}
which can be substituted into the Klein-Gordon equation \ref{eq:5.2-CosmoKG}. Applying the second slow-roll condition, the second-derivative term $\ddot{\phi}$ can be neglected, so one is left, in general, with a non-linear first order differential equation, which can be solved in many relevant cases to yield a first time-dependent approximation $\phi_1(t)$ for the scalar field. For example, for $V(\phi)=\lambda\phi^b$ with $\lambda,b\in\mathbb{R}$ the solution is 
\begin{equation} \phi_1(t) = \left[\left(\frac{b}{2}-2\right)\left(b\sqrt{{\lambda M_{Pl}^2}/{3}}\,t+A\right)\right]^{1/(2-\frac{b}{2})}, \label{eq:5.4-phisolutionpolynomial} \end{equation}
with $A$ being an integration constant that must obey certain conditions that follow from the slow-roll conditions, namely that the constant contribution to the potential $V(\phi_1)$ is large compared to the time-dependent one (we will show this explicitly below for a particular example). This solution $\phi_1$ can then be substituted back into the Friedmann equation \ref{eq:5.2-CosmoFriedmann1} to give the next approximation $H_1(t)$ to the Hubble parameter and therefore to the York parameter.\footnote{One could repeat the procedure iteratively but in general this is not necessary for an approximate description of the homogeneous cosmological background and would in any case not yield higher degrees of accuracy due to our neglect of the contribution of other matter.} The function $T(t)$ obtained can now be inverted to find $\phi$ in terms of $T$ and give a `York-time only' account of the evolution during inflation.

To illustrate this, consider a particular example of an inflationary potential, namely the case of so-called large-field inflation already considered above, $V(\phi)= m^2\phi^2$. This simple choice helps to illustrate the method of obtaining a York-time description since unlike many other cases it leads to straightforward algebra (although it is not a choice of potential favoured by current observation \citep{Planck2015_Overview}). One finds,
\begin{equation} \phi_1(t) = \mp\sqrt{\tfrac43M_{Pl}^2m^2}\,t + \phi_0, \qquad\text{for }\phi_0\gtrless0 \label{eq:5.4-phi1solution}\end{equation}
where $\phi_0$ is the value of $\phi$ at $t=0$, which must be chosen such that the approximation is valid at this time. We will justify the condition on the sign, which arises due to the appearance of `$\sqrt{\phi^2}$' in the Klein-Gordon equation, shortly. The first slow-roll condition then says
\begin{equation} \tfrac23 M_{Pl}^2m^2\ll m^2\left(\phi_0+\sqrt{\tfrac43M_{Pl}^2m^2}\,t\right)^2, \label{eq:5.4-1stSRCinExample}\end{equation}
which must be satisfied throughout inflation, in particular at $t=0$, that is,
\begin{equation}\phi_0^2\gg \tfrac23 M_{Pl}^2. \label{eq:5.4-t0SRC}\end{equation}
The constant contribution $\phi_0$ dominates the scalar field until $t\approx(\frac43M_{Pl}^2m^2)^{-\frac12}$, by which point the approximation has ceased to be valid, presumably corresponding to a time well after the end of inflation. While the approximation remains valid the sign of $\phi$ is just the sign of $\phi_0$, hence the sign dependence in equation \ref{eq:5.4-phi1solution}.

In this example the correction to $\rho$ due to the kinetic energy contribution is constant, 
\begin{equation} \frac{1}{3M_{Pl}^2}\cdot\tfrac12\dot{\phi}^2 = \tfrac29 m^2, \label{eq:5.4-KEcontribution}\end{equation}
while time dependence arises via the potential term,
\begin{equation} V(\phi_1(t)) = m^2\phi_1^2 = m^2\left(\phi_0+\sqrt{\tfrac43M_{Pl}^2m^2}\,t\right)^2.\end{equation}
This is consistent with the approximation since the time-independent contribution dominates by virtue of equation \ref{eq:5.4-1stSRCinExample}. Substituting these back into the Friedmann equation gives
\begin{equation} H_1^2 = H_0^2\left[\left(1\mp\phi_0^{-1}\sqrt{\tfrac43 M_{Pl}^2m^2}\right)^2 + H_1^{-2}\cdot\tfrac29m^2\right],\end{equation}
where
\begin{equation} H_0^2 = \frac{1}{3M_{Pl}^2}m^2\phi_0^2 \end{equation}
is the initial, constant zeroth-order approximation to the Hubble parameter as given by equation \ref{eq:5.4-FriedmannZerothOrder}. From this one obtains the expression for York time as a function of cosmological time, 
\begin{equation} T(t) = -2M_{Pl}^2\cdot H_0\left(1+\frac{m^2}{9H_0^2}-\sqrt{\tfrac43 M_{PL}^2m^2}\frac{t}{|\phi_0|}\right).\end{equation}
This expression may be straightforwardly inverted in order to express $\phi_1$ as a function of $T$ rather than $t$,
\begin{equation} t = \frac{|\phi_0|}{\sqrt{\frac43 M_{Pl}^2m^2}}\left(1+\frac{m^2}{9H_0^2}-\frac{|T|}{2M_{Pl}^2H_o}\right). \end{equation}

The primary characteristic of inflation is the rapid (approximately exponential) growth of the background scale factor $a$. In the York-time reduced-Hamiltonian formalism $a$ ceases to be a dynamical variable however since the `volume' $a^3$ is the momentum conjugate to the time parameter (up to a sign). As we discussed in section \ref{sec:classFriedmann}, the volume (up to some suitable normalisation) is therefore given by the numerical value of the reduced York-time Hamiltonian, which for a spatially flat cosmology is given by eq.\ \ref{eq:5.3-k0Hamiltonian}. The York-time derivative $a^\p$ of the scale factor obtained was given by eq.~\ref{eq:5.3-Yubble}, which read
\begin{equation} \frac{a^\prime}{a} = -\frac{T}{3T^2-4M_{Pl}^2V(\phi)}. \tag{\ref{eq:5.3-Yubble}} \end{equation}
Recall that the expression in the denominator is proportional to the total kinetic energy as is apparent from the Friedmann equation (\ref{eq:5.2-CosmoFriedmann1}). In the example above it is constant and so one can easily calculate the number of e-folds that correspond to some inflationary interval $(T_i,T_f)$,
\begin{equation} N_e = \int_{T_i}^{T_f}dT\;\frac{a^\prime}{a} = \tfrac12 \kappa \left(T_i^2-T_f^2\right), \label{eq:5.4-efolds}\end{equation}
where
\begin{equation} \kappa \equiv \frac{3}{3 T^2-4M_{Pl}^2V(\phi)} = \text{ const.} \label{eq:5.4-kappadef}\end{equation}
has been defined for convenience. For other choices of potential the solution $\phi_1(t)$ is in general not a linear function of $t$ and so the kinetic energy is not constant. However, even then one can take the solution \ref{eq:5.4-phisolutionpolynomial} and expand it, dropping terms of second and higher order in $t$ since these are small compared to the constant and linear terms by the first slow-roll condition. As a result the non-constant contribution to the kinetic energy is negligible even compared to the already small constant contribution. Equation \ref{eq:5.4-efolds} holds, therefore, more generally. 

Equation \ref{eq:5.3-Yubble} is then trivially solved, giving
\begin{equation} a(T) = a_0 e^{-\frac{\kappa}{2}T^2}, \label{eq:5.4-aduringinflation}\end{equation}
where the integration constant $a_0$ hypothetically denotes a normalisation scale corresponding to the value of $a$ in the infinite $t$-future ($T=0$), although such an interpretation is not appropriate since the inflationary description only holds for a limited duration.

A numerical estmate for the parameter $\kappa$ can be derived. For the example considered above with the inflaton potential $V(\phi)=m^2\phi^2$, the empirical estimate for the value of the Hubble parameter during inflation, $3.6\times10^{-5}M_{Pl}$ from ref.~\citep{Planck2015_ConstraintsOnInflation} (see section \ref{sec:cosmhist}), can be used to find an inequality (using equation \ref{eq:5.4-t0SRC}) for the kinetic-energy correction \ref{eq:5.4-KEcontribution},
\begin{equation} K.E. \ll 10^{-9}, \end{equation}
from which in turn one can derive an inequality for the constant $\kappa$,
\begin{equation} \kappa \gg 10^8.\end{equation}

The expression \ref{eq:5.4-aduringinflation} (with the definition \ref{eq:5.4-kappadef}) shows how the growth of the scale factor during inflation is characterised by the potential. We have also solved (approximately) for the behaviour of the scalar field during that period. In chapter \ref{chap:perttheoprelims} we will use inflation as the background cosmology on which to study perturbations, in particular, the `freezing out' of tensor modes.


\chapter{A cosmological extension}\label{chap:cosmext}

\textit{In this chapter we argue that York time as a physically fundamental notion of time suggests a natural extension of the cosmological timeline. We explain how the existence of this extension may have implications for the possible matter content of the universe, such as viable inflaton potentials. The contents of this chapter were published in ref.~\citep{Roser2015CosmExtension}.}

\section{An end to time?}\label{sec:timeends?}

Throughout this thesis we are concerned with developing an understanding of the universe, in particular, of cosmology, in terms of a description based on York time as a physical time, and with associated formal developments. In a spatially flat universe --- and we have observational reasons to believe that we live in such a universe --- cosmological history apparently ends at some finite time $T_{end}$, either equal to zero or even earlier at some small negative number (in the presence of a positive cosmological constant $\Lambda$ or an effective cosmological constant taking the form of a positive $T\rightarrow T_{end}$ limit of the matter field potential $V(\phi)$). This point corresponds to cosmological time $t\rightarrow\infty$ in the standard description, where the Hubble parameter approaches zero or a constant proportional to $\Lambda$.

Our hypothesis is that York time is physically fundamental. A valid question is therefore: Why should time end at $T_{end}\leq0$? We leave it to philosophical discourse to determine whether or not it is possible or even meaningful for time to `end' at some instant --- that is, if instants of universal history and the York timeline form a bijection --- and what the metaphysical implications are. Either way, there is no \emph{a priori} reason to assume that time indeed ends at some particular instant $T_{end}$, equal to zero or otherwise. Undoubtedly the point has certain \emph{dynamical} significance and given that time is physical it is not obviously absurd that the end of time should have a dynamical origin. In fact, this is in part what we will explore in this chapter. However, this is a matter of fact that can only be determined \emph{a posteriori}, following analysis and comparison to the observed universe.

In this chapter we explore the consequences of extending the York timeline beyond $T_{end}$, that is, beyond $t=\infty$. The usual description in terms of cosmological time $t$ fails to describe this era. In the $t$-description the transition corresponds to a coordinate singularity. The idea is analogous to a Schwarz\-schild black hole, where a new region of space-time is revealed by the use of Kruskal-Szekeres coordinates, extending to the `other side' of the singularity (where the Kruskal-Szekeres variable usually denoted by `$V$' takes negative values), a region not described by either Schwarzschild or Eddington-Finkelstein coordinates \citep[Ch.\ 31]{MisnerThorneWheeler1973}.\footnote{Arguably there is another, simpler example: the extension of dynamics beyond a black hole's event horizon in the first place. The idea that spacetime extended beyond the horizon of a Schwarzschild black hole into the region $r<2M$ took several decades to become apparent \citep{Eddington1924,Lemaitre1933,Synge1950}. However, a crucial difference here is that this extension concerns the removal of a mere coordinate singularity. No physical quantities become infinite (and geodesics crossing the surface $r=2M$ are traversed in finite proper time). In the York-time extension, as we shall see, physical quantities may become infinite depending on the particulars of the matter content.}

Our reasons for pursuing this line of inquiry are not, however, purely philosophical. In fact, there are concrete reasons (described in section \ref{sec:reasonforextension}) why such a cosmological extension is necessary for a consistent quantisation. However, unfortunately there is (as far as I can see) no viable way to explore the dynamics or even the existence of this `other side' of cosmological history in any direct empirical way.

The basic hypothesis here is that the dynamical equations in York time found via the Hamiltonian reduction remain valid beyond $T_\infty$. We restrict our analysis to the homogeneous isotropic (`Friedmann-Lema\^itre') universe explored in chapter \ref{chap:Friedmann}. In chapter \ref{chap:perttheoprelims} we will explore perturbations and it would undoubtedly be interesting to consider their evolution as the universe transitions to the `other side'. However, this is left to future work. 

At least two questions require answering. First, how do the dynamical entities --- in our case the scalar field --- evolve during the transition? Second, what evolution can we expect on the other side? The second question is easier to answer and we will do so in section \ref{sec:dynamicsonotherside}. The first is more difficult and will be explored in section \ref{sec:transition}.

\section{The necessity of a cosmological extension} \label{sec:reasonforextension}

Why should one take the extension seriously? There may be some purely philosophical arguments: that time should not end at some finite value, or that the mere mathematical possibility of the extension should at least prompt us to allow for the possibility of its physicality. However, there is a considerably more compelling, \emph{physical} reason to consider it more than mere fiction.

The model employed in this chapter and chapter \ref{chap:Friedmann} above is that of a homogeneous isotropic universe. This is, of course, an approximation. Let us consider the existence of inhomogeneities in this universe, in particular local singularities such as a Schwarzschild black hole formed at some finite time from a collapsing cloud of dust (see \cite[sec.\ 32]{MisnerThorneWheeler1973} and references therein for an introduction). Such a scenario can be described by smoothly `gluing' together parts of a closed collapsing Friedmann universe shrinking to zero size (the collapsing dust cloud), a Schwarzschild region of space-time (the near exterior of the cloud) and the chosen Friedmann cosmology (the distant rest of the universe). When described in terms of constant-mean-curvature surfaces it is found \citep{BrillCavalloIsenberg1980} that these surfaces smoothly connect between the different regions.\footnote{Interestingly, a description in terms of a conventional time parameter $t$ allows only for continuous surfaces, but not smooth ones.} Furthermore, $K\rightarrow\infty$ ($T\rightarrow\infty$) as one approaches the singularity formed by the fully collapsed dust cloud, independently of the choice of exterior cosmology.

If the cosmology is closed, all constant-mean-curvature slices inside the cloud and Schwarzschild regions connect smoothly to exterior cosmological slices \citep{QadirWheeler1985}. This is one of the features of York-time slicing explored in chapter \ref{chap:Yorktime}. If the cosmology is flat, only slices up to $K=0$ ($T=0$) connect up smoothly and slices with $K>0$ only exist in disconnected patches around the local singularities \citep[sections\ VI, VII]{BrillCavalloIsenberg1980}. An observer falling into the black hole reaches the singularity in finite proper time and therefore crosses all constant-mean-curvature surfaces up to $T=\infty$ in finite proper time. If York time has fundamental physical significance, reaching the singularity constitutes arriving at the end of the universe at $T\rightarrow\infty$ no matter the exterior cosmology. Therefore, without the extension, different areas of the universe would end at different times: at $T_{end}=0$ far away from the singularity, but at $T_{end}=\infty$ in its vicinity.

Classically this does not immediately cause a contradiction, at least not for physical reasons. However, quantisation of the theory is now a problem. The quantum theory is necessarily non-local --- this was one reason motivating the idea of a physically fundamental space-time split in the first place --- and non-local effects are incompatible with partial spatial slices of this kind. 

For example, consider a pair of entangled particles, one of which enters the Schwarz\-schild radius while the other remains in some distant part of the universe. When the falling particle reaches $T=0$ its partner encounters the end of time. What happens to the quantum state at this point is entirely unclear. At $T=0$ the configuration space (or Hilbert space) would suddenly change its dimensionality since the space on which the dynamical entities such as matter fields are defined shrinks to a set of disconnected finite patches. This does not allow for a consistent definition of the quantum theory. Furthermore, since the falling particle crosses into the region $T>0$ in finite proper time this scenario cannot be dismissed as unphysical. Our theory is furthermore supposed to be fundamental, so its validity cannot be limited to the exteriors of black holes, for example.

Therefore there is a choice: Either York time is rejected as the fundamental time parameter, or the consistency of quantisation demands the existence of the extension. In this thesis we are interested in exploring the consequences of considering York time as physically fundamental and so we require the cosmological extension.

\section{Dynamics on the other side} \label{sec:dynamicsonotherside}

The equations \ref{eq:5.3-phiprime}, \ref{eq:5.3-pphiprime} governing $\phi$ and its conjugate momentum $p_\phi$ are invariant under temporal reflection about $T=0$ (with the direction of time unchanged, so that $dT$ does not transform). Furthermore, the sign of $\phi^\prime$ is the same as the sign of $p_\phi$. Therefore the set of solutions for the interval $T\in(0,\infty)$ is exactly the same set of trajectories as for the interval $T\in(-\infty,0)$ except that the latter are traced out `backwards' and the sign of the momentum is flipped. A solution $(\phi(T),p_\phi(T))$ for $T<0$ may be used to define a corresponding solution $\tilde{\phi}(T), \tilde{p}_\phi(T)$ for $T>0$ via initial data at $T_0>0$ given by 
\begin{equation}\tilde{\phi}(T_0)=\phi(-T_0),\quad\tilde{p}_\phi(T)=-p_\phi(-T),\end{equation} 
as can be easily verified. 

The equivalence between equations \ref{eq:5.3-phiprime} and \ref{eq:5.3-pphiprime} together with $H=a^3$, and the conventional Friedmann-Lema\^itre and Klein-Gordon equations (eqs.~\ref{eq:5.2-CosmoFriedmann1}, \ref{eq:5.2-CosmoFriedmann2}, \ref{eq:5.2-CosmoKG}) implies that we should also be able to understand this symmetry in terms of the latter, even though the cosmological time parameter $t$ does not cover the period $T>0$. Instead we are able to define another analogous cosmological time parameter $\tilde{t}$ such that $\tilde{t}\rightarrow-\infty$ as $T\rightarrow0_+$ and $\tilde{t}\rightarrow0$ (or some other finite value) as $T\rightarrow\infty$. This parameter $\tilde{t}$ does not describe the period $T<0$. Equations \ref{eq:5.2-CosmoFriedmann1}, \ref{eq:5.2-CosmoFriedmann2} and \ref{eq:5.2-CosmoKG} are invariant under reversal of the direction of time, $dt\rightarrow-dt$. One can therefore see that if $t$ is mapped to $-\tilde{t}$ as part of the correspondence between solutions then these equations remain satisfied for the time variable $\tilde{t}$ instead of $t$.

However, note that in general for more complicated matter content there is no reason to expect that the universe will simply `turn around' and trace out its trajectory backwards, even if there is a one-one correspondence between solutions on `our side' and the extension.

\section{Transition behaviour and classification of potentials} \label{sec:transition}

York time $T$ is proportional to the Hubble parameter, so that $T\rightarrow0$ as $t\rightarrow\infty$ only if $\frac{\dot{a}}{a}\rightarrow0$, which is the case, for example, in the presence of a free scalar field or a field with a potential which goes to zero sufficiently fast as the end of conventional cosmological history is approached. Specifically, note that $\frac32M_{Pl}^{-2}T^2-2V>0$ as we saw above. With $T\rightarrow0$, this implies that if $V\geq0$ everywhere, then $V\rightarrow0$ faster than $\frac34M_{Pl}^{-2}T^2$ if $T$ is to go to zero. Otherwise, in the presence of a positive cosmological constant, for example, $T$ does not reach zero in finite cosmological time. While in that case there are still corresponding solutions on the `other side', there is an intermediate interval $T\in(T_\infty,-T_\infty)$ where the dynamics is not defined.

\begin{figure}[bth]
 \includegraphics[width=\linewidth]{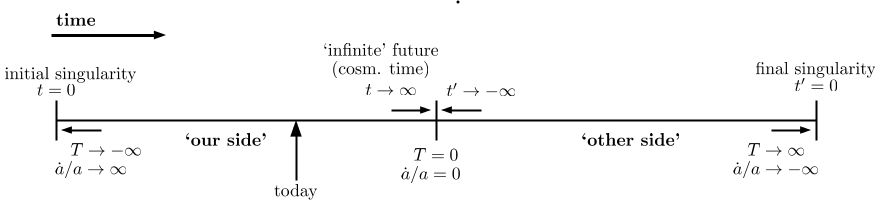}
 \caption[The extended York timeline with $\Lambda =0$]{Overview of corresponding values of cosmological time $t$ and York time $T$ in the case of a free scalar field. Only the left half of the time line is described by conventional cosmological time for a flat universe. The right hand side is however a natural extension if the York parameter is taken seriously as a fundamental time parameter. The finite `starting' and `end' values of $t$ and $t^\prime$ respectively are up to convention (due to cosmological-time translation invariance).}
  \label{fig:8.4-free}
\end{figure}

We are interested in identifying the class of potentials for which there is a smooth transition from the $T<0$ to the $T>0$ region. Specifically, what we mean by this is that 
\begin{itemize}
 \item the dynamical equations \ref{eq:5.3-phiprime}, \ref{eq:5.3-pphiprime} describe the behaviour of the scalar field(s) for the entire real line and there is no intermediate interval around $T=0$ where $\frac32M_{Pl}^{-2}T^2<2V(\phi)$,
 \item the energy density associated with the scalar field and the field value $\phi$ itself remain finite everywhere.
\end{itemize}
The first condition is equivalent to the question in the cosmological-time picture whether or not there are solutions to eq.\ \ref{eq:5.2-CosmoFriedmann1} for all values of $\dot{a}/a$. For example, the dynamics of the free field ($V(\phi)=0$) can be analytically solved,\footnote{The relation $\phi_{free}^\prime\sim|T|^{-1}$ illustrates a feature of the York time picture: More and more `happens' in a finite time interval the closer we are to $T=0$. As we discussed in section \ref{sec:cosmhist} a similar observation was made about \emph{cosmological time} regarding the very \emph{early} history (close to $t=0$) in the 1960s in \citep{Misner1969c}, who suggested that $-\ln a$ (or alternatively the logarithm of the homogeneous temperature, which turns out to roughly equivalent) might be a more appropriate choice of temporal parameter to describe the history of the universe. This is further exemplified in our language to describe the early universe in terms of `epochs' such as the `Planck', `Grand Unifying' and `Inflationary' epochs, each of which is described by a vastly different order of magnitude of duration. Another comparable choice is the parameter $\ln t$ as had been advocated by Milne another twenty years earlier (cited in \citep{Misner1969c}).}
\begin{align} 
 \phi_{free}^\prime &= sign(p_\phi^{free})\cdot\big(\tfrac32M_{Pl}^{-2}T^2\big)^{-\frac12},\qquad\qquad p_\phi^{free} = \text{const.} \notag\\
		    \Rightarrow\quad\phi_{free} &= \tfrac{\sqrt{3}}2M_{Pl}\ln|T|
\end{align}
for all $T<0$ (fig.\ \ref{fig:8.4-free}). The energy density is $\sim\dot{\phi}^2\sim T^2\rightarrow0$ and therefore obviously finite. However, $\phi$ diverges logarithmically with its sign equal to the sign of the initial value $p_\phi^{free}$. 

\begin{figure}[bth]
 \includegraphics[width=\textwidth]{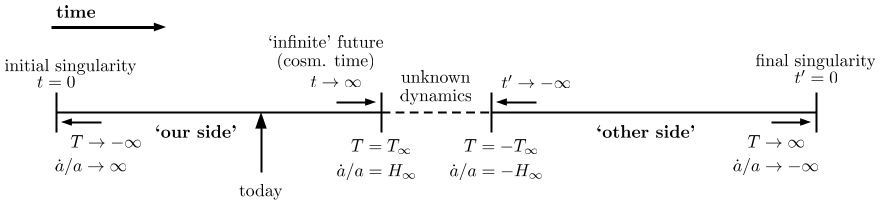}
 \caption[The extended York timeline with $\Lambda>0$]{Overview of corresponding values of cosmological time $t$ and York time $T$ in the case of a scalar field with a positive cosmological constant. The left section of the line is described by conventional cosmological time $t$, the right one by a similar parameter $t^\prime$. In the middle section, $T_\infty<T<-T_\infty$, the dynamics of the scalar field is undefined, raising philosophical questions.}
 \label{fig:8.4-cosm-const}
\end{figure}

On the other hand, consider the simple case of a positive cosmological constant, $V(\phi)=2M_{Pl}^2\Lambda$, $\Lambda>0$, and no other potential term. In terms of cosmological time, this is associated with an eternally expanding universe with $\dot{a}/a$ asymptotically approaching a finite value proportional to $\Lambda$ (de~Sitter expansion). In the York-time picture, $\frac32M_{Pl}^{-2}T^2-2V(\phi)>0$ only up until $T_\infty=-\frac16M_{Pl}^2\Lambda$ and again after $-T_\infty=+\frac16M_{Pl}^2\Lambda$. In the intermediate period the dynamics is ill-defined since $\phi$ was taken to be a real field (fig.\ \ref{fig:8.4-cosm-const}). While the dynamical equations can indeed be applied to $T>-T_\infty$, the transition is not smooth. We will provide a discussion of the meaning and implication of such an intermediate period below.

A non-trivial example of a potential which leads to a well-defined transition, but with diverging field value, is given by $V(\phi)=V_0e^{-\lambda\phi/M_P}$, where $V_0$ and $\lambda$ are constant parameters. Potentials of this form may arise in the effective four-dimensional dynamics induced by Kaluza-Klein theories, for example. A discussion of this potential in terms of conventional cosmological time $t$ is given in \citep{FerreiraJoyce1998} and references therein, in particular \citep{Halliwell1987}. This model can be solved exactly. Make the ansatz $\phi(T)=\phi_0+\alpha\ln|T|$, then realising that $dV/d\phi$ is monotonic one can infer that there is a time after which there is no further turning point, so that $sign(p_\phi)$ remains fixed and equation \ref{eq:5.3-phiprime} effectively decouples from the momentum. Solving the equation with the proposed ansatz one obtains the solution
\begin{equation}\label{eq:EXTENSION.4-exponentialpotentialsolution}
 \phi(T) = \frac{M_{Pl}}{\lambda}\ln\left(\frac{8V_0M_{Pl}^2}{(6-\lambda^2)}\frac{1}{T^2}\right),
\end{equation}
which diverges as $T\rightarrow0$. However, 
\begin{equation}\label{eq:EXTENSION.4-exponentialpotentialpotential}
 V(T) = \frac{(6-\lambda^2)T^2}{8V_0M_{Pl}^2}
\end{equation}
approaches zero in this limit, as required for the existence of a well-defined transition. Note that due to the fact that $T$ is physically meaningful there is no time-translation invariance (see \citep{RoserValentini2014a}).

A simple example of a fully smooth transition (with $\phi$ remaining finite) is provided by a scalar field with a positive mass term, $V(\phi)=m^2\phi^2$, $m^2>0$ (which is a version of large-field inflation \citep{MartinEtAl2013_EncyclopaediaInflationaris}, albeit one not favoured by the most recent data \citep{Planck2015_Overview}). One does not need to solve the equations in order to analyse the behaviour as $T\rightarrow0$. In the late-time limit one expects oscillatory behaviour with an amplitude decreasing over time. Indeed, $\phi$ begins by rolling down the potential (if its initial conditions are such that it rolls up, then eq.\ \ref{eq:5.3-pphiprime} ensures that $p_\phi$ will pass through zero at some point, a turning point, and $\phi$ changes direction) and passes through the minimum, at which point $p_\phi^\prime$ changes sign (eq.\ \ref{eq:5.3-pphiprime}). The momentum $p_\phi$ will therefore pass through zero eventually, so that a turning point results and $\phi$ reverses sign. This repeats, resulting in oscillatory behaviour. The amplitude of the oscillations decreases since $\frac32M_{Pl}^{-2} T^2$ monotonically decreases, so the turning point (recall that these can be identified by the condition $\frac32M_{Pl}^{-2}T^2-2V(\phi)=0$) are reached at a lower value of $V(\phi)$ and therefore at a lower value of $\phi^2$ than during the previous cycle. It is noteworthy that in the cosmological-time picture appeal to `Hubble friction' must be made (the Hubble parameter appears as a frictional coefficient in eq.\ \ref{eq:5.2-CosmoKG}), whereas in the York-time equations derived via the Hamiltonian reduction the decreasing amplitude is encoded in the explicit time dependence.

During the oscillations the field $\phi$ remains finite since it is bound by the field value at the most recent turning point from above and below. Both potential and kinetic energy density furthermore approach zero, so the finiteness of the energy density $\rho_\phi$ is guaranteed. In fact, we know that for oscillatory solutions that $\rho_\phi\sim a^{-3}$ (see \citep{FerreiraJoyce1998}), so even the total energy remains finite if initially normalised.

Let us now attempt a more general characterisation. We restrict our discussion to potentials bounded from below. For consider a potential without a lower bound. If there is a local minimum the field may, initial conditions permitting, become trapped in the well and the potential may be locally approximated by another potential bounded from below. If on the other hand no such minimum exists or the initial conditions are such that the field does not become trapped inside one, then $V(\phi)$ reaches arbitrarily negative values, so that $\frac32M_{Pl}^{-2}T^2-2V(\phi)>\epsilon$ as $T\rightarrow0$ for some finite $\epsilon>0$. Therefore $|\phi^\prime|$ is bounded for all $T$ greater than some reference time $T_r$. Since the interval $(T_r,0)$ is finite, $\phi$ remains finite also. The `transition' is perfectly smooth. However, it is also not a novelty since it is reached in finite cosmological time \citep{GiamboMiritzisTzanni2014}. That is, the universe recontracts even in the conventional description. No new temporal region is revealed through the use of York time.

Therefore, assume there exists $\alpha\in\mathbb{R}$ such that $\alpha=\inf V(\phi)$. Table \ref{table:6.4-TofB} summarises the implications for the transition for different cases.

 
 {\hbadness=10000 
\begin{table} 
\centering
 \resizebox{\linewidth}{!}{
\begin{tabular}{|p{0.15\linewidth}|m{0.10\linewidth}|m{0.25\linewidth}|m{0.25\linewidth}|m{0.25\linewidth}|}\hline
   \multicolumn{1}{|c|}{\multirow{2}{*}{\textbf{Value of $\inf V(\phi)$}}} & \multicolumn{1}{|c|}{\multirow{2}{*}{\textbf{Examples}}} & \multicolumn{1}{|c|}{\textbf{Value of $T_\infty$}} 
	  & \multicolumn{2}{|c|}{\textbf{Smoothness of transition}} \\\cline{4-5} 
    & & \multicolumn{1}{|c|}{\textbf{(Existence of Transition)}} & \multicolumn{1}{|c|}{\textbf{Field value finite? ($|\phi|<\infty$)}} 
	  & \multicolumn{1}{|c|}{\textbf{Energy density finite?}} \\\hline 
   $\inf V(\phi) < 0$ 
	& As below but with\newline cosm.\ const.\newline ($\Lambda<0$)
	& \textbf{None.} $T_\infty$ does not exist, the universe recontracts in finite cosmological time. No extension can be made. The null energy condition is violated.
	& \begin{center}\textbf{No real transition.}\newline (Extension does not exist.)\end{center}
	& \begin{center}\textbf{No real transition.}\newline (Extension does not exist.)\end{center} \\ \hline
   $\inf V(\phi) = 0$ 
	& AI, HI,\newline MHI, RGI,\newline SBI
	& $\mathbf{T_\infty=0}$ \textbf{(transition exists)}, unless there is a local minimum with $V(\phi_{min})>0$ in which the field is trapped, in which case behaviour as if $\inf V(\phi)>0$.
	& \textbf{Possibly:}  Yes iff there is an infinite number of turning points (i.e.\ oscillating solutions, requires existence of a minimum) (e.g.\ HI, MHI, RGI, SBI). Otherwise $\phi\rightarrow\pm\infty$ at least logarithmically (e.g.\ AI).
	& \textbf{Yes.}\newline Example: oscillating solutions in $V\sim|\phi|^n$: $\rho_\phi\sim a^{-m}$, $m=6n/(n+2)$ \citep{FerreiraJoyce1998}  \\\hline
   $\inf V(\phi) > 0$ 
	& As above but with\newline cosm.\ const.\newline ($\Lambda>0$)
	& $\mathbf{T_\infty<0}$ \textbf{(intermediate epoch of ill-defined dynamics)}. C.f.\ model with a cosmological constant.
	& \begin{center}\textbf{No transition.}\end{center} (Epoch of ill-defined dynamics before other side reached.)
	& \begin{center}\textbf{No transition.}\end{center} (Epoch of ill-defined dynamics before other side reached.)\\\hline
\end{tabular}
 }
\caption[Transition behaviour for potentials bounded during the transition into the extended York-time cosmology]{\textbf{Transition behaviour for potentials bounded from below.} The second entry in each line gives examples of potentials with the infimum as given in the first column from the most favoured inflation potentials based on the 2013 Planck data as identified in \citep{MartinEtAl2013} (although here we do not differentiate between subcategories of a particular type of potential based on parameter-value ranges), namely Arctan Inflation (AI), Higgs Inflation (HI), Mutated Hilltop Inflation (MHI), Radion Gauge Inflation (RGI) and Supergravity Brane Inflation (SBI). The third column identifies if York time reaches $T=0$, the condition for a transition to the other side to exist without an intermediate epoch of ill-defined dynamics. The last two columns provide information on the smoothness of the transition.}
\label{table:6.4-TofB}
\end{table}
 }

 \section{Discussion}\label{sec:othersidediscussion}
The matter content of the real universe is considerably more complicated than a single scalar field. However, during inflation a scalar field may have constituted the dominant constituent of the matter content. Furthermore, with all other contributions to the energy density falling to zero in the late-time limit of the universe, a non-zero contribution from a scalar field potential may become once again significant. Therefore considering a scalar-field model in order to explore the `other side' is appropriate.

The question of the significance of the transition properties of such potentials arises. Indeed, one may dismiss the cosmological extension presented here as purely mathematical speculation without any physical significance at all. However, if one considers the possibility that York time may indeed play the role of a fundamental time parameter that forms the basis of quantisation (where, as we showed, the extension is required for consistency), then the existence of models (potentials) with an intermediate period around $T=0$ of ill-defined dynamics requires consideration. One may take one of the following stances:\footnote{Not including more radical attitudes such as the view that during the intermediate epoch the dynamical evolution of matter and space cannot be derived from a variational principle, or that our mathematical tools are in some way insufficient to provide a description of the dynamics at these times. Whether or not such a view is viable is a philosophical question outside the scope of this thesis.}
\begin{enumerate}
 \item Potentials which lead to such an intermediate period are ruled out by a theory in which York time is considered fundamental. Conversely, if observational evidence suggests that inflation, for example, is driven by such a potential, then this would falsify a theory in which York time is taken to play a physically fundamental role.
 \item Time may end at some finite value, so there is no contradiction in regarding York time as physically meaningful but without taking the extension seriously. The argument of section \ref{sec:reasonforextension} is avoided by postulating some different, non-equivalent and at this stage unknown quantisation procedure which allows time to end at different values in different regions of space, or a change in the classical theory that avoids the `patchy' time slices for $T>0$. The resulting quantum theory would likely be not equivalent to standard quantisation and observational corroboration may be possible.
\end{enumerate}
The second point of view may be found viable for example in the context of Shape Dynamics \citep{GomesGrybKoslowski2011, BarbourKoslowskiMercati2013ProbOfTime,Mercati2014} (which we touched on on multiple occasions in part \ref{part1} of this thesis), where what is presumed physical is a sequence of configurations (that is, a trajectory in configuration space) without any physically fundamental time. 
There a preferred time parameter is obtained by the requirement that the dynamics be expressed in terms of dimensionless, relational quantities only \citep{BarbourKoslowskiMercati2013ProbOfTime}. The time parameter is not physically fundamental but instead supervenes on the properties of the system. Therefore there is no \emph{a priori} imperative that this time parameter should constitute a surjection from the configuration-space trajectory onto the real line. 
Furthermore, Shape Dynamics differs from general relativity in its description of black holes \citep{Gomes2014,GomesHerczeg2014}, so the theory may escape the argument outlined in section \ref{sec:reasonforextension}. In fact, in our opinion this view is the strongest reason \emph{not} to believe in the physicality of our proposed extension. Note though that Shape Dynamics is a different physical theory, which, while locally equivalent to general relativity in the constant-mean-curvature gauge, does differ in the global structures it allows and so is at least in principle\footnote{The observer would have to travel through the event horizon of a black hole in order to establish a difference.} observationally distinguishable. Hence, despite its similarities, it is not the focus of this paper. Meanwhile the view adopted here, ascribing a more fundamental role to time, remains viable at least until there is sufficient observational evidence to strongly favour theories (of quantum gravity, for example) which rely on a purely relational notion of time.

If we adopt the first point of view then a theoretical framework in which York time is fundamental imposes a restriction on inflationary potentials (or, more accurately, the complete matter content of the universe) and is therefore experimentally testable, provided cosmological flatness ($\Omega_k=0$) is conclusively established. At the time of writing the list of observationally unfalsified inflationary potentials is still too long to make any definite conclusions in this context \citep{MartinEtAl2013} and there is no obvious correlation between potentials allowing for a smooth transition and those favoured by data on inflation. However, it will be interesting to see whether `York-time friendly' potentials are favoured or not as data improves. It is, however, important to bear in mind that some inflaton potentials are merely effective (derived from some underlying supersymmetry or string model, for example) and their form may not stay the same throughout the entire history of the universe. A more careful analysis of individual models is necessary.

At this stage, whether or not we should give credence to predictions concerning the unobservable epoch $T>0$ depends on the existence of other evidence for York time as a physically fundamental time parameter. In this thesis we have argued that there are certainly are theoretical reasons. If dramatic progress is made in the development of some theory of quantum gravity that relies on York time, and observational evidence emerges that corroborates this theory, then this would be reason to take the extension seriously. Indeed it would be difficult to escape its necessity.


\chapter{Cosmological perturbation theory with York time: preliminaries}\label{chap:perttheoprelims}

\textit{In this chapter we lay the foundations for the development of the Hamiltonian-reduced York-time cosmological perturbation theory to be developed in the next chapter. We discuss some formal properties relating to the Poisson structure by looking at an anisotropic minisuperspace model and we consider some physical processes such as the `freezing out' and `Hubble re-entry' of cosmological perturbations in the York-time description. The contents of section \ref{sec:ClassKasner} were published in ref.~\citep{Roser2015a}, while sections \ref{sec:modefreezing} and \ref{sec:modereentry} were contained in ref.~\citep{RoserValentini2016}.}

\section{The importance of cosmological perturbation theory}\label{sec:pertintro}

Very broadly speaking, the study of cosmology can be divided into two areas: the study of the homogeneous `background', concerning global properties of the universe, and the study of small-scale inhomogeneities. The former is logically prior to the latter in the sense that the evolution of the background (or at least certain quantities describing aspects thereof) has to be sufficiently well understood before one can embark upon the study of small inhomogeneities on that background. However, \emph{ex hypothesi} there is no `backreaction' of the inhomogeneities on the cosmological background. The background value of some physical quantity (such as the value of some matter field $\phi$, say) is chosen to be the global average, for example, so that if one were to find that the local change in the field $\delta\phi(x)$ is positive everywhere, it would be a sign that the `wrong' background was chosen.

It is not guaranteed \emph{a priori} that this model of cosmology --- background plus perturbations --- adequately describes our actual universe. In fact, on sufficiently small scales (such as on the scale of the Earth-Moon system) it does not. Given what we know there is however strong evidence that this model does provide a viable approximation of the early universe, as well as today on sufficiently large scales ($\geq$ gigaparsecs). The evolution of small perturbations into large non-linear\footnote{`Non-linear' meaning that linearised (first-order) forms of the dynamical equations of gravity are not good approximations to describe the evolution.} inhomogeneities such as planets, galaxies and black holes (`structure formation') is a significant area of research, albeit one that we cannot do justice here.

The full equations of general relativity are highly non-linear and therefore in general difficult to solve analytically. Numerical relativity, the analysis of gravitational dynamics via computer simulation has become its own field of study in modern theoretical physics. Perturbation theory offers a method to reduce the complexity of the equation and therefore allows one to find approximate analytic solutions, at least for certain regimes. The fact that the early universe is such a regime is a fortunate `coincidence'.  

Other than through structure formation, perturbation theory may be tested experimentally via studies of the cosmic microwave background (CMB). Data from missions such as WMAP (Wilkinson Microwave Anisotropy Probe \citep{WMAP2007}) and the Planck spacecraft \citep{Planck2015_Overview} have substantially increased our understanding of the early universe within the last decade. The CMB provides a `picture' of the universe at ca.~379,000 years after the Big Bang when photons decoupled (photon-matter interactions became rare), leaving the universe transparent. The relative late time of this event limits what is known quantitatively about the universe prior to that, which is for the most part inaccessible.\footnote{A proposed method to probe the universe at earlier times is by measurements of the cosmic neutrino background (CNB). Neutrinos decoupled about one second after the big bang and would therefore provide a much earlier snapshot of the universe. However, technologically accurate measurements of the CNB are difficult at this point, although near-future experiments are not completely inconceivable \citep{FaesslerEtAl2013}.} In order to test some theory of the early universe, one is therefore forced to evolve the universe forward to the point of photon decoupling, which in itself brings theoretical and numerical uncertainties.

Finding discrepancies between predicted data of the quantised York-time reduced-Hamiltonian theory and more conventional (quantised) general relativity or other approaches to quantum gravity, remains work to be carried out in the future. In this thesis we contend with developing the formal structure of the reduced-Hamiltonian description of cosmological perturbations and gaining an understanding in terms of York time of key physical processes such as the `freezing out' of (Fourier modes of) perturbations during inflation. This is the task undertaken in this chapter and the next. A feature of the reduced theory that is relevant here is the non-canonical Poisson structure of the reduced variables, which we study in isolation in section \ref{sec:ClassKasner}. In sections \ref{sec:modefreezing} and \ref{sec:modereentry} we study the `freezing' and `Hubble re-entry' of modes relating to scalar and tensor perturbations. In chapter \ref{chap:perttheoformalism} we perform the Hamiltonian reduction with respect to York time in the perturbative regime.

\section{Classical anisotropic cosmology: Exploration of the Poisson structure}\label{sec:ClassKasner}

In chapter \ref{chap:Friedmann} we developed in detail the classical theory of the cosmological background for a homogeneous and isotropic universe in terms of York time. There was only one canonical pair of variables associated with the geometric degrees of freedom, the scale $a$ and its conjugate momentum $p_a$. These variables were eliminated in the Hamiltonian reduction in favour of physical time $T$ and its associated Hamiltonian. All \emph{anisotropic} degrees of freedom disappeared due to the symmetry assumptions, as did resultingly the non-canonical aspect of the Poisson structure that we discussed in section~\ref{sec:Poisson}. 

In this section we focus on understanding the consequences of the non-canonical Poisson brackets for the classical and quantum theory. In particular we wish to isolate this feature of our theory as far as possible from other mathematical aspects of our theory and we therefore choose the simplest conceivable non-trivial model exhibiting this structure, an anisotropic minisuperspace model. In chapter~\ref{chap:perttheoformalism} we will develop perturbation theory with the full Poisson brackets~\ref{eq:4.3-ggPB} - \ref{eq:4.3-pipiPB}. Since the perturbation theory is truly a field theory (the geometric variables have spatial dependence) we will encounter a number of new complications. It is therefore extraordinarily useful to first study the meaning and implications of the Poisson structure in isolation in the context of a finite-dimensional model.

The spatial metric $g_{ab}$ and its associated momentum $\pi^{ab}$ each have six independent components. Their determinant and trace respectively encode the overall scale and its evolution, the ratios of the diagonal entries relate to the relative scale in different spatial directions and the three off-diagonal elements concern the notion of angles. (Recall section~\ref{sec:PoTinGR} where we introduced the notion of a metric as a quantity defining the scalar product of two vectors in order to see this.) Of course, a change in the coordinates used (or a change in the frame of reference) changes what constitutes the spatial directions and can therefore mix diagonal and off-diagonal elements. However, in the case of a homogeneous minisuperspace model it makes sense to choose a metric representation that is itself homogeneous. This is only possible in the case of a spatially flat cosmology and this is what we choose to develop here. The coordinate system used in this case is just the Cartesian one.\footnote{In spatially open and closed cosmologies one would have to use hyperspherical coordinates, analogous to spherical polar coordinates in Euclidean space. In flat space such a representation is also possible, but the Cartesian one is simplest.} In order to ensure maximum simplicity, we choose to set all off-diagonal metric elements to zero, so that there are only three geometric degrees of freedom, essentially corresponding to how space is `stretched' along the three Cartesian axes. 

The resulting universe has little to do with the actual universe where no global anisotropy is observed.\footnote{A recent study based on the Planck data calculated that the odds that there is a preferred direction in the CMB are roughly $121,000:1$ against \citep{SaadehEtAl2016}.} Furthermore, the model developed here exhibits no stability around the configuration-space point corresponding to isotropy, as we will see. As a description of reality, the model is utterly abysmal. However, much can be learned from its mathematical features. In general General Relativity (without Hamiltonian reduction) the model is known as the `Kasner models' \citep{MisnerThorneWheeler1973}.

Introduce variables $Q_i\equiv g_{ii}$ and $P^i\equiv\pi^{ii}$. The metric determinant is $g=Q_1Q_2Q_3$ and the trace of the momentum is $\pi=Q_iP^i$. The York parameter is then the usual $T=2\pi/3\sqrt{g}$ and its conjugate momentum is $P_T=-\sqrt{g}$. The reduced variables are
\begin{equation} q_i = g^{-\frac13}Q_i,\qquad p^i=g^\frac13(P^i-\tfrac13\pi Q^i),\end{equation}
where $Q^i \equiv 1/Q_i$ denotes the $ii$-component of the inverse metric $g^{ij}$. For convenience we also introduce the inverse of the reduced metric variables, $q^i\equiv 1/q_i$. Because of the `compression' of two indices ($g_{ij},\pi^{ij}$ and $\tilde{g}_{ij},\tilde{\pi}^{ij}$) into one ($Q_i,P^i$ and $q_i,p^i$), we employ the following summation convention for the purposes of this chapter: Indices are summed over if they appear at least once as an upper and a lower index each. Indices may appear multiple times as either upper or lower indices only however without implying a summation.

The Poisson brackets derived from the general ones for $\tilde{g}_{ij}$ and $\tilde{\pi}^{ij}$ (eqs.~\ref{eq:4.3-ggPB} - \ref{eq:4.3-pipiPB}) are exactly those of the simple finite-dimensional model considered in the same section, namely
\begin{align} 
 \{q_i,p^j\} &= \delta_i^j-\frac{q_i}{3q_j}\tag{\ref{eq:4.3-modelqpPB}} \\
 \{q_i,q_j\} &= 0 \tag{\ref{eq:4.3-modelqqPB}} \\
 \{p^i,p^j\} &= \frac{p^i}{3q_j}-\frac{p^j}{3q_i}. \tag{\ref{eq:4.3-modelppPB}}
\end{align}
In virtue of their definition the reduced variables obey the constraints
\begin{align} q_1q_2q_3&=1, \label{eq:7.2-scaleconstraint} \\ q_ap^a &=0. \label{eq:7.2-tracefreeconstraint} \end{align}
The tracelessness \ref{eq:7.2-tracefreeconstraint} of $p^a$ ensures that the first constraint \ref{eq:7.2-scaleconstraint}, the scale-free condition, is preserved. In fact, the constraints are both first class. Furthermore, the constraints \ref{eq:7.2-scaleconstraint} and \ref{eq:7.2-tracefreeconstraint} are preserved by any motion generated by the momenta via this Poisson structure \ref{eq:4.3-modelqpPB}, \ref{eq:4.3-modelppPB}. That is, for a small vector $\epsilon_b$, the constraints are preserved under the transformation
\begin{equation} q_a\rightarrow q_a+\epsilon_b\{q_a,p^b\},\qquad p^a\rightarrow p^a+\epsilon_b\{p^a,p^b\} \end{equation}

At this point one may ask if the non-canonical structure of the Poisson brackets is necessary following the Hamiltonian reduction, or if it is only the result of a poor choice of coordinates and that a suitable coordinate transformation of the reduced variables will result in standard canonical Poisson brackets. However, no such variables exist. One way to see this is algebraically by performing a general coordinate transformation, demanding that the new coordinates satisfy canonical commutation relations and showing that no such coordinate transformations can be found. A more geometric way to see that the coordinates must be non-canonical is to look at the phase space constructed. Since the momenta all generate motion within the two-dimensional constraint surface the space spanned by the momenta is not the cotangent space of the kinematical configuration space, which is three-dimensional. The underlying reason for this is that as part of the Hamiltonian reduction the momenta are \emph{constructed} to generate motion within the constraint surface.

In terms of $q_a,p^a$ the Hamiltonian constraint takes the form
\begin{equation}\label{eq:4.3-anisotropicHamiltonianConstraint} 
  0=\mathcal{H}=2M_{Pl}^{-2}\left[ -\frac{\pi^2}{6\sqrt{g}}+\frac{1}{\sqrt{g}}q_a^{2}p^{a2}\right]=2M_{Pl}^{-2}\left[\tfrac38T^2P_T-\frac{1}{P_T}q_a^2p^{a2}\right].
\end{equation}
The physical Hamiltonian associated with York time is given by $H_{phys}=-P_T(q_a,p^a,T)$, where $P_T(q_a,p^a,T)$ is the function obtained when solving the Hamiltonian constraint for $P_T$ in terms of the other variables. In the full theory the analogous equation is a difficult elliptic equation with no known general solution. Here however it is a simple quadratic, yielding
\begin{equation}\label{eq:7.2-ClassicalHamiltonian} H_{phys}\equiv-P_T=\pm\left[\frac{8}{3T^2}q_a^2p^{a2}\right]^\frac12.\end{equation}
As discussed in chapter \ref{chap:Friedmann} the choice of sign is not physical. For any given physical trajectory corresponding to one sign choice there is a corresponding solution for the other sign choice, characterised by $q_a(T)\rightarrow q_a(T)$, $p^a(T)\rightarrow-p^a(T)$. Since the physical interpretation of the numerical value of $H_{phys}$ is that of `volume' however and volume is conventionally defined as positive, we again assume the positive sign in eq.~\ref{eq:7.2-ClassicalHamiltonian}.

The Hamiltonian is of geodesic form $N(T)(G_{ij}p^ip^j)^\frac12$. The solutions may therefore be understood geometrically as geodesics in configuration space with respect to the configuration-space metric $G_{ij}\propto q_iq_j\delta_{ij}$. Geodesic form is, in fact, a general feature of York-time reduced Hamiltonians.

Motion generated by the Hamiltonian, $q_a\rightarrow q_a+\delta T\{q_a,H\}$, $p^a\rightarrow p^a+\delta T\{p^a,H\}$, also maintains the constraints, although the tracelessness constraint \ref{eq:7.2-tracefreeconstraint} is required to show that the scale-free condition \ref{eq:7.2-scaleconstraint} is preserved. Specifically, the equation of motion are
\begin{align} 
 q_a^\prime  &= \{q_a,H\} = \sqrt{\frac{8}{3T^2\cdot q_k^2p^{k2}}}\cdot q_b^2p^b\delta^b_a \label{eq:7.2-qeom}\\
 p^{a\prime} &= \{p^a,H\} = -\sqrt{\frac{8}{3T^2\cdot q_k^2p^{k2}}}\cdot q_bp^{b2}\delta^a_b. \label{eq:7.2-peom}
\end{align}

Explicit solutions to equations \ref{eq:7.2-qeom}, \ref{eq:7.2-peom} may be found by inspection using the fact that $p^{b\prime}=-q^bp^b\delta^{ab}q_a^\prime$. One finds solutions
\begin{equation} q_a(T)=(-4/3T)^{2(s_a-\frac13)}, \qquad p^a(T) = (s_a-\tfrac13)(-4/3T)^{-2(s_a-\frac13)}, \label{eq:7.2-classicalsolutions} \end{equation}
with constant parameters $s_a$ satisfying $s_1+s_2+s_3=1$, $s_1^2+s_2^2+s_3^2=1$. These solutions are exactly the Kasner models. In order to see this and to get a better intuition of the relation of York time $T$ and standard cosmological time $t$ in these models, recall that in the Kasner models $g=t^2$ and the general fact that $T$ was defined as$-4/3$ times the fractional rate of change of volume, so that
\begin{equation} T= -\frac{4}{3t}. \end{equation}
This makes it apparent that \ref{eq:7.2-classicalsolutions} are indeed the Kasner solutions \citep{MisnerThorneWheeler1973}. The value of $H_{phys}$ is given by $-P_T=\sqrt{g}$, so that $H_{phys}=t$ --- cosmological time is just the numerical value of the physical Hamiltonian.

The fact that one obtains exactly the same solutions provides further illustration of the consistency of the reduced formalism.

A cosmological constant may be included in the above formalism, leading to the substitution $(8/3T^2)\rightarrow(\frac83T^2-2\Lambda)^{-1}$ in $H_{phys}$. The solutions of the equations of motion are then 
\begin{align}
 q_a(T) &= \gamma_a \left\vert T+\sqrt{T^2-\frac{16}{3}\Lambda}\right\vert^{+2(s_a-\frac13)} \label{eq:7.2-qsolution}\\
 p^a(T) &= \gamma_a^{-1} (s_a-\tfrac13) \left\vert T+\sqrt{T^2-\frac{16}{3}\Lambda}\right\vert^{-2(s_a-\frac13)} \label{eq:7.2-psolution}
\end{align}
where the parameters $s_a$ satisfy the same condition as in the Kasner model, $\gamma_a$ are constants chosen to satisfy eq.\ \ref{eq:7.2-scaleconstraint} and $T$ is restricted to $\frac38T^2>2\Lambda$.

The model developed in this section is not a good description of our universe. Global degrees of isotropy have not been detected. However, it helped to illustrate the mathematical interplay of the Poisson brackets and the constraints, insights that are useful in the perturbation theory to be developed in chapter \ref{chap:perttheoformalism}. However, it is first useful to gain some intuition for the mechanisms accounting for certain physical processes in the early universe when described in terms of York time. This is the subject of the rest of this chapter.

\section{Mode freezing during inflation}\label{sec:modefreezing}

Inflation originated as an explanation for various puzzles relating to cosmological initial conditions such as the horizon and flatness problems \citep{Guth1981}. Since then, however, its greatest success has arguably been to explain the near scale-invariant power spectrum of cosmological perturbations. Observations of the cosmic microwave background continue to narrow down the range of possible inflaton potentials \citep[for example, ref.\ ][]{Planck2015_Overview}.\footnote{Despite its broad appeal,  there are rival theories to inflation aiming to explain our cosmological observations, such as, for example, bouncing cosmologies \citep[see][for a recent review]{BrandenbergerPeter2016}.} Cosmological perturbation theory is one of the cornerstones of modern cosmology, accounting for both structure formation and CMB inhomogeneities. Understanding the evolution of perturbations in terms of York time is therefore essential if $T$ is to be considered the physically fundamental time parameter.

Section \ref{sec:inflation} provided a description of inflation using the York time parameter. We found that inflation occurs over a very short York-time interval $\Delta T$. In fact, in the `na\"ive' de~Sitter picture of inflation no York time would pass at all. Instead one has to employ a higher-order description in which the universe is only almost de~Sitter.

The rapid expansion of space during inflation leads to the `freezing out' of (Fourier modes of) cosmological perturbations as physical wavelengths grow significantly faster than the Hubble radius. This in turn leads to the scale-invariant power-spectrum. The freezing-our process is well understood in the conventional picture in terms of cosmological time $t$. In this section, we analyse the dynamics in terms of $T$ via `translation' of the relevant equations. The mathematical mechanism by which freezing out occurs is rather distinct from the usual, although the physics is ultimately equivalent, of course.

A more detailed analysis would lead us to consider scalar and tensor perturbations separately. For our present purposes however it suffices to consider the single sufficiently general equation of motion for a Fourier mode with wave number $k$,
\begin{equation} \frac{d^2y_k}{d\eta^2} + \left(k^2-\frac{1}{a}\frac{d^2a}{d\eta^2}\right)y_k = 0, \label{eq:7.3-ModeEqInEta}\end{equation}
where $\eta$ denotes conformal time. The variable $y_k$ here might refer to a particular tensor mode after rescaling by the scale factor. That is, we may parameterise the two independent degrees of freedom of the tensor perturbation as
\begin{equation} \delta g_{ij} = a^2\left( h_1 e_{ij}^1 + h_2 e_{ij}^2 \right),\end{equation}
where $\{e^1_{ij},e^2_{ij}\}$ form an orthonormal basis of the tensor perturbations of the spatial metric $g_{ij}$, then rescale to $y=\frac14M_{Pl}^2 ah_1$ (and similar for $h_2$) and perform a Fourier transform to arrive at equation \ref{eq:7.3-ModeEqInEta} (the equation of motion derived from the Einstein-Hilbert action). A similar equation arises in the case of scalar perturbations (scalar metric and scalar-field perturbations, for example), at least approximately in certain scenarios (most importantly for us, in quasi-de~Sitter scenarios), when expressed in appropriate variables (rescaling the field $\phi$ by the scale factor $a$) and with an appropriate gauge choice \citep{MukhanovFeldmanBrandenberger1992}. Equation \ref{eq:7.3-ModeEqInEta} is therefore rather general and will form the basis of our analysis.

The foliation is considered fixed. That is, we do not consider the back-reaction of perturbations on the foliation and the constant-mean-curvature condition may therefore be violated at the level of perturbations. However, at least for tensor perturbations there is no such back-reaction anyway since the perturbations are traceless (Tr$(e^1_{ij})=$Tr$(e^2_{ij})=0$), as a result of which the volume element $\sqrt{g}$ and the extrinsic curvature $K$ are unperturbed. The York-time analysis performed in this section is therefore exact for tensor perturbation and approximate for scalar perturbations (since there both equation \ref{eq:7.3-ModeEqInEta} as well as the foliation is approximate).

In a few cases equation \ref{eq:7.3-ModeEqInEta} can be solved exactly, but in general one can consider approximations in the super-Hubble and sub-Hubble regimes in which the wavelength of a mode is respectively much larger and much smaller than the Hubble radius, or equivalently $k\ll aH$ and $k\gg aH$ respectively. In the former case one can neglect the term in $k^2$ in equation \ref{eq:7.3-ModeEqInEta}, so that the equation has the general approximate solution (in terms of conformal time $\eta$),
\begin{equation}y_k=c_1a+c_2a\int d\eta\,a^{-2}. \label{eq:7.3-solutionforsuperHubbleineta}\end{equation}
The first term is static, while the second is dynamic and decays fast, so that only the static term remains relevant. In other words, the modes are `frozen'. For large values of $k$ on the other hand the equation is approximately that of a harmonic oscillator and such modes are evolving. During inflation, of course, a large range of modes pass from the sub-Hubble to the super-Hubble regime.

Equation \ref{eq:7.3-ModeEqInEta} may be transformed to York time, giving
\begin{align} 0 &= y^{\prime\prime}_k+\left(\frac{1}{T}+2\frac{a^\prime}{a}-\frac{a^{\prime\prime}}{a}\right)y_k^\prime
		 +\left[4M_{Pl}^4\frac{k^2}{a^2T^2}\frac{a^{\prime2}}{a^2}-\frac{a^\prime}{a}\left(\frac{1}{T}+2\frac{a^\prime}{a}\right)\right]y_k, \label{eq:7.3-ModeEqInT}
\end{align}
where primes denote derivatives with respect to $T$ as usual and we used the relative time lapse,
\begin{equation} \DERIV{\eta}{T} = -\frac{2M_{Pl}^2a^\prime}{Ta^2},\end{equation}
obtained via the relationship between $T$ and $H$ (see section \ref{sec:handsonYTcosm}). One can study equation \ref{eq:7.3-ModeEqInT} for distinct eras by choosing an appropriate function $a(T)$. During inflation the scale factor is given by equation \ref{eq:5.4-aduringinflation}, so that equation \ref{eq:7.3-ModeEqInT} takes the form
\begin{equation} 0 = y_k^{\prime\prime}-\left(T^{-1}+\kappa T\right)y_k^\prime + \kappa^2\left[4M_{Pl}^4\cdot\frac{k^2}{a^2}-2T^2\right]y_k, \label{eq:7.3-ModeEqInTInflation}\end{equation}
with $\kappa$ given by expression \ref{eq:5.4-kappadef}. Defining
\begin{align}\beta(T)&\equiv-(T^{-1}+\kappa T) \label{eq:7.3-betadef}\\ 
	     \omega^2_k(a,T)&\equiv\kappa^2[4M_{Pl}^4k^2/a^2-2T^2],  
\end{align}
equation \ref{eq:7.3-ModeEqInTInflation} may be cast into the form of a damped-oscillator equation,
\begin{equation} 0= y_k^{\prime\prime}+\beta(T)y_k^\prime + \omega_k^2(a,T) y_k\end{equation}
(noting that $\beta(T)>0$ since $T<0$ while the universe is expanding). The coefficients are time-dependent, so this oscillator-like analysis does not provide solutions for $y_k$ over an extended period. But it does give an indication of the qualitative behaviour of a mode at some particular instant during inflation. Furthermore, inflation lasts for only a short York-time interval $\Delta T$ as compared to $|T|$, so that the evolution of the coefficients is, in fact, negligible apart from the time dependence entering via the scale factor.

The sign of $\omega_k^2$ now determines the nature of the evolution. If $\omega_k^2>0$, then $\omega_k$ is a real frequency and the evolution corresponds to damped oscillations. If $\omega_k^2<0$, the frequency is imaginary, corresponding to decay and hence a freezing of the mode. Physically the $k$-dependent term contributing to $\omega_k^2$ may be expressed in a more illuminating manner in terms of the ratio $\tilde{k}$ of the actual wave number $k$ with the wave number $aH$ of the mode instantaneously crossing the Hubble radius,
\begin{equation} \tilde{k} = \frac{k}{aH},\qquad  aH = -\tfrac12 M_{Pl}^{-2} Ta,\end{equation}
so that,
\begin{equation} \omega_k^2(a,T)= \left(\tilde{k}^2-2\right)\,\kappa^2T^2. \end{equation}
Super-Hubble modes satisfy $\tilde{k}\ll 1$, so that $\omega_k^2<0$ and the modes do not oscillate. In the sub-Hubble regime, $\tilde{k}\gg1$, the frequency is real, $\omega_k^2>0$, and the modes oscillate provided they are not overdamped (which is not the case, as we will see).\footnote{The fact that physically the regimes are identified by $\tilde{k}\ll1$ and $\tilde{k}\gg1$ while mathematically the critical value is $\sqrt{2}$ rather than $1$ is of no significance provided we restrict analysis to modes sufficiently far into the sub-Hubble or super-Hubble regime. Astrophysically speaking, $\sqrt{2}\approx1$.}

We may study the effect of the damping term $\beta(T)$ by identifying whether a particular mode is overdamped or underdamped.\footnote{The possibility of critical damping may be neglected since the coefficients are time dependent and therefore no mode is critically damped for more than an instant.} The relevant quantity to consider is
\begin{align} D_k(a,T) &\equiv \beta^2(T)-4\omega_k^2(a,T) \notag\\
		       &= -16\kappa^2M_{Pl}^4\frac{k^2}{a_0^2}e^{\kappa T^2} + 9\kappa^2T^2+2\kappa +T^{-2},
\end{align}
where all terms are strictly positive except the first, which is strictly negative. Underdamping and overdamping correspond to $D_k<0$ and $D_k>0$ respectively. Since inflation occurs during a small York-time interval $\Delta T$, where $\Delta T/|T|\ll1$, one may approximate $T$ to be constant except for the $T$-dependence (entering via the scale factor) in the exponential, which decreases during inflation by a factor $e^{2N_e}$, $N_e$ being the number of e-folds given by equation \ref{eq:5.4-efolds}. Because the exponential term changes by a factor $N_e^2$ during the course of inflation, a set of modes with $k^2$ ranging over orders of magnitude given by this factor pass from being underdamped to being overdamped. This is, of course, conditional on the sign of $\omega_k^2$ not changing before the modes become overdamped. However, under reasonable assumptions they do. That is, the sign of $\omega_k^2$ changes before a mode becomes overdamped.

To see this last point, consider a mode with wave number $k$ such that at some initial York time $T_i$ near the beginning of inflation $\omega_k^2(a(T_i),T_i)>0$ and $D_k(a(T_i),T_i)<0$. The frequency changes from real to imaginary ($\omega_k^2=0$) roughly when crossing the Hubble radius, $\tilde{k}=\sqrt{2}\approx1$, that is, after an interval $\Delta T_{cross}$ given by
\begin{equation} e^{2\kappa T_i\Delta T_{cross}} = 2/\tilde{k}_i^2 ,\end{equation}
where $\tilde{k}_i$ refers to the value of $\tilde{k}$ corresponding to the mode $k$ at time $T_i$. Here only the leading-order change to the exponential has been considered, $\kappa T^2\approx \kappa T_i^2+2\kappa T_i\Delta T$, which is once again possible since $\Delta T/|T|\ll1$. Explicitly,
\begin{equation} \Delta T_{cross} = \left(2\kappa T_i\right)^{-1}\ln \left(2/k_i^2\right). \end{equation}
Meanwhile the mode passes from one damping regime to the other when $D_k=0$, or
\begin{equation} e^{2\kappa T_i\Delta T_{CD}}=\frac{9\kappa^2 T_i^2+2\kappa+T_i^{-2}}{16 M_{Pl}^4\kappa^2 k^2/a_i^2}, \label{eq:7.3-expdeltaTCD}\end{equation}
which may be further approximated by dropping the second and third term in the numerator if $\kappa^2T_i^2$ dominates. 

A numerical estimate for a given choice of inflaton potential makes this clear. For the example considered in section \ref{sec:inflation} with the inflaton potential $V(\phi)=m^2\phi^2$, the empirical estimate for $H$ during inflation (from recent Planck data, \citep{Planck2015_ConstraintsOnInflation}) of $H\lesssim 3.6\times10^{-5}$ (in reduced Planck units) can be used to find an inequality (using equation \ref{eq:5.4-t0SRC}) for the kinetic-energy correction \ref{eq:5.4-KEcontribution},
\begin{equation} K.E. \ll 10^{-9}, \end{equation}
from which in turn one can derive an inequality for the constant $\kappa$,
\begin{equation} \kappa \gg 10^8.\end{equation}
This implies that the two expressions in the damping term $\beta(T)$ (equation \ref{eq:7.3-betadef}) are of roughly equal magnitude, while the term $\kappa^2T_i^2$ does indeed dominate in equation \ref{eq:7.3-expdeltaTCD}.

One then has
\begin{equation} \Delta T_{CD} \approx \left(2\kappa T_i\right)^{-1}\ln \left(9/4\tilde{k}_i^2\right), \end{equation}
which is always strictly greater than $\Delta T_{cross}$. Thus the frequency will change from real to imaginary before the transition from underdamping to overdamping.

While it may be possible to construct conditions such that the above approximations are not valid and for which this result would not hold, the result does hold under conditions corresponding to inflation as it is presumed to occur in our actual universe. Hence the freezing-out process occurs, mathematically speaking, as a result of the change of sign of $\omega_k^2$ rather than a transition to overdamping.

This mathematical mechanism is fundamentally different from the mechanism in the conformal-time description, namely the solution \ref{eq:7.3-solutionforsuperHubbleineta} applying to the super-Hubble regime together with modes entering this regime during inflation. The physical evolution is, of course, the same. The perspective on these processes from the York-time point of view, is however new.

\section[Re-entry of the Hubble radius in the radiation-dominated universe]{\texorpdfstring{Re-entry of the Hubble radius in the radiation-\\dominated universe}{Re-entry of the Hubble radius in the radiation-dominated universe}}\label{sec:modereentry}

According to our best understanding of cosmic history inflation was followed by a period of `reheating' during which the inflaton field decayed into the known particles of the standard model, although details of this process are uncertain. The energy density of the universe became subsequently dominated by radiation and during this period the scale factor behaved as $a\sim t^\frac12$. The Hubble radius grew again as $\sim t$, allowing modes to `re-enter' and evolve after having previously been frozen. In this section we compare the York-time and conformal-time descriptions of the `Hubble re-entry' process of Fourier modes of perturbations, much like we did for their `freezing out' in the last section.

During the radiation-dominated era York time relates\footnote{Any explicit relationship between $t$ and $T$ can be modified by translation in $t$ since the cosmological equations in $t$ are time-translation invariant. However, the York-time theory is not $T$-translation invariant since $T$ has a physical meaning and the equations are explicitly $T$-dependent. The relationship given here therefore depends on the appropriate choice of time origin, namely that $t=0$ corresponds to $T=-\infty$.} to cosmological time as $T=-M_{Pl}^2/t$, so that the scale factor behaves as
\begin{equation} a(T) = a_{Pl}\frac{M_{Pl}}{|T|^\frac12}, \end{equation}
where $a_{Pl}$ is a proportionality constant formally referring to the scale factor at Planck York-time, $T=-M_{Pl}^2$.

The physical Hubble radius $H^{-1}$ (known as an explicit function of time since $H=-\frac12 M_{Pl}^{-2}T$) may be compared with the evolution of the physical wavelength of a mode,
\begin{equation} \lambda_{phys}(k) = a \lambda(k) = a_0\left(\frac{t}{t_0}\right)^\frac12\lambda = a_0\left(\frac{|T_0|}{|T|}\right)^\frac12\lambda,\end{equation}
where $\lambda$ is the comoving wavelength and $a_0$, $t_0$ and $T_0$ refer to the values of the corresponding quantities at some reference time (usually today). The Hubble radius grows faster than physical wavelengths and modes do indeed enter the Hubble radius during the radiation-dominated era.

The equation of motion for this period takes the form
\begin{equation} 0=y_k^{\prime\prime} + \frac{1}{2T} y_k^\prime + \frac{M_{Pl}^2k^2}{a_{Pl}^2|T|^3}y_k.\end{equation}
This can be solved exactly,
\begin{align} y_k(T)&=c_1\cdot\sqrt{|T|}e^{-i\frac{2M_{Pl}|k|}{a_{Pl}\sqrt{|T|}}}	
		     +c_2\cdot i\sqrt{|T|}\frac{a_{Pl}}{4M_{Pl}|k|}e^{i\frac{2M_{Pl}|k|}{a_{Pl}\sqrt{|T|}}}, \label{eq:7.4-yksolutionsradiation}
\end{align}
where $c_1$ and $c_2$ are arbitrary constants. The pre-factor of $\sqrt{T}$ results from `absorbing' a factor of $a$ into the variable $y$. The actual dynamical evolution is in the phases. A mode becomes `unfrozen' when its phase changes sufficiently rapidly. With $d/dT[\arg y_k]\sim|k||T|^{-\frac32}$, once unfrozen the speed of evolution only increases as $T\rightarrow0$ from below. Contrast this with the description in conformal time $\eta$. During the radiation-dominated era it follows from $a=a_0\sqrt{t/t_0}$ that $T=-M_{Pl}^2 t^{-1}$ and $dt\sqrt{t_0/t} = a_0 d\eta$ (so that $\eta=2\sqrt{tt_0}/a_0$ up to an additive constant). Hence the solutions \ref{eq:7.4-yksolutionsradiation} in terms of conformal time are
\begin{align} y_k(\eta) &= \bar{c}_1 \cdot \eta^{-1}e^{-i\frac{|k|}{a_{Pl}}\eta} \notag\\ 
			&\qquad +\bar{c}_2\cdot i\eta^{-1}\frac{a_{Pl}}{4M_{Pl}|k|} e^{i\frac{|k|}{a_{Pl}}\eta}, \end{align}
where $\bar{c}_1$ and $\bar{c}_2$ are arbitrary coefficients (related to $c_1$ and $c_2$ respectively by a constant factor). We see that the phase of each contribution is linear in $\eta$, so that the conformal `speed' of evolution $\deriv{\eta}(\arg y_k)$ is constant.

If the radiation-dominated period is taken to have begun at some point in time $\eta_r=2\sqrt{t_rt_0}/a_0=2M_{Pl}^2(a_0\sqrt{|T_r||T_0|})^{-1}$ when the scale factor had a value $a_r$, then the condition $\tilde{k}\ll1$ for a mode to be super-Hubble may be expressed as
\begin{equation} |k|\gg|T|^\frac12\cdot\pi a_r|T_r|^\frac12/M_{Pl}^2 = |T|^\frac12\cdot\text{const.}\end{equation}
When the inequality becomes an approximate equality,
\begin{equation} |k|\approx |T|^\frac12\cdot\pi a_r|T_r|^\frac12/M_{Pl}^2,\label{eq:7.4-modesstartevolving}\end{equation}
one expects the mode to evolve and indeed this is consistent with the solutions \ref{eq:7.4-yksolutionsradiation}.

Once a mode $k $ has become unfrozen ($\arg(y_k)\not\approx 0$), what are the time scales $\Delta \eta_k$ and $\Delta T_k$ for its evolution? In terms of conformal time $\eta$ the requirement for evolution to occur is 
\begin{equation} |k|\Delta\eta \approx a_0\eta_r, \end{equation}
while in terms of York time the condition \ref{eq:7.4-modesstartevolving} becomes $\Delta \sqrt{T} \approx\frac{|k|M_{Pl}^2}{a_r\sqrt{|T_r|}}$, or
\begin{equation} \Delta T_k \approx |T|\left( 1-\frac{|k|^2}{\left(|k|+a_rM_{Pl}^{-2}|T_r|^\frac12|T|^\frac12\right)^2}\right).\end{equation}
In the short-wavelength limit ($k\rightarrow\infty$), the York-time scale becomes small, $\Delta T_k\rightarrow0$ and so evolution is fast, as expected. Similarly, for modes in the super-Hubble regime ($k\rightarrow0$), which have not re-entered, one finds $\Delta T_k\rightarrow|T|$, that is, the York-time scale approaches the full duration of the remaining history of the universe.

The description in this section and the last were not in terms of a reduced-Hamiltonian formalism but through the transformation of the dynamical equations themselves from $t$ to $T$. This has allowed us to develop an intuition for the form of some of the mathematical mechanisms that play a role in the York-time description. We started with the equations of established conventional perturbation theory. In contrast, in the next chapter we being `from scratch' with the ADM action applied to a perturbative regime and perform the Hamiltonian reduction with respect to York time explicitly in order to arrive at the fundamental equations of York-time cosmological perturbation theory.

\chapter{Cosmological perturbation theory with York time: Hamiltonian reduction}\label{chap:perttheoformalism}

\textit{In this chapter we develop classical cosmological perturbation theory with York time using a perturbative approach to Hamiltonian reduction. The contents of this chapter formed part of ref.~\citep{Roser2015b}.}

\section[Principles of cosmological perturbation theory via York-time Hamiltonian reduction]{\texorpdfstring{Principles of cosmological perturbation theory via\\ York-time Hamiltonian reduction}{Principles of cosmological perturbation theory via York-time Hamiltonian reduction}}

The developments in sections \ref{sec:modefreezing} and \ref{sec:modereentry} gave us a `hand-on' understanding of examples of physical processes when viewed from the perspective of York time. This was achieved by the transformation in the time parameter from cosmological time $t$ to York time $T$ for a known background expansion $a(t)$. In this section we finally develop the reduced-Hamiltonian formalism for cosmological perturbation theory. The insights we gained in section \ref{sec:ClassKasner} about the non-canonical Poisson structure will be useful. In section \ref{sec:quantperts} we quantise the theory developed here.

Hamiltonian reduction applied to perturbation theory differs from the reductions carried out for minisuperspace models (sections \ref{sec:classFriedmann}, \ref{sec:ClassKasner}) in a number of ways: First, variables have a spatial dependence (they are field variables and not homogeneous), so that the momentum conjugate to $T$ gives a physical Hamiltonian \emph{density}, while the physical Hamiltonian itself is recovered by integration over all space. Second, the Hamiltonian density is derived perturbatively and therefore not exact. Third, the foliation is fixed at the level of the background but not initially at the level of perturbations. So a \emph{gauge choice} has to be made. For reasons we will discuss, the choice that implements York time exactly at the perturbative level (rather than only approximately) is not only theoretically the most desirable but also algebraically advantageous. However, in principle other choices exist.

Quantising the theory developed here does not give a true theory of quantum gravity, which would have to be non-perturbative. Analogous to the question that arose in the case of the quantisation of cosmologies, a caveat of the perturbative approach of this section is that it is not obvious that the quantisation of the perturbatively developed classical theory is physically equivalent to a perturbative limit of the full quantum theory. In much of the literature on quantised perturbations (irrespective of any York-time reference) this is tacitly assumed. However, in the absence of a complete theory of quantum gravity, all such work is contingent on that final theory being compatible in this way. Fortunately we know that the observed universe is well described in terms of a homogeneous isotropic `background' together with perturbations described by linearised field equations. While the reduced Hamiltonian at which we arrive is not fundamental and the quantised theory does not constitute a fundamental theory of quantum gravity, the theory's experimental predictions may nonetheless be compared to the observed universe. Whether corroborated or falsified, the result would narrow the search for the fundamental theory.

This chapter is without doubt the one heaviest on algebraic developments. For better readability some of the details have been banned into `sidebars' (boxes) and the reader is welcome to skip them on a first reading.

The method we employ here in order to derive the cosmological York-time perturbation theory is as follows. One first splits the geometric and matter variables and momenta appearing in the Hamiltonian constraint into homogeneous isotropic background variables and perturbations. For matter content we choose a set of scalar fields $\phi_a$ for simplicity. One can then solve the Hamiltonian constraint to zeroth order and derive a Hamiltonian for the background dynamics, essentially the result found in chapter \ref{chap:Friedmann}. The zeroth-order equations of motion may then be solved and one substitutes their solutions as functions of York time back into the Hamiltonian constraint, which one can then solve to second order in the perturbation variables in order to arrive at a reduced Hamiltonian that is also second order in those variables. Note that first-order quantities in general vanish if the background equations of motion are satisfied, so that the Hamiltonian describing the perturbations is second order only and therefore gives linear equations of motion.

Before we can write down the perturbative expansion of the dynamical variables and the resulting Poisson structure explicitly, the matter of foliation-related gauge freedom must be addressed. The issue is well known from conventional cosmological perturbation theory \citep{MukhanovFeldmanBrandenberger1992,Mukhanov2005}: writing down some perturbative expansion of the three-metric and matter fields does not by itself separate truly physical perturbations from apparent perturbations due to the choice of coordinate system. Furthermore, small adjustments in the coordinate system (the `gauge choice') allow one to move the physical content of the perturbation expansion between the different variables. Ultimately one is interested in the two physical degrees of freedom, manifest in gauge-independent quantities such as the Bardeen variables \citep{Bardeen1980}.

When performing perturbation theory with York time the situation is similar, although the gauge freedom is limited to three-diffeomorphisms on the spatial slice since the foliation itself is fixed if \emph{exact} York slicing is chosen. It is possible to make a different choice of foliation (at the perturbative level) while still using York time as the time parameter and performing the Hamiltonian reduction. In the case of a homogeneous isotropic background the zeroth-order (background) foliation is exactly that of the standard Friedmann description, although the slices are differently parameterised, and one has the full gauge freedom of standard cosmological perturbation theory available. There is no bar to developing perturbation theory in this framework.\footnote{The findings of sections \ref{sec:modefreezing} and \ref{sec:modereentry} correspond to a description in terms of York time without choosing the exact constant-mean-curvature slicing, although in the case of tensor perturbations exactness of slicing was recovered in virtue of the perturbation type.}

However, it turns out that in this case the resulting expressions occurring in the dynamical equations, for example, are algebraically considerably more complicated. The primary reason for this is that the partial\footnote{Freedom to choose coordinates on each slice, that is, to perform a spatial three-diffeomorphism, remains. Only the foliation itself is fixed} gauge choice of what we will call the `York gauge' --- implementing the York slicing on the perturbative as well as the background level --- imposes conditions that ensure that the determinant of the metric $g\equiv\det(g_{ab})$ and the trace of the conjugate momentum $\pi\equiv Tr(\pi^{ab})$ remain unperturbed, or equivalently that the linear change in the metric $\delta g_{ab}$ and in the momentum $\delta \pi^{ab}$ are both traceless at first order. A change in these quantities would correspond to a shift in York time $T$, perturbing the foliation. These conditions eliminate a large number of terms in various stages of the perturbative expansion. Furthermore, the terms through which scalar fields couple to the geometry are exactly those proportional to perturbations of the volume element $\rootg$, so that the York gauge not only simplifies the algebra but also \emph{decouples matter and geometry}, at least to leading order. The York gauge is also the `correct' gauge to use if York time really is considered the physical time parameter.

Working in the York gauge from the beginning, we perturbatively expand the reduced variables $\gtilde_{ab}$, $\pitilde^{ab}$ only (rather than the original canonical variables $g_{ab}$, $\pi^{ab}$), and the variables describing the scalar field. Throughout this chapter we denote zeroth-order (background) terms by an overbar. We define the perturbation variables $h_{ab}(x)$, $\nu^{ab}(x)$ in terms of the reduced variables $\tilde{g}_{ab}$ and $\tilde{\pi}^{ab}$ (see section \ref{sec:Poisson} ) as
\begin{align}
 \gtilde_{ab}&\equiv\gamma_{ab}+h_{ab}, \label{eq:8.1-defofh}, \\
 \pitilde^{ab}&\equiv\tilde{\bar{\pi}}^{ab} + \nu^{ab} = \nu^{ab} \label{eq:8.1-defofnu} ,
\end{align}
where $\gamma_{ab}$ is the scale-free metric of the background (in the case of a flat universe this is just the identity) and the last equality holds because the scale-free part of the background does not evolve, so its conjugate momentum $\tilde{\bar{\pi}}^{ab}$ vanishes. That is, the reduced momentum is perturbation only. The associated Poisson brackets to first order are
\begin{align} 
\{h_{ab}(x),\nu^{cd}(y)\} &= \Big[ \delta_a^{(c}\delta_b^{d)} - \tfrac13\gamma_{ab}\gamma^{cd} + \tfrac13\gamma^{cd}h_{ab} \notag\\
      &\hspace{0.2\linewidth}-\tfrac13\gamma_{ab}h^{cd}  +\text{h.o.}\Big]\;\delta^3(x-y), \label{eq:8.1-hnuPB} \\
\{\nu^{ab}(x),\nu^{cd}(y)\} &= \frac13\left[\gamma^{cd}\nu^{ab}-\gamma^{ab}\nu^{cd}\right]\;\delta^3(x-y). \label{eq:8.1-nunuPB}
\end{align}
Indices of $h_{ab}$ and $\nu^{ab}$ are raised and lowered by $\gamma^{bc}$ and $\gamma_{bc}$ respectively and `h.o.' stands for higher-order terms. These equations are derived from the general Poisson brackets \ref{eq:4.3-ggPB}, \ref{eq:4.3-gpiPB} and we used the fact that the expansion of the inverse metric is (to first order) $\gtilde^{ab}=\gamma^{ab}-h^{ab}$, with the scale-free inverse background metric $\gamma^{ab}$ defined via $\delta_a^c=\gamma_{ab}\gamma^{bc}$.

By construction the physical Hamiltonian obtained in the reduction is quad\-ratic because the perturbation expansion is to second order, although the terms contain time-dependent coefficients determined by the solution of the background dynamics. Later on both background and perturbations will be quantised. In order to obtain trajectories for the background dynamics, that is, functions of time which will constitute the coefficients of the perturbation terms, we employ the de~Broglie-Bohm pilot-wave formulation of quantum mechanics \citep{deBroglie1928,Bohm1952,Holland1993}, where such trajectories are part of the fundamental ontology. De~Broglie-Bohm trajectories have previously been used as a mathematical tool in cosmological perturbation theory in \citep{Pinto-NetoSantosStruyve2012,Pinto-NetoSantosStruyve2014}, although with conventional cosmological time rather than York time. The fact that the Hamiltonian is a sum of quadratic terms means that the quantisation procedure is straightforward (there are no `square-root' operators at the perturbative level) once a representation of the commutator algebra derived from the non-canonical Poisson structure is found. The time-dependent coefficients do however make finding solutions to the quantum dynamics more difficult. 

The fact that the variables are not canonical does not hinder quantisation, as will become clear in the following chapters. However, the fact that the momentum-momentum bracket does not vanish implies that the quantum theory cannot possess a momentum but only a position representation, a matter we discuss in section \ref{sec:quantKasner} below. The fact that the quantum theory has a preferred basis, the position representation, may be taken to provide hints for answering foundational questions in quantum theory. For example, formulations in which the position representation has a special status, such as the de~Broglie-Bohm formulation, would not have to explain why the position representation should be taken as fundamental.

\section{The perturbative Hamiltonian constraint}\label{sec:pertHamiltonianconstraint}

As in the case of minisuperspace models, we begin with the ADM action for general relativity (see section \ref{sec:PoTinGR} and ref.~\citep{ADM1962}) minimally coupled to a set of scalar fields $\phi_A$ with momenta $\pphi{A}$,\footnote{The choice to consider multiple scalar field rather than just one is not of further importance. The zeroth-order terms found correspond to a mild generalisation of the minisuperspace model of chapter \ref{chap:Friedmann}, where we looked at a single scalar field.}
\begin{equation} \label{eq:8.2-ADMaction}
 S = \int dt\,d^3x\;\left[\gdot_{ab}\pi^{ab}+\phidot_A\pphi{A} -N_i\mathscr{H}^i-N\mathscr{H}\right],
\end{equation}
where 
\begin{align}
 \mathscr{H}&= -M_{Pl}^2\rootg R+\frac{1}{M_{Pl}^2\rootg}\,(g_{ac}g_{bd}-\tfrac12 g_{ab}g_{cd})\pi^{ab}\pi^{cd} \notag\\
		 &\hspace{0.2\linewidth}+ \frac{1}{2\rootg}\pphisq+\frac{\rootg}{2}g^{ab}\phi_{A,a}\phi_{A,b} + \rootg V(\phi), \label{eq:8.2-GeneralHamiltonianConstraint} \\
 \mathscr{H}^a &= -2\nabla_b\pi^{ab} + \pphi{A} g^{ab}\de_b\phi_A \label{eq:8.2-GeneralMomentumConstraint}
\end{align}
constitute the Hamiltonian and momentum constraints respectively for this choice of matter content. Summation over repeated field indices ($A$, etc.) is assumed. As before $R$ denotes the scalar three-curvature of space, $M_{Pl}$ the Planck mass, $\pphisq=\sum_A\pphi{A}^2$ and $V(\phi)$ is a currently arbitrary potential of the scalar fields $\phi=\{\phi_1,\phi_2,\dots\}$. We suppress the spatial and temporal arguments of the field quantities where they are unambiguous in order to avoid notational clutter.

The goal is to solve the Hamiltonian constraint, $\mathscr{H}=0$, for $P_T = -\sqrt{g}$, the momentum conjugate to the York time parameter $T\equiv\frac{2\pi}{3\rootg}$. One therefore performs the change of variables $\{g_{ab},\pi^{ab}\}\rightarrow\{\gtilde_{ab},\pitilde^{ab},T,P_T\}$ (see chapter \ref{chap:choiceoftime}). In the perturbative case considered here we instead make the change $\{g_{ab},\pi^{ab}\}\rightarrow\{h_{ab},\nu^{ab},T,P_T\}$ with $h_{ab}$ and $\nu^{ab}$ defined by eqs.\ \ref{eq:8.1-defofh} and \ref{eq:8.1-defofnu} respectively. The fields are expanded as
\begin{equation} \phi_A=\phibar_A +\delta\phi_A,\qquad\qquad\pphi{A}=\pphibar{A}+\delta\pphi{A}.\end{equation}
We keep terms up to second order in the geometric and matter perturbation variables. While not difficult, this leads to some lengthy expressions.

The scaling of the matter terms in \ref{eq:8.2-GeneralHamiltonianConstraint} is straightforward since their only scale-dependence is in the prefactor $g^\frac12\sim P_T$ or $g^{-\frac12}\sim P_T^{-1}$. The curvature terms obtained in the expansion of $R$ are homogeneous in their scale dependence, $R\sim g^{-\frac13}\sim P_T^{-\frac23}$ and only the momentum terms are inhomogeneous in their $T,P_T$-dependence. The latter fact follows from the expression obtained when writing $\pi^{ab}$ in terms of $\tilde{\pi}^{ab}$,
\begin{align}\pi^{ab} &= g^{-\frac13}\pitilde^{ab}+\tfrac13\pi g^{-\frac13}\gtilde^{ab} \notag\\
		      &= \mathcal{H}_T^{-\frac23}\nu^{ab}+\tfrac13\left(\tfrac32T\mathcal{H}_T\right)\cdot \mathcal{H}_T^{-\frac23}
			   \left(\gamma^{ab}-\gamma^{ac}\gamma^{bd}h_{cd}+\gamma^{ac}\gamma^{de}\gamma^{fb}h_{cd}h_{ef}\right),
\end{align} 
the last equality holding to second order. The Hamiltonian constraint equation becomes the equation determining $H_T$,
\begin{align}\label{eq:8.2-HCEq}
 0 &= \mathcal{H}_T\Big(-\tfrac{3}{8M_{Pl}^2}\cdot T^2+\QU{V}\Big) + \mathcal{H}_T^\frac13\Big(\QU{R}+\QU{\nabla\phi}\Big) +M_{Pl}^{-2} Th_{ab}\nu^{ab} \notag\\ 
	 &\hspace{0.4\linewidth} +\mathcal{H}_T^{-1}\Big(M_{Pl}^{-2}\gamma_{ac}\gamma_{bd}\nu^{ab}\nu^{cd}+\QU{p_\phi}\Big),
\end{align}
where we have introduced the shorthand $\QU{X}$ to denote the terms in the perturbative expansion after factoring out the scale dependence which are derived from the term containing $X$ in the Hamiltonian constraint \ref{eq:8.2-GeneralHamiltonianConstraint}. With the York gauge applied these are given by
\begin{align}
 \QU{R} &= -M_{Pl}^{2}\tilde{\Rbar}-M_{Pl}^{2}\widetilde{\delta R}^\one -M_{Pl}^{2}\widetilde{\delta R}^\two  \label{eq:8.2-Rterms}\\
 \QU{p_\phi} &= \tfrac12\bar{p}_\phi^2+\pphibar{A}\delta\pphi{A}+\delta\pphisq. \label{eq:8.2-pphiterms}\\ 
 \QU{\nabla\phi} &= \tfrac12\gamma^{ij}\delta\phi_{A,i}\delta\phi_{A,j} \label{eq:8.2-delphiterms}\\
 \QU{V} &= V(\phibar)+\delta\phi_A\left.\PD{V}{\phi_A}\right|_{\phibar} 
		    +\tfrac12\delta\phi_A\delta\phi_B\left.\frac{\partial^2V}{\partial\phi_A\partial\phi_B}\right|_{\phibar}.\label{eq:8.2-Vterms}
\end{align}
For the algebraic details of the expansion as well as the form of the first and second-order curvature-perturbation terms $\widetilde{\delta R}^\one$, $\widetilde{\delta R}^\two$, see the box below. Throughout this chapter we use superscripts `$(n)$' to denote the $n$th-order contribution to the preceding quantity. At this point we have retained the first-order terms, although these will cancel when the background equations of motion are satisfied. This is a general result following from the fact that the equations are derivable via an extremisation principle.

{
\begin{mdframed}
  \setlength{\parindent}{10pt}

\small
\bigskip
\textbf{Expansion of relevant terms up to second order}\bigskip

Prior to application of the York gauge (that is, if York time is only implemented at the background level and one retains the full gauge freedom of standard perturbation theory), expressions \ref{eq:8.2-Rterms}-\ref{eq:8.2-Vterms} are
\begin{align*}
 [\lqu R\rqu] &=-\frac1\tk\left(\tilde{\Rbar}+\widetilde{\delta R}^\one +\eta^\one\tilde{\Rbar}+\widetilde{\delta R}^\two +\eta^\one\widetilde{\delta R}^\one +\eta^\two\tilde{\Rbar} \right) \\
 [\lqu\nabla\phi\rqu] &= \frac12\gamma^{ij}\delta\phi_{A,i}\delta\phi_{A,j} \\
 [\lqu p_\phi\rqu] &= \frac12\Big(\pphibar{A}^2-\eta^\one\pphibar{A}^2+2\pphibar{A}\delta\pphi{A} \notag\\
		    &\hspace{0.2\linewidth}+\big(-\eta^\two+(\eta^\one)^2\big)\pphibar{A}^2-2\eta^\one\pphibar{A}\delta\pphi{A}+\delta\pphi{A}^2\Big) \\ 
 [\lqu V\rqu] &= V(\phibar)+\eta^\one V(\phibar)+\delta\phi_A\left.\PD{V}{\phi_A}\right|_{\phibar} \notag\\ 
		    &\hspace{0.2\linewidth}+\eta^\two V(\phibar) +\eta^\one\delta\phi_A\left.\PD{V}{\phi_A}\right|_{\phibar}
			  +\frac12\delta\phi_A\delta\phi_B\left.\frac{\partial^2V}{\partial\phi_A\partial\phi_B}\right|_{\phibar}, \label{eq:8.2-Vtermsgeneral}
\end{align*}
where
\begin{align*} \eta^\one &= \tfrac12 h_{ab}\gamma^{ab} \\
	      \eta^\two &= \tfrac18(h_{ab}\gamma^{ab})^2 - \tfrac14 h_{ac}h_{bd}\gamma^{ab}\gamma^{cd}
\end{align*}
are the first and second order fractional perturbation in the metric. In the York gauge these are set to zero. Note that the expansions of $[\lqu p_\phi\rqu]$ and $[\lqu V\rqu]$ 
contain mixed terms, which would lead to a coupling between matter and geometric perturbation in the linearised equations of motion. However, in the York gauge the perturbation is set to zero, eliminating exactly those terms.

The expressions for the perturbative expansion of $\tilde{R}$ are
\begin{align*}
 \tilde{\Rbar} &= \gbar^\frac13\Rbar=\gbar^\frac13\gbar^{ij}\Rbar_{ij} = \gamma^{ij}R_{ij}, \\
 \widetilde{\delta R}^\one &= \gbar^\frac13\delta R^\one = \gbar^\frac13\delta(g^{ij}R_{ij})^\one = \gamma^{ij}\delta R^\one_{ij} + (-\gamma^{ik}h_{kl}\gamma^{lj})\Rbar_{ij}, \\
 \widetilde{\delta R}^\two &= \gbar^\frac13\delta R^\two = \gbar^\frac13\delta(g^{ij}R_{ij})^\two \notag\\
		       &= \gamma^{ij}\delta R^\two_{ij} + (-\gamma^{ik}h_{kl}\gamma^{lj})\delta R^\one_{ij}   + (-\gamma^{ik}h_{kl}\gamma^{lm}h_{mn}\gamma^{nj})\Rbar_{ij}.
\end{align*}
with
\begin{align*}
  \delta R^\one_{ij} &= \de_k\delta\Gamma^{\one k}_{ij}-\de_i\delta\Gamma^{\one k}_{jk} \\
  \delta R^\two_{ij} &=	  \de_k\delta\Gamma^{\two k}_{ij}-\de_i\delta\Gamma^{\two k}_{jk}
		+\delta\Gamma^{\one k}_{ij}\delta\Gamma^{\one l}_{kl}-\delta\Gamma^{\one k}_{il}\delta\Gamma^{\one l}_{jk},
\end{align*}
where
\begin{align*}
 \delta\Gamma^{\one q}_{rs} &= \tfrac12\gamma^{qt} (h_{tr,s}+h_{ts,r}-h_{rs,t} ) \\
 \delta\Gamma^{\two q}_{rs} &= -\gamma^{qu}h_{uv}\delta\Gamma^{\one v}_{rs}
\end{align*}
denote the perturbation in the Levi-Civita connection.

\end{mdframed}
}

\section{The physical Hamiltonian and the equations of motion}

After multiplication by $\mathcal{H}_T$ eq.~\ref{eq:8.2-HCEq} is a sextic equation in $\mathcal{H}_T^\frac13$, which in general cannot be solved analytically. As previously anticipated, the procedure now is to first solve this equation at zeroth order and use the solution as the basis for solving it up to second order perturbatively. At each order only a new linear equation has to be solved. Since the background is homogeneous the zeroth-order terms\footnote{At zeroth order (the background) the universe is homogeneous, so that the Hamiltonian itself is just equal to the Hamiltonian density (up to a constant multiplicative factor). Hence we write $H_T^\zero$ rather than $\mathcal{H}_T^\zero$, consistent with the notation in chapter~\ref{chap:Friedmann}.} have no spatial dependence and the equation is a simple polynomial,
\begin{equation} 0=\Big(-\tfrac{3}{8M_{Pl}^2}\cdot T^2+V(\phibar)\Big)H_T^{\zero2}-M_{Pl}^2\tilde{\Rbar}H_T^{\zero\frac43}+\tfrac12p_\phi^2 .\label{eq:8.3-BGHCEq}\end{equation}
Since the $\curlyH^0$ term in \ref{eq:8.2-HCEq} was second order only, the zeroth-order equation does not have a corresponding term and therefore after multiplication by $H_T^{\zero-2}$ takes the form of a depressed cubic in $u\equiv H_T^{\zero-\frac23}$, which has solutions
\begin{equation} u=(A+\sqrt{A^2-C^3})^\frac13 + (A+\sqrt{A^2-C^3})^\frac13,\end{equation}
with 
\begin{equation} A= \frac{1}{\pphisq}\left(\tfrac{3}{8M_{Pl}^2}\cdot T^2-V(\phibar)\right), \qquad C= \frac{1}{6\pphisq}M_{Pl}^2\tilde{\Rbar} .\end{equation}
These are, of course, exactly the solutions found in chapter~\ref{chap:Friedmann}, generalised to more than a single scalar field. For the purposes of our model we now make the assumption that the universe is flat at zeroth order, just as we did before. Again this choice is motivated by the two facts that (1) the algebra in what follows is significantly less convoluted, even at the level of the background and that (2) observation suggests that the universe is flat on sufficiently large scales within experimental uncertainty. Having chosen the global geometry of the background we can also fix the background frame of reference, making the obvious (inertial) choice $\gamma_{ab}=diag(1,1,1)$. However, for the most part we will retain reference to $\gamma_{ab}$ explicitly.

With the assumption of flatness one has $\tilde{\Rbar}=0$ and the background equation \ref{eq:8.3-BGHCEq} is trivially solved,
\begin{equation} H_T^\zero = \pm\left[\frac{\frac12\bar{p}_\phi^2}{\frac{3}{8M_{Pl}^2}\cdot T^2-V(\phibar)}\right]^\frac12. \label{eq:8.3-flatBGHCEq}\end{equation}
The ambiguous sign has no physical effect. Recall that a change in the choice of the sign in \ref{eq:8.3-flatBGHCEq} leads to the same set of physical trajectories with the exception that the sign of corresponding momenta are swapped (see chapter~\ref{chap:Friedmann}). That is, the set of solutions of the dynamical equations obtained from the Hamiltonian density $H_T$ in \ref{eq:8.3-flatBGHCEq} with a positive sign is related to the set of solutions of $H_T$ with a negative sign via a reflection in phase space. The physical interpretation of the numeric value of the Hamiltonian is that of volume (since $H_T=\rootg$, see chapter~\ref{chap:choiceoftime}), which suggests that $H_T\geq0$ is the more physically meaningful choice, and indeed we will adopt this convention here too. 

For a single scalar field, the dynamics of this spatially constant Hamiltonian density was discussed in chapter~\ref{chap:Friedmann}, where integration over a conformal normalisation volume was assumed. The details depend on the choice of $V(\phi)$. Note that there are no remaining geometric degrees of freedom at the background level since the scale variable $g$ and its conjugate momentum $\pi$ have been eliminated in favour of $T$ and $P_T=-H_T$. In the case of a set of free fields the momenta $\pphi{A}$ are constant and the fields simply evolve according to the Hamiltonian density \ref{eq:8.3-flatBGHCEq} for $V=0$, leading to
\begin{equation} \phi_A^\prime = \frac{\pphi{A}}{\sqrt{\frac{3}{4M_{Pl}^2}\cdot\pphisq T^2}},\end{equation}
so that, for an expanding universe ($T<0$), 
\begin{equation} \phi_A(T) = \frac{\pphi{A}}{\sqrt{\frac{3}{4M_{Pl}^2}\cdot\pphisq}}\ln|T|.\end{equation}
The volume of the universe is the numerical value of the background Hamiltonian density obtained by substituting the solution back into expression \ref{eq:8.3-flatBGHCEq}, up to a constant determined by the coordinate volume chosen for normalisation in the case of a flat universe as discussed above (the volume is well-defined without normalisation for a closed universe, though here the Hamiltonian takes a more complicated form).

For general $V(\phi)$ one obtains solutions ($\phi_A(T),\pphi{A}(T)$) by an appropriate method of solving Hamilton's equation for the Hamiltonian obtained after integrating $H_T^\zero$ over a (comoving) normalisation volume. Assuming Hamilton's equations for the background to be satisfied one can show that the first-order terms in eq.\ \ref{eq:8.3-BGHCEq} (note that these are matter terms only) cancel out. 

With $H_T^\zero$ found, one can substitute the result back into eq.\ \ref{eq:8.2-HCEq} and proceed to first order. This is trivially $[\mathcal{H}_T^\frac13]^\one=0$ since there are no first-order terms. The second order equation may be solved to yield the $[\mathcal{H}_T^\frac13]^\two$ from which one then obtains the second-order contribution to the Hamiltonian density,
\begin{align} \mathcal{H}_T^\two(x) 
 &=F(T)\cdot\Bigg[ M_{Pl}^{-2}\gamma_{ik}\gamma_{jl}\nu^{ij}\nu^{kl} + H_T^\zero\cdot M_{Pl}^{-2} Th_{ij}\nu^{ij} \notag\\
	&\hspace{0.15\linewidth} -H_T^{\zero\frac43}\cdot M_{Pl}^2\widetilde{\delta R}^\two(h_{ab}) +\tfrac12\delta\pphisq
						-H_T^{\zero\frac43}\cdot\tfrac12\gamma^{ij}\delta\phi_{A,i}\delta\phi_{A,j} \notag\\ 
	&\hspace{0.15\linewidth} + H_T^{\zero2}\cdot\tfrac12\delta\phi_A\delta\phi_B\left.\frac{\partial^2V}{\partial\phi_A\partial\phi_B}\right|_{\phibar} \Bigg] \label{eq:8.3-Hdens2}
\end{align}
where
\begin{equation} F(T)\equiv H_T^{\zero-1}\cdot\left[ \tfrac34\cdot M_{Pl}^{-2} T^2-2V(\phibar)\right]^{-1}
			=\left[2(\tfrac38\cdot M_{Pl}^{-2} T^2-V(\phibar))\pphibar{A}^2\right]^{-\frac12}. \label{eq:8.3-FofT}\end{equation}
The details of this procedure are spelled out in the box on the perturbative approach to solving the Hamiltonian constraint to second order. The Hamiltonian determining the perturbative dynamics is
\begin{equation} H_{pert}=\int_{\mathscr{V}}\,d\text{vol}\;\mathcal{H}_T(x), \label{eq:8.3-Hpert}\end{equation}
where $\mathscr{V}$ is the chosen normalisation coordinate volume. (Hence a change in the choice of coordinates on the slice at this stage would imply a change in the limits, leaving value of the Hamiltonian overall unchanged. However, having chosen some coordinates initially one now has that $d$vol$=1\cdot d^3x$ since the notion of scale has been extracted during the Hamiltonian reduction.)

{
\begin{mdframed}
  \setlength{\parindent}{10pt}

\small
\bigskip
\textbf{Perturbative approach to solving the Hamiltonian constraint to second order}\bigskip

Writing $x\equiv \mathcal{H}_T^\frac13$, eq.~\ref{eq:8.2-HCEq}, after multiplication by $\mathcal{H}_T=x^3$, has the form
\begin{equation} 0 = ax^6 + bx^4 + cx^3 + d, \tag{8.A}\label{xequation}\end{equation}
where, separating different orders visually,
\begin{alignat*}{4}
 &a= &&-\frac{3}{8M_{Pl}^2}\cdot T^2+ V(\phibar)  \quad&& +\delta\phi_A\left.\PD{V}{\phi_A}\right|_{\phibar} 
		  \quad&& +\tfrac12\delta\phi_A\delta\phi_B\left.\frac{\partial^2V}{\partial\phi_A\partial\phi_B}\right|_{\phibar} \\
 &b= && &&-M_{Pl}^2\widetilde{\delta R}^\one \quad&&-M_{Pl}^2\widetilde{\delta R}^\two +\tfrac12\gamma^{ab}\delta\phi_{A,a}\delta\phi_{A,b} \\
 &c= && && && M_{Pl}^{-2} T h_{ab}\nu^{ab} \\
 &d= &&  \tfrac12\pphibar{A}^2 \quad&& +\pphibar{A}\delta\pphi{A} \quad&& +\delta\pphi{A}^2 + M_{Pl}^{-2}\gamma_{ac}\gamma_{bd}\nu^{ab}\nu^{cd}.
\end{alignat*}
Here we assumed a spatially flat background, so that $\tilde{\Rbar}=0$. As in the text we let $a^\zero$ denote the zeroth order term of $a$ and so on. Similarly, we expand the sought function order by order, $x=x^\zero+x^\one+x^\two$. 

One first solves the zeroth-order equation,
\begin{equation*} 0 = a^\zero x^{\zero6}+d^\zero,\end{equation*}
so that 
\begin{equation} x^{\zero6} = -\frac{d^\zero}{a^\zero} = \frac{\frac12\bar{p}_\phi^2}{\frac{3}{8M_{Pl}^2}\cdot T^2-V(\phibar)}, \tag{8.B}\label{x0solution}\end{equation}
consistent with the result in chapter~\ref{chap:Friedmann}. The first order equation is trivial since $a^\one$, $b^\one$ and $d^\one$ vanish when the background equations are satisfied, hence $x^\one=0$. Then the second-order equation is
\begin{equation*} 0=a^\two x^{\zero6}+6a^\zero x^{\zero5} x^\two+b^\two x^{\zero4}+c^\two x^{\zero3}+d^\two, \end{equation*}
an equation linear in $x^\two$, giving after substitution of the coefficients and the background solution \ref{x0solution},
\begin{align*}
 x^\two &=  \Bigg[ \frac12\delta\phi_A\delta\phi_B\left.\frac{\partial^2V}{\partial\phi_A\partial\phi_B}\right|_{\phibar}x_0^6
		    -\left(M_{Pl}^2\widetilde{\delta R}^\two+\frac12\gamma^{ij}\delta\phi_{A,i}\delta\phi_{A,j}\right)x_0^4  \notag\\
	&\hspace{0.05\linewidth} +M_{Pl}^{-2} Th_{ij}\nu^{ij}x_0^3+\delta\pphi{A}^2+M_{Pl}^{-2}\gamma_{ik}\gamma_{jl}\nu^{ij}\nu^{kl} \Bigg]
				\Big[\left(\tfrac{3}{8M_{Pl}^{2}} T^2-V(\phibar)\right)x_0^5\Big]^{-1}.
\end{align*}
The Hamiltonian density at second order is then
\begin{equation} \mathcal{H}_T=(x^\zero+x^\two)^3 = x^{\zero3} + 3x^{\zero2}x^\two. \tag{8.C}\label{fullHsolution}\end{equation}

\end{mdframed}

}

From the perturbation Hamiltonian one can obtain the equations of motion,
\begin{align}
 \PD{h_{ab}}{T}  &=\{h_{ab}(x),H_{pert}\} \notag\\
		 &= F(T)\left[2\cdot M_{Pl}^{-2}\nu_{ab}+H_T^\zero\cdot M_{Pl}^{-2} Th_{ab}\right], \label{eq:8.3-heom}\\
 \PD{\nu^{ab}}{T}&=\{\nu^{ab}(x),H_{pert}\} \notag\\
		 &= F(T)\bigg[-H_T^\zero\cdot M_{Pl}^{-2} T\nu^{ab}   
		 -2H_T^{\zero\frac43}\cdot M_{Pl}^{2}\left(\delta^{(ab)(cd)}-\tfrac13\gamma^{ab}\gamma^{cd}\right)
			    \gamma^{ij}h_{ic,jd} \label{eq:8.3-nueom} \bigg]
\end{align}
where $\delta^{(ab)(cd)}\equiv\delta_m^{(a}\delta^{b)}_n\gamma^{mc}\gamma^{nd}$. The term in the last line is derived from the perturbation of the scale-free curvature, $\de\widetilde{\delta R}^\two/\de h_{ij}$, and is discussed in the box discussing the expansion of curvature terms at the end of this section. 

Since the Hamiltonian density contains only second-order terms, only zeroth-order contributions to the Poisson brackets \ref{eq:8.1-hnuPB}, \ref{eq:8.1-nunuPB} contribute to first-order terms in equations \ref{eq:8.3-heom}, \ref{eq:8.3-nueom}. That is, the momentum-momentum Poisson bracket \ref{eq:8.1-nunuPB} is effectively canonical, while the position-momentum bracket remains non-canonical due to the `$-\frac13\gamma_{ab}\gamma^{cd}$ term. For the purposes of first-order perturbation theory the relevant Poisson structure is therefore determined by the background only.

It is easy to confirm using eqs.~\ref{eq:8.3-heom}, \ref{eq:8.3-nueom} that the tracelessness of $h_{ab}$ and $\nu^{ab}$ is indeed conserved as contraction of eqs.~\ref{eq:8.3-heom} and \ref{eq:8.3-nueom} with $\gamma^{ab}$ and $\gamma_{ab}$ respectively reveals. Therefore the constraints $\gamma^{ab}h_{ab}=0$ and $\gamma_{ab}\nu^{ab}=0$ are first class.

The equations for the scalar fields are
\begin{align}
 \PD{\delta\phi_A}{T}(x)   &=\{\delta\phi_A(x),H_{pert}\} = F(T)\cdot\delta\pphi{A}(x) \label{eq:8.3-deltaphiEOM}\\
 \PD{\delta\pphi{A}}{T}(x) &=\{\delta\pphi{A}(x),H_{pert}\} \notag\\
			   &= -F(T) \bigg[H_T^{\zero \frac43}\int d^3y\; \gamma^{ij}\de_i\left(\delta^3(x-y)\de_j\delta\phi_A(y)\right) \notag\\
			   &\hspace{0.4\linewidth}+H_T^{\zero 2}\frac{\de^2V}{\de\phi_A\de\phi_C}\delta\phi_C(x) \bigg] \label{eq:8.3-deltapphiEOM}
\end{align}
It is a noteworthy feature that these equations have decoupled from the geometric degrees of freedom. This is a particular feature of the York gauge since matter-field perturbations would otherwise couple with perturbations in the local scale, that is, the metric determinant. In the box at the end of the last section we discussed what the matter-geometry mixed terms are and how they are eliminated by the York gauge. This result does not generalise to any form of matter, however. In particular, tensor fields coupling to the metric would lead to mixed terms not eliminated by the choice of gauge and therefore a geometric-matter interaction even at the linear level.

By construction equations \ref{eq:8.3-heom} and \ref{eq:8.3-nueom} (as well as \ref{eq:8.3-deltaphiEOM} and \ref{eq:8.3-deltapphiEOM}) are linear and therefore may be Fourier analysed. The behaviour of the solutions strongly depends on that of the time-dependent factors $F(T)$ and $H_T(T)$, which is in turn dependent on the choice of potential $V(\phi)$. For an analysis of the behaviour of the background, in particular at late times $T\rightarrow0^-$ and for candidates of inflationary potentials, see our discussion in chapter~\ref{chap:cosmext}. If we consider a free field, explicit functions of time for $F(T)$ and $H_T^\zero(T)$ can easily be written down since $\pphi{A}(T)=\pphi{A}(T_0)$ is constant and fixed by the boundary conditions at some time $T_0$:
\begin{align}
F(T) 		 &= \sqrt{\tfrac43M_{Pl}^2}|T|^{-1}, \label{eq:8.3-FofTfreefield} \\
H_T^\zero(T) &= \sqrt{\tfrac43 M_{Pl}^{2}\pphisq(T_0)}|T|^{-1} = \sqrt{\pphisq(T_0)}\cdot F(T). \label{eq:8.3-H_BGfreefield}
\end{align}

For the Fourier analysis we first expand the perturbation variables,
\begin{equation} h_{ab}(x) = \int d^3k\;\xi(k)_{ab}e^{ik\cdot x},\hspace{0.1\linewidth} \nu^{ab}(x) = \int d^3k\;\mu^{ab}(k) e^{ik\cdot x}.\end{equation}
Eq.~\ref{eq:8.3-heom} defines the relationship between $h^\prime_{ab}$ and the momenta $\nu^{cd}$ and, after acting with $\int d^3x\,e^{-il\cdot x}$ on both sides, yields
\begin{equation}\xi_{ab}^\prime (k) = F(T) \cdot M_{Pl}^{-2}\left[ 2\gamma_{ac}\gamma_{bd}\mu^{cd}(k)+H_T^\zero T\xi_{ab}(k) \right] \label{eq:8.3-xieom}\end{equation}
and is independent of the chosen mode. Eq.~\ref{eq:8.3-nueom} contains the actual dynamics and leads to $k$-dependent terms,
\begin{align} \mu^{ab\prime}(k) 
    &= F(T)\Big[ -H_T^\zero\cdot M_{Pl}^{-2} T\mu^{ab}(k) \notag\\ 
    &\hspace{0.15\linewidth} + 2H_T^{\zero\frac43}M_{Pl}^{2}\left(\delta^{(ab)(cd)}-\tfrac13\gamma^{ab}\gamma^{cd}\right)\gamma^{ij}\xi_{ic}k_jk_d \Big], \label{eq:8.3-mueom}
\end{align}

{
\begin{mdframed}
  \setlength{\parindent}{10pt}

\small
\bigskip
\textbf{Terms relating to perturbations in the curvature}\bigskip

Derivation of the classical equations of motion involves the term 
\begin{align}\{\nu^{ab}(y),\int d^3x\,\widetilde{\delta R}^\two(x)\} 
	  &= \int d^3x\;\{\nu^{ab}(y),\widetilde{\delta R}(x)\} \notag\\
	  &= \int d^3x\; \{\nu^{ab}(y),h_{cd}(x)\}\;\PD{\widetilde{\delta R}^\two}{h_{cd}}. \tag{8.D} \label{PBnudeltaR2}
\end{align}
For the flat background assumed in the text and expressed in a frame with Cartesian coordinates, one has
\begin{align*} \widetilde{\delta R}^\two 
  &= \gamma^{ij}\delta R_{ij} + (-\gamma^{ik}h_{kl}\gamma^{lj})\delta R^\one_{ij} \notag\\
  &= \gamma^{ij}\big(\de_k\delta\Gamma^{\two k}_{ij} - \de_i\delta\Gamma^{\two k}_{jk} 
	      + \delta\Gamma^{\one k}_{ij}\delta\Gamma^{\one l}_{kl} -\delta\Gamma^{\one k}_{il}\delta\Gamma^{\one l}_{jk} \big) \notag\\
  &\hspace{0.05\linewidth} -\gamma^{ik}h_{kl}\gamma^{lj}\big(\de_k\delta\Gamma^{\one k}_{ij}-\de_i\delta\Gamma^{\one k}_{jk} \big).
\end{align*}

Since $\widetilde{\delta R}^\two$ is first order, the Poisson bracket only contributes at zeroth order and is therefore spatially constant and may be taken outside the integral. The remaining term is then, abbreviating $\delta^{(3)}\equiv \delta^{(3)}(x-y)$,
\begin{align} &\int d^3x\;\delta^{(3)}\PD{\widetilde{\delta R}^\two}{h_{cd}}(y) \notag\\
 &\quad= \int d^3y\;\Bigg[ \tau^{\one k.cd}_{ij.k}[\delta^{(3)}]\gamma^{ij} - \tau^{one k.cd}_{jk.i}[\delta^{(3)}]\gamma^{ij} 
	    + \sigma^{\zero k.cd}_{ij}[\delta^{(3)}\delta\Gamma^{\one l}_{kl}] \notag \\
	&\hspace{0.2\linewidth} +\sigma^{\zero l.cd}_{kl}[\delta^{(3)}\delta\Gamma^{\one k}_{ij}] -\sigma^{\zero k.cd}_{il}[\delta^{(3)}\delta\Gamma^{\one l}_{jk}]\gamma^{ij} \notag\\
	&\hspace{0.2\linewidth}	-\sigma^{\zero l.cd}_{jk}[\delta^{(3)}\delta\Gamma^{\one k}_{il}]\gamma^{ij} 
				- \gamma^{im}\delta^{cd}_{mn}\gamma^{nj}(\de_k\delta\Gamma^{\one k}_{ij}-\de_i\delta\Gamma^{\one k}_{jk}) \notag\\
	&\hspace{0.2\linewidth} -\tau^{\zero k.cd}_{ij.k}[\delta^{(3)}h^{ij}] + \tau^{\zero k.cd}_{jk.i}[\delta^{(3)}h^{ij}]\Bigg] \tag{8.E}\label{dR2dh}
\end{align}
with
\begin{align*}
 \sigma^{\zero q.ab}_{rs}[f] &= -\tfrac12\gamma^{qt} (\delta^{ab}_{tr}\de_s + \delta^{ab}_{ts}\de_r - \delta^{ab}_{rs}\de_t ) f \\
 \sigma^{\one q.ab}_{rs}[f] &= -\gamma^{q(a}\delta\Gamma^{\one v)}_{rs} \cdot f 
				    + \tfrac12\gamma^{vt}\gamma^{qu}(\delta^{ab}_{tr}\de_s + \delta^{ab}_{ts}\de_r - \delta^{ab}_{rs}\de_t ) (h_{uv}f) \\
 \tau^{\zero q.ab}_{rs.i}[f] &=  \tfrac12\gamma^{qt}(\delta^{ab}_{tr}\de_i\de_s + \delta^{ab}_{ts}\de_i\de_r-\delta^{ab}_{rs}\de_i\de_t ) f \\
 \tau^{\one q.ab}_{rs.i}[f] &= \delta^{ab}_{uv}\gamma^{qu}\de_i(\delta\Gamma^{\one v}_{rs} f) - \tfrac12\delta^{ab}_{uv}\gamma^{qu}\gamma^{vt}(h_{tr,si}+h_{ts,ri}-h_{rs,ti})\cdot f \notag\\
			    &\hspace{0.05\linewidth} +\tfrac12\gamma^{qu}\gamma^{vt} (\delta^{ab}_{tr}\de_i\de_s + \delta^{ab}_{ts}\de_i\de_r - \delta^{ab}_{rs}\de_i\de_t ) (h_{uv} f),
\end{align*}
defined as
\begin{align*}
 \int d^3x\;\sigma^{\zero q.ab}_{rs}[f] &= \int d^3x\;\PD{\Gamma^{\one q}_{rs}}{h_{ab}} f \\  
 \int d^3x\;\sigma^{\one q.ab}_{rs}[f] &= \int d^3x\;\PD{\Gamma^{\two q}_{rs}}{h_{ab}} f \\
 \int d^3x\;\tau^{\zero q.ab}_{rs.i}[f] &= \int d^3x\;\pd{h_{ab}} \left[\de_i\Gamma^{\one q}_{rs} \right] f \\  
 \int d^3x\;\tau^{\one q.ab}_{rs}[f] &= \int d^3x\;\pd{h_{ab}} \left[\de_i\Gamma^{\two q}_{rs} \right] f,
\end{align*}
where $f$ is any appropriate functions of the spatial coordinates. The mismatch of notational superscripts $(n)$ was chosen in order to remain consistent with their role of identifying the order of perturbation, which is reduced by one as a result of the differentiation with respect to $h_{ab}$.

Expression \ref{dR2dh} contains a number of total derivatives as well as a significant degree of symmetry and evaluates to the simpler expression
\begin{equation*} \int d^3x\;\delta^{(3)}\PD{\widetilde{\delta R}^\two}{h_{cd}}(y) = -2\delta^{(cd)(kt)}\gamma^{ij}h_{it,jk} \end{equation*}
after all boundary terms have been dropped, which follows from assuming appropriate boundary conditions, and the constraint $\gamma^{ij}h_{ij}(x)=0$ has been used.

\end{mdframed}

}

\section{Analysis of the dynamics}

The dynamics of the perturbations is now completely defined. However, it is useful to develop an idea about some of the mechanisms that follow from this set of equations. Equations \ref{eq:8.3-xieom} and \ref{eq:8.3-mueom} were derived as part of a framework that described the full dynamics of the universe: both background and perturbations. But carefully considering the steps carried out above makes clear that the equations for $\xi_{ab}$ and $\mu^{ab}$ are independent of the \emph{specifics} of the background, so that any background evolution may be chosen and substituted into the expressions for $H_T^\zero(T)$ and $F(T)$. Hence it is possible, for example, to describe perturbations for an inflationary scenario, whose background was developed in section \ref{sec:inflation}. There one would use (with all symbols defined as above)
\begin{align} 
  H_{T\;infl}^\zero(T) &= a_0^{-3} e^{\frac32\kappa T^2} \\
  F_{infl}(T) &=a_0^{-3} e^{\frac32\kappa T^2} \cdot\bigg[ \frac{3}{4M_{Pl}^2} T^2 - 2m^2\phi_0(1-\epsilon_0+\mu T)\bigg]^{-1}
\end{align}
for the inflaton potential $V(\phi)=m^2\phi^2$.

Here we are interested in some more general features of the dynamics defined by equations \ref{eq:8.3-xieom} and \ref{eq:8.3-mueom}. In particular, for each of the equations there are two regimes corresponding to the dominance of one of the two terms inside the respective square parentheses. Consider equation \ref{eq:8.3-xieom} in the regime where
\begin{equation} H_T^\zero T\xi_{ab}(k) \gg 2\mu_{ab}(k).\end{equation}
Indices are raised an lowered using $\gamma_{ij}$ and its inverse. This regime could correspond to a very large universe (large volume $H_T^\zero$) at a sufficiently early time, for example, or an appropriately sized mode function $\xi_{ab}(k)$ versus the corresponding `momentum mode' $\mu^{ab}(k)$ for some given mode with wave number $k$. In this regime the equation becomes independent of $\mu^{ab}$, 
\begin{equation} \xi_{ab}^\prime = M_{Pl}^{-2} F(T) H_T^\zero T\cdot \xi_{ab}, \end{equation}
which is readily solved by
\begin{equation} \xi_{ab}(T) = \xi_{ab}(T_1) e^{M_{Pl}^{-2}\bigintssss dT F(T) H_T^\zero T},\end{equation}
where $T_1$ is some reference time. Each mode just evolves as a function of time depending on the background and, crucially, the details are independent of the wave number of the modes, so that the overall shape of the perturbation (given by $h_{ab}$) is unchanged. There is no genuine evolution, corresponding to frozen modes. During an inflationary scenario, for example, $H_T^\zero$ grows dramatically while $T$ itself does not change significantly, so that a large number of modes (modes with a large spread of ratios $\xi_{ab}(k)/\mu_{ab}(k)$) enter this regime and freeze out.

In the opposite regime,
\begin{equation} H_T^\zero T\xi_{ab}(k) \ll 2\mu_{ab}(k),\end{equation}
the resulting equation
\begin{equation} \xi_{ab}^\prime (k) = 2 F(T) \cdot M_{Pl}^{-2} \gamma_{ac}\gamma_{bd}\mu^{cd}(k) \end{equation}
has essentially the usual form `velocity $\propto$ momentum', up to an overall time-dependent scaling factor that depends on the background. This equation implies that the dynamics are not frozen out ($\mu^{ab}=0$ would contradict the assumption about the regime) and therefore corresponds to genuine evolution. Just like in the case of a particle system where the corresponding equation is `$\dot{x}=p/m$' the real content of the dynamics is found in the equation describing the evolution of the momentum.

The regimes for the equation \ref{eq:8.3-mueom} describing the `momentum modes' are not identical to the regimes of equation \ref{eq:8.3-xieom} but instead correspond to long and short wavelengths (small and large $|k|$ respectively). In the long wavelength regime one has
\begin{equation} |T|\mu^{ab}(k) \gg M_{Pl}^4 2H_T^{\zero\frac13}\left(\delta^{(ab)(cd)}-\tfrac13\gamma^{ab}\gamma^{cd}\right)\gamma^{ij}\xi_{ic}k_jk_d,\end{equation}
which means that the equation \ref{eq:8.3-mueom} has the approximate form
\begin{equation} \mu^{ab\prime}(k) = M_{Pl}^{-2} F(T)H_T^\zero |T| \cdot \mu^{ab} \end{equation}
and therefore solutions (in an expanding universe, where $T<0$)
\begin{equation} \mu^{ab}(T) = \mu^{ab}(T_1) e^{-M_{Pl}^{-2}\bigintssss dT F(T) H_T^\zero T}, \end{equation}
that is, all long-wavelength momentum modes scale alike (independent of the exact value of $k$) as a function of time whose form is dependent on the background only and which furthermore matches the (inverse of the) time evolution of the frozen $\xi_{ab}$ modes. In other words, in the long-wavelength regime the momentum modes do not experience real evolution.

On the other hand, for short wavelengths (large $|k|$) the second term dominates and the equation approximates to
\begin{equation} \mu^{ab\prime}(k) = F(T)H_T^{\zero\frac43} M_{Pl}^2\Big(\xi^{aj}k_Jk^b+\xi^{bj}k_jk^a-\tfrac23\gamma^{ab}\xi^{ij}k_ik_j\Big), \end{equation}
which corresponds to genuine evolution since there is linear dependence on the variable $\xi^{ab}$. This is furthermore an indication that one could describe the system once again in terms of a damped harmonic oscillator with time dependent terms (compare section \ref{sec:modefreezing}).  

The system described by these equations is complicated and a full analysis likely requires numerical tools. We will not explore the classical dynamics further here. Instead, in the next part of this thesis we proceed with the development of the quantum theory derived by canonical quantisation of the theory developed here.

\part{Canonical quantum cosmology with York time}\label{part3}
 
\chapter{Quantum Friedmann-Lema\^itre cosmology}\label{chap:QuantFriedmann}

\textit{In this chapter we canonically quantise the cosmological minisuperspace model developed in chapter \ref{chap:Friedmann}. We discuss some of the features of quantum theories derived via canonical quantisation from York-time Hamiltonian-reduced classical theories in the context of this model. The contents of this chapter were published as part of ref.~\citep{RoserValentini2014a}.}

\section{Quantum theory of the free scalar field}\label{sec:quantumFriedmannfree}

Whether or not the application of the quantisation recipe to gravity leads to a viable theory of quantum gravity remains to be seen. However, our theory of gravity obtained via Hamiltonian reduction from the ADM formalism does not suffer from the problem of time and so canonical quantisation is at least possible. A fundamentally different quantisation procedure will be proposed in chapter \ref{chap:traj} and applied to cosmology in chapter \ref{chap:cosmtraj}.

Ultimately empirical data must determine the correct form of quantum gravity. With such measurements however unavailable at the time of writing and, arguably, in the foreseeable future, theoretical motivations may be considered. The application of canonical quantisation to the metric field is motivated by the success of its application to matter fields, specifically the fields of the Standard Model. Canonical quantisation of gravity is therefore plausible \emph{if} the gravitational field is sufficiently `matter-field-like'. In the aside on proper time in section \ref{sec:YTproperties} we touched on some arguments suggesting that gravity ought to be considered a dynamical field like matter fields and that it is special only in that it has dynamics which allows for a geometric representation. Specifically, it couples to all other fields in a specific way, namely via so-called minimal coupling. In this picture the geometric aspect is however not fundamental. If, on the other hand, the converse is true and the gravitational field tensor $g_{ab}$ (or $\supfour g_{\mu\nu}$) really is just a mathematical representation of actual spatial (or spatio-temporal) geometry, then treating it like a matter field with regards to quantisation is ad hoc.\footnote{This is one example where philosophical reasoning and the details of one's realist stance are tied to how one proceeds mathematically.}

Conversely, the success of different treatments of gravity with regards to quantisation may give an indication of which `view' is correct. This requires the pursuit of all lines of enquiry. Quantising the 3-metric field (with a particular, theoretically motivated choice of foliation) is one such line. The simple minisuperspace model considered here is a first step along that line, although it does, of course, have direct physical application given the observed approximate homogeneity and isotropy of our universe.

To begin, we wish to construct a quantum theory using the classical theory developed in chapter \ref{chap:Friedmann} as a starting point. For illustrative purposes, we first consider the free theory (with $V(\phi)=0$). In the free theory the sign of the momentum is unchanging, so that classically we may replace $|p_\phi|\rightarrow p_\phi$,\footnote{This assumption was made inadvertently in ref.~\citep{RoserValentini2014a} where we first reported these results. This treatment is however technically incorrect: That the momentum is not changing sign is a result of the classical dynamics and not a defining feature of the theory. The more accurate way to proceed would be the method applied in the next chapter and briefly discussed in the box below, where a square-root Hamiltonian is interpreted via diagonalisation of its square. Fortunately, the results obtained with the present method are however encouraging. The pseudo-classical behaviour of the quantum trajectories (see below) has also been deduced by \citet{John2015}, although without reference to York time and with a slightly expanded notion of quantum trajectory.} up to a sign (though recall that the sign of the Hamiltonian is arbitrary anyway). The classical Hamiltonian is then 
\begin{equation}H=-\sqrt{\tfrac23}M_{Pl}\frac{p_\phi}{-T}. \label{eq:9.1-classHamiltonian}\end{equation} 
We promote the canonical variables to operators, $\phi\rightarrow\hat{\phi}$ and $p_\phi\rightarrow\hat{p}_\phi=-i\pd{\phi}$, to act on a wavefunction $\Psi(\phi,T)$ and as our law of evolution we take the Schr\"odinger equation:
\begin{equation} \label{eq:FRIEDQUANTfreeSchroedinger} i\PD{\Psi}{T}(\phi,T) = \hat{H}\Psi(\phi,T) = -i\sqrt{\tfrac23}M_{Pl}\frac{1}{T}\PD{\Psi}{\phi}(\phi,T).\end{equation} 

The Hamiltonian, being proportional to $\hat{p}_\phi$, is a time-dependent generator of translations in the configuration space of $\phi$. Thus the wavefunction is shifted in its entirety and the evolution is real (in the sense that if $\Psi$ only takes real values at one time $T_0$, it takes real values at all times). To see this explicitly, first consider the evolution of $\Psi$ over a small time interval $T_0\rightarrow T_0+\delta T$. To first order in $\delta T$ we have
\begin{align}\Psi(\phi,T_0+\delta T)
  &= \Psi(\phi,T_0)+\delta T\PD{\Psi}{T}(\phi,T_0) \notag\\
  &= \Psi(\phi,T_0)-\sqrt{\tfrac23}M_{Pl}\frac{\delta T}{T_0}\PD{\Psi}{\phi}(\phi,T_0) \notag\\
  &= \Psi\left(\phi-\sqrt{\tfrac23}M_{Pl}\frac{\delta T}{T_0},T_0\right).
\end{align}
Thus during the interval $\delta T$ the wavefunction is shifted in the direction of decreasing $\phi$ (since $T_0<0$ by assumption). 
For $T_0\rightarrow T=T_0+\Delta T$ with finite $\Delta T$, it is readily shown that
\begin{equation}\label{eq:FRIEDQUANTfreeWFsolution} \Psi(\phi,T)=\Psi\left(\phi-\sqrt{\tfrac23}M_{Pl}\ln\frac{T}{T_0}, T_0\right) \end{equation}
satisfies eq.\ \ref{eq:FRIEDQUANTfreeSchroedinger}.

Eq.\ \ref{eq:FRIEDQUANTfreeWFsolution} implies that for $T\rightarrow-\infty$ the wavepacket is located at $\phi\rightarrow+\infty$ and for $T\rightarrow0$ at $\phi\rightarrow-\infty$, matching the classical evolution of $\phi$. Since the shape of $\Psi$ in $\phi$-space remains unchanged, the probability current $j(\phi,T)$ will be uniform in $\phi$ (though $T$-dependent). This will be seen explicitly below when we consider the de~Broglie-Bohm trajectories.

In standard quantum mechanics (without trajectories) the `classicality' of the quantum evolution may be seen at the level of expectation values. For an appropriately narrow wave function, we can apply Ehrenfest's theorem and obtain 
\begin{equation} \deriv{T}\left\langle\phi\right\rangle = \sqrt{\tfrac23}M_{Pl}T^{-1}, \qquad \deriv{T}\left\langle p_\phi\right\rangle = 0, \end{equation}
that is, the classical evolution, but for expectation values. Using trajectories, we will show that the assumption of a narrow wave packet is, in fact, unnecessary in order to obtain the classical-like behaviour.

First, however, consider now the role of the scale factor $a$. Previously we argued that the numerical value of the Hamiltonian is to be interpreted as the volume of the universe (section \ref{sec:meaningofHamiltonian}). This is also true for the quantum theory.

In a more usual context the classical interpretation of the value of the Hamiltonian is that of energy. In the quantum theory this translates into an energy spectrum, given by the eigenvalues of the Hamiltonian operator (with corresponding eigenstates). Analogously, it is natural here to interpret the eigenvalues of the Hamiltonian as elements of a `volume spectrum' with corresponding `volume eigenstates'. In the quantum theory it is therefore plausible to characterise the volume (or scale factor) associated with a particular state via the expectation value of the Hamiltonian\footnote{If $a$ were a dynamical variable of our Hamiltonian system (as it was in the unreduced model), this equation would denote a constraint imposed at the level of expectation values. However, in our case $a$ is not such a variable and is instead introduced as a definition.}
\begin{equation} a^3\equiv\oppr{\Psi}{\hat{H}}{\Psi}.\label{eq:FRIEDQUANTexpectationvaluedef}\end{equation}

In order to see that this makes sense, consider the eigenstates of the Hamiltonian. It is easy to confirm that these are given by $\Psi_k(\phi)=N_ke^{-ik\phi}$, where $k\in\mathbb{R}$ labels the eigenstate and $N_k$ is a normalisation constant. The corresponding eigenvalues, which we write as `$a_k^3$' since they are to correspond to a volume, are 
\begin{equation} a_k^3 = \sqrt{\tfrac23}M_{Pl}T^{-1}k.\end{equation}
Thus the eigenvalue depends linearly on the wavenumber of the eigenstate and inversely on York time $T$. Since $k$ can take any real value, the spectrum is continuous. Note that the time dependence is the same for all eigenstates and therefore also for arbitrary linear superpositions. Since the set of $\Psi_k$ form a complete basis the time dependence is therefore identical for all states. Consider a linear superposition $\Psi(\phi)=\sum_kA_k\Psi_k(\phi)=\sum_kA_kN_ke^{-ik\phi}$, where $A_k\in\mathbb{C}$ are arbitrary coefficients. The volume expectation value (\ref{eq:FRIEDQUANTexpectationvaluedef}) is
\begin{align} \oppr{\Psi}{\hat{H}}{\Psi} 
    &= \int d\phi\; \Psi^\ast\hat{H}\Psi  \notag\\
    &= \sqrt{\tfrac23}M_{Pl}\frac{1}{T}\sum_{k.k^\prime}\int d\phi\;A_{k^\prime}^\ast\Psi_{k^\prime}^\ast kA_k\Psi_k \notag\\
    &= \sqrt{\tfrac23}M_{Pl}\frac{1}{T}\sum_k k|A_k|^2
\end{align}
where in the last step we used the orthonormality of the eigenfunctions. The volume expectation value is thus inversely proportional to $T$. The chosen state $\Psi$ only determines an overall factor $\sum_k k|A_k|^2$.

Finally consider the time evolution of $a^3$ for an arbitrary state $\Psi$. From $da^3/dT = d\oppr{\Psi}{\hat{H}}{\Psi}/dT$, and using eq.\ \ref{eq:FRIEDQUANTexpectationvaluedef}, we see that 
\begin{equation}\frac{a^\prime}{a}=-\frac{1}{3T}\label{freeaprimea}\end{equation}
independently of $\Psi$. Note that this is exactly the equation giving the evolution of $a$ in the classical theory. 

If we consider quantum trajectories as they appear in the de~Broglie-Bohm formulation these will be geometrically parallel in the extended configuration space ($\phi$-$T$-space). The velocity (and therefore guidance equation for the trajectories) can be read off from eq.\ \ref{eq:FRIEDQUANTfreeWFsolution} to be 
\begin{equation} v_\phi(T)=\sqrt{\tfrac23}M_{Pl}\frac{1}{T}.\end{equation}
This result could also have been obtained by evaluating $\pd{T}|\Psi|^2$ using the Schr\"odinger equation in order to arrive at the continuity equation for the probability density. This result matches eq.\ \ref{eq:5.3-phiprimefree}, the equation for the classical trajectories. Thus the de~Broglie-Bohm and classical trajectories are identical. This may have been expected for a free theory. However, below we show that a similar result holds even in the presence of a potential, which one would not naturally expect.

The use of quantum trajectories allows for an alternative method to calculate the evolution of $a$ via the identification of `local expectation values' $a^3\equiv\operatorname{Re}[\psi^*\hat{H}\psi/|\psi|^2](\phi,T)$ (evaluated along the trajectory), in the sense of \citet[sec.\ 3.5]{Holland1993}. However, in this thesis we choose to employ the identification of $a^3$ via eq.\ \ref{eq:FRIEDQUANTexpectationvaluedef}.


{
\begin{mdframed}
  \setlength{\parindent}{10pt}

\small
\bigskip
\textbf{A more careful treatment of `square-root' Hamiltonian}\bigskip

Above we treated the `square-root' Hamiltonian via the simplification of considering the quantisation of the Hamiltonian \ref{eq:9.1-classHamiltonian} rather than the more fundamentally correct expressions,
\begin{equation*} H_T = -\sqrt{\tfrac23} \frac{M_{Pl}}{-T} \sqrt{p_\phi^2}.\end{equation*}
This modification is adequate for the classical theory where the momentum $p_\phi$ does not change sign. In fact, we will use it again below in section \ref{sec:quantumFriedmannpotential} even in the presence of a potential where the applicability of this modification of the Hamiltonian is more limited (see our discussion of `turning points' in chapter \ref{chap:Friedmann}). The fact that it is classically adequate follows from the classical dynamics and so, strictly speaking, this should not be assumed in the quantum theory. This raises the question how to interpret the operator resulting from the quantisation of the classical phase-space function $\sqrt{p_\phi^2}$. One might be tempted to write
\begin{equation*} \widehat{\sqrt{p_\phi^2}} = \sqrt{\hat{p}_\phi^2} = \sqrt{-\partial_\phi^2}, \end{equation*}
(with $\partial_\phi=\pd{\phi}$) which is not problematic but does not offer a resolution either since one is left to interpret the mathematical meaning of the square root of a second derivative.

The way to deal with this difficulty is by interpreting the expression via diagonalisation. In general, if $\ket{h}$ is an eigenfunction of $\hat{h}$ with eigenvalue $h$, $\hat{h}\ket{h}=h\ket{h}$, then it is also an eigenfunction of $\hat{f}\equiv\hat{h}^2$ with eigenvalue $h^2$, 
\begin{equation*} \hat{h}^2\ket{h}=h^2\ket{h}.\end{equation*}
The converse is in general not true: there may be eigenfunctions of $\hat{f}$ which are not eigenfunctions of $\hat{h}$. However, if $\hat{f}$ is diagonalisable with a complete set of eigenstates (this follows from Hermiticity), then these also form a set of eigenstates for $\hat{h}=\hat{f}^\frac12$ with square-rooted eigenvalues.

Regarding Hermiticity, if $\hat{h}$ is Hermitian, then so is $\hat{f}$. Again, for arbitrary operators the converse is not true: If $\hat{A}^2=\hat{B}$ and $\hat{B}$ is Hermitian, then it is not guaranteed that $\hat{A}$ is Hermitian. However, if $\hat{A}=\hat{B}^\frac12$ is defined in terms of the diagonalised operator, then $\hat{A}$ will be Hermitian too (up to a subtlety that we will establish below for the particular case of the quantisation of the anisotropic universe in section~\ref{sec:quantKasner}, where we will deal with the square-root operator more carefully than in the present chapter. The subtlety is that $\hat{B}$ must be positive semi-definite, that is, it must have only non-negative eigenvalues, so that their square roots are real). That is, not every square root of a Hermitian operator is Hermitian, but one can always find one via diagonalisation.

Let us consider the implications for the minisuperspace model considered in this chapter. Define the square of the Hamiltonian operator,
\begin{equation*} \hat{F} \equiv\hat{H}_T^2 = \tfrac23 M_{Pl}^2 T^{-2} \hat{p}_\phi^2\end{equation*}
and find its eigenfunctions $\psi_k$, satisfying
\begin{equation*} \hat{F}\psi_k(\phi) = \tfrac23 M_{Pl}^2 T^{-2} (-\partial^2_\phi)\psi_k(\phi) = g\psi_k(\phi).\end{equation*}
This is solved by the eigenfunctions
\begin{equation*} \psi_k(\phi) = Ne^{ik\phi} \end{equation*}
with eigenvalues
\begin{equation*} g = \tfrac23 M_{Pl}^2 T^{-2} (-k^2).\end{equation*}
The prefactor $N$ is a normalisation constant. In general $k$ could be any complex number. However, the functions are only bounded for $k\in\mathbb{R}$, so viable eigenfunctions are restricted to real values of $k$.\footnote{The functions must be bounded if it is to be possible to interpret $|\psi_k|^2$ as a probability distribution. It is not completely clear if the possibility of such an interpretation is a necessary feature of a theory describing the evolution of a single isolated universe, especially if the universe were represented by a de~Broglie-Bohm trajectory and the role of the $\psi_k$ is fundamentally that of a guiding wave. So relaxing this condition may be a defensible option in the present case. However, complex $k$ imply complex volume eigenvalues. (This essentially relates to the fact that Hermiticity of the Hamiltonian is not established for unbounded functions, a mathematical detail that is rarely relevant in conventional quantum mechanics, where the existence of a probability interpretation is assumed.)} The eigenvalues of the actual Hamiltonian are then 
\begin{equation*} h = -\sqrt{\tfrac23}M_{Pl}\frac{|k|}{T} \end{equation*}
and are therefore unchanged from what was established in the text using the simplified square-root-free Hamiltonian.

In order to derive the de~Broglie-Bohm velocity $v$ one needs the probability current \citep{StruyveValentini2009}, given via the continuity equation
\begin{equation*} \PD{|\Psi|^2}{T} - \partial_\phi (|\Psi|^2 v) = 0. \end{equation*}
Using the Schr\"odinger equation and its conjugate one arrives at an expression for the velocity:
\begin{equation*} v = -|\Psi|^{-2}\partial_\phi^{-1}\bigg[ i\left((H\Psi)^\dagger\psi-i\psi^\dagger(H\psi)\right)\bigg].\end{equation*}
In general the inverse operator `$\nabla^{-1}$' cannot be evaluated explicitly and some other manipulation is necessary for an explicit result. However, in the one-dimensional case the operator $\partial_\phi^{-1}$ simply denotes integration with respect to $\phi$. The integration constant corresponds to the general divergence-free term that may be added to the de~Broglie-Bohm velocity without affecting the probability density and its evolution. The application of $\hat{H}$ to a general state $\Psi$ is evaluated via the decomposition of $\Psi$ into volume eigenfunctions,
\begin{equation*} \Psi = \sum_k c_k \psi_k, \end{equation*}
with the sum being over all values of $k$ for which there is a non-vanishing complex coefficient $c_k$. The action of the Hamiltonian on $\psi_k$ is known.


\end{mdframed}
}

\section{Quantum theory with a potential}\label{sec:quantumFriedmannpotential}

The quantisation of the homogeneous isotropic minisuperspace model with a non-vanishing scalar-field potential proceeds along similar lines. Here (as we did in ref.~\citep{RoserValentini2014a}) we will also make the simplification of replacing the `square-root of the squared momentum' operator by the momentum itself. Unlike in the case of the free theory where the momentum does not change sign classically, it may do so in the case of a potential, so that the applicability of the quantised theory developed in this section is limited. A proper treatment should proceed along the lines described in the box above. In the next chapter, we will give an example of the proper treatment of a square-root Hamiltonian in the form of the quantisation of the anisotropic minisuperspace model of section~\ref{sec:ClassKasner}.

The recipe of quantisation is ambiguous for a non-zero potential $V(\phi)\neq 0$ due to the necessity of choosing a factor ordering in the Hamiltonian \ref{eq:5.3-k0Hamiltonian}. Different choices will lead to different dynamics. It turns out that the symmetric choice
\begin{align} 
  \hat{H}&=-\half\left( \hat{U}^{-\half}\hat{p}_\phi+\hat{p}_\phi \hat{U}^{-\half} \right) \notag\\
	 &=+\half\left(\frac{i}{\sqrt{\frac{3}{2M_{Pl}^2} T^2 -2V(\phi)}}\pd{\phi}+\pd{\phi}\frac{i}{\sqrt{\frac{3}{2M_{Pl}^2} T^2 -2V(\phi)}}\right)
	 , 
\end{align}
results in a Hermitian Hamiltonian, as may be readily verified.\footnote{Other orderings, while not in general Hermitian, are however PT-symmetric for even potentials $V(\phi)$, and so they may conceivably be considered as other viable choices, provided the treatment is adjusted accordingly (with appropriately modified inner products). See \citet{BenderIntro2005} for an introduction to PT-symmetric Hamiltonians in quantum theory. As always, ultimately the right choice is determined by observation.} The term $U$ is a shorthand defined by equation~\ref{eq:5.3-defofU}.
At the end of this section we will also show that this ordering has the special property that the resulting de~Broglie-Bohm trajectories match those of the classical theory just as they did in the free case, although the applicability of this result is limited given that the modified Hamiltonian without the square-root operator is used. 

Application of Ehrenfest's theorem gives
\begin{align} \deriv{T}\left\langle\phi\right\rangle &= \left\langle\frac{1}{\sqrt{\frac{3}{2M_{Pl}^2} T^2-2V(\phi)}}\right\rangle \notag\\
	      \deriv{T}\left\langle p_\phi\right\rangle &= -\left\langle\frac{1}{(\frac{3}{2M_{Pl}^2}T^2-2V(\phi))^\frac{3}{2}}\PD{V}{\phi}p_\phi\right\rangle
			    + \half\left\langle\left[p_\phi,\frac{1}{(\frac{3}{2M_{Pl}^2}T^2-2V(\phi))^\frac{3}{2} }\PD{V}{\phi}\right]\right\rangle,\label{eq:FRIEDQUANTEhrenfesteqs}
\end{align}
where the equation for $\phi$ has the form of the classical evolution equation, while the momentum equation differs by the second expression. For a state that is sufficiently `classical', that is, appropriately localised in both position and momentum (while consistent with the uncertainty principle) such that for a relevant functions $f(\phi)$ it is the case that 
\begin{equation}\langle f(\phi,p_\phi)\rangle\approx f(\langle\phi\rangle,\langle p_\phi\rangle), \label{eq:FRIEDQUANTclassicalitycondition}\end{equation} 
we have for $F(\phi)=(\frac{3}{2M_{Pl}^2}T^2-2V(\phi))^{-\frac{3}{2}}\PD{V}{\phi}$ that
\begin{equation} \langle[p_\phi,F]\rangle = \langle(p_\phi F-Fp_\phi)\rangle = \langle p_\phi F\rangle-\langle Fp_\phi\rangle=0,\end{equation}
having used the condition \ref{eq:FRIEDQUANTclassicalitycondition} for the choice $f(\phi,p_\phi)=p_\phi\cdot F(\phi)$ in the last equality. Thus the final term in the second equation of (\ref{eq:FRIEDQUANTEhrenfesteqs}) vanishes. Condition \ref{eq:FRIEDQUANTclassicalitycondition} also allows us to replace the remaining expressions on the right-hand sides of eqs.\ \ref{eq:FRIEDQUANTEhrenfesteqs} by the appropriate functions of $\langle\phi\rangle$ and we obtain classically evolving expectation values since the equations have the same form as the classical equations fo $\phi^\prime$ and $p_\phi^\prime$. Below we will show the recovery of classicality more fully at the level of trajectories.

Let us consider the evolution of the scale factor $a$ given by (\ref{eq:FRIEDQUANTexpectationvaluedef}). Proceeding as for the free theory, we first find the `volume eigenfunctions'. These are
\begin{equation}\psi_k^T = N_k\left(1-\frac{4M_{Pl}^2V(\phi)}{3T^2}\right)^\frac{1}{4} e^{-ik\int d\phi\,\left(1-\frac{4M_{Pl}^2V(\phi)}{3T^2}\right)^\half}, \label{eq:FRIEDQUANTeigenfunctionswithpotential}
\end{equation}
as is easy to verify. Unlike the eigenfunctions for the free theory, which were time independent (only the eigen\emph{values} carried time dependence), the functions here are time dependent also. \footnote{Indeed, it is easy to see why this must be so. Suppose we wish to find time-independent eigenfunctions $\chi_k(\phi)$, satisfying $\hat{H}\chi_k=w_k(T)\chi_k(\phi)$, where $w_k(T)$ are the corresponding eigenvalues, which may depend on time $T$. Evaluating $\hat{H}\chi_k$, the condition on $\chi_k$ would be that
\begin{equation}w_k(T) = iU^{-\half}\left(\PD{\ln\chi_k}{\phi}-\half U^{-1}\PD{U}{\phi}\right)\end{equation}
is only a function of time $T$ (since the eigenvalue $w_k(T)$ cannot have dependence on $\phi$). But no time-independent $\chi$ can satisfy this condition since $U(\phi,T)$ is time dependent.}

Note that we could have rescaled the eigenfunctions $\psi_k^T$ by an arbitrary function of time. This ambiguity is resolved by demanding that for early times $-T\gg1$ the eigenfunctions match those of the free theory. This is a reasonable condition since we have already established that the early-time dynamics for the trajectories is well approximated by that of the free theory.

The eigenvalues corresponding to these eigenfunctions (\ref{eq:FRIEDQUANTeigenfunctionswithpotential}) are 
\begin{equation}a_k^3=\sqrt{\frac23}M_{Pl}\frac{k}{T},\end{equation}
that is, they match those of the free theory. However, the eigenfunctions $\psi_k$ are not normalisable over the entire real line, $\phi\in(-\infty,\infty)$ unless the $V(\phi)$ falls off sufficiently fast as $\phi\rightarrow\pm\infty$, which is not the case for many commonly investigated potentials, such as $V(\phi)=M^4(\phi/M_{Pl})^2$ considered above.

Time-dependent eigenfunctions of a time-dependent Hamiltonian do not, in general, evolve into eigenfunctions (that is, a general eigenstate $\psi_k^{T_0}$ of $H(T_0)$ does not evolve into an eigenstate $\psi_{k^\prime}^{T_1}$ of $H(T_1)$) and few general results are known. We will postpone detailed analysis of appropriate models to future work. 

Instead, let us consider a general superposition $\Psi(T_0)=\sum_k A_k(T_0) \psi_k^{T_0}(T_0)$ at time $T_0$. For later times, we write $\Psi(T)=\sum_kA_k(T)\psi_k^T(T)$, where we must include the time-dependence of the coefficients $A_k(T)$ in order to account for the fact that eigenfunctions of $H(T_0)$ do not evolve into eigenfunctions of $H(T)$ for some $T>T_0$. Setting $a^3\equiv \oppr{\Psi}{H}{\Psi}$, we find
\begin{equation} a^3= \sum\limits_k|A_k(T)|^2|N_k|^2\int d\phi \sqrt{1-\frac{4M_{Pl}^2V(\phi)}{3T^2}}\cdot\sqrt{\frac23}M_{Pl}\frac{k}{T}.\end{equation}
Differentiation with respect to $T$ gives
\begin{equation}
 \frac{a^\prime}{a} = -\frac{1}{3T}\left(1-\frac{ \int d\phi\; 2V(\phi)/\sqrt{\frac{3}{2M_{Pl}^2}T^2-2V(\phi)} }{ \int d\phi\,\sqrt{\frac{3}{2M_{Pl}^2}T^2-2V(\phi) }}\right)
			+\frac{ \sum_k |N_k|^2\cdot k\cdot \deriv{T}|A_k(T)|^2 }{ 3\sum_k |N_k|^2\cdot k\cdot |A_k(T)|^2 }.
\end{equation}
We note that in this expression only the last term depends on the quantum state. In general, the two integrals may be divergent and care must be taken when evaluating their ratio. In practice, some regularisation method may have to be applied. The result for the free theory is recovered in the case when $V(\phi)=0$.

Consider finally the de~Broglie-Bohm trajectories. From the Schr\"odinger equation for this Hamiltonian $\hat{H}$ it is straightforward to derive the continuity equation for the probability density,
\begin{equation} \PD{|\Psi|^2}{T} = \pd{\phi}\left(U^{-\half}|\Psi|^2\right) \end{equation}
from which we read off the guidance equation for the de~Broglie-Bohm trajectories,
\begin{equation} v_\phi = -U^{-\half} = -\frac{1}{\sqrt{\frac{3}{2M_{Pl}^2}T^2-2V(\phi)}}. \end{equation}
This is indeed the classical velocity $\phi^\prime$ for an arbitrary potential $V(\phi)$.

It is interesting to note that the velocity $v_\phi$ is independent of the state $\Psi$ and only depends on the position in $\phi$-configuration space. This is a result of the first-order form of the Hamiltonian and contrasts to more conventional second-order systems (where the velocity might depend on the phase of $\Psi$, for example).

The identical law for the classical and de~Broglie-Bohm trajectories implies that our considerations for the classical evolution of the universe apply to the quantum evolution, too. This enables us to develop a perturbation theory that is entirely quantum, using the fact that the homogeneous background evolves just as if it were classical and independently of the quantum state.


\chapter{Quantum York-time cosmological perturbation theory}\label{chap:quantcosmpert}

\textit{The goal of this chapter is the canonical quantisation of the reduced-Hamiltonian cosmological perturbation theory developed in chapter \ref{chap:perttheoformalism}. As a preliminary step, in order to discuss the unconventional commutator structure we first consider the canonical quantum theory of the anisotropic minisuperspace model of section \ref{sec:ClassKasner} before moving on to perturbation theory proper. Section \ref{sec:quantKasner} was published as part of ref.~\citep{Roser2015a}, while section \ref{sec:quantperts} formed part of ref.~\citep{Roser2015b}.}

\section{Quantisation, representations and the anisotropic homogeneous quantum universe}\label{sec:quantKasner}

Whether or not following the canonical quantisation recipe is appropriate for arriving at a quantum theory of gravity is yet to be determined. Rather than theoretical speculation about what formal features we expect a quantum theory of gravity to have, our approach is to examine the theory resulting from an application of the standard quantisation procedure to the classical reduced-Hamiltonian theory. If the quantum theory is found to be wanting with regards to internal consistency or empirical adequacy, another approach will have to be proposed.

The quantisation procedure is not completely straightforward. In the homogeneous case we were able to follow the canonical-quantisation procedure without difficulty (the only `unusual' feature was having a strongly time-dependent Hamiltonian and its `square-root' form). In contrast, in the present anisotropic model and in the general case the non-canonical Poisson brackets must lead to a modified commutator-bracket structure.

It is therefore worth isolating this feature of the quantum theory by exploring the canonical quantisation of the anisotropic minisuperspace model of section \ref{sec:ClassKasner} before proceeding to the quantisation of perturbation theory itself. While empirically inadequate with regard to our actual universe, useful insights about the structure of the quantum theory are gained.

\subsection{Quantisation with non-canonical Poisson brackets}

The natural generalisation of the usual bracket `promotion', schematically expressible as
\begin{equation} \text{`` } \{q,p\}=\mathbb{1}\text{ ''}\qquad\longrightarrow\qquad \text{`` }[\hat{q},\hat{p}]=i\hbar\hat{\mathbb{1}} \text{ ''}, \end{equation}
is that the right-hand side of the Poisson equations is also promoted to operators and multiplied by a factor $i\hbar$,
\begin{equation} \text{`` } \{q,p\} = f(q,p) \text{ ''}\qquad\longrightarrow\qquad \text{`` }[\hat{q},\hat{p}]=i\hbar\widehat{f(q,p)} \text{ ''}\end{equation} 
and similar for other non-commuting brackets. Few works consider quantisation in general for non-canonical Poisson structure. One notable exception is ref.~\citep{Thiemann2007}. In general this promotion procedure leads to factor-ordering ambiguities however since the right-hand side may contain (classically) products of $p$s and $q$s, or other non-Poisson-commuting combinations (such as $p$s and $p$s in our case). This implies that there is, strictly speaking, an ambiguity due to factor order in the function denoted by $f(q,p)$ above. The ambiguity is however `small' since all possible orderings are related by adding or subtracting a commutator bracket on the right-hand side. This commutator bracket, when evaluated, introduces a factor of $\hbar$ (in addition to the overall factor of $\hbar$ present on the right), so that the ambiguity is of order $\hbar^2$, while the unambiguous part of the commutator expression is only of order $\hbar$. Nonetheless the quantisation recipe is indisputably ambiguous when applied to classical theories with such Poisson structures. Fortunately for our purposes there will be easy criteria to choose an appropriate order in the cases we have to consider.

The Poisson structure of the reduced variables in the general York-time reduced-Hamiltonian theory is given by eqs.~\ref{eq:4.3-ggPB} and \ref{eq:4.3-gpiPB} (and the vanishing of $\{\tilde{g}_{ab},\tilde{g}_{cd}\}$). The resulting commutator algebra is then given by
\begin{align}
 [\hat{\gtilde}_{ab}(x),\hat{\gtilde}_{cd}(y)] &= 0 \label{eq:9.1-ggCB}\\
 [\hat{\gtilde}_{ab}(x),\hat{\pitilde}^{cd}(y)]   &= i\hbar\left(\delta_a^{(c}\delta_b^{d)}-\tfrac13\hat{\gtilde}_{ab}\hat{\gtilde}^{cd}\right)\delta^3(x-y) \label{eq:9.1-gpiCB}\\
 [\hat{\pitilde}^{ab}(x),\hat{\pitilde}^{cd}(y)] &= \frac{i\hbar}{3}\left(\hat{\gtilde}^{cd}\hat{\pitilde}^{ab}-\hat{\gtilde}^{ab}\hat{\pitilde}^{cd}\right)\delta^3(x-y). \label{eq:9.1-pipiCB}
\end{align}
The operators $\hat{\gtilde}_{ab}$ and $\hat{\pitilde}^{cd}$ act on wave functionals $\Psi$.\footnote{In the presence of matter this functional also depends on the matter field configuration, such as a scalar field $\phi(x)$. Since this is not important for our purposes here, we do not include it here in order to avoid obscuring the notation. Of course, $\Psi$ is furthermore dependent on York time $T$, which we also suppress at this stage.} The inverse-metric operator $\gtilde^{ab}$ has the obvious meaning, satisfying $\gtilde_{ab}\gtilde^{bc}=\delta_a^c$. For reduced notational clutter it is useful to suppress the position arguments ($x$, $y$, etc.).

By inspection one can find a `position' basis for $\Psi$, that is, a representation of the operators and the wave functional such that $\Psi$ is expressed as a functional of the reduced-metric configurations $\gtilde_{ab}(x)$ and constitutes eigenfunctions of the operator $\hat{\gtilde}_{ab}$ with the eigenvalue being the argument of $\Psi$,
\begin{equation} \hat{\gtilde}_{ab}\Psi[\gtilde_{ij}] = \gtilde_{ab}\Psi[\gtilde_{ij}].\end{equation}
The momentum operator acts as
\begin{equation}\hat{\pitilde}^{ab}\Psi[\gtilde_{ij}] = 
		    \left[-i\hbar\left(\delta^{(ab)}_{(cd)}-\tfrac13\gtilde^{ab}\gtilde_{cd}\right)\frac{\delta}{\delta\gtilde_{cd}}\right]\Psi[\gtilde_{ij}]. \label{eq:10.1-posreppi}\end{equation}
One can verify that this is indeed a representation of the commutators \ref{eq:9.1-ggCB} - \ref{eq:9.1-pipiCB}. 

However, there is no momentum representation. This follows directly from the fact that momenta do not commute with each other, eq.~\ref{eq:9.1-pipiCB}. A momentum representation would require functionals $\Psi[\pitilde^{ij}]$ such that 
\begin{equation} \hat{\pitilde}^{ab}\Psi[\pitilde^{ij}] = \pitilde^{ab}\Psi[\pitilde^{ij}], \end{equation}
but then
\begin{equation} \label{eq:9.1-Pidontcommute}
\hat{\pi}^{ab}\hat{\pi}^{cd}\tilde{\Psi}[\pi^{ij}(x)]=\pi^{ab}\pi^{cd}\tilde{\Psi}[\pi^{ij}(x)]
	  =\pi^{cd}\pi^{ab}\tilde{\Psi}[\pi^{ij}(x)]=\hat{\pi}^{cd}\hat{\pi}^{ab}\tilde{\Psi}[\pi^{ij}(x)] ,
\end{equation}
contradicting the commutator relation \ref{eq:9.1-pipiCB}. The box below discusses the physical and formal meaning of this and draws contrasts with loop quantum gravity.

The position representation identically satisfies the quantum constraint demanding the vanishing of the trace of the momentum for physical states,
\begin{equation} \hat{\gtilde}{ab}\hat{\pitilde}^{ab} \Psi_{phys}[\gtilde_{ij}] = 0.\end{equation}
The `scale-free' constraint,
\begin{equation} \hat{\gtilde}\Psi[\gtilde_{ij}] = \hat{\mathbb{1}}\Psi[\gtilde_{ij}], \end{equation}
on the other restricts $\Psi$ to be non-vanishing on the surface defined by the condition $g=1$. This may be more clearly understood in the context of a finite-dimensional model, such as the anisotropic cosmological model considered in this chapter.

{
\begin{mdframed}
  \setlength{\parindent}{10pt}

\small
\bigskip
\textbf{Momentum observables: \newline\indent Reduced-Hamiltonian quantisation versus Loop Quantum Gravity}\bigskip

The origin of the non-Hermiticity of the momenta may be understood as follows. In the usual formal canonical quantisation procedure one constructs a commutative $C^*$-algebra of complex functions and the action of complex vector fields on the configuration space (with its representation respecting an involution) in order to ensure that all phase-space functions become `observables' in the quantum theory. For example, in Loop Quantum Gravity this is implemented via the shift to the loop variables, which have the appropriate structure. There one has a kinematical (i.e.\ prior to `knowledge' of the constraints) phase space in the form of a cotangent bundle over the configuration space and can proceed accordingly. Loop quantum gravity is perhaps the most advanced approach to canonical quantum gravity (a phrase borrowed from ref.~\citep{Koslowski2013}), with a rigorously defined quantisation and Hilbert space. However, the introduction of loop variables comes at the cost of having configuration variables that (arguably) lack a clear physical interpretation.

The present theory, as constructed via the Hamiltonian reduction, differs qualitatively from the procedure used for Loop Quantum Gravity. The metric-derived variables $g_{ab}$ (or $q_i$ in the anisotropic minisuperspace model) and associated momenta do not lead to the full structure of a $C^*$-algebra. The set of position operators form their own (smaller) $C^*$-algebra (with $g_{ab}=g_{ab}^*=g_{ab}$ or $q_a^*=q_a$, and $F(g_{ab})^*=\bar F(g_{ab})$ or $F(q^a)^*=\bar F(q^a)$ for the involution) and this ensures the existence of the position basis (e.g.\ theorem 29.2.2 in ref.~\citep{Thiemann2007}). What is not present here is the involution of the vector fields on configuration space. This leads to the momenta not being `observables' and neither is the operator form of a general phase-space function involving the momenta (although some functions may be, including, fortunately, the Hamiltonian).

In contrast with conventional canonical quantisation such as employed in the construction of Loop Quantum Gravity, the reduced momenta do not span the cotangent space of the kinematical configuration variables $g_{ab}$ (or $q_i$) since instead they maintain the constraint $g=1$ (or $q_1q_2q_3=1$) \emph{by construction}, acting only tangentially to the constraint surface. The Poisson structure also ensures that the other constraint, $g_{ab}\pi^{ab}=0$ (or $q_ip^i=0$) is maintained by any generator constructed from the momenta and variables. In the quantum theory this is in effect the reason the analogous quantum constraint holds identically at the operator level.

Fortunately, the lack of momentum observables does not make the quantum theory unfeasible. Physically speaking, as has been pointed out by Feynman and Hibbs \citep[ch.\ 5]{FeynmanHibbs1965} and repeatedly been emphasised by Bell \citep[e.g.\ p.\ 196]{Bell1987} all physical observation is ultimately that of positions, such as time-of-flight measurements for particle momenta or simply the position of an apparatus pointer that is appropriately coupled to the system. 

Elaborating on this point, note that quantum theories may be viewed as theories of \emph{positions}. In general, this requires one to view the position basis as preferred\footnote{Some formulations, such as de~Broglie-Bohm theory, do this at the fundamental level, whereas in the Everettian formulation, for example, an (approximate) position basis may be preferred as a result of decoherence.}, although for our purposes this is already the case in light of the Poisson structure, as argued in the text. Particular types of position measurements can be interpreted as `momentum measurements'. For example, in the standard quantum mechanics of a single particle with the wavefunction initially localised in position space, the statistical spread of position-measurement outcomes after a fixed time interval corresponds to the probability distribution of potential momentum-measurement outcomes. The initial uncertainty in position becomes negligible at sufficiently long times of flight (see ref.~\citep[sec.\ 5-1]{FeynmanHibbs1965} for a detailed analysis). Accepting that the observation of momentum is only a particular type of position measurement, it becomes clear that a quantum theory is not strictly required to have this structure. The absence of self-adjoint momentum operators implies that the theory lacks certain formal properties in contrast with conventional quantum theories, but it is not physically inconsistent or unviable. 

\end{mdframed}
}

\subsection{Application to the anisotropic homogeneous universe}

Applying this quantisation scheme to the anisotropic cosmological model developed in section \ref{sec:ClassKasner}, we promote the (reduced) variables $q_i$ and $p^i$ to operators that are understood to act on kinematically allowed quantum states $\Psi_{kin}$. `Kinematically allowed' encompasses all states that are expressible as suitable normalisable functions over the Hilbert space that corresponds to the classical configuration space defined by the three position variables $q_1,q_2,q_3$. Much smaller is the space of physically allowed states --- states to which a physical meaning can be given and which are not mere mathematical artifacts. These states must also satisfy the quantum constraints, obtained by operator-promotion of the classical constraint equations. That is, a constraint $\phi(q_i,p^j)=0$ acts on states as $\phi(\qhat_i,\phat^j)\Psi_{phys}=0$, selecting physically meaningful states $\Psi_{phys}$. 

In the present case the constraint equations were given by eqs.~\ref{eq:7.2-scaleconstraint} and \ref{eq:7.2-tracefreeconstraint} with the resulting quantum constraints
\begin{align}
 \qhat_1\qhat_2\qhat_3\Psi_{phys} &= \mathbb{1} \Psi_{phys} \label{eq:9.1-quantumscaleconstraint}\\
 \qhat_a\phat^a\Psi_{phys} &= 0. \label{eq:9.1-quantumtraceconstraint}
\end{align}
The Poisson algebra is promoted to a commutator algebra of operators:
\begin{align}
 \label{eq:9.1-qpcommutator} [\qhat_a,\phat^b] &= i\hbar\left(\delta_a^b-\tfrac13\qhat_a\qhat^b\right), \\
 \label{eq:9.1-ppcommutator} [\phat^a,\phat^b] &= \frac{i\hbar}{3}\left[\phat^a\qhat^b-\phat^b\qhat^a\right]
\end{align}
The momentum operators do not commute, so as in the general case the theory does not possess a momentum representation. However, there is a `position' representation (a representation defined by the fact that the operators $\qhat_a$ act by multiplication of the argument, $\qhat_a\Psi(q_i)=q_a\Psi(q_i)$), given by
\begin{align}
 \qhat_a\Psi_{kin}(q_i) &= q_a\Psi_{kin}(q_i), \\
 \phat^b\Psi_{kin}(q_i) &= \left[-i\hbar\left(\delta^b_a-\tfrac13q^bq_a\right)\pd{q_a}\right]\Psi_{kin}(q_i).
\end{align}
It is easy to verify that these do indeed satisfy the commutation relations \ref{eq:9.1-qpcommutator}, \ref{eq:9.1-ppcommutator}. 

We are left to consider the meaning of the constraints. The tracelessness requirement in operator form \ref{eq:9.1-quantumtraceconstraint} is satisfied identically in this representation, so that it does not restrict the space of physical wavefunctions. On the other hand, the absence of absolute scale, taking the form of the self-adjoint quantum constraint \ref{eq:9.1-quantumscaleconstraint} implies that physical wavefunctions must vanish off the two-dimensional surface $\Sigma$ given by $q_1q_2q_3=1$. On the full configuration space physical wavefunctions are therefore discontinuous at $\Sigma$. One may be worried that this is problematic given that one encounters expressions of the form $\partial/\partial q_i\;\Psi_{phys}$, which are derivatives across the discontinuity. However, derivatives in the position representation only appear in the form of linear combinations as given by the momenta, which are directional derivatives tangent to the constraint surface. This is, of course, the quantum result corresponding to the fact that classically momenta generate motion within the constraint surface. Specifically, considering a basis for the tangent space to $\Sigma$ at an arbitrary point by writing $q_3=(q_1q_2)^{-1}$ and calculating the corresponding tangent vectors,
\begin{align}
 T_1 \equiv\left(\PD{q_1}{q_1},\PD{q_2}{q_1},\PD{q_3}{q_1}\right) &= \left(1,0, -1/q_1^2q_2\right), \\
 T_2 \equiv\left(\PD{q_1}{q_2},\PD{q_2}{q_2},\PD{q_3}{q_2}\right) &= \left(0,1, -1/q_1q_2^2\right),
\end{align}
the momenta can be expressed in terms of this basis,\footnote{Furthermore, one notes that the momenta are not linearly independent operators.}
\begin{align}
 \phat^1 &=\left(\tfrac23 T_1-\frac{q_2}{3q_1}T_2\right)\cdot\nabla \\
 \phat^2 &=\left(-\frac{q_1}{3q_2}T_1+\tfrac23 T_2\right)\cdot\nabla \\
 \phat^3 &=\left(-\tfrac13q_1^2q_2T_1-\tfrac13q_1q_2^2T_2\right)\cdot\nabla.
\end{align}
The evolution of the value of $\Psi_{phys}$ at some point $\vec{q}$ is therefore `blind' to the properties of $\Psi_{phys}$ outside the constraint surface, regardless of the form of the Hamiltonian.

\subsection{The quantum theory of the anisotropic universe}

The dynamics of the quantum theory is determined by the quantised, time-dependent Hamiltonian,
\begin{equation} \hat{H} = \sqrt{\frac{8}{3T^2}}\widehat{\sqrt{q_i^2p^{i2}}}. \label{HamiltonianOperator}\end{equation}
Since classically the numerical value of the Hamiltonian is `volume', in the quantum theory $\hat{H}$ defines a `volume spectrum' with volume eigenfunctions and eigenvalues rather than the more conventionally encountered energy spectrum (section \ref{sec:meaningofHamiltonian}).

The obvious difficulty with $\hat{H}$ is the appearance of a `square-root' operator, whose meaning is not initially clear. Another question is the factor-ordering ambiguity in the radicand. Regarding the latter, one can show using the commutation relations that changing the ordering of $(\qhat_a\qhat_a\phat^a\phat^a)$ either has no effect or merely corresponds to adding or subtracting a constant $\frac43\hbar^2$ or $\frac83\hbar^2$. A different ordering choice therefore corresponds to a shift in the Hamiltonian eigenvalues but does not change the eigenfunctions. Where necessary, we will therefore with minimal loss of generality assume the ordering `$qqpp$', also expressible in the form
\begin{equation} \qhat_a\qhat_a\phat^a\phat^a\Psi_{phys}(q)=\left(q_a^2\del^{a2}-\tfrac13q_aq_b\del^a\del^b+\tfrac23q_a\del^a\right)\Psi_{phys}(q). \end{equation}

The square-root operator could be dealt with formally by assuming the existence of a series expansion,
\begin{equation} \hat{H}=(8/3T^2)^\frac12\sum_{n=0}^\infty w_{n,a}(q)(\del^a)^n,\end{equation}
which acts straightforwardly on a state when expressed as a Fourier transform,\footnote{Note however that the Fourier expansion does not constitute a integral over momentum eigenstates since no momentum representation exists. In fact, the relation of the Fourier components $k^c$ and the momenta can be seen via the action of the latter on the former,
  \begin{equation} \phat^a\psi(q)=\hbar\int d^3k\; \tilde{\psi}(k)\cdot(\delta^a_b-\tfrac13q^aq_b)k^b\,e^{ik^cq_c}
				 =\hbar k^a\psi(q)-\frac{\hbar}{3}\int d^3k\; \tilde{\psi}(k)\cdot q^aq_b k^b\,e^{ik^cq_c}. 
  \end{equation} 
}
\begin{equation} \psi(q)=\int d^3k\; \tilde{\psi}(k)e^{ik^cq_c},\end{equation}
giving
\begin{align} \hat{H}\left(\tilde{\psi}(k)e^{ik^cq_c}\right)
 &= \int d^3k\;\sum_{n=0}^\infty w_{n,a}(q)\,(ik^a)^n\cdot\tilde{\psi}(k)e^{ik^cq_c} \notag\\
 &= \int d^3k\;\sqrt{q_a^2(ik^a)^2-\tfrac13q_aq_b(ik^a)(ik^b)+\tfrac23q_a(ik^a)}\;\cdot\tilde{\psi}(k)e^{ik^cq_c}. 
\end{align}
This formal result does not seem however very practical or insightful.

A more productive approach is to deal with the square-root explicitly. We intend to interpret $\hat{h}=(\qhat_a^2\phat^{a2})^\frac12$ in the manner described in the general discussion in the box at the end of section~\ref{sec:quantumFriedmannfree}: Since analysis of $\hat{H}^2$ is a lot easier than $\hat{H}$ itself (as there is no square root and so the interpretation of the operator expression is clear) interpretation of the operator via diagonalisation offers the way forward. Hermiticity and positive semi-definiteness need to be established for this to be viable however. 



The operator $\hat{f}\equiv\hat{h}^2$ is Hermitian only on $\Sigma$, not on the full, unconstrained configuration space. In order to establish its Hermiticity it is useful to perform a change of coordinates $(q_1,q_2,q_3)\rightarrow(u\equiv q_1,v\equiv q_2, w\equiv q_1q_2q_3)$, so that $w=1$ describes the constraint surface, facilitating the integration. With $\partial_u=\partial^1-(w/u^2v)\partial^3$, $\partial_v=\partial^2-(w/uv^2)\partial^3$, $\partial_w=(1/uv)\partial^3$, the momenta are
\begin{align}
 \phat^1 &=-i\hbar\left(\tfrac23\del_u-\frac{v}{3u}\del_v\right) \\
 \phat^2 &=-i\hbar\left(\tfrac23\del_v-\frac{u}{3v}\del_u\right) \\
 \phat^3 &= \frac{i\hbar}{3}\left(\frac{u^2v}{w}\del_u + \frac{uv^2}{w}\del_v\right),
\end{align}
the volume element is $dq_1dq_2dq_3=(uv)^{-1}dudvdw$ and the operator $\hat{f}$ is
\begin{equation}\hat{f}\equiv\qhat_a\qhat_a\phat^a\phat^a = -\frac{2\hbar^2}{3}\Big[u^2\del_u^2+v^2\del_v^2-uv\del_u\del_v+v\del_v+u\del_u\Big]. \label{finuv}\end{equation}
In terms of these coordinates the Hermiticity of $\hat{f}$ on the constraint surface,
\begin{equation}\int_\Sigma \psi_{phys}^\dag (\hat{f}\chi_{phys}) =\int_\Sigma (\hat{f}\psi_{phys})^\dag\chi_{phys},\end{equation}
is easily shown. The momenta themselves are not Hermitian and therefore do not constitute `observables' in the conventional sense (see box above for a formal discussion). For example,
\begin{equation} \int\limits_\Sigma \psi_{phys}^\dag(\phat^1\chi_{phys}) 
    = \int_\Sigma (\phat^1\psi_{phys})^\dag\chi_{phys} - \int_0^\infty\int_0^\infty du\;dv\;\frac{1}{uv}\cdot\frac{i\hbar}{u}\psi_{phys}^\dag\chi_{phys}. 
\end{equation}

The reason we chose to examine the simplest model in which the Poisson brackets are non-trivial is that it allows us to explore their quantisation without having to worry about too many cumbersome notational or other details. An added benefit of this model is that the Hamiltonian eigenequation, that is, the `volume eigenspectrum' can be solved exactly. This is because $\hat{f}$ is a homogeneous operator (eq.\ \ref{finuv}) whose eigenfunctions may be readily found by inspection,
\begin{equation} \phi_{n,m}(u,v)=A_{n,m}u^nv^m, \end{equation}
with eigenvalues
\begin{equation} \hat{f}\phi_{n,m}(u,v) =-\frac{2\hbar^2}{3}\big[n^2+m^2-nm\big]\phi_{n,m}(u,v).\end{equation}
Na\"ively $m,n\in\mathbb{C}$, although the values will be restricted shortly. With the interpretation of the square-root operator discussed above, the Hamiltonian eigensolutions are
\begin{equation} \hat{H}\phi_{n,m}(u,v) = h_{n,m}(T)\;\phi_{n,m}(u,v),\qquad h_{n,m}(T)=i\;\frac{4\hbar}{3|T|}\sqrt{n^2+m^2-nm}.\end{equation}
The eigenfunctions $\phi_{m,n}(u,v)$ are, however, not normalisable on $u,v\in(0,\infty)$ for any values $n,m$. But they are bounded (and, in fact, of constant magnitude) for purely imaginary $n,m$, and divergent for all other values. That is, let $n=i\beta$, $m=i\delta$, $\beta,\delta\in\mathbb{R}$. Then
\begin{equation} h_{i\beta,i\delta}(T) = -\;\frac{4\hbar}{3|T|}\sqrt{\beta^2+\delta^2-\beta\delta}, \end{equation}
where $\beta^2+\delta^2-\beta\delta\geq0$ always, so that the eigenvalues are real. Thus the reality of eigenvalues is equivalent with the non-divergence of the eigenfunctions. This is closely analogous to the `plane wave' eigenfunctions of a free particle in basic particle quantum mechanics. There the eigenfunctions are of the form $\exp(ikx)$ for real $k$ and therefore not normalisable either but bounded with constant magnitude. Imaginary values of $k$ are excluded even though they solve the eigenvalue equation because they entail divergent eigenfunctions. The analogy can be made more apparent by writing $\phi_{i\beta,i\delta}=e^{i(\beta\ln u+\delta\ln v)}$. The eigenvalues are sign-definite, that is, either all positive or all negative, depending on the choice of sign in the Hamiltonian. As we did in the classical theory we can choose `volume' to be positive, so that the Hamiltonian is a positive-semidefinite operator. We see furthermore that there is a unique minimum-volume state, $\phi_{0,0}$, which has constant eigenvalue zero, somewhat analogous to a `vacuum' state.

One can also derive uncertainty relations. However, since the momenta are not observables the relevance of these relation is questionable and their meaning is obscure.\footnote{And not just in the way in which even in canonical systems the meaning of the uncertainty relation remains full of controversy.} Nonetheless we include them for completeness. One finds
\begin{equation} \sigma_{q_i}\sigma_{p^j} \geq \frac{\hbar}{3}\qquad (i=j) \end{equation}
but the other non-zero relations are not as well-behaved and state-dependent,
\begin{align}
 \sigma_{q_i}\sigma_{p^j} &= \infty, \qquad (i\neq j)\qquad \text{ for H-eigenstates}\\
 \sigma_{p^i}\sigma_{p^j} &= 0 \text{ or } \infty \text{ depending on symmetry properties of chosen state}.
\end{align}

Including a cosmological constant in the quantum theory is relatively straightforward since only the time-dependent pre-factor of the Hamiltonian changes, with the resulting Hamiltonian eigenvalues changing accordingly. The eigenfunctions remain the same.

The model developed in this section is not a good approximation to our universe. However, the analysis of the Poisson brackets, in particular, how it guarantees motion on the constraint surface and leads to a position representation that automatically satisfies one of the quantum constraint while maintaining the other independent of the Hamiltonian, will help us understand the analogous structure in the more complicated case of cosmological perturbations, developed in the next section.

\section{Canonical quantisation of York-time cosmological perturbation theory}\label{sec:quantperts}

The primary physical object in a quantum theory is the quantum state, which is represented as a complex function (or functional) defined on the configuration space of the variables of the classical theory and on which operators derived from the classical variables and momenta act. It evolves according to the Schr\"odinger equation defined by the operator-promoted Hamiltonian. In general the state may also be expressed as a function of other variables, such as the momenta. This is not the case here, as we discussed in the box in section \ref{sec:quantKasner}, since the classical momenta do not Poisson commute among themselves, a feature they retain when `promoted' to commutator brackets between operators. Instead due to the asymmetric nature of the $\gtilde_{ab}$ and $\pitilde^{ab}$ only a `position' (that is, field-value) representation can be constructed. If sufficient evidence were found to lend credence to York time as a physically fundamental time, then this asymmetry may be taken to suggest that position is a physically preferred basis. 

In the particular case of the perturbation theory, the `position' variables are $h_{ab}$, while the momenta are $\nu^{ab}$. The promotion of their Poisson structure (given by eqs.~\ref{eq:8.1-hnuPB}, \ref{eq:8.1-nunuPB}), yields
\begin{align} 
[\hat{h}_{ab}(x),\hat{\nu}^{cd}(y)] &= i\hbar\Big[ \delta_a^{(c}\delta_b^{d)} - \tfrac13\gamma_{ab}\gamma^{cd} + \tfrac13\gamma^{cd}\hat{h}_{ab} \notag\\
      &\hspace{0.2\linewidth}-\tfrac13\gamma_{ab}\hat{h}^{cd}  +\text{h.o.}\Big]\;\delta^3(x-y), \label{eq:10.2-hnuCB} \\
[\hat{\nu}^{ab}(x),\hat{\nu}^{cd}(y)] &= \frac{i\hbar}{3}\left[\gamma^{cd}\hat{\nu}^{ab}-\gamma^{ab}\hat{\nu}^{cd}\right]\;\delta^3(x-y). \label{eq:10.2-nunuCB}
\end{align}
From this commutator algebra --- or alternatively from the general framework discussed at the beginning of this chapter --- one obtains the representation of the operator-promoted perturbation variables $\hat h_{ab}$ and $\hat\nu^{ab}$ for a wave functional in the position basis,
\begin{align}
 \hat h_{ab}\Psi(h_{ij})  &= h_{ab}\Psi(h_{ij}) \label{eq:10.2-posreph} \\
 \hat\nu^{ab}\Psi(h_{ij}) &= -i\hbar\big(\delta^{ab}_{cd}-\tfrac13(\gamma^{ab}-\gamma^{am}h_{mn}\gamma^{nd})(\gamma_{cd}+h_{cd})\big)\fd{h_{cd}} \Psi(h_{ij}) \notag\\
			  &= -i\hbar\big(\delta^{ab}_{cd}-\tfrac13\gamma^{ab}\gamma_{cd}-\tfrac13\gamma^{ab}h_{cd}+\tfrac13h^{ab}\gamma_{cd} +\text{h.o.}\big)
			  \fd{h_{cd}}\Psi(h_{ij}).    \label{eq:10.2-posrepnu}
\end{align}

Since the matter variables are canonical and furthermore commute with the geometric variables, their representation is not restricted to a particular basis. While one could imagine `mixed' bases such that $\Psi=\Psi(\gtilde_{ab},\pphi{A})$ (or $\Psi=\Psi(h_{ab},\delta\pphi{A})$ in the present case), here we assume the full position basis where $\Psi=\Psi(\gtilde_{ab},\phi_A)$ (or $\Psi=\Psi(h_{ab},\delta\phi_A)$), although since our focus in on the geometric degrees of freedom, this is not important for our discussion.

The quantum dynamics is determined by the operator form of the Hamiltonian given by \ref{eq:8.3-Hdens2}-\ref{eq:8.3-Hpert}. However, the Hamiltonian involves explicit functions of time obtained from solutions of the classical dynamics of the background. For the purposes of the quantum theory one could choose to adopt the same form, contending with quantised perturbations on a classical background. Here we are interested in a cosmological theory that is fully quantum and we therefore also quantise the background. Standard quantum mechanics does not, however, provide solutions that could be substituted for the classical ones since there are no trajectories in configuration space, as the object of interest is the state or wavefunction. We resolve this conundrum by employing the trajectories obtained in de~Broglie-Bohm theory \citep{deBroglie1928,Bohm1952,Holland1993}, similar to methods employed in conventional (unreduced, York-time unrelated) cosmological perturbation theory \citep[e.g.][]{PeterPinhoPintoneto2007,PeterPintoNeto2008}. While non-equilibrium de~Broglie-Bohm theory when applied to cosmological applications may lead to a distinct phenomenology \citep{Valentini2010InflCosm,ValentiniColin2015a,Valentini2015a, UnderwoodValentini2015} it agrees perfectly in its statistical predicitons with standard quantum mechanics in its equilibrium form and can therefore be employed as a mathematical tool even if one is unwilling to make any form of ontological commitment. The fact that de~Broglie-Bohm theory regards the position basis as privileged is furthermore consistent with its special status in the present formalism. Whether or not this consistency adds credence to de~Broglie-Bohm theory as the fundamentally correct (with respect to ontology) way to formulate quantum mechanics, is another discussion.

We fully solved the background dynamics for a single scalar field in the last chapter, where we found that the quantum trajectories do, in fact, match the classical ones (although this is not generally the case and even here certain provisos regarding our quantisation procedure apply). Furthermore, the Hamiltonian expectation values --- physically the volume of the universe up to a constant, analogous to the notion of energy in quantum mechanics of more conventional systems --- also match the evolution of the classical volume (this is true without the provisos). A similar quantum-classical correspondence was established using de~Broglie-Bohm theory for other appropriate matter content with standard cosmological time and outside the reduced-Hamiltonian picture in \citep{John2015}. While such results are encouraging, the equality of classical and quantum trajectories is not necessary for the method described here to be applied. However, the fact that the two do match for certain scenarios may explain why quantising only the perturbations and leaving the background classical, as is often done, is empirically successful. For the case of a free field we found $\phi(T)\sim\ln|T|$ and $\oppr{\Psi}{\hat H_{BG}}{\Psi}\sim |T|^{-1}$ with the factors of proportionality depending on the chosen initial state (see section \ref{sec:quantumFriedmannfree}).

Having solved the background one has explicit functions of time $\phibar_A(T)$ and $\pphibar{A}(T)$, and the quantum analogue of $H_T^\zero(T)$ is given by the expectation value of the background Hamiltonian. This fully determines the perturbation Hamiltonian in the quantum theory.

The dynamics is further defined by the presence of the constraints implied by the York gauge, that is, the choice of gauge such that the foliation is parameterised by York time exactly rather than merely at zeroth order. Classically this condition implied that $h_{ab}$ and $\nu^{ab}$ are traceless, which according to the Dirac quantisation procedure leads to constraints on the set of physical states $\Psi_{phys}$,
\begin{align}
 \gamma^{ab}\hat h_{ab} \Psi_{phys} &= 0 \label{eq:10.2-qhconstraint} \\
 \gamma_{ab}\hat\nu^{ab}\Psi_{phys} &= 0. \label{eq:10.2-qnuconstraint}
\end{align}
In the position representation the first of these implies that $\Psi_{phys}$ must vanish off the classical constraint surface defined by $h_{ij}=0$, reducing the dimensionality of the physical configuration space by one per spatial point.\footnote{Recall that the true number of physical degrees of freedom is, of course, further reduced by the 3-conformal-diffeomorphism invariance.} From the linearity of the Schr\"odinger equation it follows that the vanishing of $\Psi_{phys}$ off the surface is consistent with its time evolution. This result is exactly analogous to what we found in our analysis for the anisotropic minisuperspace model of section \ref{sec:quantKasner}.

The physical wave functional is therefore necessarily discontinuous in the full (unconstrained) configuration space. This is nonetheless not in conflict with the action of the momenta $\hat\nu^{ab}$ as derivatives since they act tangential to the constraint surface, just as their classical generator counterparts and analogous to their nature in the minisuperspace model. Here one could also change variables (separately at each point in space) in order to express all operators in terms of tangential derivatives only, as we did above, although we will not do so here since no further insight is gained.

The momentum-trace constraint \ref{eq:10.2-qnuconstraint} does not restrict the set of possible physical states any further since it is identically satisfied by the choice of representation of $\hat\pitilde^{ab}$ even in the general case (eq.~\ref{eq:10.1-posreppi}).

The momentum operator is Hermitian only when considered to zeroth order (meaning the first two terms in \ref{eq:10.2-posrepnu}), where its form is fully determined by the background. At this order a momentum representation exists, although note that the representation of the momentum in the position basis is still not the canonical one. The classical analogue is that the Poisson brackets only contribute to the linear perturbation equations at zeroth order as discussed above. The Hamiltonian may be made Hermitian by choosing the symmetric ordering for the mixed term,
\begin{equation} F(T)H_T^\zero\cdot M_{Pl}^{-2} T\;\cdot\tfrac12(\hat h_{ij}\hat\nu^{ij} + \hat\nu^{ij}\hat h_{ij}).\end{equation} 

The reason Hermiticity cannot be established at all orders is in part that in the derivation of the Hamiltonian it was assumed that the perturbation variables $h_{ab}$ and $\nu^{ab}$ are sufficiently small for a second-order Hamiltonian to adequately describe the dynamics. The statement that the momentum or Hamiltonian operators be Hermitian however is, algebraically speaking, a claim involving the functional integral over all allowed functions $h_{ij}$, including those where individual components may be large. Therefore one would have to apply finite (and adequately small) limits to the functional integral, or use an appropriate attenuation functional, representing the notion that physical wavefunctionals must become small for functions $h_{ij}(x)$ with large upper bounds. However, even then one can at best hope to establish Hermiticity approximately. This issue does not depend on whether or not one restricts the functional integration to the constraint surface, whereas in the non-perturbative finite-dimensional model developed in section~\ref{sec:quantKasner} Hermiticity can be established if one does make that restriction.

In practice one might ignore these issues and consider expression \ref{eq:10.2-posrepnu} to leading order only, so that the Hamiltonian is Hermitian and a well-defined probability current exists. One consequence of this is however that the geometric momentum operator $\hat\nu^{ab}$ is therefore only a quantum observable at leading order. At this order, however, the quantum theory is well defined. Even at higher order however there is no issue with the momentum not being an observable as we discussed in the box in section~\ref{sec:quantKasner}. Momentum is only approximately observable.

The quantum theory developed here is consistent and at this stage not obviously unreasonable. However, a full understanding can only be achieved once a number of topics have been described within this framework, such as detailed solutions (possibly through numerical work), the notion of a vacuum state, what kind of quantum effects might be detectable, and so on. This will comprise a significant amount of work that is left for future inquiry. Here we instead return to a more fundamental issue and question whether there are alternatives to the canonical quantisation method.

\part{Quantum cosmology without the wavefunction}\label{part4}
\chapter{Trajectory-Weyl quantisation}\label{chap:traj}

\textit{In this chapter we propose a new quantisation procedure that does not rely on a `wavefunction' and may be specifically suited to the quantisation of cosmological theories. The approach combines aspects from two different lines of inquiry: the quantisation based on ensembles of trajectories and quantum mechanics understood as a dynamical Weyl geometry on configuration space. This chapter closely resembles the contents of ref.~\citep{Roser2015_TrajGeometry}.}

\section{Introduction to trajectory-based quantisation}\label{sec:trajintro}

Formulating a consistent theory of quantum gravity requires a modification of either our classical theory of gravity or our method to arrive at a quantum theory from a classical theory, or possibly both. In the past chapters we considered almost exclusively the former, introducing the notion of a physically preferred time and exploring the consequences of doing so. In this final part we finally consider the latter, a modification of the quantisation process. We reported the contents of this chapter in ref.~\citep{Roser2015_TrajGeometry}. The schematic developments in chapter \ref{chap:cosmtraj} have not yet been published in any form.

Among the conceptual troubles of formulating a theory of quantum gravity is, for example, the interpretation of the wavefunction of the universe, in particular with regards to probabilistic notions. The scheme proposed here does not include any form of wavefunction. However, the idea presented here is still in its infancy and this final major part of the present thesis should therefore be understood as a starting point for a future line of investigation rather than a fully developed proposal for a theory of quantum gravity. In fact, in this chapter we present the proposed formalism without any reference to gravity at all and focus almost entirely on the quantisation of non-relativistic theories with finite-dimensional configuration spaces. In the next chapter we discuss the application of this scheme to gravity and cosmology, although results are at this point only schematic. The proper extension of this approach to field theories involves certain technical complications, which we will address in future work. 

The present proposal is based on two recent but entirely independent developments in the foundations of quantum theory. The first is a number of `trajectory-only' formulations of quantum theory \citep{Sebens2014, HallDeckertWiseman2014, Tipler2010, Bostrom2014, Poirier2010, SchiffPoirier2011, Schmelzer2011a}, though also see ref.~\citep{Holland2005}, intended to recover quantum mechanics without reference to a physical wavefunction. Two related approaches which however lead to an experimentally distinguishable theory are the real-ensemble formulations of refs.~\citep{Smolin2011,Smolin2015}. The other recent development of importance to us is the insight that the non-classical nature of quantum dynamics may be understood as a geometrical phenomenon, as has been shown for non-relativistic particle mechanics \citep{Santamato1984, Carroll2007, NovelloSalimFalciano2009}, relativistic particle mechanics \citep{Santamato1984b,Santamato1985} and for the particle Dirac equation \citep{SantamatoMartini2011,SantamatoMartini2012,SantamatoMartini2014}. In this chapter we demonstrate that these two ideas form a natural union, and that this union may indeed help to overcome a technical problem of the trajectory-only approach that has remained largely, though not entirely, unaddressed.

We will show how quantum theory may be understood as a natural extension of classical mechanics by allowing the configuration-space geometry to be dynamical. Here we focus on the non-relativistic theory. In previous approaches to such trajectory-only formulations quantum mechanics was recovered by postulating a `force' term that was Newtonian in the sense that it appeared on the right-hand side of `$m\ddot{x} = $' (but was not Newtonian in any other sense of the word). In its simplest formulation \citep{Sebens2014} this term was chosen to match the quantum potential of Bohm \citep{Bohm1952, Holland1993} with the substitution $|\psi|\rightarrow\sqrt\rho$. That is, the Born probability amplitude $|\psi|$ was replaced by the square root of the system density $\rho$ in configuration space. This eliminates any reference to a wave function $\psi$, which is not part of such a trajectory-only theory. However, it does require that $\rho$ is a physically real quantity. This means it requires the reality of \emph{all} dynamically allowed configuration-space trajectories, making it a theory of many worlds.\footnote{Some authors in their presentation have therefore likened aspects of the many-trajectory approach to Everettian `many-world' theory. We find this more confusing than helpful since the two have very little in common with the exception of being described as theories of `many worlds', a phrase with rather distinct meanings in the two theories.}

We take a different approach and do not postulate such a force. Instead, we begin by reconsidering aspects of classical mechanics, constructing an action that leads to a simultaneous determination of all dynamically possible paths. Mathematically this is almost trivial given the action for a single trajectory. However, it prepares the next step, the introduction of a geometrical term in the action. This term is nothing but the curvature of the configuration space. So the action to be associated with quantum mechanics will be the sum of two terms: The classical matter action plus a curvature. Formally it is similar to the Einstein-Hilbert action, although this analogy should not be taken too literally since the latter concerns the curvature of ($3+1$)-spacetime, not configuration space.

The action we obtain is effectively that of Santamato \citep{Santamato1984}, but unlike Santamato's it does not rely on the minimisation of expectation values. Nonetheless a substantial part of our derivation is taken from \citep{Santamato1984}. Our action differs from that of \citep{NovelloSalimFalciano2009} in its functional form, although it turns out to be numerically equal on shell.

The trajectories we find turn out to be exactly those of de~Broglie-Bohm theory \citep{Bohm1952, Holland1993, deBroglie1928} for an ensemble in equilibrium ($\rho=|\psi|^2$).
Equilibrium de~Broglie-Bohm theory exactly recovers all experimental predictions of quantum mechanics, and so the theory presented here does so too. There is, however, a subtlety , which we will address.

The dynamical geometry of the configuration space is a so-called \emph{Weyl geometry} \citep{Weyl1918}, of which we will give a brief review in section \ref{Weyl}. We will find that the geometry has singularities, namely at points which in conventional quantum mechanics (or de~Broglie-Bohm theory) correspond to nodes in the wavefunction ($\psi=0$). This will lead us to refine our approach: We will move from trajectories \emph{in} the configuration-space manifold to a manifold \emph{of} trajectories. This refinement is crucial to what is to follow, as well as strongly preferable for philosophical reasons. However, for reasons of presentational clarity we will only make this change in our perspective once the basic tenets of our theory are established. 

As the final step in our recovery of the phenomenology of quantum mechanics we address the aforementioned `subtlety', namely the quantised nature of angular momentum, manifest in the fact that loop integrals of the form $\oint \vec{p}\cdot\vec{dl}$ may be non-zero, yet only take values $2\pi\hbar\cdot m$, $m\in\mathbb{Z}$. In standard de~Broglie-Bohm theory this result follows from the multi-valuedness of the phase of the wavefunction and the fact that the phase is undefined at nodes. In a na\"ive trajectory-only theory it is entirely unclear why such a result should hold, or how the integral could be non-zero at all. This was first pointed out by Wallstrom \citep{Wallstrom1994} in connection with hydrodynamical approaches and has remained largely unaddressed.\footnote{Strictly speaking only the first of these questions is the one asked by Wallstrom in \citep{Wallstrom1994}. However, I will use the phrase `Wallstrom's objection' more broadly and include also the second where applicable.}

Coming from the perspective of de~Broglie-Bohm, the idea of `many de~Broglie-Bohm world' theory may be considered a precursor to the trajectory-only theory. Here an infinite ensemble of worlds evolves according to the laws of standard de~Broglie-Bohm theory, guided by an ontological wave function. The notion appeared inadvertently as a result of the `many-worlds-in-denial' objection to de~Broglie-Bohm theory \citep{Deutsch1996,BrownWallace2005} and was discussed and criticised by Valentini \citep{Valentini2010}, whose position was subsequently challenged by Brown \citep{BrownReplyToValentini2010}. Since there is a physical wave function in this theory, Wallstrom's problem does not arise. The concept is also discussed by Sebens \citep{Sebens2014}, who calls the theory `prodigal QM', although he considers it with only a finite number of trajectories.

Let us now turn to many-trajectory theory proper. It is instructive to outline the approaches of other authors briefly and review how they have addressed the Wallstrom objection, before summarising the solution presented in this paper. This short survey is not intended to be comprehensive or detailed. I am however not aware of any such survey on the subject in the existing literature (with the exception of ref.~\citep{Roser2015_TrajGeometry}, of course, which contains significant overlap with the present chapter).
 
The idea of formulating quantum mechanics in terms of the flow of an ensemble of configuration-space particles is sometimes traced back to ideas of Madelung \citep{Madelung1926}. However, this is a rather anachronistic reading of Madelung and historically inaccurate \citep{BacciagaluppiValentini2009}: Madelung imagined the electron as a charged fluid in three-space and he did not make the leap to \emph{configuration space} as the arena of the dynamics, nor did he consider individual trajectories within the fluid to have any ontological significance. His approach is therefore more akin to that of Schr\"odinger than having any resemblance to the trajectory-based approached with which we are concerned here. 

Sebens \citep{Sebens2014} arrives at the trajectory-only theory by starting from the point of view of de~Broglie-Bohm, introducing an ensemble of `worlds' distributed according to $|\psi|^2$ and then eliminates $\psi$ in favour of the ensemble density. However, he suggests that this is only a continuum limit of a more fundamental ontology with only a finite (though large) number of worlds. As such, quantum mechanics is considered to be only approximately correct, though sufficiently accurate for a difference to be experimentally undetected at this point. Sebens addresses Wallstrom's objection by introducing an appropriate quantisation condition, although stating that this is preliminary and that for now the quantised nature of these loop integrals is an empirically discovered feature of our universe. 

The approach of Hall et al.\ \citep{HallDeckertWiseman2014} is similar to that of Sebens, except that they discuss in more detail the nature of the inter-world interaction in the case of a finite number of worlds and indeed provide a concrete example for a one-dimensional model. They also apply the theory to a series of quantum phenomena and use it as a tool to numerically approximate ground state wavefunctions. They do not address Wallstrom's objection however.

Tipler \citep{Tipler2010} (but also see his earlier work \citep{Tipler2006}) begins with classical mechanics with an infinity of trajectories and then introduces the `smoothing potential' $U$ (equal to the quantum potential once the function $R$ appearing in the definition of $U$ has been made) as the unique correction to the Hamilton-Jacobi equation that avoids the crossing of trajectories and thereby ensures the smoothness of Hamilton's principal function $S$ and, in his words, prevents `god from playing dice', referring to the ambiguity in evolution where trajectories would cross in the absence of the smoothing potential. He likens this addition to Maxwell's introduction of the displacement current in Amp\`ere's law. The constant $\hbar$ must exist based on dimensional grounds. He also discusses the recovery of Born probabilities and the uncertainty principle in terms of his theory. However, despite employing an example involving angular momentum he does not address Wallstrom's issue. Indeed, $S$ is a real valued function by construction ---the classical Hamilton principal function--- and therefore any non-zero value of angular momentum remains to be explained.

A critique of Tipler's proposal can be found in Bostr\"om \citep{Bostrom2014}. Bostr\"om's approach shares features with that of Sebens, Hall et al.\ and Tipler, but he is very careful about discussing the philosophical and logical foundations required for the construction of his theory. Somewhat peculiarly though the existence of a wave function is one of the postulates of his theory, although he makes clear that `existence' here is meant in a mathematical sense (that is, such a function is definable) and not in some ontological sense. This does help his theory to overcome Wallstrom's objection, although the theory may be seen as somewhat unsatisfactory given the reliance on a wave function, ontological or not. Unlike Sebens and Hall et al., Bostr\"om argues in favour of a continuous infinity of worlds rather than a large but finite number and he briefly discusses the measure-theoretic foundations for such a picture, in particular in response to criticism by Vaidman \citep{Vaidman2014}.

The approach of Schiff and Poirier \citep{Poirier2010, SchiffPoirier2011} is different in that their formulation does not rely on the introduction of a quantum potential in the usual form. Instead, they consider higher-order time derivatives of the variables to appear in the expressions for the Lagrangian and energy of their theory. Certain constraints on their form then allow the identification of a term that effectively constitutes (at least numerically) the quantum potential as the simplest option (although at least in the time-dependent case reference to a density of trajectories in configuration space cannot be avoided). The authors show how the function $S$ can be recovered as the action integral along a trajectory, but why this function should indeed be phase-like in its multivaluedness remains unaddressed.

Schmelzer's \citep{Schmelzer2011a, Schmelzer2011b} construction is somewhat similar to Tipler, Sebens and Hall et al., although he discusses a number of foundational points not addressed by them and, crucially, provides a detailed discussion of Wallstrom's objection together with a proposal to overcome it. He shows that a regularity postulate for the Laplacian of the density $\Delta\rho$ suffices to restrict the value of the loop integrals $\oint \vec{p}\cdot\vec{dl}$ to take values equal to $\pm\hbar$. The postulate is equivalent to demanding that the nodes of the wave function are simple zeros. Since other integer values are however possible in standard quantum theory such situations must correspond to the presence of multiple zeros within the loop. For this to be viable one must consider quantum mechanics to be merely an approximation to some underlying `subquantum' theory, whose exact nature is unknown, but which may be empirically distinguishable, at least in principle. Therefore, while not proposing an experimentally distinct theory, Schmelzer's reasoning may nonetheless be experimentally assessible, although hard to ever rule out since, if standard quantum mechanics holds up, one is always able to change the regime (scale) on which differences are expected to become detectable. Despite this departure from quantum mechanics, Schmelzer does resolve the Wallstrom objection in his framework. The regularity postulate itself is justified by another principle concerning the nature of the elusive subquantum theory, which he calls the `principle of minimal distortion', together with a recognition of the role of $\Delta\rho$ in the energy-balance equation associated with the trajectories, although he acknowledges that this specification is `speculative in character' \citep{Schmelzer2011b} because it concerns the unknown subquantum theory.

Smolin \citep{Smolin2011,Smolin2015} has made two proposals for \emph{real}-ensemble formulations, meaning formulations of quantum mechanics in terms of ensembles of identical systems in a single (our) world. As such they do not belong to the family of many-trajectory theories discussed so far. However, Smolin's theories address Wallstrom`s objection and are therefore of interest to our purposes, although we cannot, of course, do his approaches justice here. His first formulation \citep{Smolin2011} is based on a stochastic dynamics governing how systems copy their beables onto those of similar system (in addition to a further deterministic dynamical law concerning a `phase' associated with each system). Wallstrom's objection is avoided in virtue of the form of the action, in which the phase variables appear as phases, not their real-valued gradient. There are other issues with this approach however, notably what the author calls `phase alignment', for which a mechanism is suggested but which, in the author's own words, ``is ad hoc and can probably be improved on''. His second formulation \citep{Smolin2015} does not rely on a stochastic mechanism but the `principle of maximal variety', which effectively drives systems towards greater distinctiveness (a term which is given a quantitative definition) with regards to their relation to the rest of the universe. Mathematically this is achieved via a potential-energy term that is a function of the ensemble properties, in particular the aforementioned `distinctiveness'. In the appropriate limit quantum mechanics can be recovered, although one can expect experimental deviations from quantum mechanics for appropriately rare systems, such as macroscopic objects. Motivation for this idea is found in Leibniz' Principle of the Identity of Indiscernibles. Smolin addresses Wallstrom's issue by introducing unimodular complex beables to take the role of momenta, rather than real-valued phases. This is a relatively straightforward solution, although it may be criticised on the basis that it was introduced specifically to solve this problem and without independent motivation.

Our solution to the Wallstrom's objection consists of a slight generalisation of our action, bringing it into the most general form that still retains the same equations of motion. While this modification is, of course, prompted by the necessity to address Wallstrom's objection, we emphasise that the new term is by no means ad hoc. Rather, it is the most general total derivative that can be consistently added to the Lagrangian and the addition of the total derivative in the first place is connected to the open-endpoint variation employed in the construction (see sec.\ \ref{classical}). 

At the end of this chapter we discuss the viability of this quantisation scheme to field theory and gravity. The detailed implementation thereof is left to future work.


\section{Relevant aspects of classical mechanics}\label{classical}

In this section we discuss two somewhat unconventional aspects of classical mechanics that will be relevant when we introduce our action for quantum trajectories below. The first of these is variation of an action without fixing the end point, apparently leading to the condition that the momenta $p_i$ vanish. However, if the Lagrangian is specified only up to a total derivative of some unknown function $S$ (this does not change the equations of motion), then the condition obtained is $p_i=\partial_iS$. This use of open-endpoint variation is due to Santamato \citep{Santamato1984}. Tipler \citep{Tipler2010} uses open-endpoint variation in a similar though not identical manner in his derivation of the Hamilton-Jacobi equation. The second point of discussion is the simultaneous variation of multiple trajectories in configuration space by summing (or integrating) the actions of each individual system. At first this remark may appear almost trivial, yet it will present a natural starting point for our derivation of quantum trajectories later on.

\subsection{Open-endpoint variation and total derivatives}
Let us assume that in our classical theory the trajectories $q(t)$ traced out in a configuration space $\mathcal{C}$ may be obtained by extremisation of an action,
\begin{equation} I_1 = \int dt\; L(q(t),\dot{q}(t),t) \label{eq:11.2-classActionBasic}\end{equation}
where $q=\{q^1, \dots, q^n\}$ are the configuration variables and $L$ is the Lagrangian. A dot denotes differentiation with respect to the time parameter $t$. In the usual approach one fixes the two end points of the trajectory, that is, the configuration of the system at two times $t_0$ and $t_1$: $q_0=q(t_0)$ and $q_1=q(t_1)$, where $t_1>t_0$. Since the value of $I_1$ (and $L$ and the intermediate points $q(t)$) depends on the choice of end points, we should strictly write
\begin{equation} I_1(q_0,t_0;q_1,t_1) = \int_{t_0}^{t_1} dt\; L\big(q(q_0;q_1;t),\dot{q}(q_0;q_1;t),t\big).\end{equation}
This is usually not done since it is cumbersome notation and the choice of end points $q_0$ and $q_1$ does not affect the equations of motion. Only the functional form of the Lagrangian is of significance. Rarely is the numerical value of the action integral of interest.

The variation $\delta q$ at $t_0$ and $t_1$ is assumed to vanish, and so there is no boundary term when performing the partial integrations that lead to the equations of motion. But consider now the case where we only set $\delta q=0$ at $t_0$, but leave it open at $t_1$.\footnote{We could have chosen to leave both open, or only the initial variation. The reasoning that follows would be similar.} We therefore write
\begin{equation} I_1(q_0,t_0;t_1) = \int_{t_0}^{t_1} dt\; L\big(q(q_0;t),\dot{q}(q_0;t),t\big).\label{eq:11.2-classActionWithInitial}\end{equation}
In that case variation yields a non-zero boundary term, proportional to the variation at $t_1$:
\begin{align}\notag \delta I_1(q_0,t_0;t_1) =& \PD{L}{\dot{q}^i}\big(q(t_1,q_0), \dot{q}(t_1,q_0),t\big) \cdot\delta q^i(t_1,q_0)  \\
				 &\hspace{-24pt}-\int\limits_{t_0}^{t_1}dt\, \left[\pd{t}\PD{L}{\dot{q}^i}\big(q(t,q_0),\dot{q}(t,q_0),t)\big) 
				    - \PD{L}{q^i}\big(q(t,q_0),\dot{q}(t,q_0),t\big) \right] \cdot\delta q^i(t,q_0).
\end{align} 
The variation $\delta q(t_1,q_0)$ at the end point is independent of the variation $\delta q(t,q_0)$ at intermediate times. The vanishing of the boundary term is not a result of the vanishing of the integrand along the path. Rather it is the result of the applications of Hamilton's principle, demanding that $\delta I_1=0$ for any choice of path and endpoint. As a result each of the two terms must vanish separately. Hence an extremisation of $I_1$ implies that the term multiplying $\delta q(t_1,q_0)$ must vanish, as does the expression inside the square brackets. The latter gives the usual Euler-Lagrange equation, while the former implies that 
\begin{equation} 
    p_i\big(q(t_1,q_0), \dot{q}(t_1,q_0),t\big)\equiv\PD{L}{\dot{q}^i}\big(q(t_1,q_0), \dot{q}(t_1,q_0),t\big) = 0.\label{eq:11.2-RawBoundaryEq.}
\end{equation}
The momenta at the end point vanish. In fact, the appearance of such an end-point constraint is a general feature of variational problems with open boundary conditions \citep[chap.~2]{Lanczos1949}.. 

In the context of Lagrangian mechanics this condition appears rather undesirable, as it seems to suggest that `ignorance' of the final position of our system implies that its momenta must vanish (and if our action principle is be extremised for \emph{any} choice of $t_0$ and $t_1$ the implication is that the momenta would have to vanish at all times). Nonetheless open-endpoint variation has found to be useful in the elimination of unphysical (gauge) degrees of freedom via a procedure called `best-matching' \citep[Sec. 4]{Barbour2003SIGPD}, which we briefly discussed in chapter \ref{chap:problemoftime}. Here we employ the method for another purpose however.

A slight generalisation of our Lagrangian alleviates any concern we might have regarding the boundary equation \ref{eq:11.2-RawBoundaryEq.}. It is well known that the equations of motions are unchanged if a total time derivative of an arbitrary function $S(q,t)$ of the system variables and time is added to the Lagrangian:
\begin{equation} \bar{I}_1(q_0,t_0;t_1) = \int_{t_0}^{t_1} dt\; \left[L\big(q(q_0;t),\dot{q}(q_0;t),t\big)-\DERIV{S}{t}\big(q(q_0,t),t\big)\right].\end{equation}
The relative minus sign is for later convenience. Note that $\DERIV{S}{t}=\PD{S}{t}+\dot{q}^i\partial_iS$. Performing the variation as before, the boundary equation is now
\begin{equation}
  p_i\big(q(t_1,q_0), \dot{q}(t_1,q_0),t\big)-\partial_iS\big(q(t_1,q_0),t\big) = 0, \label{eq:11.2-BoundaryEq.}
\end{equation}
where $p_i$ is defined as above (that is, in terms of the original Lagrangian without the total time derivative). Since this is to hold for any interval $[t_0,t_1]$, we have in general that $p_i=\partial_iS$. This equation can be seen as determining $S$ up to an arbitrary additive function of time.\footnote{If we really wish to add the most general total derivative to the Lagrangian (which still does not change the equations of motion, of course), then another subtlety arises regarding the function $S$. We will not go into it here though but reserve it for section \ref{Wallstrom}, where it will be of relevance.} It only restricts the momenta $p_i$ in that these must be the gradient of \emph{some} function.

Of course, $S$ turns out to have all the properties of a generator of those canonical transformations that represent the time evolution of the system. In a more usual approach we might have introduced it as such. However, here we see that such a function can also be introduced in the Lagrangian picture as an arbitrary addition to the Lagrangian that leaves the equations of motion unchanged. For example, using $\dot{q}^ip_i=\dot{q}^i\partial_iS=\DERIV{S}{t}-\PD{S}{t}$, together with $H(q,\nabla S,t)\equiv\dot{q}^i\partial_iS-L(q,\dot{q},t)$ allows one to obtain the usual Hamilton-Jacobi equation for our function $S$.

Returning briefly to the use of open-endpoint variations in the context of 'best-matching', it is worth noting that Barbour's and his collaborators' use \citep{Barbour2003SIGPD} and ours is perfectly consistent, provided we demand that the function $S$ depend only on physical and not on non-physical gauge degrees of freedom. Then the momenta associated with the non-physical variables vanish, while those associated with physical ones satisfy $p_i=\partial_iS$. In fact, we believe this to be a promising line of investigation for the construction of a fully relational quantum theory from first principles. This might then be extended to form a natural quantisation scheme of Shape Dynamics (see chapter \ref{chap:problemoftime}) to produce a promising candidate for a viable theory of quantum gravity.

\subsection{Action principles for multiple simultaneous trajectories}
The extremisation of $I_1$ (or $\bar{I}_1$) determines dynamically allowed configuration-space trajectories of a single system, starting at $q_0$ at time $t_0$. In the absence of a fixed end point, a second boundary condition (such as the initial velocity $\dot{q}(t_0)$) must be given in order to pick out a single trajectory uniquely.

Suppose now that instead of one such system we have $N$ systems of the same type, located at positions $q_{0(1)},q_{0(2)},\dots$ at start time $t_0$, with $q_{0(J)}=\{q_{0(J)}^1,\dots,q_{0(J)}^n\}$, $J=1,\dots,N$. The usual thing to do would be to extremise $\bar{I}_1(q_{0(J)},t_0;t_1)$ for each $q_{0(J)}$. Each time we would get the same equations of motion and only the initial conditions differ. That is, we have a different action for each system (taking different numerical values), although all the actions are identical in their functional form. This latter fact is the reason that in practice we would only derive the equations of motion once and then merely introduce the boundary conditions when solving them.

However, it is also possible to extremise a single action in order to obtain all $N$ trajectories at once (provided sufficiently many boundary conditions are specified). To do so we use the action
\begin{equation} I_N(q_{0(1)},\dots,q_{0(N)},t_0;t_1)=\frac{1}{N}\sum\limits_{J=1}^N I_1(q_{0(J)}).\end{equation}
The factor $1/N$ is not necessary for finite $N$ but is introduced here for consistency with what is to follow. The variations $\delta q(t,q_{0(J)})$ for each $J$ are independent of one another and so this action yields $N$ sets of formally identical Euler-Lagrange equations. In addition we have $N$ sets of boundary equations \ref{eq:11.2-BoundaryEq.}, one for each system.


We can generalise this to an infinite number of systems. For a countable infinity we take the limit,
\begin{equation} I_\infty = \lim_{N\rightarrow\infty} I_N(q_{0(J)};J\in\mathbb{N}).\end{equation}
Here the factor $1/N$ is important since otherwise the numerical value of the action would in general be infinite and thus not extremisable.\footnote{Alternatively we could have written down an infinite sum explicitly, but then we must use coefficients $\alpha_J$, which converge to zero sufficiently quickly as $J\rightarrow\infty$, so that the sum $\sum_{J=1}^\infty \alpha_J I_1(q_{0(J)})$ is finite. However, this would give different `weights' to different systems.} Here we will not consider the countable case any further. 

For a continuous (uncountable) infinity of system we must define a measure $\mu(q_0)$ over the configuration space, representing the density of systems at $q_0$ at time $t_0$. Since $\mu$ is defined over the whole configuration space (with $\mu=0$ in places where there are no systems), the starting positions $q_0$ take on the role of \emph{coordinates}. Let us therefore make the purely notational replacement $q_0\rightarrow x$. Instead of the discrete label $J=1,2,\dots$, systems are now labelled by the \emph{value} of their initial position $x$ (or $q_0$). The total action for this infinite ensemble of systems is now an integral over all starting positions $x$ with measure $\mu(x)$, which we assume to be normalisable. The action is therefore
\begin{align} I_{Ens} 
 &= \int_\mathcal{C} d^nx\,\mu(x) I_1(x,t_0;t_1) \notag\\
 &= \int_\mathcal{C} d^nx\,\mu(x) \int_{t_0}^{t_1} dt\, \left[L\big(q(t,x),\dot{q}(t,x), t\big)-\DERIV{S}{t}\big(q(x,t),t\big)\right] \label{eq:11.2-ClassicalEnsembleAction}
\end{align}

Let us reiterate: $x$ labels the initial position at time $t_0$ and therefore functions as a coordinate for an ensemble whose distribution in $\mathcal{C}$ forms a continuum (or at least may be approximated by one). The dynamical variable itself is $q(t,x)$, that is, the variables are labelled by their initial position $q(t_0,x)=x$. The integral of $x$ over the configuration space $\mathcal{C}$ is not an integral of the dynamical variable but of the initial positions. (It would be nonsensical to integrate over all $q$ and at the same time attempt to arrive at dynamical laws for $q$.) 

Note that $\mu(x)$ is a function over the configuration space describing the ensemble density at time $t_0$, not over the variables $q$. This measure turns the integral over configuration space into an integral over all systems in the ensemble. If the ensemble is allowed to evolve for some time, and then the systems are `relabelled' at some later time $t_0^\prime$ by their instantaneous positions $q(t_0^\prime,y)=y$, the new density function $\mu^\prime(y)$ will of course not be identical with the original $\mu(x)$.

One may furthermore show using standard results of classical mechanics that the measure $\mu$ obeys a continuity equation,
\begin{equation} \PD{\mu}{t}+\partial_i\left(\dot{q}^i\mu\right) =0. \label{eq:11.2-ClassicalCE1}\end{equation}
This follows from the fact that the momentum $p_i$ is given by the spatial gradient of a function $S$, which ensures that trajectories can never cross (or begin or end) and furthermore are continuous on the configuration space.

We are now in a position to make the step towards the quantum theory.

\section{Review of Weyl geometry}\label{Weyl}
In this short section we present the main ideas and relevant results of a Weyl geometry. 
The primary source (and in our opinion still the best one despite its age) is found in the mathematical sections of Weyl`s own work \citep{Weyl1918}.\footnote{The sections on physical interpretation in Weyl's paper are also of extraordinary interest, although it seems that the (rather beautiful) idea to identify the Weyl one-form with the electromagnetic 4-potential ultimately fails. See the Einstein-Weyl correspondence following Weyl's other 1918 paper \citep{LorentzEtAl1923}.} Here we will remain brief and only quote the results relevant to us. 

In order to avoid confusion we emphasise that a Weyl geometry is a concept that is distinct from those of the perhaps more widely known `Weyl transformations', `Weyl-invariant theories' and so on. The latter concepts are applicable to Riemannian geometries. For example, a Weyl-invariant theory is a theory determined by an action principle which is invariant under a local rescaling $g_{\mu\nu}(x)\rightarrow\Omega^2(x)g_{\mu\nu}(x)$ (plus matter-field rescaling depending on their conformal weights) and has found application in some spacetime theories of gravity and matter.

A Weyl geometry on the other hand is a generalisation of a Riemannian geometry in the sense that it is concerned with manifolds equipped with a metric $g_{ij}$ but also further geometric degrees of freedom not encoded in the metric. They key difference to a Riemannian geometry is that `length' is not comparable across finite distances. In Weyl's words, it is a `pure infinitesimal geometry'. Consider a vector $A^k$ at some point $P$ in the manifold, and suppose the vector is parallel-transported along some curve $C$ to a point $P^\prime$, where the transported vector is denoted by $A^{\prime k}$. In the case of a Riemannian geometry the direction of the vector is dependent on the connecting curve $C$. However, its length is not. The length $|A^\prime|=\sqrt{g_{ij}A^{\prime i}A^{\prime j}}$ depends only on the local metric $g_{ab}(P^\prime)$. In this sense length is path-independent and can be compared across finite distances in a Riemannian manifold. 

This is not the case if the geometry is a Weyl geometry. Here the length $|A^\prime|$ of the transported vector $A^k$ is also path dependent. For reasons described in \citep{Weyl1918} it is arguably natural to demand that for an infinitesimal transformation $P\rightarrow P+dP$, or in coordinates $x^i+dx^i$, the change in length is linear. That is, for a general length $\ell$, we find
\begin{equation} \ell+d\ell = \ell+ \ell\phi_kdx^k, \end{equation}
where $\phi_k$ is an arbitrary one-form and the full geometric properties of the manifold are given by the pair $(g_{ij},\phi_k)$. The two sets of variables $g_{ij}$ and $\phi_k$ are \emph{a priori} independent from one another, so that they may be varied independently in the application of an action principle, for example.

In general, an infinitesimal transport results in a change in the vector given by $\delta A^k=\Gamma^k_{ij}dx^iA^k$, where the coefficients $\Gamma^k_{ij}$ are a priori arbitrary. However, this definition of $\Gamma^k_{ij}$, together with $\delta |A|^2 = \delta (g_{ij}A^iA^j)$, implies that
\begin{equation} \Gamma^k_{ij} = \half g^{kl}\left(\partial_ig_{jl}+\partial_jg_{il}-\partial_lg_{ij}\right)
				    +\half g^{kl}\left(g_{li}\phi_j + g_{lj}\phi_i - g_{ij}\phi_l\right), \label{eq:11.3-Weylconnection}
\end{equation}
If $\phi_k\equiv0$, this expression reduces to the usual Christoffel symbol and the geometry is Riemannian. That is, the vectors undergo no change in length as they are transported along a curve.

Curvature is defined exactly as in the Riemannian case as a commutator of covariant-derivative operators, $R^i_{jkl}A^j=\nabla_l\nabla_kA^i-\nabla_k\nabla_lA^i$, leading to the usual expression in terms in the connections $\Gamma^k_{ij}$, although their expression \ref{eq:11.3-Weylconnection} differ from the Riemannian case. After some algebra we see that the scalar curvature (the one pertinent to our action) is given by
\begin{equation}
 R = R_{Riem} + (n-1)(n-2)\phi_k\phi^k-\frac{2}{\sqrt{g}}(n-1)\partial_k(\sqrt{g}\phi^k).
\end{equation}
Also note that the covariant derivative (defined in the usual way) of the metric does not vanish:
\begin{equation}
 \nabla_kg_{ij} = g_{ij}\phi_k.
\end{equation}
This implies that if one chooses coordinates such that locally the metric is $g_{ab}=diag(1,1,\dots)$ it is \emph{not} the case that the connection coefficients $\Gamma^a_{bc}$ vanish.\footnote{Contrast this to General Relativity, where $\nabla g_{\mu\nu}=0$, so that Lorentz frames (where $g_{\mu\nu}=diag(-1,1,\dots)$ locally) are inertial frames (where $\Gamma^\alpha_{\mu\nu}=0$) and vice versa (see \citep{MisnerThorneWheeler1973})}

One can in general consider a joint local transformation of the two geometric degrees of freedom, $g_{ij}\rightarrow\omega g_{ij}$, $\phi_k\rightarrow\phi_k-\partial_k\ln\omega$, under which it is reasonable to demand that a physical theory be invariant, given that such a transformation does not change the local notion of length and angles, or how to transport them. These gauge transformation are clearly similar to the notion of `Weyl transformations' in a Riemannian geometry since the metric $g_{ij}$ transforms in the same manner, but the latter does not have the separate notion of $\phi_k$, that is, path-dependent transport properties of length. 

Accordingly, one can make a gauge choice by choosing $\omega(x)$ with some given initial values $g_{ij}$ and $\phi_k$. In particular, if the Weyl geometry is \emph{integrable}, meaning $\phi_k$ is the gradient of a scalar function, then it is possible to choose $\omega$ such that after the transformation $\phi_k=0$ and the geometry is Riemannian. If on the other hand the metric can be expressed $g_{ij}(x)=\lambda(x)\bar{g}_{ij}$ everywhere with $\bar{g}_{ij}$ constant, then one can choose a gauge such that $g_{ij}$ is spatially constant and only $\phi_k$ has spatial dependence, so that the curvature is only a function of $\phi_k$. This latter scenario will be the case relevant to us below.

For our considerations below we do not consider the Riemannian part of the geometry, that is, the part determined by the metric $g_{ij}$, to have any dynamical meaning. Ultimately the metric geometry of the configuration space may find some application but at this preliminary stage we contend with illustrating that the Weylian part (determined by $\phi_k$) of the configuration-space geometry can be understood to encode the difference between classical and quantum dynamics. In \citep{Santamato1984} Santamato chose a similar route. He chose to include explicit reference to the metric $g_{ij}$ ``[o]nly for the sake of generality, as well as in view of further extension to spinning particles'' [p.\ 216]. For the sake of algebraic simplicity we will not include reference to $g_{ij}$ in this work and leave a generalisation and extension entirely to future inquiry.

\section{Quantum trajectories from an action principle}\label{sec:trajquantum}
In this section we allow the geometry of the configuration space to become dynamical. To this end, we treat the Weyl one-form $\phi_k$ as a dynamical variable and add to the classical ensemble action $I_{Ens}$ a curvature term with an appropriately chosen constant of a form closely resembling conformal coupling. Extremising this action we obtain the equation of motion for $q$ (now with an extra non-classical term stemming from the curvature term), the continuity equation (from variation of the function $S$) and the relationship between the geometry and the measure $\mu$. We show how these equations reproduce quantum behaviour, up to a subtlety that we will address in section \ref{Wallstrom}.

\subsection{Preliminary remarks}
In our discussion in section \ref{classical} the density $\mu(x)$ did not play any dynamical role since in the extremisation of the ensemble action $I_{Ens}$ the total minimum was found as the weighted sum of the minima of each trajectory. That is, the minimisation of one trajectory was independent from the minimisation of any other. We will now introduce a term that effectively constitutes an interaction between trajectories. More precisely, we add the curvature of the configuration space to our Lagrangian (with a suitably chosen coupling constant that essentially turns out to correspond to $\hbar^2$ up to a numerical factor resembling a conformal coupling). This curvature is that of a Weyl geometry (see section \ref{Weyl} and \citep{Weyl1918}) and its Weyl one-form $\phi_k$ is considered to be a dynamical variable. The role of the `Riemannian' part of the curvature (due to the metric $g_{ij}$) is yet to be investigated. Here it is of secondary importance and $g_{ij}$ is not considered dynamical for our current purposes. 

Much of the mathematical structure appearing in this construction is again due to Santamato \citep{Santamato1984} and our notation is close to (though not identical with) his. However, Santamato's approach is founded on the stochastic approach of Nelson and Madelung, not a many-trajectories theory. Santamato's quantum action also involves the minimisation of an expectation value, which one might find ontologically questionable.\footnote{How can we make sense of the concept prior to having a theory of measurement and without the notion of quantum states, for example?} In our case, the minimisation principle remains unchanged from the classical minimisation of $I_{Ens}$. Merely an interaction term is added in the form of the curvature term. This curvature term constitutes an interaction between `neighbouring' trajectories since it contains configuration-spatial derivatives.

Note that our (and equivalently Santamato's) action is numerically identical to the action used in \citep{NovelloSalimFalciano2009} after use of $\DERIV{S}{t}=\PD{S}{t}+\dot{q}^i\PD{S}{q^i}$, $p_i=\PD{S}{q^i}$ and $H=\dot{q}^ip_i-L$. However, we consider our action to be a more natural starting point as a natural extension of the classical action. The approach of \citep{NovelloSalimFalciano2009} also differs in three other regards: Firstly, the authors are committed to consider a geometry on space rather than configuration space and their approach therefore only applies to a single particle (although without this commitment their formalism easily generalises). Secondly, the variation of their curvature term is done in the style of Palatini, that is, by considering the connection $\Gamma^a_{bc}$ as independent variables. Here on the other hand we will vary the curvature $R$ as a function of the Riemannian metric $g_{ij}$ and Weyl one-form $\phi_k$, although this latter difference is primarily aesthetic. The final difference is that they presuppose the Weyl geometry to be integrable, whereas in our formalism (just as in Santamato's) this is a result that follows from the equations of motion.

\subsection{The quantum action and its variation}

We now append the classical Lagrangian by a curvature term $\gamma(n)\lambda^2R[g,\phi]$, where $\lambda$ is a coupling constant that is ultimately fixed by observation. The term $\gamma(n)=\frac{1}{8}\frac{n-2}{n-1}$ is a numerical constant dependent on the dimensionality of the configuration space, chosen for later convenience.\footnote{Note that this expression for $\gamma$ is introduced with the knowledge of hindsight. Without this hindsight, we might have absorbed any such constant into $\lambda^2$.} That is, the action for the Trajectory-Weyl theory (TWT) is
\begin{equation}  \label{eq:11.4-QMAction}
  I_{TWT}= \int\limits_\mathcal{C}d^nx\, \sqrt{g}\mu(x) \int\limits_{t_0}^{t_1} dt\,\Bigg( L_{class}(q,\dot{q},t) - \DERIV{S}{t}(q,t) + \gamma(n)\lambda^2R[g_{ij}(q,t),\phi_i(q,t)]\Bigg),
\end{equation}
where
\begin{equation}
 R[g_{ij}(q,t),\phi_i(q,t)] \equiv R_{Riem}[g_{ij}] + (n-1)\left[ (n-2)\phi_i\phi^i-\frac{2}{\sqrt{g}}\partial_i\left(\sqrt{g}\phi^i\right) \right] \label{eq:11.4-WeylR}
\end{equation}
as obtained above. Note that we have included $\sqrt{g}$ in the volume element inside the integral since we allow for a general metric $g_{ij}$. The dependence of the variables on the initial position $x$ has been suppressed. 

In what follows, we assume for simplicity that the configuration-space $\mathcal{C}$ is flat in the Riemannian sense ($R_{Riem}=0$) and the $\sqrt{g}$ is merely determined by coordinates. For the purposes of illustration one may choose the coordinates to be Cartesian, so that $\sqrt{g}=1$. This fixes the Weyl gauge in the manner described at the end of section \ref{Weyl}. The only geometric degree of freedom is $\phi_k$, which, we emphasise, is a variable independent of $g_{ij}$.

We take note though that in classical mechanics $\mathcal{C}$ may have a non-zero $R_{Riem}$ for systems with certain types of constraints which lead to a reduced configuration space (a $(n-l)$-dimensional surface with non-vanishing curvature within the $n$-dimensional configuration space, where $l$ is the number of independent constraints). Whether or not the `quantisation' method proposed here could (or should) be applied to a classical reduced configuration space in the presence of constraints is to be investigated in the future. 

The similarity of \ref{eq:11.4-QMAction} with the Einstein-Hilbert action is obvious and may in part be the superficial appeal of this approach. However, the analogy is minimal. Our action is defined in configuration space, not physical space(-time). Furthermore, because we do not worry about tensorial properties relating to a theory's general covariance (as one would in general relativity with matter fields) but use fixed coordinates we do not have to generalise the derivative operators in the Lagrangian to covariant derivatives. Covariant generalisations are left to future work.

Let us for the sake of specificality consider the classical starting point to be an $N$-particle system in three dimensions ($n=3N$) with particle masses $m_I$, $I=1,\dots N$. The masses can be absorbed in the metric $g_{ij}$ on $\mathcal{C}$ in the usual way. For example for two particles of masses $m_1$ and $m_2$, we would take $g_{ij}=diag(m_1,m_1,m_1,m_2,m_2,m_2)$ (in Cartesian coordinates), enabling us to write terms in the form $\sum_I\half m_1\dot{\vec{q}}_I\cdot\dot{\vec{q}}_I = \half g_{ij}\dot{q}^i\dot{q}^j = \half \dot{q}^i\dot{q}_i$ and $\sum_i\frac{1}{2m_I} \vec{p}_I\cdot\vec{p}_I = \half g^{ij}p_ip_j=\half p^ip_j$, and so on, being careful to keep track of the natural index position for the dynamical quantities. See \cite{Lanczos1949} or any other book on analytic mechanics for further details. Hence there is no need to include explicit mass terms in what follows.

\paragraph{Variation of $q$: The law of motion}
Since $R$ has only dependence on $q$ and not $\dot{q}$, the boundary equation is unchanged, $p_i=\partial_iS$, and the Euler-Lagrange equation simply acquires an extra term $\gamma\lambda^2\PD{R}{q^i}$:
\begin{align} 0 &= \deriv{t}\PD{L_{class}}{\dot{q}^i}\big(q(t,x),\dot{q}(t,x),t\big)-\PD{L_{class}}{q^i}\big(q(t,x),\dot{q}(t,x),t\big) \notag\\
		&\hspace{120pt}-\gamma\lambda^2\PD{R}{q^i}\Big[g_{ij}(q(t,x),t),\phi_i(q(t,x),t)\Big] \label{eq:11.4-QuantumEOM}
\end{align}
The new term will ultimately be interpretable as a `quantum potential' $Q$ that matches Bohm's, as we will see. The equation then takes the form of the classical equation of motion with $Q$ appearing in the role of a new potential term, at least formally.

\paragraph{Variation of $\phi$: Relationship between $\phi$ and $\mu$}
We note that only the new term carries dependence on the Weyl one-form $\phi$. Hence we wish to minimise
\begin{align} I_R 
 &= (n+1)\int\limits_\mathcal{C} d^nx\sqrt{g}\mu\,\left[(n-2)\phi_i\phi^i-\frac{2}{\sqrt{g}}\partial_i(\sqrt{g}\phi^i)\right] \notag\\
 &= (n+1)\int\limits_\mathcal{C} d^nx\,\left[\mu\sqrt{g}(n-2)\phi_i\phi^i+\sqrt{g}\phi^i\cdot2\partial_i\mu\right].\label{eq:11.4-OnlyRAction}
\end{align}
where we have partially integrated in order to arrive at the last line. The boundary term drops out since $\mu$ is normalisable and therefore vanishing at infinity. For better readability arguments have been suppressed. 

Following \citep{Santamato1984}, a little algebra reveals that the term to be extremised can be written as
\begin{align} I_R
 &= -\frac{n-1}{n-2}\int\limits_\mathcal{C}d^nq\,\mu\sqrt{g}g^{ij}\partial_i(\ln\mu)\partial_j(\ln\mu) \notag\\
 &\qquad	+\frac{n-1}{n-2}\int\limits_\mathcal{c}\mu\sqrt{g} g^{ij}\big((n-2)\phi_i+\partial_i(\ln\mu)\big)\big((n-2)\phi_j+\partial_j(\ln\mu)\big).
\end{align}
This brings $I_R$ into a form that is independent of derivatives of $\phi_i$. Furthermore, the first term contains no dependence on $\phi_i$ and can therefore be ignored for the purposes of extremisation.

The extremisation is easy since the integrand is positive definite (with $\sqrt{g}$ and $\mu$ being positive definite and the rest being of the square form $g^{ij}W_iW_j$ with $W_i=(n-2)\phi_i+\partial_i(\ln\mu)$). Hence the the minimum must be bounded from below by zero and indeed it is clear that this minimum is obtained by 
\begin{equation} \phi_i = -\frac{1}{n-2}\partial_i(\ln\mu). \label{eq:11.4-phisolution}\end{equation}

This shows that the Weyl geometry is \emph{integrable}, that is, there is a scalar function $f$ on $\mathcal{C}$ such that $\phi_i=\partial_if$. Hence a local scale may be defined on $\mathcal{C}$ and the final length of a vector transported between two points in $\mathcal{C}$ is path-independent. As we stated an integrable Weyl geometry allows a gauge choice such that $\phi_i=0$ (that is, $f=$ const.) everywhere. However, in our case we already fixed the gauge such that the Riemannian metric $g_{ij}$ is constant. 

\paragraph{The continuity equation}

The continuity equation may be obtained in the same manner as for the classical case, using the fact that the momentum is given by a spatial derivatives at all times as a result of the boundary equation \ref{eq:11.2-BoundaryEq.}. The presence of the curvature term has no impact. It takes the form
\begin{equation} \label{eq:11.4-CE} \PD{\mu}{t}+\partial_i\left(\mu\dot{q}^i\right) = 0. \end{equation}

Had we not fixed the coordinates to be Cartesian, then the equation would include additional factors of $\sqrt{g}$ \citep[see][]{Santamato1984},
\begin{equation} \pd{t}\left(\sqrt{g}\mu\right)+\partial_i\left(\sqrt{g}\mu\dot{q}^i\right) = 0, \end{equation}
The appearance of $\sqrt{g}$ with $\mu$ corresponds to the fact that $\mu$ is a scalar density of weight one. 

From the meaning of $\mu$, introduced in the process of taking the continuum limit for a configuration-space ensemble, we should have expected such a continuity equation to hold. It confirms that $\mu$ may be interpreted as a density. This equation furthermore shows that $\mu$ is conserved along a trajectory.

\subsection{Equivalence with equilibrium de~Broglie-Bohm theory}
Substituting the solution \ref{eq:11.4-phisolution} into the expression \ref{eq:11.4-WeylR} for the curvature, we find that (by simple algebra, and using $\gamma=\frac{1}{8}\frac{n-2}{n-1}$)
\begin{align} R
 &= R_{Riem} +\frac{n-1}{n-2}\left[\partial_i(\ln\mu)g^{ij}\partial_j(\ln\mu)+\frac{2}{\sqrt{g}}\partial_i\left(\sqrt{g}g^{ij}\partial_j(\ln\mu)\right)\right]
			      \notag \\
 &= R_{Riem} +\frac{1}{2\gamma\sqrt{\mu}}\partial_i\left(\sqrt{g}g^{ij}\partial_j\sqrt{\mu}\right).
\end{align}

With this, the Hamilton-Jacobi equation derived from the Lagrangian $L-\DERIV{S}{t}+\gamma\lambda^2R$ becomes
\begin{align} 0 
  &= \PD{S}{t} + H(q,\nabla S,t) \notag\\
  &= \PD{S}{t} + H_{class}(q,\nabla S,t)- \gamma\lambda^2R_{Riem}  -\frac{\lambda^2}{2}\frac{\partial_i(\sqrt{g}g^{ij}\partial_j\sqrt{\mu})}{\sqrt{\mu}},
	\label{eq:11.4-QHJE}
\end{align}
where we used the fact that the momentum conjugate to $\phi_i$ vanishes identically since $\dot{\phi}_i$ does not appear in the expression for $R$.

The reader familiar with de~Broglie-Bohm theory in Bohm's second-order formulation will recognise the last term on the right as the quantum potential (at least for a broad class of particle systems), 
provided $\lambda^2 = \hbar^2$ (which ultimately is just the experimental discovery of the value of a constant, or, viewed differently, the discovery of a property of the system of units used) and that $\mu$, the configuration-space measure, can be interpreted as the equilibrium density $|\Psi|^2$. But the measure $\mu$ was introduced as describing the local (in $\mathcal{C}$) density of systems of our ensemble in the continuum limit and so this is indeed the natural interpretation. 

In order to obtain the evolution of the ensemble as a whole, eq.\ \ref{eq:11.4-QHJE} must now be solved simultaneously with the continuity equation (eq.~\ref{eq:11.4-CE}) for $\mu$ and $S$, which in practice would be done by combining them into the complex field $\psi=\sqrt\mu e^{iS/\hbar}$ on $\mathcal{C}$. Solving for a single trajectory alone is not possible, since $R$ contains reference to $\mu$, which is determined by the collection of all trajectories.

The equations \ref{eq:11.4-QHJE} and \ref{eq:11.4-CE} match, of course, the usual quantum Hamilton-Jacobi and continuity equations known from equilibrium de~Broglie-Bohm theory, with the difference that no reference to any wavefunction $\psi$ is made and the `quantum potential' contains $\sqrt{\mu}$ instead of $|\psi|$, with the consequence that there is no state of quantum non-equilibrium.\footnote{The term 'non-equilibrium' here is meant in the sense of de~Broglie-Bohm, that is, a state where $\rho\neq|\psi|^2$. This is impossible here since $\psi$ is not an ontological entity in its own right but only definable in terms of $\rho$ and $S$ (up to an unphysical overall phase factor). Contrast this to `many-de~Broglie-Bohm worlds' theory \citep{Sebens2014,Valentini2010}, where such $\rho$ and $|\psi|$ are a priori independent of one another.}

Note that de~Broglie's first-order dynamics\footnote{The distinction between the theories of de~Broglie and Bohm is often overlooked. The equilibrium state of de~Broglie's dynamics is stable (an ensemble in quantum non-equilibrium, $\rho\neq|\psi|^2$, will approach equilibrium), while in Bohm's theory there is a a more general form of non-equilibrium (where de~Broglie's relation $\vec{p}=\nabla S$ is violated), which is unstable \citep{ColinValentini2014}.} also emerges from this formalism (at least for certain Lagrangians), namely as the condition $p_i=\partial_iS$ derived from the boundary equation resulting from the open-endpoint variation.

Thus the equations derived here are equivalent with those of de~Broglie-Bohm theory. It appears that therefore the predictions of quantum mechanics have also been recovered. However, this conclusion would be incorrect, as we will discuss in the next section.

\section{Nodes and the phase-like nature of \texorpdfstring{$S$}{S}}\label{Wallstrom}
In this section we discuss the inequivalence between the present theory as it stands and de~Broglie-Bohm quantum mechanics. The issue is related to the respective single versus multi-valuedness of $S$. However, we show how the phase-like behaviour of $S$ can be recovered in our theory using a further generalisation of our Lagrangian. To do so, we first discuss the concept and nature of `nodes' in our theory and their relationship to the geometry. This requires a shift in our perspective: from a configuration-space manifold \emph{with} trajectories to a manifold \emph{of} trajectories.

\subsection{Wallstrom's criticism}
The problem with the above theory, and indeed any many-trajectories theory with or without a geometric foundation \citep{Sebens2014,HallDeckertWiseman2014,SchiffPoirier2011} is that the space of solutions of such a theory is smaller than that of de~Broglie-Bohm quantum mechanics.\footnote{Alternatively a many-trajectories theory may be constructed using the velocities or momenta themselves as the variables and not making reference to a function $S$. In this case the space of solutions is not smaller than that of de~Broglie-Bohm but instead larger since angular momenta can take continuous values. Hence there are two versions of Wallstrom's criticism applicable to different sorts of formulations.} While on a small local patch in configuration space the equations have the same solutions, a difference in the possible \emph{global} properties of the solutions arises. To our knowledge this was first noted by Wallstrom \citep{Wallstrom1994} in the context of the Madelung hydrodynamic approach.

The function $S$, when introduced as the phase of the (physically real) wavefunction $\psi=|\psi|e^{iS/\hbar}$, is multi-valued, $S\sim S+m\cdot2\pi\hbar$, $m\in\mathbb{Z}$. Furthermore, $S$ is \emph{undefined} at nodes of the wavefunction (points where $\psi=0$). These two facts together imply that there may be `states' (functions $\psi$) such that a closed loop $L$ may be found such that
\begin{equation} \oint_L \partial_iS\,dl^i \neq 0, \label{eq:11.5-non-integrability}\end{equation}
where $dl$ is a line element in the configuration space. Specifically this is possible if and only if $L$ is \emph{non-contractible}, in the sense that it is impossible to deform $L$ continuously into a point without passing through a node. Such loops exist in the presence of $(n-2)$-dimensional nodes (points in two dimension, lines in three dimensions, etc.). 

Contrast this to the case where $S$ is introduced as a scalar function on $\mathcal{C}$ \citep[such as in ref.\ ][]{SantamatoMartini2014}. Here $S$ is defined even where $\mu=0$ and is a real field on $\mathcal{C}$, so that the integral \ref{eq:11.5-non-integrability} vanishes for \emph{any} loop.

Solutions to the Schr\"odinger equation that display this non-integrability of $\partial_iS$ arise, for example, in the case of stationary states of hydrogen-like atoms with non-zero angular momentum. It follows that the present theory and other, similar many-trajectories theories do not allow for non-zero angular momentum states and therefore do not match the predictions of quantum mechanics.

Wallstrom's criticism has remained largely unaddressed by authors proposing many-trajectory theories, as we discussed in the short survey above. The solution we will construct here is superficially similar to Smolin's \citep{Smolin2015}, who introduces a unimodular complex variable $w$ (a `complex factor beable') in place of the momentum. However, in our case the goal will be achieved via a generalisation of the total derivative appearing in the Lagrangian, not by revising the mathematical representation of momenta.

\subsection{Nodes}
The necessity of having nodes present in cases where this mismatch between our `trajec\-tory-Weyl theory' (TWT) and quantum mechanics arises leads us to investigate the nature of nodes in the former. The term `node' in TWT is, of course, not a good choice of terminology given that no waves are involved. However, it is clear what is meant: A point $P$ in $\mathcal{C}$ is a node if $\mu(P)=0$. 
In de~Broglie-Bohm theory this does not necessarily imply that there is no trajectory that passes through the node. Whether a consistent law of motion exists for this system depends on whether the phase function $S(q)$ or at least the velocity field can be analytically extended through the node. This is trivial if the wave function is real (e.g. the $n=2$, $\ell=1$, $m=0$ state of hydrogen), in which case the particle placed at the node remains at the node. On the other hand, the $n=2$, $\ell=1$, $m=1$ state does not allow for such an analytic extension. There can be no continuous velocity field that includes the node and no trajectory can exist there within the framework of the theory. More generally, a trajectory exists (in a mathematical sense) at the node if and only if the loop integrals $\oint_L \partial_iS\,dl^i$ around the node vanish. Whether or not this extra `world' exists in an ontological sense is however irrelevant for the dynamics of all other `worlds' in the case of `many-de~Broglie-Bohm-worlds' theory. Furthermore, adding a single trajectory does not change the local system density.

Above we obtained the result $\phi_i=-\frac{1}{n-2}\partial_i(\ln\mu)$ (eq.\ \ref{eq:11.4-phisolution}). At a node, where $\mu=0$, this becomes undefined. That is, the mathematical notion of vector transport and the idea of `scale' fails to make sense at nodes. So nodes correspond to a certain kind of geometric singularity.

Let us examine the behaviour of the Weyl form $\phi_i$ in the immediate vicinity of a node. Suppose $\mu\sim r^{2m}$ as we approach the node, where $r$ is the distance from the node and $m$ is some (not necessarily integer) positive power. The factor of two in the exponent is for later convenience.

Approaching the node radially, that is, moving in the negative-$r$ direction, we see that
\begin{equation} \phi_{-r}=-\frac{2}{n-2}\pd{(-r)}\ln r^{2m} = \frac{2m}{n-2}\frac{1}{r}, \end{equation}
which shows that the length of a vector being transported radially towards the node by a small distance $\delta r$ grows by a factor proportional to $r^{-1}$. Since the Weyl geometry is integrable, we can also look at the length of a vector itself, which goes as
\begin{equation} \ell\sim -\ln r, \end{equation}
and so $\ell\rightarrow\infty$ as $r\rightarrow0$.

\subsection{From a manifold with trajectories to a manifold of trajectories}

It is now time for a shift in our thinking. We began by considering systems moving in the configuration-space manifold $\mathcal{C}$ with time, forming trajectories in the `configuration-space-time' ($\mathcal{C}\,\otimes$ time, sometimes called the `extended configuration space' \citep{Lanczos1949}). However, the dynamics of the theory imply that geometric properties are undefined at certain points in $\mathcal{C}$, namely those points not occupied by a system. This suggests that instead of considering systems particles (or `corpuscles', or `worlds') moving \emph{in} a manifold, we should consider the manifold \emph{of} all system particles (or of all trajectories if we think of it in a configuration-space-time picture), which we shall call $\mathcal{T}$. The configuration space merely becomes a way to label systems within our continuously infinite ensemble, namely by using the physical properties (the value of the variables $q_i$). In the presence of nodes, $\mathcal{T}$ is not simply connected.

The view taken is therefore the following. The physical world in its entirety consists of an infinite collection of similar systems (described by the same set of variables, that is, corresponding to points in the same configuration space) whose properties are specified in the form of $n$ variables $q^i$. These variables determine the continuity properties of the collection, thereby forming an $n$-dimensional manifold $\mathcal{T}$. Since the number density of systems is not uniform across the $q^i$, this manifold, in general, has non-zero curvature. The manifold $\mathcal{T}$ may be embedded in a configuration space $\mathcal{C}$ of the same dimension via a Weyl geometry, on which points are labelled by the $q^i$. Continuous subsets of $\mathcal{C}$ with a dimension smaller than $n$ might not have an inverse image (are not mapped to in the embedding) and form nodes, illustrating that the original manifold might not be simply connected. In practice, we work in $\mathcal{C}$, labelling systems by $q^i$ and tracking their local number density via the function $\mu:\mathcal{C}\rightarrow\mathbb{R}_{\geq0}$.

The function $S$ introduced in the action as an arbitrary function is defined on $\mathcal{T}$, not $\mathcal{C}$. Following the embedding in $\mathcal{C}$ it is therefore natural that $S$ is not defined at nodes, since these are exactly those points that are \emph{not} in the image of $\mathcal{T}$. Of course, by itself this does not answer Wallstrom's criticism. While the presence of nodes is necessary for a mismatch between na\"ive TWT and quantum mechanics to occur, it is not sufficient. In quantum mechanics $S$ is multi-valued and the presence of nodes merely allows one to find closed paths changing $S$ continuously from one value at a point to another at the same point.

Before coming to the second part of our solution to Wallstrom's problem, return briefly to the question of the existence of trajectories at nodes. In TWT, unlike in `many-de~Broglie-Bohm worlds' theory, the manifold of system particles is considered fundamental. This manifold may or may not have a non-trivial topology, and what functions $S$ (or `$U+W$' as introduced in the next section) are possible is contingent on that topology. One result of this is that a scenario in which there is a node in the wave function (as described by the standard theory) through which $S$ can be analytically extended may or may not correspond to a simply connected topology for $\mathcal{T}$. But whether or not there is a system particle at the node matters---unlike in the case of `many-de~Broglie-Bohm worlds'---since it changes the topology of the manifold and therefore the global properties of functions definable on it.

\subsection{The most general time derivative \texorpdfstring{$dS/dt$}{dSdt}}

We now discuss a possible solution to Wallstrom's problem. We consider the most general total time-derivative $dS/dt$ that can be added to the Lagrangian. This is, in fact, the sum of two time derivatives, one of which takes values on $\mathbb{R}$ as usual, the other one is a derivative of a function that takes values from the \emph{closed} one-dimensional manifold $S^1$, not $\mathbb{R}$.  

In order to see this, let us take a step back and consider how $S$ was introduced. We added an arbitrary total derivative $\DERIV{S}{t}$ to the action. The only requirement was that $\DERIV{S}{t}\in\mathbb{R}$. This requirement may seem almost not worth stating since it is generally assumed that terms in the action take values over $\mathbb{R}$ or sometimes $\mathbb{C}$. However, the implication of $\DERIV{S}{t}\in\mathbb{R}$ is \emph{not} that $S$ itself must take values on $\mathbb{R}$. Instead we could define $S:\mathcal{T}\rightarrow S^1$, that is, let $S$ map from $\mathcal{T}$ (or equivalently $\mathcal{C}\setminus\{$nodes$\}$) to the $\emph{closed}$ one-dimensional manifold, the `circle' $S^1$. The spaces $S^1$ and $\mathbb{R}$ share the same tangent space $\mathbb{R}$. Furthermore, they are the \emph{only} spaces that do so and our Lagrangian will therefore indeed be the most general.

It turns out that the resultant function `$S$' has the required properties to resolve Wallstrom's problem. 
A critic might be inclined to say that we `cheated' by giving $S$ the required phase-like property by fiat. 
Yet without wishing to discuss the philosophical issues too deeply, this is no different than any other introduction of mathematical objects in quantitative physics. The choice of what particular type of object should represent some proposed physical entity is founded on the properties that entity is supposed to have, inferred from observed phenomena and, arguably, certain theoretical guiding principles (such as generality, simplicity, symmetries, etc.). For example, in standard quantum mechanics we invent a mathematical object called the `wavefunction' that takes values from $\mathbb{C}$ on configuration space (or, more abstractly, some Hilbert space whose physical meaning is more obscure). 
Similarly, we have chosen the correct mathematical structure to address Wallstrom's objection and match the predictions of the theory to empirical facts.

We could justify the step of adding to the action a total time-derivative of a function $S:\mathcal{T}\rightarrow S^1$ rather than one mapping to $\mathbb{R}$ by its ultimate observational success (such as in measurements of atomic angular momentum). However, since the requirement $\DERIV{S}{t}\in\mathbb{R}$ gives two options, $S:\mathcal{T}\rightarrow\mathbb{R}$ and $S:\mathcal{T}\rightarrow S^1$, and neither changes the equations of motion, a priori we should add \emph{both} terms in order to achieve maximal generality. The added term is therefore not an \emph{ad-hoc} addition, but a natural generalisation. Let us therefore introduce a function $U:\mathcal{T}\rightarrow\mathbb{R}$, while the symbol $W$ will be used for the function $W:\mathcal{T}\rightarrow S^1$. These take the place of the symbol '$S$'. Our action becomes
\begin{align}  \label{eq:11.5-QMActionWithDoubleTDs}
  I_{TWT}^\star&=\int_{t_0}^{t_1} dt\, \int_\mathcal{C}d^nx\, \sqrt{g}\mu(x) \Bigg( L_{class}(q,\dot{q},t) + \gamma(n)\lambda^2R[g_{ij}(q,t),\phi_i(q,t)] \notag\\
	       &\hspace{216pt}- \DERIV{U}{t}(q,t) - \DERIV{W}{t}(q,t) \Bigg).
\end{align}
The equation of motion (\ref{eq:11.4-QuantumEOM}) remains unchanged, as does the result for the Weyl geometry (\ref{eq:11.4-phisolution}). The one difference is in the boundary equation, which now reads
\begin{equation} p_i = \partial_i W + \partial_i U. \end{equation}
All the previous derivations to recover de~Broglie-Bohm trajectories hold provided we replace $S$ by `$W+U$'. The expression is in quotation marks since addition is not defined for two functions taking values in different spaces, even if both are one-dimensional, and is therefore not to be interpreted as literal addition. However, $S=`U+W'$ never occurs itself, only its derivatives do and among them addition is well-defined (since they all take values in $\mathbb{R}$).

Now since $U$ is a globally defined real field, we have 
\begin{equation}\oint_L\,\partial_i U\cdot dl^i =0\end{equation}
for any closed loop, so there is no contribution from $U$ to closed-loop integrals of $\nabla S$. Meanwhile,
\begin{equation}\oint_L\,\partial_i W\cdot dl^i = z\cdot h,\qquad z\in\mathbb{Z},\end{equation}
where $h$ is defined to be the value obtained when the image of the loop $L$ under $W$ winds around the manifold $S^1$ exactly once. The numerical value of $h$ depends on how the manifold $S^1$ is parametrised (how it is represented in terms of real numbers, ultimately a choice of units). The value of $h$ is the same for all nodes in the manifold.

The constant $h$ can be thought of as the `circumference' of $S^1$ if we visualise the manifold as a circle embedded in $\mathbb{R}^2$. It functions as a `constant of conversion' between lengths on $S^1$ and on $\mathbb{R}$. For dimensional reasons it takes units of angular momentum if dimensionful variables are used. What is left to be addressed is the relationship between the associated radius, $\hbar=h/2\pi$, and the geometric `coupling constant' $\lambda$.

The proportionality of the geometric coupling constant $\gamma\lambda^2$ follows from dimensional analysis. The functions $U$ and $W$ must have the same dimension as the action, length squared over time ($L^2T^{-1}$). The only fundamental scale in the theory is the `radius' $\hbar$, which therefore must also have dimensions $L^2T^{-1}$. But the dimensions of the curvature $R$ are $L^{-2}$, so $\gamma\lambda^2$ must have the dimensions of $\hbar^2$, allowing us to set $\lambda=\hbar$, leaving only the numerical constant $\gamma$. 

The constant $\hbar$ is the one fundamental scale provided by the theory. It is remarkable that this one fundamental scale is, in fact, sufficient to compare all terms in the Lagrangian (that is, to let them have the same dimensionality). It is even possible to go one step further and demand that the action be dimensionless. This can be achieved by rescaling $I_{TWT}$ by a factor $\hbar^{-1}$. In this case $U$ and $W$ are dimensionless, too, and the closed manifold onto which $W$ maps is simply parametrised by angular values.

So only the numerical value of $\gamma$ is left to be discussed. The value $\frac{1}{8}\frac{n-2}{n-1}$ was, of course, chosen with the hindsight of experimental outcomes (matching those of quantum mechanics). However, it is itself distinguished: Its $n$-dependence is the unique choice such that the resultant equations of motion do not explicitly depend on $n$. Only the geometric properties of the manifold do (eq.~\ref{eq:11.4-phisolution}). The coupling constant resembles a `conformal coupling` as familiar from field theory in Riemannian geometries.

\section{Discussion of trajectory-Weyl quantisation}\label{sec:trajdiscussion}

In this chapter we showed how the phenomenology of non-relativistic quantum mechanics may be recovered starting with an `ensemble' of systems continuously distributed in configuration space. Thus our theory is one of many worlds, though very much distinct from the notion of `world' in (Everettian) `Many-Worlds' Theory. By including a curvature term in the action the trajectories turned out to match those of equilibrium de~Broglie-Bohm theory. Our construction was closely modelled along that of \citep{Santamato1984}, although our starting point was different. Nor is the idea of recovering quantum dynamics via an infinite ensemble of systems, covering all possible configurations, completely new, although it is a relatively recent development.

Our purpose here was to show that the two ideas, a dynamical Weyl geometry together with a `many-trajectory' theory, form a natural union. In order to emphasise this point we showed how Wallstrom's objection, a key problem with `quantum' theories not based on a wavefunction, can be overcome via a generalisation of the total time derivative in the Lagrangian. 

In our procedure we demoted the configuration space to a purely mathematical concept, with the physical reality corresponding to a manifold of configuration-space particles (or `worlds'), or --- in a configuration-space-time picture --- of trajectories. Other than the fact that this manifold, unlike the configuration space, has topological properties that allow the existence of non-contractible loops and thereby forms the first step in our solution to Wallstrom's objection, it may also be philosophically preferable: In standard quantum mechanics the wavefunction is defined on configuration space and it is unclear what this space is without appeal to some previous classical theory with a certain set of dynamical variables. In contrast, the present trajectory-Weyl theory has no such ambiguous ontology. Functions such as $S$ (or $U$ and $W$) are defined on the set of all worlds, each of which is physically concrete. If the manifold allows for non-contractible loops then the $S^1$-valued function $W$ may yield non-zero quantised values for the integral $\oint \partial_iW\cdot dl^i$. The appearance of $W$ in the generalisation of the Lagrangian thus completes the solution to Wallstrom's objection.

The development of this wavefunction-free method of arriving at a quantum theory from a classical theory was motivated by the failure of the canonical quantisation scheme in gravity, and from conceptual issues with interpreting the wavefunction of the universe, for example. Our motivation has therefore been entirely negative, namely the failure of another quantisation scheme. There is another positive, \emph{philosophical} reason however for this kind of quantisation. In fact, even in the absence of any empirical evidence for quantum effects a philosopher might propose a scheme similar to ours because of conceptual trouble with the \emph{classical} theory.

Consider a philosopher and scientist at the end of the late nineteenth century. Looking at the known laws of physics, he\footnote{Gender inequality at the time implies that chances are this scholar would likely have been male.} realised that two pieces of information are required for a physical description of the universe: its laws (in the form of equations) and boundary conditions (selecting the actually instantiated solution). Leaving aside the possibly deeper questions of why the laws are what they are, he wonders not only why nature chose some particular boundary conditions rather than another, but also why one set of boundary conditions must be selected --- that is, somehow ontologically preferred --- in the first place. He finds the demand of classical physics to arbitrarily choose a single solution from the set of all solutions of the laws of nature unacceptable. That is, he dislikes the categorisation of worlds into impossible worlds (those not satisfying the laws of nature), possible but non-actual worlds (solutions to the laws of nature that differ from the actual world), and possible actual worlds (the one world we live in). 

Suppose, he says, there is no distinction between actual and non-actual possible worlds. All initial conditions are equally real.\footnote{\label{footnote:11.Lewis} The philosophical position proposed here is somewhat similar to \emph{modal realism} (most prominently held by David Lewis \citep{Lewis1986Book}), which states that all possible worlds are equally real. There `possible' means not logically inconsistent (while worlds with laws of physics different from ours are considered `possible'). The view of our fictional polymath here is more restricted in that `possible' is understood as `possible given that the laws of physics are just as they are in our world.' We will return to this discussion briefly in the conclusion (chapter \ref{chap:conclusions}). } We would be unable to observe worlds other than our own (or they would be part of our world by definition), so there is no empirical reason either for or against their reality. His intellectual adversaries bring up the principle of parsimony (`Ockham's razor') but he replies confidently that parsimony was exactly the reason to abandon the superfluous notion of `possible but not actual' worlds. However, he soon realises that an implication of classical mechanics understood in this sense is that individual world trajectories cross and that configuration-space position alone does not uniquely identify a world (one has to include velocities, or momenta). Dissatisfied by this underdetermination he formalises this problem mathematically in the Hamilton-Jacobi description and adds the demand that the Hamilton-Jacobi equation be $C^2$ (doubly differentiable) everywhere as this would fix the problem. Imposing this constraint, he arrives at trajectory-based quantisation. 

The development from Hamilton-Jacobi theory to quantum trajectories in this manner was, in fact, proposed by Tipler \citep{Tipler2010}, as we discussed. The preceding philosophical argument has to my knowledge not been made in this exact form, though it is closely tied to the reasoning of Lewis \citep{Lewis1986Book} and others on the reality of all possible worlds (see footnote \ref{footnote:11.Lewis}).

While the reasoning is somewhat tentative and may need to be spelled out more rigorously, it does in any case serve as an example of how purely philosophical reasoning --- so often dismissed by physicists --- can lead to empirically adequate theories. Of course, if physicists were to adopt this reasoning, it would undoubtedly no longer be considered `philosophical'.\footnote{Coincidentally, the fact that once a particular line of philosophical reasoning has consequences for a physical theory it becomes adopted by physicists and loses its status of being `philosophical', and that as a result philosophers are considered by some to never contribute anything useful to physics, has been pointed out in a popular article by Hossenfelder \citep{Hossenfelder2016PostOnPhilosophy}.}

The scientific test of the theory is, of course, another. Can this quantisation scheme provide a path toward an adequate theory of quantum gravity?

\chapter{Quantum cosmology from trajectory-Weyl quantisation?}\label{chap:cosmtraj}

\textit{In this chapter we sketch the application of the wavefunction-free quantisation scheme developed in the last chapter to gravity. Detailed technical developments are left for future work.}

\section{Extension of the trajectory-Weyl approach}\label{sec:extendtraj}

The developments of the last chapter have provided us with a fundamentally new method of arriving at a quantum theory from a classical theory. For a finite-dimensional particle system with the usual action we have exactly recovered the empirical predictions of equilibrium de~Broglie-Bohm theory and hence standard operational quantum mechanics. It is however not guaranteed that this equivalence should hold for any type of system, in particular, for systems with classical equations that have a fundamentally different form, such as those resulting from Hamiltonian reduction. In the case of quantum gravity and cosmology it is not even clear what the `standard' approach is (there is none). The phenomenological equivalence with canonical quantisation for conventional actions is, in a sense, coincidental.\footnote{Of course, without this coincidence we would never have discovered the possibility of trajectory-Weyl quantisation. More precisely, if the fundamental ontology of the universe is really such an infinite ensemble of systems individually described by classical language and if the dynamics were not equivalent with canonical quantisation, then canonical quantisation and standard operational quantum theory would never have been developed, at least not in the present form.}

The geometric approach (without a trajectory ontology) has been extended to relativistic particle quantum mechanics \citep{Santamato1984b,Santamato1985} (in the sense of a non-field-theoretic picture, where the Klein-Gordon equation is considered to hold for a wavefunction, not for a classical field) and the particle Dirac equation \citep{SantamatoMartini2011,SantamatoMartini2014}, starting from the configuration space of a classical rigid body (although without attention to Wallstrom's criticism). The trajectory ontology could however be incorporated into their method. Extension to field theory proper is technically more difficult. The primary reason is that the field-theoretic configuration space (the space of all allowed functions $\phi(x)$ in scalar field theory, for example, or (conformal) superspace in the case of gravity) is infinite-dimensional. The standard measure on this space is therefore not defined but only Gaussian measures are available \citep{CorichiCortezQuevedo2002,CorichiCortezQuevedo2003}. This complication is likely manageable. An open question is the proper definition of a geometry (specifically, the curvature term that enters the action) in the case of an infinite-dimensional space. A rigorously constructed theory will require considerable technical and formal development, an endeavour left for future work. In the next section (\ref{sec:cosmwithtraj}) we sketch some of the expressions that are the metric-field counterparts of the equations in chapter \ref{chap:traj}.

Wallstrom's problem exists in field theory, too. Recall that the presence of non-contractible loops requires the existence of nodes of co-dimension two. In field theory this occurs in states corresponding to particular linear combinations of two field modes, analogous to circular polarisation. For example, if only two such modes $k_1,k_2$ in Fourier space are non-zero, then the infinite-dimensional wave functional can be written in the form of a two-dimensional wavefunction. For example,
\begin{equation}\Psi(k_1,k_2) = R(|k_1^2+k_2^2|)\;[\cos k_1 + i\;\sin k_2],\end{equation}
where $R(|k_1^2+k_2^2|)$ is the `radial' part of $\Psi$ and satisfies $R(0)=0$, so that there is a node on the `$\infty-2$' dimensional `line' $k_1=0=k_2$. The existence of such states implies that Wallstrom's objection also applies to the infinite-dimensional field theory, not just to simple non-relativistic particle mechanics.

The success of such the trajectory-Weyl approach to quantum field theory is yet to be determined, although one may be hopeful. Ultimately the goal is, of course, its application to gravity. Multiple preliminary investigations need to be undertaken here: the quantisation of time-reparameterisation invariant systems such as those described in chapter \ref{chap:problemoftime}, the quantisation of relational theories such as the precursors to Shape Dynamics as described above, a series of minisuperspace models, which may provide some insight regarding the scheme's application to gravity, in particular in the context of cosmology, the question whether a fully covariant quantisation is possible with this method or if a particular choice of physical time such as York time has to be made, or if we should disregard general relativity all together in favour of Shape Dynamics, and so on. Undoubtedly this work would easily take up another thesis, likely more. Here we contend with laying the foundation for what may ultimately turn out to be a viable contender for a theory of quantum gravity.


\section{Application to gravity and cosmology}\label{sec:cosmwithtraj}

Applying the above quantisation scheme to gravity --- presumably to the spatial metric field for a chosen slicing such as the York-time (constant-mean-curvature) foliation --- introduces a number of technical issues. In addition, it is notationally cumbersome and some development in this area will be necessary for the theory to be viably analysed. Let us sketch some of the steps toward this wavefunction-free theory of quantum gravity.

The basic dynamical variables (the `$q^i$' above) in the case of gravity are the possible configuration of the metric field $q_{ab}(x)$. We reserve the symbol $g_{ab}(x)$ (or $(q_0)_{ab}(x)$) for the analogue of the initial variable value or `coordinate' that was called $x$ (or initially $q_0^i$) above, while $x$ in this chapter refers to points of space. The configuration space is the infinite-dimensional function space of all possible functions $q_{ab}(x)$ which are suitably well behaved, the precise meaning of which will have to be determined but will presumably involve being symmetric in the indices $(ab)$ and smooth almost everywhere, and possibly satisfying other properties relating to finiteness and continuity. One way to `translate' the finite-dimensional quantities from the last chapter to the functional quantities of the present one is to treat $x$ itself like an additional index, albeit one that takes values not from a finite subset of $\mathbb{N}$ but from $\mathbb{R}^3$ in the case of a flat cosmology and from $S^3$ in a closed one. Sums over indices become integrals over space, in addition to the finite sums over the two standard indices of the metric. One plausible proposal for a simplifying notation would be to write $q_{ab(x)}$ rather than $q_{ab}(x)$ together with an `Einstein integration convention', integrating over recurring `$(x)$'. However, we will not use this here but leave this to be investigated further in future work.

A technical issue that arises in the case of the flat cosmology is that integrals over $\mathbb{R}^3$ are, in general, not finite even if the integrand is finite everywhere. One will therefore require some form of normalisation. This could take the form of a suitably infinitesimal pre-factor cancelling the infinity or the introduction of a coordinate `box' of space, giving finite integration limits. The infinite limit of this box could then be taken at the level of measurable quantities.\footnote{Alternatively, the theory might be investigated with a suitably chosen box size, such as the size of the observable universe (of order $10-100$ gigaparsecs today). The latter idea would correspond to choosing a coordinate limit such that the line integral across the box would give a physical distance of this order, using a classically plausible choice of metric configuration. However, since here classical notions of distance are introduced, this method appears less desirable than taking the infinite limit.} 

The expression for the classical action analogous to equation \ref{eq:11.2-classActionBasic}, using York time $T$, would read
\begin{equation} I_1 = \int dT \; L\Big[q_{ab}(x)(T),q_{ab}^\p(x)(T),T\Big), \qquad L =\int d^3x\;\mathcal{L}\Big(q_{ab}(x)(T),q_{ab}^\p(x)(T),T\Big).\end{equation}
Here we used the convention to use square brackets for dependence on functional arguments and mixed brackets `$[ ...)$' if both functional as well as conventional arguments occur. The expression $q_{ab}(x)(T)$ is to be understood as the variable $q$ with finite indices $ab$, continuous `index' $(x)$, as a function of time $T$. We see how the notation quickly becomes difficult to read and proposed convention of placing $(x)$ in the super or subscript, for example, may help with clarity.

The analogue of the action written with the initial `coordinate' $g_{ab}(x)\equiv (q_0)_{ab}(x)$ included (equation \ref{eq:11.2-classActionWithInitial}) is
\begin{align} &I_1\Bigg[(q_0)_{ab}(x),T_0;T_1\Bigg) = \notag\\
		&\qquad\qquad\bigintsss_{T_0}^{T_1} dT\; L\Bigg[ q_{ab}(x)\Big[(q_0)_{ab}(x);T\Big),q^\p_{ab}(x)\Big[(q_0)_{ab}(x)\Big[(q_0)_{ab}(x);T\Big),T\Bigg). 
\end{align}
The need for more concise yet clear notation is evident. Application of the open-endpoint principle leads to vanishing of the momenta just as above. Spelled out (analogous to eq.~\ref{eq:11.2-RawBoundaryEq.}),
\begin{align} &\pi^{ab}(x)\Bigg[q_{ab}(x)\Big(T_1,(q_0)_{ab}(x)\Big],q_{ab}^\p(x)\Big(T_1,(q_0)_{ab}(x)\Big], T\Bigg) \notag\\
	      &\qquad\qquad\equiv\frac{\delta L}{\delta q_{ab}(x)} \Bigg[q_{ab}(x)\Big(T_1,(q_0)_{ab}(x)\Big], q_{ab}^\p(x)\Big(T_1,(q_0)_{ab}(x)\Big], T\Bigg) =0.
\end{align}

The ensemble minimisation principle above involved an integral of the full configuration space together with a suitable measure $\mu$. The corresponding expression (analogous to eq.~\ref{eq:11.2-ClassicalEnsembleAction}) is, formally,
\begin{equation} I_{Ens} = \int_C \mathcal{D}g_{ab}(x)\;\mu\Big[g_{ab}(x)\Big]\; I_1\Big[g_{ab}(x),T_0;T_1\Big).\end{equation}
In close analogy with the developments in chapter \ref{chap:traj} we have now abandoned the notation $(q_0)_{ab}(x)$ in favour of the `coordinate' $g_{ab}(x)$. The term $\mathcal{D}g_{ab}(x)$ denotes the coordinate integration measure (itself an issue, to be addressed below), with the physical, dynamical measure being given by $\mu$ as above.

This raises the second major technical problem. Measures on function spaces are difficult to define. For example, there is no notion of a basic Lebesque measure. Only Gaussian measures exist \citep{CorichiCortezQuevedo2002,CorichiCortezQuevedo2003}. The way forward here would likely be to first repeat the finite-dimensional trajectory-Weyl quantisation with such Gaussian measures only (in standard quantum mechanics a Gaussian rather than Lebesque measure over the configuration/Hilbert space changes the position representation of operators, for example, but is also possible). Then one could construct a field-theoretic analogy of \emph{that}. The issue is, I believe, manageable at least in principle, even if complicated.

Another difficulty is the correct notion of the analogue of a Weyl geometry in a function space. The na\"ive expression for length transport\footnote{The way to define the length of a vector in a function space in the first place is not obviously unique either.} is
\begin{equation} \ell +d \ell = \ell + \ell\int d^3x\;\phi^{ab}(x)\mathcal{D}g_{ab}(x).\end{equation}
This expression alone involves two technical issues: First, the integral is over all space, which may be infinite. This can be dealt with via a nominal coordinate box. The second issue is that a small shift $\mathscr{D}g_{ab}$ (not to be confused with the functional volume element $\mathcal{D}g_{ab}$) in the function space requires a metric $M$ on this function space in the first place.\footnote{The deWitt supermetric $G_{abcd}$ together with integration over all space (since the deWitt supermetric is map from two metric values \emph{at a point} to a number, while $M$ is to be a map from two metric \emph{functions}) comes to mind, although this would require some a priori justification.} This problem is exactly that of the coordinate integration measure mentioned earlier. There is no unique or obvious way to choose this metric. A further complication (though one that can be dealt with) arises with the need to identify metric configurations that correspond to the same geometry (that is, are related via three-diffeomorphisms). 

One then has to define a notion akin to curvature for the function space. I am not currently aware of how far such notions have been developed by mathematicians. A significant degree of function-space theory will be necessary to make trajectory-Weyl quantisation rigorous for the metric field. What the resulting dynamics look like will be seen.

A more basic development is necessary first if the system to be quantised in this manner was obtained by Hamiltonian reduction, specifically with York time as the physical time parameter. The developments of chapter \ref{chap:traj} were in a Lagrangian style (a term was added to the action), while the reduced-Hamiltonian theory is by its very nature a Hamiltonian formalism. Na\"ively, on might propose deriving a physical (or `reduced') Lagrangian from the reduced Hamiltonian via a Legendre transform, schematically (for a finite-dimensional model):
\begin{equation} L_{phys} = q^{\p a}p_a-H_{phys}(q,p,T).\end{equation}
This is however problematic since the dynamics in the reduced-Hamiltonian theory is not determined only by the physical Hamiltonian but also, crucially, by the Poisson structure. Equations of motion derived via minimisation from a Lagrangian principle would be incorrect since the non-canonical Poisson structure would be lost.

Instead this requires the development of the proposed quantisation scheme in the Hamiltonian picture. Another option however could be to quantise in the trajectory-Weyl style \emph{prior} to Hamiltonian reduction. In this case the quantisation scheme needs to be extended to include time-reparameterisation invariant systems (with general lapse functions $N$) and the procedure of Hamiltonian reduction would need to be extended to include terms involving the configuration-space curvature. Developments in fundamental analytical mechanics appear to form the foundation of this approach to quantum gravity, no matter which line of inquiry is pursued.

A short-term goal may be the quantisation of cosmological minisuperspace models (such as those of chapter \ref{chap:Friedmann} and section \ref{sec:ClassKasner}), which only have a finite number of dimensions. In this case the homogeneous gravitational and matter degrees of freedom would take the role of the variables $q^i$ of chapter \ref{chap:traj}. Even this development however requires the quantisation scheme to be expressed in a Hamiltonian-compatible form, at least if connection with a reduced-Hamiltonian picture is to be made. An open question here is one we already touched upon, then with regards to canonical quantisation. It equally arises in the case of geometry-trajectory quantisation: is quantising the cosmology equivalent to `cosmologising' the quantum theory? That is, is the dynamics one obtains by quantising a minisuperspace model the same as those obtained by imposing the selected minisuperspace symmetries onto the full quantum theory? 

The goal of this chapter was not to present a full theory of quantum gravity. Indeed, this would be beyond the scope of any single thesis. Instead the goal was to lay out as clearly as reasonably possible a research programme based on the quantisation method developed in chapter \ref{chap:traj}, possibly in conjunction with the physical-time reduced-Hamiltonian approach that has formed the main part of this work.

\part*{} 
\addtocontents{toc}{\protect\vspace{30pt}}
\chapter{Conclusion}\label{chap:conclusions} 

An ideal candidate for a fundamental theory of quantum gravity would have the following features: 
\begin{compactenum}
\item It should have a self-consistent mathematical structure and its axioms or principles should be independent of any non-quantum-gravitational regime, such as a low-energy or large-scale limit.
\item It should provide a clear ontological picture of what sorts of entities constitute the world at the fundamental level.
\item In the appropriate respective limits it should recover an empirically adequate classical theory of gravity (such as general relativity or Shape Dynamics) and the quantum field theory of the standard model.\footnote{The subject here is a fundamental theory of quantum gravity that includes appropriate matter content, not a quantum theory of pure gravity (unless is itself purely geometrical, for example, described via a Kaluza-Klein mechanism).}
\item It should make clear predictions about in principle empirically accessible quantities for regimes in which both quantum as well as gravitational effects are important (such as the very early universe) and that are outside the scope of our current best fundamental theories.
\end{compactenum}
I would believe that these criteria are for the most part uncontroversial. That the theory be consistent is obvious and that it avoid reference to other regimes is a requirement for the theory to be fundamental and not, for example, to suffer from an analogue of the measurement problem, which in quantum mechanics arises as the result of the reliance of the theory on external regimes (such as classical observers) to which the theory does not apply. 

The second point (concerning ontology) might not be considered important by workers without realist sympathies. This is ultimately a question of the philosophy of science, the full extent of which we cannot discuss in the confines of this thesis. However, providing a clear picture of ontology is arguably necessary for the theory to have explanatory rather than merely predictive value. At least historically this has been a feature of successful theories in physics, including some formulations of quantum mechanics.

The third point is a necessity if the theory is to be empirically adequate. It may be possible that the limit of the fundamental theory leads to a classical theory of gravity, for example, that is structurally different from known theories such as general relativity or Shape Dynamics, but still agrees with observation. In such a case the theory of quantum gravity would provide a new way `for free' to understand classical gravity. But even then general relativity and Shape Dynamics would form effective theories in the low-energy large-scale regime.

The final point is necessary for the theory to be falsifiable independent of the empirical status of its effective form in the low-energy regimes of classical gravity and quantum field theory. If the theory has open parameters these should be determinable by experiment but not in such a way that they can be made to fit \emph{any} set of observations, lest the theory lose its scientific value.

A criterion that could be added to the list is that the theory should be derivable from a small number of well-defined physical principles. Indeed such a property would be desirable. However, it is not obvious that nature has to behave in a manner that allows for the reduction of its structure to simple principles or axioms. While many theories in the past have satisfied this criterion (indeed, a substantial effort is made today to axiomatise the structure of quantum theory, information-theoretically or otherwise), it is not a given that there is a fundamental theory of quantum gravity that is reducible in this manner.

How does the quantisation (canonical or otherwise) of reduced-Hamiltonian theories of gravity, in particular our theory based on York time fare? The mathematical structure of our theory is effectively that of quantum mechanics applied to a conformal geometry as its physical degrees of freedom, together with an appropriate matter content. In the canonical theory one has a wavefunction and (in the de~Broglie-Bohm approach) a trajectory in configuration space (recall that position space is considered physically preferred not only in de~Broglie-Bohm theory, but also by the commutator structure of the quantised reduced-Hamiltonian theory). The constraints that arise as part of the reduction fit neatly into this framework (section \ref{sec:quantKasner} above) and there is no sign of any internal inconsistency.\footnote{Proving that a theory is actually consistent is in general very difficult. It is a lot easier, usually, to show that an inconsistent theory is just that, while in the case of a consistent theory one must point to the absence of obvious inconsistencies.} Crucially, the theory is also consistent with time evolution, which was not the case in the unreduced theory of general relativity in the ADM description, whose quantisation led to a frozen universe. In the trajectory-Weyl approach a number of technical questions remain to be explored (see chapter \ref{chap:cosmtraj}), and only once these are overcome the consistency of the theory can be fully assessed. However, at least in the finite-dimensional model this method of quantisation did not lead to any obvious inconsistencies.

The ontology of our theories is also clear. In the canonical theory the universe is made up of a complex entity mathematically described by a wavefunctional --- a complex function over spatial geometries --- and possibly (in the de~Broglie-Bohm picture) a concrete `world' represented by a moving point (or `particle') in the configuration space. Whether or not this is a satisfactory ontological picture is a different question (mostly independent of gravity and dependent more generally on quantum theory). But either way, it is a clear proposal for an ontology. In the trajectory-Weyl picture, the universe instead consists of an infinite continuity of `classical-like' worlds (each representable by a single point in configuration space), evolving dynamically in time dependent on one another. There is no wavefunction. In both theories time has a status analogous to standard classical and quantum mechanics, although it is physically meaningful (since the dynamics is explicitly time dependent). While this does not answer the question `what is time?' this issue is no more pressing than the same question asked in the context of classical mechanics.

The limits of the theories are expected to satisfy the third criterion. A classical limit may be obtained, for example, via Ehrenfest's theorem or by analysis of the quantum trajectories in conditions representing a classical-like regime.\footnote{There is no complete consensus, it seems, on what constitutes a rigorous derivation of the classical limit of a quantum theory. Frequent approaches include considering the limit of the theory as $\hbar\rightarrow0$ (which makes no physical sense since $\hbar$ is a constant and showing that the theory leads to certain dynamics in this limit has no bearing on the physical emergence of the macroscopic classical world) or introducing `sufficiently classical states' (which begs the question how these states arise dynamically in the case of systems that are expected to behave approximately classical, such as people and planets).} Quantum field theory is expected to be recovered in the case where matter fields dominate and the gravitational field is approximately static flat space. An issue that arises here is that in such a scenario no York time passes, so quantum field theory can never be obtained exactly but must be always be understood as a limit (to some extent the situation is analogous to the York-time description of inflation in section~\ref{sec:inflation}). In fact, given the intrinsically geometrical nature of York time, completely new insights into `Minkowski' quantum field theory may be gained by analysing this limit. A non-trivial amount of work will be necessary to develop this. 

Finally, the description of the early universe in York-time geometrodynamics is qualitatively no different from later eras: time passes and geometric and matter quantities evolve. In fact, the universe is infinitely old and all time but the very last finite interval belong to the so-called Planck era. Our theory requires further developments in order to identify what the measurable remnants of this era are. This may be via the quantum perturbation theory developed in this thesis, or take the form of non-linear (that is, non-perturbative) effects. Relating the theory to actual experimental measurements and distinguishing it empirically will require more work. The scientific value of the theory will rest, at least to a substantial part, on these developments.

Suppose then that quantised York-time geometrodynamics (canonical or trajectory-Weylian) or perhaps some other still unknown approach to quantum gravity turns out to be an ideal theory in the sense of the criteria above, and further presume that this theory becomes well corroborated by experiment. That is, \emph{the} theory of quantum gravity has been found. What is left for fundamental physics?

The search for quantum gravity is not the only big outstanding question in theoretical physics: What is the cosmological constant and why is it what it is? Did inflation really occur and what are the actual mechanisms responsible for it if it did? If it did not, what explains the CMB homogeneity and scale-invariant power spectrum? What about the flatness problem? What are the details of the reheating mechanism? In what ways, if any, does the standard model have to be modified? What is the fundamental (field?) theory describing matter physics at high energies and justifies the renormalisation of known field theories, which are understood to be only effective at sufficiently low energies? The list goes on.

One would hope that a full theory of quantum gravity might provide answers to a number of these questions, for example, explain the apparent value of the cosmological constant and provide a natural mechanism for inflation. However, even then physics is left with more fundamental questions: Why are there three spatial and one temporal dimension (at least at the classical level)? What explains the value of constants in physics, such as the parameters of the standard model? Why are the symmetries of space and time --- or whatever structure underlies them at the fundamental level --- what they are?

One can ask questions that go even deeper: Is there an underlying reason why the laws of physics are what they are? That is, why do the equations of quantum gravity take one form rather than another? Even with the laws in place, how did the universe `decide' what the boundary conditions should be? In a very modest sense, the trajectory-geometry approach to quantisation might be considered to contribute to that last question. No single trajectory is ontologically preferred and no initial configuration is selected. However, as the theory stands currently boundary conditions still need to be applied to the density of trajectories (or the measure on the space). 

This line of questioning leads naturally to a desire for a sort of necessitarianism --- the idea that the universe is exactly as it is because it could not possibly have been otherwise, neither laws nor boundary conditions. That is, there is an underlying reason why this is the only consistent way for the universe to be. The idea of necessitarianism has its own history going back at least Anthony Collins in the early eighteenth century \citep{Collins1717}, although the subject has been firmly one of philosophy and arguably not physics.\footnote{Coincidentally, Collins had an extensive correspondence on a variety of philosophical questions with the very same Samuel Clarke who famously argued with Leibniz over the nature of space, time and motion \citep{LeibnizClarkeCorrespondence}.} I do not doubt though that if mathematical or logical reasons are found that explain the boundary conditions of the universe, for example, the programme would very quickly be adopted by physicists. The implication of this, the idea that all of physics would be reducible to pure reason (although its discovery would in practice be guided by observation) is not new. However, at this point in time it is extraordinarily remote.  Nonetheless, I would argue, it is worth pursuing these questions. Who knows where they might lead.


\bibliographystyle{abbrvnat}	
\bibliography{Bibloi}	



\end{document}